\documentclass[%
reprint,
amsmath,amssymb,
aps,
rmp,
floatfix,
]{revtex4-2}
\usepackage{graphicx}
\usepackage{dcolumn}
\usepackage{bm}
\usepackage[T1]{fontenc}
\usepackage{color}

\usepackage{enumitem}

\definecolor{darkred}{rgb}{0.6,0,0}
\usepackage{hyperref}
\newcommand\br{{\bf r}}

\newcommand\bq{{\bf q}}
\newcommand{\bfr}{{\bf r}}
\newcommand{\bfF}{{\bf F}}
\newcommand{\fp}{f_{\rm p}}
\newcommand{\vp}{v_{\rm p}}
\newcommand{\ellp}{\ell_{\rm p}}

\newcommand{\Gvec}[1]{\boldsymbol{#1}}
\newcommand{\cF}{\mathcal{F}}

\newcommand{\grad}{\nabla}

\newcommand\bfu{{\bf u}}
\newcommand\bffp{{\bf f}_{\rm p}}
\newcommand\bfJ{{\bf J}}
\newcommand\bfp{{\bf p}}

\newcommand\bfeta{{\boldsymbol{\eta}}}

\newcommand\blue[1]{{\color{blue} #1}}

\usepackage[normalem]{ulem}

\usepackage{tikz}
\newcommand\OO[1]{{${\rm O}^{(#1)}$}}

\begin{document}
	
	\preprint{APS/123-QED}
	
	\title{Inclusions, Boundaries and Disorder in Scalar Active Matter}

	\author{Omer Granek}
	\affiliation{Department of Physics, Technion -- Israel Institute of Technology, Haifa 32000, Israel}        

	\author{Yariv Kafri}
	\affiliation{Department of Physics, Technion -- Israel Institute of Technology, Haifa 32000, Israel}

	\author{Mehran Kardar}
	\affiliation{Department of Physics, Massachusetts Institute of Technology, Cambridge, Massachusetts 02139, USA}
        
        \author{Sunghan Ro}
	\affiliation{Department of Physics, Massachusetts Institute of Technology, Cambridge, Massachusetts 02139, USA}
        
	\author{Alexandre Solon}
        \affiliation{Sorbonne Université, CNRS,
          Laboratoire de Physique Théorique de la Matière Condensée
          (UMR CNRS 7600), 4 Place Jussieu, 75252 Paris Cedex 05,
          France}

	\author{Julien Tailleur}
	\affiliation{Department of Physics, Massachusetts Institute of Technology, Cambridge, Massachusetts 02139, USA}

 \begin{abstract}
     Active systems are driven out of equilibrium by exchanging energy and momentum with their environment. This endows them with anomalous mechanical properties that we review in this colloquium for the case of dry scalar active matter, which has attracted considerable attention. These unusual properties lead to a rich physics when active fluids are in contact with boundaries, inclusions, tracers, or disordered potentials. Indeed, studies of the mechanical pressure of active fluids and of the dynamics of passive tracers have shown that active systems impact their environment in non-trivial ways, for example, by propelling and rotating anisotropic inclusions. Conversely, the long-ranged density and current modulations induced by localized obstacles show how the environment can have a far-reaching impact on active fluids. This is best exemplified by the propensity of bulk and boundary disorder to destroy bulk phase separation in active matter, showing active systems to be much more sensitive to their surroundings than passive ones. This colloquium aims at providing a unifying perspective on the rich interplay between active systems and their environments.
 \end{abstract}

	\maketitle

\tableofcontents{}
        \section{Introduction}

Active matter comprises entities that dissipate
energy to exert propelling forces on their
environment~\cite{ramaswamy2010mechanics,marchetti2013hydrodynamics,bechinger_active_2016,chate_dry_2020,fodor2018statistical,tailleur2022active}. From
molecular motors to bacteria and large groups of animals, active
systems are ubiquitous in biology. Furthermore, over the past two
decades, physicists and chemists have devised active
particles in the lab, paving the way toward engineering synthetic
active materials. The non-equilibrium drive at the microscopic scale
endows active materials with a plethora of collective behaviors
unmatched in equilibrium physics. The rich contrast with equilibrium has
germinated considerable experimental and theoretical research on the
subject, which has turned \textit{active matter} into a central field
of condensed matter physics.

One of the most
striking differences between active and passive systems lies in their manifestations of force and work, running counter to intuition from equilibrium thermodynamics.
While in some aspects, bulk fluids of active particles resemble equilibrium matter, the forces exerted on their confining
vessels~\cite{takatori2014swim,yang2014aggregation,solon2015pressure,junot2017active,zakine2020surface}
display a host of unusual phenomena. Examples range from ratchet
currents~\cite{di2010bacterial,sokolov2010swimming}, to anisotropic
pressure~\cite{solon2015pressure}, and long-ranged density
modulations induced by inclusions~\cite{galajda2007wall,tailleur2009sedimentation,angelani2011active,rodenburg2018ratchet,baek_generic_2018}.
More recently, it became clear that the set of results pertaining to
boundaries, inclusions and tracers, and to the forces exerted on them, have a much broader
impact in the context of disordered active
materials~\cite{morin2017distortion,toner2018hydrodynamic,duan2021breakdown,chardac2021emergence}. In
particular, disorder has been shown to play a fundamentally different
role for active systems than for equilibrium ones~\cite{dor2019ramifications,ro2021disorder,dor2022disordered}.

In this colloquium, we review recent results on scalar active matter
in the presence of boundaries, inclusions, tracers, and disorder, for
which we try to offer a coherent physical picture. Scalar systems
correspond to `dry' active matter whose only hydrodynamic mode is the
conserved density field. It offers the simplest-yet-not-too-simple
framework to study the interplay between activity and mechanical
forces. The insights gained from studying these systems can then be
used in other situations, such as `wet' active matter, where the
presence of a momentum-conserving solvent plays an important
role. Similarly, the interplay between activity and mechanics in polar
or nematic active fluids forms a current frontier of the field that is
beyond the scope of this colloquium.

\bigskip

We first briefly introduce scalar active matter in
Sec.~\ref{sec:scalar}, starting at the single-particle level and progressing
to collective behaviors. Then, in Sec.~\ref{sec:pressure} we discuss
the anomalous properties of the forces that active fluids exert on
confining boundaries. In Sec.~\ref{sec:obstacles}, we turn to obstacles and tracers immersed in active baths. The
results presented in these sections finally allow us to discuss the
effect of bulk and boundary disorder on active fluids in
Secs.~\ref{sec:bulkdisorder} and~\ref{sec:boundarydisorder},
respectively.

\section{Scalar Active Matter}
\label{sec:scalar}

To introduce scalar active matter at the microscopic scale, we first
review the standard models of active particles in
Sec.~\ref{sec:free}. We then discuss in Sec.~\ref{sec:int} the
different interactions between particles that have been considered, and
the conditions under which the resulting large-scale physics reduces to
the dynamics of a single density field. Finally, we describe the
collective behaviors encountered in scalar active systems in
Sec.~\ref{sec:collective}, focusing on Motility-Induced Phase
Separation (MIPS), and we review the corresponding hydrodynamic
description in Sec.~\ref{sec:mips}.

\subsection{Noninteracting active particles}
\label{sec:free}
A large diversity of active particles, each capable of dissipating
energy to self-propel, exists across scales in nature, from molecular
motors at the nano-scale to cells and macroscopic animals. In
addition, many types of artificial self-propelled particles (SPPs) are
now engineered, examples being chemically powered Janus
colloids~\cite{howse2007self}, colloidal
rollers~\cite{bricard2013emergence}, vibrated
grains~\cite{deseigne2010collective}, self-propelled
droplets~\cite{thutupalli_swarming_2011} or `hexbug' toy
robots~\cite{li_asymmetric_2013}, to name but a few. Although they
vary greatly in their details, these active particles share the
feature of being persistent random walkers. Compared to a passive
random walker, this introduces a typical scale separating ballistic
motion at a small scale from a diffusive behavior on large
scales. This scale is quantified by the persistence time $\tau$ and
the average distance traveled during this time, which is called the
persistence length $\ellp$. Three types of active particles have been
most commonly used in theoretical and numerical studies of active
matter.

Active Brownian particles (ABPs)~\cite{fily_athermal_2012} and
run-and-tumble particles (RTPs)~\cite{schnitzer1993theory} propel at a
constant speed $\vp=\mu \fp$, where $\mu$ is the particle mobility and
$\fp$ is the constant magnitude of the self-propelling force. The
orientations of ABPs change continuously due to rotational diffusion,
characterized by a rotational diffusivity $D_{\rm r}$. By contrast, RTPs
randomize their directions of motion instantaneously during `tumbles'
that occur at a constant rate $\alpha$. ABPs are a good model for
self-propelled colloids~\cite{howse2007self,ginot2018sedimentation},
while RTPs have been used to model the dynamics of swimming bacteria
like E.Coli~\cite{berg2004coli}. The spatial dynamics of RTPs and
ABPs, in their simplest form, reads
\begin{equation}
  \dot \br(t) = \mu \bffp(t)\;,\label{eq:AP}
\end{equation}
where the self-propulsion force $\bffp(t)$ can be seen as a non-Gaussian
noise of fixed magnitude $\fp$. The persistence time of an ABP in $d$
space dimensions is $\tau=\left[(d-1)D_{\rm r}\right]^{-1}$ while
$\tau=\alpha^{-1}$ for RTPs. The corresponding persistence length is given
by $\ellp=\mu \fp \tau$.

To model active particles whose propulsion forces have fluctuating
norms, a third model has been introduced in which the active force evolves according to an Ornstein-Uhlenbeck process:
\begin{equation}  \label{eq:OU}
  \tau \dot {\bf f}_p=-\bffp + \frac{\sqrt {2D_{\rm eff}}}{\mu}\boldsymbol{\eta}\;.
\end{equation}
Here, $\boldsymbol{\eta}(t)$ is a centered Gaussian white noise of
unit amplitude and independent components. This model is commonly referred to as
Active Ornstein-Uhlenbeck particles (AOUPs). It was studied long before
active matter existed as a field, as one of the simplest models of
diffusion with colored noise~\cite{hanggi1994colored}. In the context
of active matter, AOUPs were independently introduced to model the
dynamics of crawling cells~\cite{sepulveda2013collective} and as a
simplified model for which analytical progress is
tractable~\cite{szamel2014self}\footnote{Other models of active particles have also been considered in the literature, e.g. by considering underdamped Langevin equations with nonlinear friction~\cite{Romanczuk2012b}. In some limits, they yield back the standard ABP and RTP models.}. The Gaussian nature of the
self-propelled force has indeed allowed studying analytically a
variety of
problems~\cite{maggi2015multidimensional,fodor_how_2016,berthier_how_2017,wittmann2017effective1,wittmann2017effective2,woillez2020nonlocal,woillez2020active,martin_statistical_2021}. For
AOUPs, the persistence time is given by $\tau$, while the typical
propulsion force can be defined from
$ \fp^2=\langle \bffp^2 \rangle= d D_{\rm eff}/(\mu^2 \tau)$, where $\langle\cdot\rangle$ is an average over histories. The persistence
length can then be computed as
\begin{equation}\notag
  \ellp\equiv \Big\langle [\bfr(t)-\bfr(0)]\cdot \frac{\bffp(0)}{|\bffp(0)|}\Big\rangle \underset{t\to\infty} \sim \mu \tau\langle  |\bffp(0)|\rangle = \sqrt{ \frac{2 D_{\rm eff} \tau}{\pi}}\,.
\end{equation}

Despite their different dynamics, ABPs, RTPs and AOUPs all lead to
force auto-correlation functions that decay exponentially in time:
\begin{equation}
  \label{eq:vv-ABP-RTP}
  \langle f_{ \mathrm{p},i}(t) f_{ \mathrm{p},j}(0)\rangle=\delta_{ij}\frac{\fp^2}{d}e^{-t/\tau}\;,
\end{equation}
with $f_{ \mathrm{p},k}$ denoting the spatial components of $\bffp$. At large
scales, all  lead to  diffusive dynamics with an effective diffusion
coefficient $D_{\rm eff}=\mu^2 \fp^2\tau /d$. For ABPs and RTPs, this can be
written as $D_{\rm eff}=\ellp^2/(d\tau)$. For AOUPs, the scale of the active
force is proportional to the number of space dimensions and the fluctuations
of $\fp$ lead to an extra contribution to the diffusivity, so
that $D_{\rm eff}=\pi \ellp^2/(2\tau)$.

\begin{figure}
  \includegraphics[width=\columnwidth]{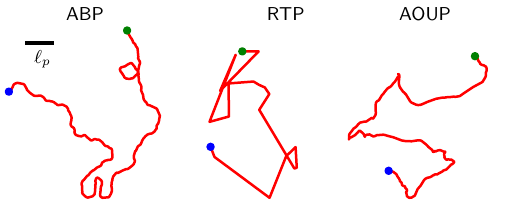}

\includegraphics{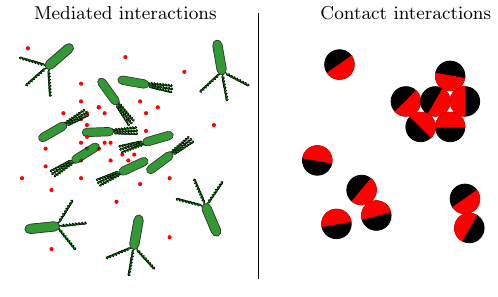}
\if{\beginpgfgraphicnamed{Fig1}
  \begin{tikzpicture}[scale=1]
    \def\w{0.05}
    \draw (10.25,12) node {Mediated interactions};
    \draw (15,12) node {Contact interactions};
\draw (10,10) node{\includegraphics[width=\w\textwidth,angle=144]{greenrun.pdf}};

        \fill[red] (10.3,9.5) circle (1pt);
        \fill[red] (9.8,10.4) circle (1pt);
        \fill[red] (9.8,10.4) circle (1pt);
        \fill[red] (9.6,10.3) circle (1pt);
        \fill[red] (10.4,9.6) circle (1pt);
        \fill[red] (9.9,9.8)  circle (1pt);
        \fill[red] (10.2,9.6) circle (1pt);
        \fill[red] (10,9.8)   circle (1pt);
        \fill[red] (10,9.5)   circle (1pt);
        \fill[red] (10.6,10)  circle (1pt);
        \fill[red] (9.5,10.2) circle (1pt);
        \fill[red] (9.2,10.3) circle (1pt);
        \fill[red] (9.5,9.8)  circle (1pt);
        \fill[red] (9.6,9.7)  circle (1pt);
        \fill[red] (9.6,9.9)  circle (1pt);
        \fill[red] (9.5,9.5)  circle (1pt);
        \fill[red] (9.8,9.7)  circle (1pt);
        \fill[red] (9.6,10.1) circle (1pt);
        \fill[red] (9.8,9.7)  circle (1pt);
        \fill[red] (10.3,10.2)circle (1pt);
        
        \fill[red] (9,9)      circle (1pt);
        \fill[red] (9.9,10.3) circle (1pt);
        \fill[red] (10.5,9.1) circle (1pt);
        \fill[red] (10.5,10.5)circle (1pt);
        \fill[red] (8.9,9.7)  circle (1pt);
        \fill[red] (10.8,10.4)circle (1pt);
        \fill[red] (8.9,9.4)  circle (1pt);
        \fill[red] (11,9.2)   circle (1pt);
        \fill[red] (9.5,10.8) circle (1pt);
        \fill[red] (9.5,10.5) circle (1pt);
        \fill[red] (10.1,8.8) circle (1pt);
        \fill[red] (10.6,10.3)circle (1pt);
        \fill[red] (9.5,9.2)  circle (1pt);

        \fill[red] (8.5,9.2)  circle (1pt);
        \fill[red] (8.6,11.4)  circle (1pt);
        \fill[red] (10.25,11.25)  circle (1pt);
        \fill[red] (11.25,11)  circle (1pt);
        \fill[red] (11.75,10.1)  circle (1pt);
        \fill[red] (9,8)  circle (1pt);
        \fill[red] (11,8.2)  circle (1pt);

        \draw (10.1,10.5) node{\includegraphics[width=\w\textwidth,angle=88]{greenrun.pdf}};
          \draw (10.6,10.8) node{\includegraphics[width=\w\textwidth,angle=132]{greenrun.pdf}};
        \draw (11,9.6)    node{\includegraphics[width=\w\textwidth,angle=179]{greenrun.pdf}};
        \draw (10.6,9.8)  node{\includegraphics[width=\w\textwidth,angle=338]{greenrun.pdf}};
        \draw (9.2,10.1)  node{\includegraphics[width=\w\textwidth,angle=172]{greenrun.pdf}};
        \draw (9.4,9.5)   node{\includegraphics[width=\w\textwidth,angle=354]{greenrun.pdf}};
        \draw (10.2,9.3)  node{\includegraphics[width=\w\textwidth,angle=347]{greenrun.pdf}};
        
        \draw (9,11) node{\includegraphics[width=\w\textwidth,angle=288]{greentumble.pdf}};
        \draw (12,11)  node{\includegraphics[width=\w\textwidth,angle=347]{greentumble.pdf}};
        \draw (10.5,8)  node{\includegraphics[width=\w\textwidth,angle=327]{greentumble.pdf}};
        \draw (9.1,8.5)  node{\includegraphics[width=\w\textwidth,angle=73]{greentumble.pdf}};
        \draw (11.5,8.7)  node{\includegraphics[width=\w\textwidth,angle=180]{greentumble.pdf}};

        \draw (12.5,7.5) -- (12.5,12);

        \begin{scope}[xshift=15cm,yshift=10cm,scale=.5]
        \foreach \x/\y/\theta in
        {
        0/1/-135,
        1/1/240,
        2/1/90,
        1.5/0.134/0,
        1.5/1.866/170,
        .5/0.134/-45,
        -1.5/-2/230,
        -1/-2.866/15,
        -3/-1/-10,
        -2.25/2.25/-145,
        2/-2.3/215,
        2.15/-2.3-.988/60,
        -2.35/-3.15/10
}
{
\filldraw[black] (\x,\y) circle (.5);
\filldraw[rotate around={\theta:(\x,\y)},red] (\x-.5,\y) arc (180:0:.5) |- cycle;

(\x,\y) circle (.5);
}
\end{scope}
       
    \end{tikzpicture}
\endpgfgraphicnamed{Fig1}}\fi

  \caption{{\bf Top}: Representative trajectories for the three types
    of active particles described in Sec.~\ref{sec:free}, all with the
    same persistence time $\tau=1$, total duration $T=20$ and
    effective diffusion coefficient $D_{\rm eff}=1$. This corresponds
    to $v_{\rm p}=\sqrt{2}$ (for ABP and RTP) and a persistence length
    $\ellp=\sqrt{2}$ indicated as a scale bar. The blue and green dots
    indicate the starting and finishing positions, respectively. {\bf
    Bottom}: Schematic representation of mediated and contact
    interactions occurring between active
    particles.}\label{fig:traj-int}
\end{figure}

Typical trajectories for the three types of particles are shown in
Fig.~\ref{fig:traj-int} (top), which highlights their differences on short
length scales. While these differences are irrelevant at the scale of
diffusive dynamics, they play an important role in the presence of
external potentials. For example, in the large persistence regime, it
has been predicted
theoretically~\cite{tailleur2008statistical,tailleur2009sedimentation,hennes2014self,solon2015active,malakar2020steady,basu2020exact,smith2022exact}
and observed experimentally~\cite{takatori2016acoustic,schmidt2021non}
that RTPs and ABPs accumulate away from the center of a confining
harmonic well. On the contrary, in a harmonic potential, AOUPs always
have a steady state given by a centered Gaussian
distribution~\cite{szamel2014self} and, somewhat surprisingly, their
dynamics obey detailed balance~\cite{fodor_how_2016}.

\subsection{Interacting scalar active matter}
\label{sec:int}

In this colloquium, we focus on scalar active matter, i.e.,\,on active
systems whose long-time and large-scale behaviors are entirely
captured by the stochastic dynamics of a density field. All dilute dry
active systems fall in this class, as do a wealth of interacting ones.

A scalar theory generically describes systems where the interactions
only depend on and impact the particle positions. This is for instance
the case for attractive or repulsive pairwise forces that play an
important role in dense active
systems~\cite{fily_athermal_2012,redner_structure_2013,stenhammar2014phase,wysocki2014cooperative,stenhammar_activity-induced_2015}. It
also applies to interactions that act only on the magnitude of the
self-propulsion velocity and not on its
direction~\cite{liu2011sequential,dalessandro_contact_2017}. Such
interactions can, for instance, be mediated by chemical signals as
encountered in assemblies of cells interacting via ``quorum sensing''
(QS). This is depicted in Fig.~\ref{fig:traj-int} (bottom left) where the
cells adapt their behaviors to the concentration of diffusing
signaling molecules. QS is generic in
nature~\cite{miller_quorum_2001} and plays an important role in
diverse biological functions ranging from
bioluminescence~\cite{nealson1970luminescence,engebrecht1984lux,fuqua1994quorum,verma2013quorum}
and virulence~\cite{tsou2010virulence} to biofilm formation and
swarming~\cite{hammer_quorum_2003,daniels_quorum_2004}. It can also be
engineered by genetic manipulation of
bacteria~\cite{liu2011sequential,curatolo2020cooperative} or using
light-controlled
colloids~\cite{bauerle_self-organization_2018,massana-cid_rectification_2022}. Integrating
out the dynamics of the mediating chemical field, particles
interacting via QS can be modeled by motility parameters that depend
on the density field, for example, through a density-dependent
self-propulsion force $\fp(\br,[\rho])$~\cite{cates2015motility}.

It is important to note that many active systems also experience
interactions that require more complex effective (`hydrodynamic')
descriptions. This is, for example, the case when the rotational
symmetry of the system is spontaneously broken. The prototypical
example is that of the Vicsek model~\cite{vicsek1995novel} where
strong-enough aligning torques between the particles lead to the
emergence of an ordered polar
phase in dense systems~\cite{marchetti2013hydrodynamics,chate_dry_2020}. A proper
hydrodynamic description of the corresponding flocking phase then
requires the inclusion of the orientation
field~\cite{toner_hydrodynamics_2005}. A wealth of other interactions
may impact the particle's orientations, like
chemotaxis~\cite{berg2004coli}---which makes cells move preferably up
or down chemical gradients---, or contact inhibition of
locomotion~\cite{stramer2017mechanisms} that may lead to cells
reverting their directions of motion upon encounters. We stress that
the disordered phases of all such systems nevertheless remain part of
scalar active matter and already exhibit non-trivial collective
behaviors~\cite{brenner1998physical,saha2014clusters,sese2018velocity,o2020lamellar,spera2023nematic}.

\subsection{Collective behavior}
\label{sec:collective}
In this Section, we  turn to  the collective behaviors that are typically encountered in scalar active matter.

\begin{figure}
\includegraphics{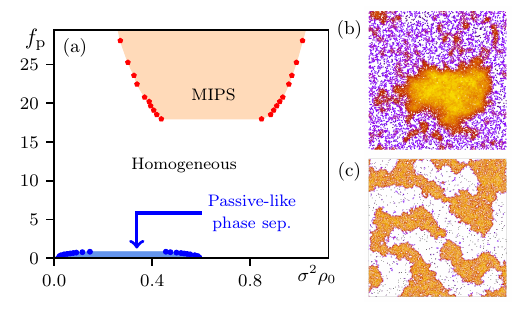}
\if{\beginpgfgraphicnamed{Fig2}
  \begin{tikzpicture}
    \def\x{4.5}
    \def\y {-.1}
    \def\yy{-1.75}

    \def\xb{2.1}
    \def\yb{-2.25}
    \def\dx{.7}
    \def\dy{.4}

   \path (0,\y) node {\includegraphics[totalheight=4.5cm]{./Phase-Diag-MIPS-LJ_v2.pdf}};

   \filldraw[white] (\xb,\yb) rectangle (\xb+\dx,\yb+\dy);

   \draw (\xb+.5*\dx,\yb+.5*\dy) node {$\sigma^2 \rho_0$};

    \draw[blue] (1.35,-1) node[align=center, text width=1.75cm] {\footnotesize Passive-like phase sep.};
    
    \draw[blue,ultra thick,->] (.5,-1) -- (-.6,-1) -- (-.6,-1.6);
    
      \draw (0.7,1 ) node[align=center, text width=1.75cm] {\footnotesize MIPS};

    \draw (0.2,-.2) node {\footnotesize Homogeneous};
    
\path (\x,1.25) node {\includegraphics[totalheight=2.5cm]{./mips.png}};

\path (\x,-1.25) node {\includegraphics[totalheight=2.5cm]{./lj.png}};
      

      \draw (-1.65,1.8) node {(a)};
      \draw (\x-1.5,1+1.1) node {(b)};
      \draw (\x-1.5,-1+.7) node {(c)};

  \end{tikzpicture}
\endpgfgraphicnamed{Fig2}}\fi

  \caption{ Simulations of ABPs interacting via a Lennard-Jones
  potential,
  $V(r)=4\epsilon\big[(\frac{\sigma}{r})^{12}-(\frac{\sigma}{r})^{6}\big]$
  for $r<2.7\sigma$ and $V(r)=0$ otherwise. {\bf (a)} Phase diagram
  obtained by varying $\fp$ and the average density $\rho_0$. {\bf
  (b)} Representative snapshot of the system in the MIPS region
  ($\fp=42$, $\rho_0=0.9$). {\bf (c)} Representative snapshot of the
  system in the passive-like phase separation, during the coarsening stage
  ($\fp=0.4$, $\rho_0=0.5$). Parameters: $\mu=1$, translational
  diffusivity $D_{\rm t}=0.4$, $D_{\rm r}=2$, $\sigma=0.89$, system size
  $300\times 300$. Data courtesy of Gianmarco
  Spera.} \label{fig:MIPS-LJ}
\end{figure}

\subsubsection{Bulk phase separation in scalar active matter}

When considering systems whose large-scale behaviors are characterized
by the conserved dynamics of a density field, possibly the simplest phase
transition corresponds to condensation and
the breaking of translational uniformity.

In many active particle contexts, condensation has been predicted
theoretically~\cite{fily_athermal_2012,redner_structure_2013,redner2013reentrant,mognetti2013living,stenhammar2014phase,wysocki2014cooperative,stenhammar_activity-induced_2015,paliwal2018chemical,spera2023nematic},
and observed
experimentally~\cite{theurkauff2012dynamic,buttinoni_dynamical_2013,palacci2013living,liu2019self,van2019interrupted},
arising from the interplay between attractive, repulsive, and
propulsion forces. When self-propulsion is
weak, attractive forces between the particles can easily
overcome activity, and  the expected equilibrium phase transitions typically survives. This is illustrated in
Fig.~\ref{fig:MIPS-LJ} using numerical simulations of self-propelled
ABPs in the presence of a translational noise and Lennard-Jones (LJ)
interactions. At ${\fp}=0$, the system undergoes equilibrium
liquid-gas phase separation. As $\fp$ increases, the phase-separated
region shrinks until activity overcomes the attractive forces, and the
system turns into a homogeneous fluid. Surprisingly, at even larger
propulsion forces, a reentrant phase transition into a phase-separated
region is observed ~\cite{redner2013reentrant,spera2023nematic}.

The mechanism underlying this reentrant phase transition is
MIPS, which is distinct from
equilibrium phase separation and does not require attraction between
the particles. Instead, MIPS results from the interplay between the
tendency of active particles to accumulate where they move slower and
their slowdown at high density due to collisions and repulsive
forces~\cite{cates2015motility}.

MIPS has been reported in a wealth of active systems and can arise
from a variety of interactions: pairwise
forces~\cite{fily_athermal_2012,redner_structure_2013,stenhammar2014phase,wysocki2014cooperative},
QS~\cite{tailleur2008statistical,liu2011sequential,cates_when_2013,bauerle_self-organization_2018,curatolo2020cooperative},
chemotaxis~\cite{o2020lamellar,zhang2021active,Hongbo2023}, or steric
hindrance~\cite{thompson_lattice_2011,whitelam2018phase,dittrich2021critical,kourbane2018exact,adachi2020universality,maggi_universality_2021,agranov2021exact}. It
has been extensively
reviewed~\cite{cates2015motility,fodor2018statistical,kurzthaler2022out,stenhammar2021introduction}
and we focus here on its simplest hydrodynamic description. 

\subsubsection{A minimal hydrodynamic description of MIPS}
\label{sec:mips}

\begin{figure}
  \includegraphics[width=.6\columnwidth]{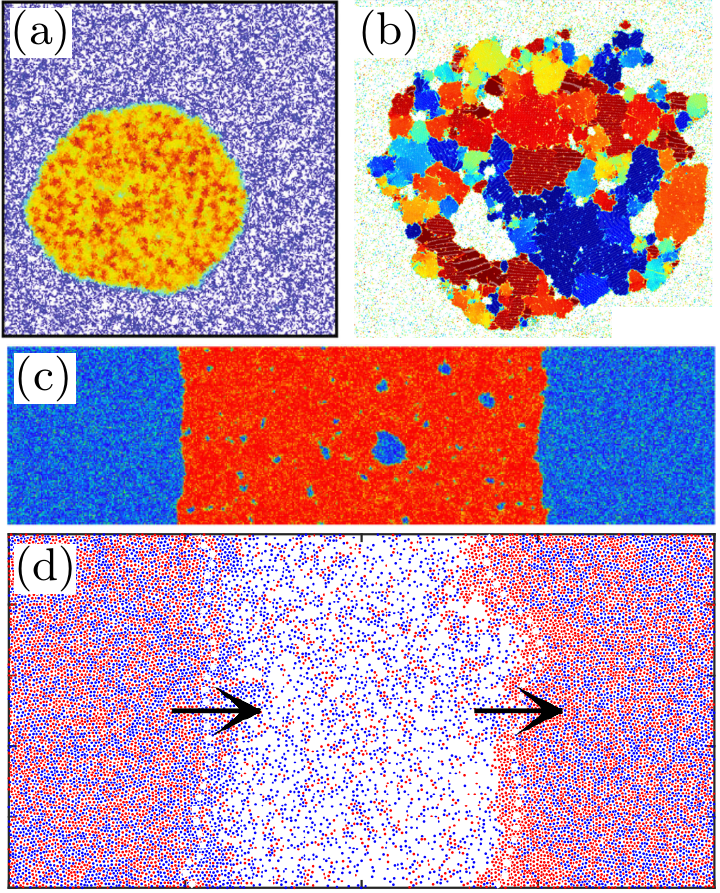}
  \caption{{\bf (a)} MIPS in ABPs interacting via quorum sensing showing a liquid-gas phase separation. Reproduced  with kind permission of The European Physical Journal (EPJ) 
    from~\cite{solon2015active}. {\bf (b)} MIPS in ABPs interacting via a stiff repulsion
    showing a mosaic of hexatic domains and gas inclusions in the
    dense phase. The color indicates the direction of the hexatic
    order. Reproduced with permission from ~\cite{caporusso_motility-induced_2020}. {\bf (c)} MIPS in
    ABPs interacting via harmonic repulsion showing a distribution of
    gas bubbles in the dense phase. Warmer colors indicate a higher
    density. Reproduced with permission from~\cite{shi_self-organized_2020}. {\bf
      (d)} MIPS in a mixture of ABPs and passive Brownian particles
    showing propagating interfaces in the direction indicated by the
    black arrows. Reproduced with permission from~\cite{wysocki_propagating_2016}.}
  \label{fig:collective}
\end{figure}

MIPS was first observed in collections of RTPs
interacting via quorum-sensing (QSAPs; see Fig.~\ref{fig:collective}(a)) whose self-propulsion speed
$v_{\rm p} =\mu \fp$ decreases rapidly enough as the local density
increases~\cite{tailleur2008statistical}. For this system, it is
possible to derive a fluctuating hydrodynamics for the density field
as
\begin{equation}\label{eq:QSAPs}
  \partial_t {\hat \rho} = \nabla \cdot [\hat \rho \mathcal{D}_{\rm eff} \nabla \mathfrak{u} + \sqrt{ 2 \hat \rho \mathcal{D}_{\rm eff}} \boldsymbol{\Lambda}]\;.
\end{equation}
In Eq.~\eqref{eq:QSAPs}, the density field is constructed from the
$N$
particle positions as
\begin{equation}
\hat \rho(\br,t)\equiv \sum_{i=1}^N \delta[\br-\br_i(t)]\;,
\end{equation}
$\mathcal{D}_{\rm eff}(\br,[\hat\rho])=v^2(\br,[\hat\rho])\tau/d$ is the large-scale
diffusivity, $\mathfrak{u}(\br,[\hat\rho])=\log[\hat\rho(\bfr) v(\br,[\hat\rho]) ]$ is a
non-equilibrium chemical potential, and $\boldsymbol{\Lambda}$ is a
centered Gaussian white noise of unit variance and delta-correlated in space. The derivation
of Eq.~\eqref{eq:QSAPs} relies on a diffusive approximation of the
microscopic run-and-tumble
dynamics~\cite{tailleur2008statistical,cates_when_2013,solon2015active}
and on stochastic calculus~\cite{dean1996langevin,solon2015active}.

In a system with a homogeneous density $\rho_0$, the chemical
potential can be written as $\mathfrak{u}=\log[\rho_0
v(\br,[\rho_0])]\equiv f'(\rho_0)$, where $f(\rho_0)$ is a Landau-like free
energy density. For all density values such that $f''(\rho_0)<0$, a
homogeneous system is linearly unstable and separates into liquid and
gas phases. This instability defines a spinodal region. To predict the
coexisting `binodal' densities and the phase diagram of the system,
one needs to consider inhomogeneous density profiles. The mapping onto
an equilibrium theory then breaks down because $\mathfrak{u}$, in general, cannot be
written as the functional derivative of a free energy
$\mathcal{F}[\rho]$. To leading order in a gradient
expansion, however, a predictive theory can be formulated to determine
the phase
diagram~\cite{solon_generalized_2018,solon_generalized_2018-1}.

Active particles interacting via pairwise forces (PFAPs) provide
another system for which an explicit coarse-graining of the
microscopic dynamics is possible. In the presence of repulsive interactions, it has
led to a predictive theory for MIPS. This case is significantly more
involved than QSAPs but a series of articles have led to the
understanding of the spinodal instability, first, and more recently to the
prediction of the binodals~\cite{fily_athermal_2012,redner_structure_2013,stenhammar2014phase,takatori2015towards,Solon_interactions,solon_generalized_2018,solon_generalized_2018-1,speck2021coexistence}.

An alternative route to account for MIPS is to follow phenomenological
approaches and construct a hydrodynamic description for a fluctuating
coarse-grained density field $\rho$ that includes all terms allowed by
symmetry~\cite{wittkowski_scalar_2014}. At the fourth order in a gradient
expansion, one obtains active model B+
dynamics~\cite{tjhung_cluster_2018}:
\begin{align}
  \label{eq:AMBrho}
  \partial_t\rho&=-\nabla \cdot \left[\mathbf{J}+\sqrt{2 D M}\Gvec\Lambda \right] ,\\
  \label{eq:AMBJ}
  \mathbf{J}/M&=-\grad\left[\frac{\delta \cF}{\delta\rho}+\lambda |\grad\rho|^2\right]+\zeta(\nabla^2\rho)\grad\rho\,,
\end{align}
where $\cF=\int d^d\bfr\, \left[f(\rho)+\kappa |\grad\rho|^2\right] $ and
$f(\rho)$ plays again the role of a Landau free energy density. All
coefficients $D$, $M$, $\kappa$, $\lambda$ and $\zeta$ depend in
principle on $\rho$. The $\lambda$ and $\zeta$ terms both break time
reversal symmetry~\cite{nardini_entropy_2017,tjhung_cluster_2018} and
the dynamics in Eqs.~\eqref{eq:AMBrho}-\eqref{eq:AMBJ} cannot be derived from a free energy as in regular model B. Active model B+ is thus a generalization
of Eq.~\eqref{eq:QSAPs}, which accounts for MIPS in a broader set of
systems. Using the methods developed
in~\cite{solon_generalized_2018,solon_generalized_2018-1}, it is
again possible to predict the coexisting densities of a fully
phase-separated system. An interesting difference with equilibrium
physics is that the interfacial terms contribute to the results~\cite{wittkowski_scalar_2014}.

\subsubsection{Beyond the simple MIPS scenario}

Interestingly, Eqs.~\eqref{eq:AMBrho}-\eqref{eq:AMBJ} predict a richer physics than
the simple MIPS scenario discussed above: for large values of
$|\zeta|$, one observes a reversed Ostwald ripening whereby small
droplets grow at the expense of larger
ones~\cite{tjhung_cluster_2018}. In the phase coexistence regime, this
leads to a dense phase that contains droplets of the dilute phase evolving with
complex dynamics. This is reminiscent of what has been observed in
simulations of PFAPs, where the droplet sizes are found to be
algebraically distributed, leading to a critical dense
phase~\cite{caporusso_motility-induced_2020,shi_self-organized_2020}
shown Fig.~\ref{fig:collective}(c).

Besides the nature of the phase-separated state, the critical
properties of MIPS have also attracted attention. Conflicting results
have been reported regarding the critical point~\cite{Speck2022rg}.
It has been measured and argued
to have exponents either in the Ising universality
class~\cite{partridge_critical_2019,maggi_universality_2021,maggi_critical_2022}
or another one~\cite{siebert_critical_2018,caballero2018bulk}, and the effect,
or absence thereof, of bubbles on the critical point remains
unclear~\cite{caballero2018bulk}.

The microscopic systems described above  also display interesting
properties beyond those captured solely by the dynamics of their density
field, and thus fall outside the scope of this colloquium. For
instance, the dense phases observed in ($d=2$) simulations of PFAPs have
revealed a rich structure: they are found either in a disordered
liquid form~\cite{fily_athermal_2012} for soft repulsive potentials,
or as a mosaic of hexatic domains with different orientations for
stiffer ones (see~\cite{redner_structure_2013,digregorio_full_2018}
and Fig.~\ref{fig:collective}(b)). Furthermore, the hexatic phase
observed in the passive limit of this system survives the addition of
activity~\cite{digregorio_full_2018} and has attracted a lot of
attention recently~\cite{digregorio20192d}. At high density, active
systems also exhibit a glassy dynamics, which has been observed
experimentally~\cite{angelini_glass-like_2011,garcia_physics_2015,klongvessa_active_2019}
and
numerically~\cite{levis_single-particle_2015,berthier_how_2017,berthier_glassy_2019,klongvessa_active_2019}.

Finally, rich physical phenomena occur in mixtures of scalar active systems. For example, mixtures of active and passive particles
interacting via pairwise repulsion can
demix~\cite{ilker_phase_2020,weber_binary_2016} or jointly undergo
MIPS~\cite{stenhammar_activity-induced_2015}, while mixed species of
QSAPs can either segregate or
co-localize~\cite{curatolo2020cooperative}. Out of equilibrium, one
expects generically that the interactions between two species are non-reciprocal. Even for PFAPs, which obey Newton's third law of action and
reaction microscopically, the large-scale description of a mixture of
active and passive particles features non-reciprocal
couplings~\cite{wittkowski_nonequilibrium_2017}. When strong enough,
the non-reciprocity gives rise to propagating patterns, as shown using
a phenomenological
theory~\cite{saha_scalar_2020,you2020nonreciprocity} and observed in
mixtures of active and passive ABPs~\cite{wysocki_propagating_2016}
(see Fig.~\ref{fig:collective}(d)), as well as for mixtures of
QSAPs~\cite{dinelli_non-reciprocity_2022}. 

All in all, most results on the collective behaviors of scalar active
matter have been established in idealized systems, invariant by
translation and endowed with periodic boundary conditions. In
equilibrium, when the correlation length is finite, obstacles and
boundaries alter the system only in their immediate
vicinity, which legitimates this approach. As we shall see in Section~\ref{sec:boundarydisorder}, the situation is very different for active system, where boundaries may have a far-reaching influence on bulk behaviors. Before we discuss this case, we first review the anomalous mechanical forces exerted by active systems on boundaries and inclusions, which are at the root of this important difference between active and passive systems. 

\section{Mechanical forces on confining boundaries}
\label{sec:pressure}

\begin{figure}
  \begin{center}
    \includegraphics[width=.8\columnwidth]{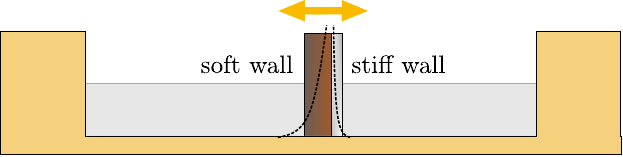}

    \includegraphics[width=.49\columnwidth]{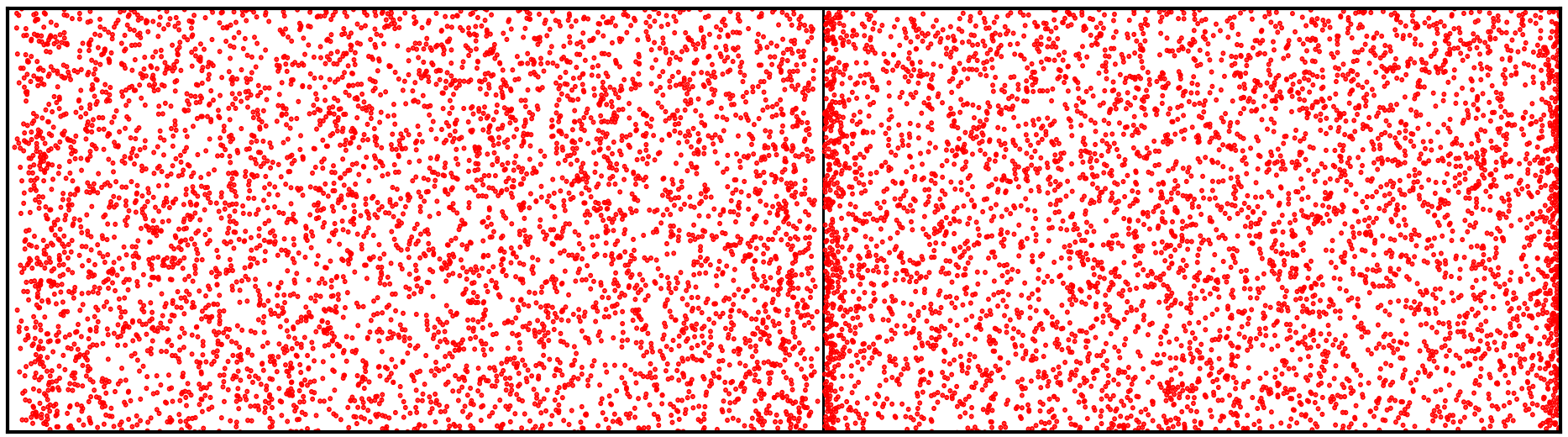}
    \includegraphics[width=.49\columnwidth]{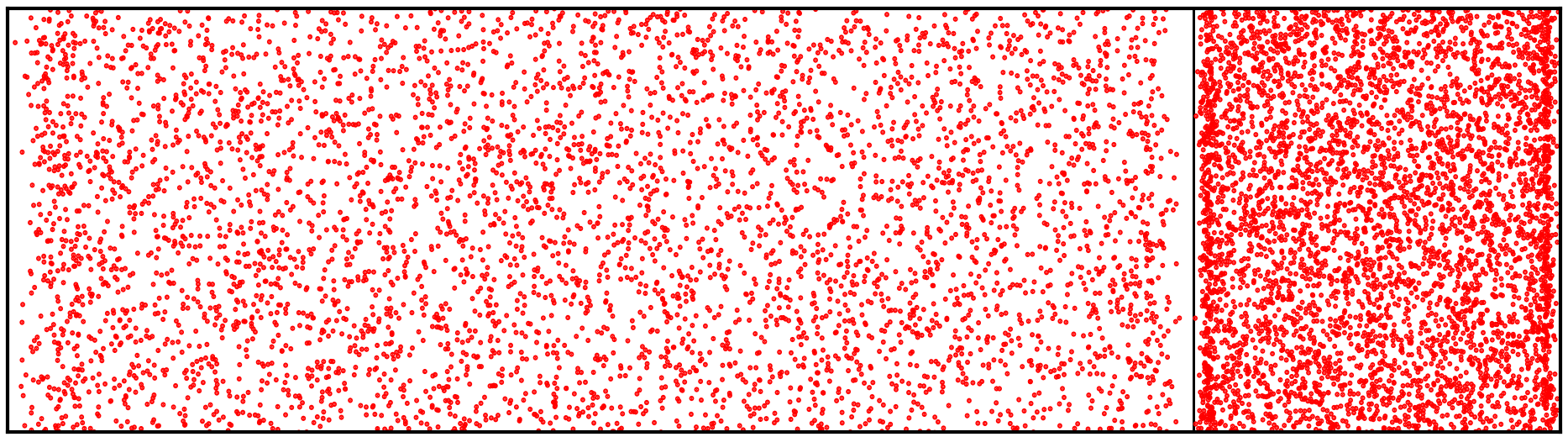}
  \end{center}

  \caption[Caption for LOF]{{\bf Top:} A mobile partition that is stiffer on one side divides the system into two compartments, each initially with equal density. Since the pressure depends on the wall's stiffness through Eq.~\eqref{eq:pressurenoeos}, each side of the partition experiences a different force from the active particles. As a result, the mobile partition moves until the forces on both sides balance. {This material was originally published in~\cite{tailleur2022active} and has been reproduced by permission of  \href{https://doi.org/10.1093/oso/9780192858313.001.0001}{Oxford University Press}. For permission to reuse this material, please visit \url{http://global.oup.com/academic/rights}}. {\bf Bottom:} Numerical simulations corresponding to the setup described
    in the top panel using either circular ABPs
    (left panel) or elliptical ABPs (right panel). The absence of an equation of state
    (EOS) in the latter case is apparent from the spontaneous
    compression of one half of the system. Reproduced with permission    from~\protect\cite{solon2015pressure}.}\label{fig:piston}
\end{figure}

In recent years, it has become evident that active systems display many
anomalous properties when they interact with boundaries. Arguably, the
simplest demonstration of this is obtained by placing an asymmetric
mobile partition in a cavity comprising a homogeneous gas of
self-propelled ellipses. One then observes a spontaneous compression
of one side of the system, in violation of the second law of
thermodynamics (see Fig.~\ref{fig:piston}). By now, it has become
clear that such phenomena can be rationalized in terms of the
anomalous mechanical properties of active fluids, which we review in
this section.

\subsection{Mechanical pressure on flat confining boundaries}

\begin{figure}
\includegraphics{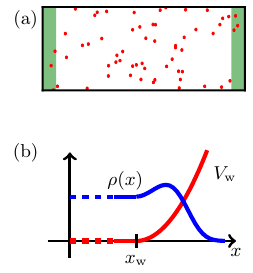}
\if{\beginpgfgraphicnamed{Fig5a}
  \begin{tikzpicture}[scale=.5 ]
    \draw (0,0) node {\includegraphics[width=.4\columnwidth]{./box.pdf}};
    \draw (-4,1) node {(a)};                               
    \begin{scope}[xshift=-2.5cm,yshift=-6.5cm,scale=.75]
      \draw[->,line width=0.05cm] (-1,0) --(7.5,0) node[below] {$x$};
      \draw[->,line width=0.05cm] (0,-.75) --(0,4) node[xshift=-.75cm] {(b)};
      \draw[red,line width=0.09cm,dashed] (0,0) -- (2,0); 
      \draw[red,line width=0.07cm] (2,0) -- (3,0); 
      \draw[red,line width=0.07cm,domain=3:6.2] plot (\x,{0.4*(\x-3)*(\x-3)}); 
      \draw (7,3) node {$V_{\rm w}$};
      \draw[blue,line width=0.07cm,dashed] (0,2) -- (2,2); 
      \draw[blue,line width=0.07cm] (2,2) -- (3,2);
      \draw[blue,line width=0.07cm,domain=3:7] plot (\x,{2*exp(.4*(\x-3)*(\x-3)-0.2*(\x-3)*(\x-3)*(\x-3))}); 
      \draw (2.5,2.7) node {$\rho(x)$};
      \draw[line width=0.05cm] (3,.3) -- (3,-.3) node [anchor=north] {$x_{\rm w}$};
    \end{scope}
  \end{tikzpicture}
\endpgfgraphicnamed{Fig5a}}\fi
\includegraphics{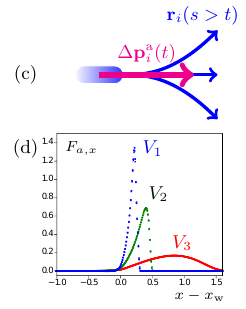}
\if{\beginpgfgraphicnamed{Fig5d}
\begin{tikzpicture}
\draw (-1.75,1) node {(d)};
\draw (0,0) node {\includegraphics[trim=14cm 0 0 0,clip,width=3.2cm]{./BoundaryLayer.pdf}};
\draw (.4,1) node {\blue{$V_1$}};
\draw (.5,.23) node {\color{green!10!black}{$V_2$}};
\draw (.9,-.6) node {\color{red}{$V_3$}};
\draw (-0.8,1) node {\scriptsize {$F_{a,x}$}};
\draw (1.2,-1.5) node {\scriptsize $x-x_{\rm w}$};

\begin{scope}[xshift=-.5cm,yshift=2.25cm]
        \draw (-1.25,0) node {(c)};                                
                  \draw[white, right color=blue!80!white, left color=blue!5!white,rounded corners=4pt] (-.4,-.15) rectangle (.4,.15);
                    \draw[->,ultra thick,blue] (.4,0) .. controls (.75,0) and (1.25,0) .. (2,.75) node[xshift=-.25cm,yshift=.25cm] {$\bfr_i(s>t)$};
          \draw[->,ultra thick,blue] (.4,0) .. controls (.75,0) and (1.25,0) .. (2,0);
          \draw[->,ultra thick,blue] (.4,0) .. controls (.75,0) and (1.25,0) .. (2,-.75);
            \draw[->,line width = 3pt,magenta] (0,0) -- (1.6,0) ;
            \draw[magenta] (.8,.35) node {$\Delta{\bf p}_i^{\rm a}(t)$};
\end{scope}
        
\end{tikzpicture}
\endpgfgraphicnamed{Fig5d}}\fi
  \caption{\textbf{(a)} A simple setup to compute the pressure exerted by an active gas on confining walls consists of a $2d$ box with periodic boundaries along $y$ and a confining potential along $x$. \textbf{(b)} Schematic representation of the density of active particles and of the confining wall potential shown in panel (a). \textbf{(c)} Despite the fact that the active force is doing an isotropic random walk, it will transmit a non-zero average momentum to the active particle between any time $s=t$ and $s=+\infty$. The corresponding `active' impulse $\Delta {\bf p}_i^{\rm a}(t)$ is therefore nonzero. \textbf{(d)} The active force densities measured for three different confining potentials is non-zero close to a confining wall. The figure shows a clear dependence on the wall potential. In the presence of an EOS, the areas under the three curves are the same. }\label{fig:pressure}\label{fig:ActiveImpulse}
\end{figure}

Consider the simplest case of a two-dimensional gas of non-interacting
active particles confined by a vertical flat wall localized at
$x=x_{\rm w}$, which we model by a repulsive potential $V_{\rm w}(x)$ that
vanishes for $x<x_{\rm w}$ and diverges at larger values of $x$. The
pressure exerted by the gas on the wall can be computed as:
\begin{equation}\label{eq:pres:def}
  P = \int_{x_b}^\infty dx\, \rho(x,y_b) \partial_x V_{\rm w}(x)\;,
\end{equation}
where $(x_b,y_b)\equiv \br_b$ corresponds to a point deep in the
bulk of the active fluid, $x_b\ll x_{\rm w}$, and $\rho(\br, t)=\langle \sum_i
\delta[\br-\br_i(t)]\rangle$ is the average number density (see
Fig.~\ref{fig:pressure}). To determine $P$, it is natural to start
from the dynamics of particle $i$, which reads
\begin{equation} \label{eq:singledynamics}
  \dot \bfr_i = \mu \bffp^i - \mu \nabla V_{\rm w}(\bfr_i)+\sqrt{2 D_{\rm t}}\bfeta_i\;,
\end{equation}
where $\bffp^i\equiv \fp \bfu(\theta_i)$ is the particle propulsion force, $\mu$ its mobility,
$D_{\rm t}$ a translational diffusivity, and $\bfeta_i$ a centered Gaussian
white noise of unit variance. The evolution of $\rho$ then satisfies a
conservation law
\begin{equation}\label{eq:dynamicsdensitylaw}
  \partial_t \rho(\br,t)=-\nabla\cdot \bfJ(\br,t)\;,
\end{equation}
where the current $\bfJ$ is given by
\begin{equation}\label{eq:pres:current}
  \bfJ=\mu \bfF_a(\br) - \mu\rho(\br)\nabla V_{\rm w}(\bfr) - D_{\rm t} \nabla \rho(\br)\;,
\end{equation}
and $\bfF_a\equiv \langle \sum_i  \bffp^i \delta(\br-\br_i)\rangle$
is the active-force density. In the steady state, the confinement by a
wall and the translational symmetry along the wall imply a vanishing
current $\bfJ=0$. Using Eqs.~\eqref{eq:pres:def} and
\eqref{eq:pres:current}, the pressure can then be written as
\begin{equation}\label{eq:pres:pres2}
  P=\frac{D_{\rm t}}\mu \rho(\br_b) +\int_{x_b}^\infty dx\, \bfF_a(x,y_b) \;.
\end{equation}
The pressure exerted by the system on the wall is thus the sum of the passive ideal-gas pressure and a contribution stemming from the active force density, which is typically non-zero close to confining walls (See Fig.~\ref{fig:ActiveImpulse}(d)).

Before considering the case of self-propelled ellipses shown in  Fig.~\ref{fig:piston}, we start with
the simpler case of ABPs that undergo isotropic rotational diffusion everywhere in space,
$\dot \theta_i=\sqrt{2 D_{\rm r}}\eta^{\rm r}_i$, where $\eta_i^{\rm r}$ is a centered Gaussian white noise of unit variance.  To
make progress, it is useful to introduce the active impulse of
particle $i$, which is the average momentum the particle will receive
in the future from the substrate on which it is pushing. For a
circular ABP, whose orientation is $\bfu[\theta_i(t)]$ at time $t$,
the active impulse can be computed as
\begin{equation}\label{eq:pres:AIdef}
  \Delta \bfp_i^a \equiv \int_t^\infty ds\, \fp \overline{\bfu[\theta_i(s)]} =\frac{\fp}{D_{\rm r}} \bfu[\theta_i(t)]\;,
\end{equation}
where the overline denotes an average over future histories, i.e., over
$\eta_i^{\rm r}(s\geq t)$, and we have used that
$\overline{\bfu[\theta_i(s)]}=\bfu[\theta_i(t)] \exp[-D_{\rm r}
  (s-t)]$. Even though the particle is undergoing a random walk, so
that its average active force is zero, the active impulse at time $t$
is non-zero because of persistence (see Fig.~\ref{fig:ActiveImpulse}(c)). Direct aglebra then shows that the time evolution of the active impulse field
$\Delta \bfp^a(\bfr)\equiv \langle \sum_i \Delta \bfp^a_i \delta(\br-\br_i)\rangle $
is given by
\begin{equation}\label{eq:pres:activeimpdyn}
  \partial_t \Delta \bfp^a(\bfr)=-\bfF_a(\br)+\nabla \cdot \sigma_a(\bfr)\;,
\end{equation}
where
\begin{equation}
-\sigma_a\equiv  \langle \sum_i \dot \br_i \otimes \Delta \bfp_i^a
\delta(\br-\br_i)\rangle- D_{\rm t} \nabla \otimes \Delta \bfp^a(\br)
\end{equation}
is a tensor that measures the flux of active impulse. In the homogeneous
isotropic bulk of the system, a direct computation shows that
$\sigma_a=\rho_b \mu \fp^2/(2D_{\rm r})\mathbb{I}_d$, where $\mathbb{I}_d$ is the identity tensor. In the steady state,
Eq.~\eqref{eq:pres:activeimpdyn} shows that the density of active
force is the divergence of the `active' stress tensor $\sigma_a$:
\begin{eqnarray}\label{eq:pres:sigmaa}
  \bfF_a(\br)&=&\nabla \cdot \sigma_a(\bfr)\;.
\end{eqnarray}
Equation~\eqref{eq:pres:sigmaa} has a simple interpretation: the
active impulse acts as a ``momentum reservoir'' for the particle. To
produce a non-zero density of active force in a region of space,
incoming and outcoming fluxes of active impulse have to
differ. Finally, the pressure takes the ideal gas law form
\begin{equation}\label{eq:pres:EOS1}
  P=\rho(\br_b)\frac{D_{\rm t}}\mu - \hat {\bf x}\cdot\sigma_a(\br_b)\cdot \hat {\bf x}=\rho_b T_{\rm eff}\;,
\end{equation}
where $\rho_b\equiv\rho(\br_b)$, $D_{\rm eff}=D_{\rm t}+(\mu\fp)^2/2D_{\rm r}$ is the large-scale diffusivity of the particle, $T_{\rm eff}\equiv D_{\rm eff}/\mu$ is an effective temperature~\footnote{We note, beyond the analogy with the ideal gas law, $T_{\rm eff}$ does not, in general, play any thermodynamic role in active systems.}, and $\hat{\bf x}$ is a unit vector in the $x$ direction. The pressure can thus be written as the sum of a passive and an active
stress, which is remarkable due to the absence of momentum
conservation in the system. Since
$\partial_t \langle \br_i \otimes \bfu_i \rangle= \langle \dot \br_i \otimes  \bfu_i
\rangle - D_{\rm r} \langle \br_i \otimes  \bfu_i \rangle$, one finds that, in a
homogeneous bulk, the active stress tensor can be rewritten as
\begin{equation}
\sigma_a=-\langle \sum_i \br_i \otimes \bffp^i
\delta(\br-\br_i)\rangle\;.
\end{equation}
This expression was introduced in \cite{takatori2014swim}, where it
was termed the ``swim
pressure''. It can also be obtained using methods developed by Irving
and Kirkwood~\cite{yang2014aggregation} or using a generalized virial
theorem~\cite{Winkler2015SoftMatter,Falasco:2016:NJP}.

It is interesting to note that when the active force density
satisfies Eq.~\eqref{eq:pres:sigmaa}, there is a relation in the
steady state between the total current ${\bf J}_{\rm tot}$ flowing through the system and
the total force ${\bf F}_{\rm tot}$ exerted by the particles on the boundary. To see this,
integrate Eq.~\eqref{eq:pres:current} over space to get that the total
current ${\bf J}_{\rm tot}\equiv \int d^2\, \br \bfJ(\br)$ satisfies:
\begin{equation}
{\bf J}_{\rm tot}=-\mu \int d^2 \br\, \rho(\br) \nabla V_{\rm w}(\br) \equiv  -\mu {\bf F}_{\rm tot} \;,
\end{equation}
where we have used that $\int d^2 \br\, \bfF_a(\br)=0$ due to
Eq.~\eqref{eq:pres:sigmaa}. A more intuitive derivation of this result
can be obtained by summing Eq.~\eqref{eq:singledynamics} over all
particles and averaging over the steady-state distribution. One then
gets
\begin{equation} \label{eq:currentforcetot}
{\bf J}_{\rm tot}=\langle \sum_i \dot\bfr_i \rangle = -\mu \langle \sum_i\nabla V_{\rm w}(\bfr_i) \rangle = -\mu {\bf F}_{\rm tot} \;,
\end{equation}
where the sum of the active forces has vanished since the dynamics of
the orientations are isotropic random walks decoupled from
$\br_i(t)$---which, as will become clear below, is the reason why
Eq.~\eqref{eq:pres:sigmaa} holds. In flux-free systems,
Eq.~\eqref{eq:currentforcetot} thus implies that boundaries cannot
exert any net total force on an active bath.

We note that Eq.~\eqref{eq:pres:EOS1} shows that the pressure is
independent of the wall potential. This is remarkable since
$\bfF_a(\br)$ depends on the choice of confining potential $V_{\rm w}(\br)$
(see Fig.~\ref{fig:ActiveImpulse}(d)). The underlying reason is that,
since the dynamics of the particle orientation is independent from all
other degrees of freedom, the total active impulse that the particle
can transfer to the wall does not depend on the wall potential.
The existence of an equation of state for the pressure can be
generalized to ABPs interacting via pairwise forces and can be used to
derive a mechanical theory for MIPS in such
systems~\cite{takatori2014swim,Solon_interactions,solon_generalized_2018,solon_generalized_2018-1,speck2021coexistence,omar2023mechanical}.

We note, however, that the derivation above does not account for the
simulations reported in Fig.~\ref{fig:piston}. One thus needs to go one
step further and account for the ellipsoidal particle shapes by
considering the torques $\Gamma$ exerted by the walls on the
particles: $\dot \theta_i=\Gamma(\br_i,\theta_i) + \sqrt{2 D_{\rm r}}
\eta_i^{\rm r}$. In this case, the time evolution of the active-force density
$\bfF_a(\br)$ is given by:
\begin{eqnarray}
  \partial_t \bfF_a(\br) &=  
\langle \sum_i f_{\rm p} \Gamma(\br_i,\theta_i) {\bf u}^\perp(\theta_i) \delta(\br-\br_i) \rangle - D_{\rm r} \bfF_a(\br)\nonumber\\
  &\qquad  -
\nabla\cdot [  \langle \sum_i \dot \br_i \otimes \bffp^i
\delta(\br-\br_i)\rangle]\;,
\end{eqnarray}
where $\bfu^\perp(\theta)=\partial_\theta \bfu(\theta)$. In the steady state, one  finds
\begin{eqnarray}\label{eq:pres:sources}
\bfF_a(\br) = \nabla \cdot \sigma_a^{\rm tf}+\langle \sum_i \frac{f_{\rm p}}{D_{\rm r}} \Gamma(\br_i,\theta_i) {\bf u}^\perp(\theta_i) \delta(\br-\br_i) \rangle\;,
\end{eqnarray}
where
$\sigma_a^{\rm tf}=-\langle \sum_i \dot \br_i \otimes \frac{\bffp^i}{D_{\rm r}}
 \delta(\br-\br_i)\rangle$ is the flux of active impulse
in the absence of torques. Equation~\eqref{eq:pres:sources} thus
splits the contribution to the density of active forces between
conserved and non-conserved parts, hence showing that wall-induced
torques can be seen as sources or sinks of active impulse. Introducing
$\psi(\br,\theta)=\langle \sum_i
\delta(\br-\br_i)\delta(\theta-\theta_i)\rangle$, the pressure can be written
as
\begin{eqnarray}\label{eq:pressure}
  P&=&\rho_b T_{\rm eff} + \Delta P_w\;,
\end{eqnarray}
where $\Delta P_w$ is a wall-dependent contribution given by
\begin{equation}\label{eq:pressurenoeos}
\Delta P_w= \frac{\mu \fp
 }{D_{\rm r}}\! \int \!\!d\theta \!\!\int_{x_b}^\infty \!\!\! dx\, \psi(x,y_b,\theta) \Gamma(x,y_b,\theta)\sin \theta\,.
\end{equation}
Importantly, the pressure is not independent of the wall anymore,
which explains the spontaneous compression of the asymmetric piston
shown in Fig.~\ref{fig:piston}. Indeed, the piston is stalled when the
pressures on both sides are equal, which requires
$\Delta P_{w_L}(\rho_L)=\Delta P_{w_R}(\rho_R)$. When the left and
right walls of the piston are different, this equality requires
different densities on both sides of the piston.

\begin{figure}
  \includegraphics[width=1\linewidth]{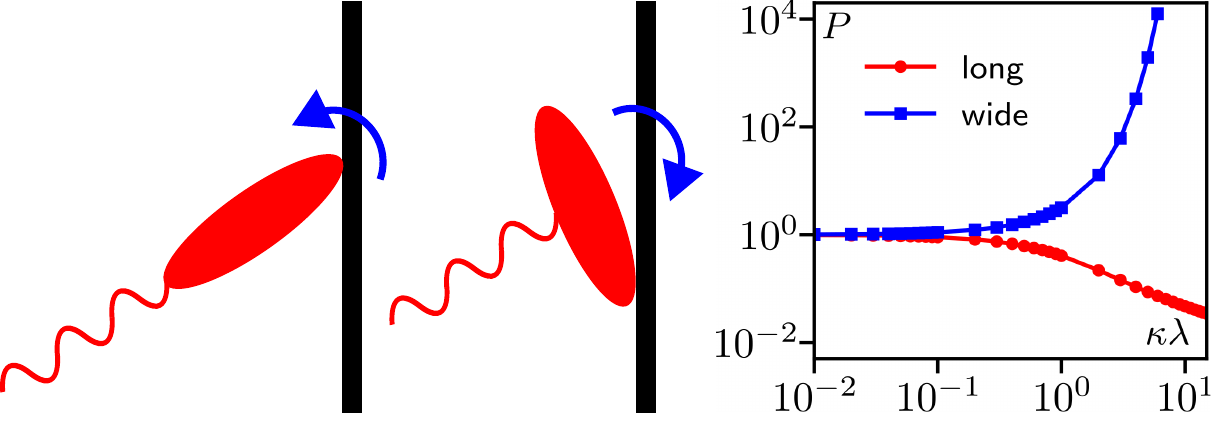} \caption{{\bf
  Left:} Schematic representation of an elliptic active particle
  hitting a wall. Depending on the aspect ratio, the collision rotates
  the particle in opposite directions, making it face parallel or
  toward the wall. {\bf Right:} Mechanical pressure measured on the
  wall for non-interacting ABPs with an aspect-ratio parameter
  $\kappa=|a^2-b^2|/8$, where $a$ and $b$ are the axis lengths of the
  ellipse. The wall is a confining harmonic potential of stiffness
  $\lambda$. Parameters: $f_{\rm p}=\mu=1$, $D_{\rm r}=0.5$, $\lambda=1$, bulk
  density $\rho=1$, system size $20\times 2$. As suggested by the left
  panel, the pressure exerted by long particles ($a>b$) is
  decreased by the wall torques, whereas that of wide particles ($a<b$)
  is enhanced.}  \label{fig:torque-wall}
\end{figure}

\begin{figure}  
  \includegraphics[width=1\linewidth]{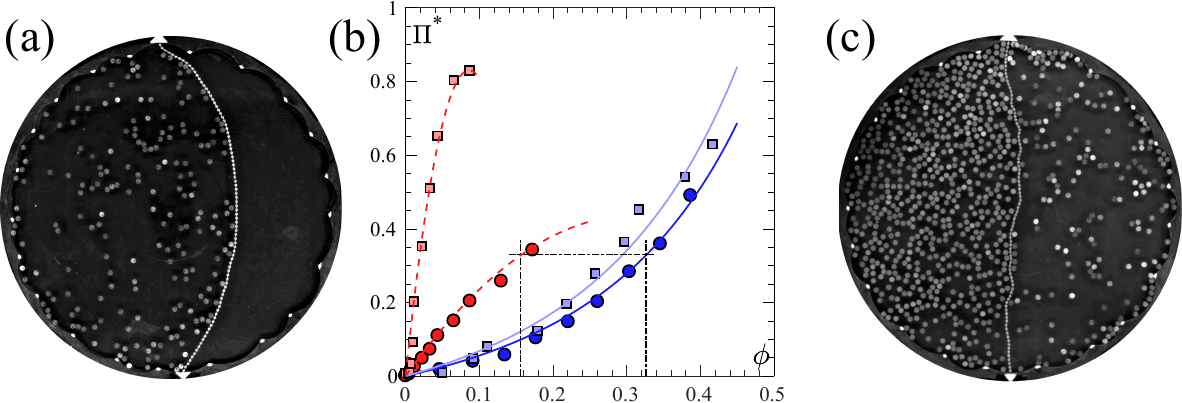} \caption{{\bf
  (a)} A flexible chain separates a system of vibrated grains into two
  cavities. The chain curvature is used to measure the pressure
  exerted by the particles confined in the left cavity, which can be
  either isotropic passive-like disks or anisotropic active
  disks. {\bf (b)} Pressure $\Pi$ measured as the packing fraction
  $\phi$ is varied. For isotropic disks (blue), $\Pi$ is in agreement
  with the equilibrium equation of states for hard disks (plain
  lines).  For active disks (red), two types of chains made with links
  of different sizes measure different pressures (square and circle
  symbols), signaling the absence of an equation of state for the
  pressure. Dot-dashed line denotes positions where active and passive
  disks are in equilibrium at different densities, as shown in {\bf
  (c)}. Reproduced with permission
  from~\cite{junot2017active}} \label{fig:Olivier}
\end{figure}

The discussion above can be understood by thinking about the most
commonly encountered active particles around us: pedestrians. Think
about a small child running towards you. Stopping the child requires
making its translational speed vanish, hence bringing the
corresponding incoming momentum flux to zero. This corresponds to the
drop of the passive momentum ($\rho D_{\rm t}/\mu$ in the case of an ideal
gas). If the child keeps running while you are holding them, you have
to absorb an additional momentum flux that the child is transferring
from the ground onto you, which corresponds to the contribution of the
active force in Eq.~\eqref{eq:pres:pres2}. Because many different
strategies can be employed to stop the child from running, the active
pressure will generically depend on the restraining adult, hence
leading to the lack of an equation of state. This is how torques lead
to a lack of equation of state, as illustrated in
Fig.~\ref{fig:torque-wall}. For torque-free ABPs, because the dynamics
of the active force is independent of all other degrees of freedom,
the momentum flux they transfer to the wall through their active force
before running away is always given by $\mu \fp^2/2 D_{\rm r}$, which leads
to an equation of state.  We note that this is an idealized limit and
that, for dry active systems, the lack of equation of state is
expected to be generic, a fact that has been
confirmed experimentally~\cite{junot2017active} for self-propelled discs (see
Fig.~\ref{fig:Olivier}). We stress that the lack of equation of state is not necessarily related to
torques induced by confining walls: Aligning interactions and motility regulation have, for
instance, also been shown to prevent the existence of an equation of
state for the pressure~\cite{solon2015pressure}.

\subsection{Curved and flexible boundaries}

\begin{figure}
\includegraphics{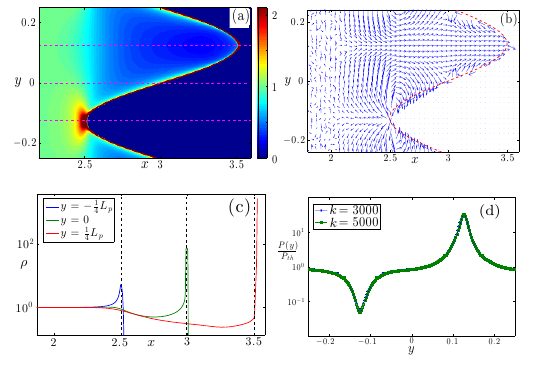}  
\if{\beginpgfgraphicnamed{Fig8}
  \begin{tikzpicture}
\def\x{4.4}
\def\y{3.1}
\def\xw{-1.25}
\def\yw{.73}
\def\dx{.15}
\def\dy{.2}
\def\ys{.22}

\path(.08,0) node {\includegraphics[width=.5225\columnwidth,height=.32\columnwidth]{Density_ABP_new_A_lines_v2.pdf}};
\path (\x,0) node {\includegraphics[width=.475\columnwidth,height=.32\columnwidth]{Current_ABP_new_B_v2.pdf}};
\path (0,-\y) node {\includegraphics[width=.494\columnwidth,height=.31\columnwidth]{DensityCross_gold_C_v2.pdf}};
\path (.1+\x,-\y) node {\includegraphics[width=.5225\columnwidth,height=.35\columnwidth]{P_Y_2lmbda_new2_v2.pdf}};


\draw (\x+1.5,-\y+1) node {\scriptsize (d)};
  \end{tikzpicture}
\endpgfgraphicnamed{Fig8}}\fi
  \caption{Density \textbf{(a)} and current \textbf{(b)} of
    non-interacting ABPs near the right edge of the system with the
    curved-wall potential given in
    Eq.~\eqref{eq:curvedwallpot}. Parameters: $f_{\rm p}=D_{\rm r}=24$,
    $D_{\rm t}=0$, $L_p=0.5$, $A=0.5$, $k=10^3$, $x_0=3$, and $\mu=1$. The
    red dashed curve corresponds to $x_{\rm w}(y)$. \textbf{(c)} Three cross
    sections of the particle density taken at the three horizontal
    dashed lines in (a). The vertical lines correspond to
    $x_{\rm w}(y)$. \textbf{(d)} Pressure normal to the wall, normalized by
    Eq.~\eqref{eq:pres:EOS1}, as a function of $y$, in the hard wall
    limit.  Reproduced with permission
    from~\cite{nikola2016active}.}  \label{fig:Density}\label{current} \label{density_cross}\label{P_y_profile}\label{fig:Curved}
\end{figure}

The anomalous mechanical properties of dry active systems are not
restricted to the lack of an equation of state for flat walls. Indeed,
even in cases where an equation of state exists, the local pressure
exerted by active fluids on confining walls generically depend on the boundary
shape~\cite{mallory2014curvature,fily2014dynamicsSM,fily2015dynamics,yan2015force,nikola2016active}. This
can be traced back to the fact that, after colliding with a wall,
active particles glide along it and accumulate in regions with higher
curvatures~\cite{mallory2014curvature,fily2014dynamicsSM}. This leads
to non-trivial density modulations and currents near the
wall~\cite{mallory2014curvature,fily2014dynamicsSM,fily2015dynamics,yan2015force}
that are illustrated in Fig.~\ref{fig:Curved} using numerical
simulations of ABPs interacting with a periodic soft-wall potential that vanishes for $x<x_w$ and is otherwise given by
\begin{equation}\label{eq:curvedwallpot}
V(\br)=\frac{k}2 [x-x_{\rm w}(y)]^2\;\text{with}\; x_{\rm w}(y)=x_0+A\sin(2\pi y/L_p)\;\,
\end{equation}
and $L_p$ the period. Importantly, Fig.~\ref{fig:Curved}(d)
shows that the pressure, measured as the force normal to the wall,
depends on the exact location along the wall.

\begin{figure}
\includegraphics[width=0.9\columnwidth]{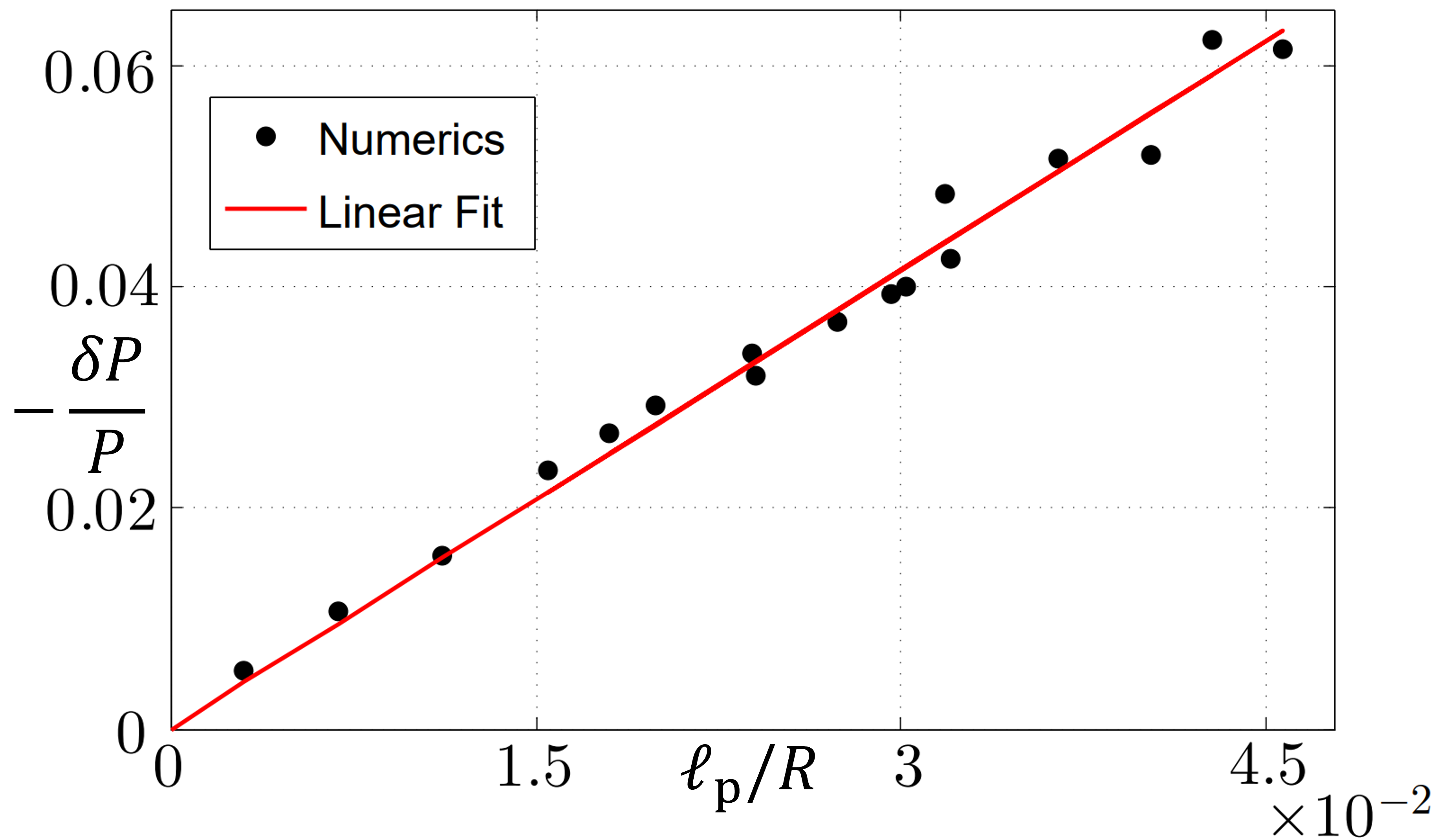}
\caption{Difference $\delta P$ between the pressure at the concave and convex
    apices of the curved-wall potential given in
    Eq.~\eqref{eq:curvedwallpot}, as the radius of curvature $R =
    L_p/(4\pi^2A)$ is varied. The red line is a linear fit of the data
    leading to a slope $\sim 1.3$. Numerical data were obtained using
    simulations of ABPs with $f_{\rm p} = 0.75$ and $\mu=D_{\rm r} =
    1$. Wall parameters in the ranges $A \in [3.1\cdot10^{-3},
    3.3\cdot10^{-2}]$ and $L_p \in [3.6, 5.5]$ were used, with $k =
    2 \cdot 10^4$. Dimensional analysis predicts a slope of
    $2$. Reproduced with permission from~\cite{nikola2016active}.  }
\label{fig:laplace}
\end{figure}

The difference between the point of highest pressure and lowest pressure, $\delta P$, is found numerically to be proportional to the
curvature $1/R$ at the tips of the sinusoidal wall, see Fig.~\ref{fig:laplace}. This is consistent with measurements of the pressure exerted by an active ideal gas on a 2d circular cavity of radius $R$~\cite{mallory2014anomalous,yan2015force,sandford2017pressure}. Moreover, recently, the corresponding finite-size correction was computed analytically and shown to be given by~\cite{zakine2020surface}:
\begin{equation}
    P(R)=P_b -\frac{\gamma}{R}+\mathcal{O}(R^{-1})\;,
\end{equation}
where $P_b=\rho_b T_{\rm eff}$ is the pressure in an infinite system at density $\rho_b$ and $\gamma$ is the fluid-solid surface tension that can be computed as
\begin{equation}\label{eq:surfacetension}
    \gamma\equiv \int_0^R dr\, \frac{\mu }2 f_{\rm p}^2\tau \rho_b-\int_0^\infty dr\, \frac{\mu }2 f_{\rm p}^2\tau \rho(r)\;.
\end{equation}
This result, which generalizes Lapace's law to active fluids, directly suggests that $\delta P \simeq 2\gamma/R$.  On dimensional grounds, one can estimate the surface tension as $\gamma \simeq -P_b \ell_{\rm p}$,
with $\ell_{\rm p}$ the
persistence length. This agrees semi-quantitatively with the measurements shown in Fig.~\ref{fig:laplace}. As shown by Eq.~\eqref{eq:surfacetension}, the negative sign of $\gamma$ is due to the tendency of active particles to accumulate at the wall. It is consistent with the overall sign of $\delta P$, which can also be
understood heuristically as the active particles accumulate at concave
regions of the wall. Note that the sign of $\delta P$ shows that active particles tend to exert forces on curved boundaries that would amplify the deformation of a flexible boundary, as discussed below.

Interestingly, for walls that are not reflection-symmetric, a net
shearing force develops parallel to the
wall~\cite{nikola2016active}. It can be understood as a consequence of
a ratchet
effect~\cite{reichhardt2013active,ai2014transport,angelani2011active}:
The breaking of time-reversal symmetry by the active particles coupled
to a breaking of an inversion symmetry leads to a steady-state current
along the wall. In turn, Eq.~\eqref{eq:pres:current} tells us that
such a current will generically be associated with a force tangential to
the wall. This explains the spontaneous rotation of
microscopic gears observed in simulations~\cite{angelani2009self} and
experiments~\cite{di2010bacterial,sokolov2010swimming}.

The dependence of the active pressure on the boundary shape is
exemplified by considering the behavior of flexible elastic objects
inside an active fluid. For concreteness, first consider a flexible
partition whose ends are held at walls at the top and bottom of a
container filled with active particles. Once a fluctuation creates a
local deformation in the filament, a finite pressure 
difference $\Delta P$ develops between its two sides due to the difference in active
forces exerted on the two sides of the apex. This tends to increase
the deformation and is opposed by the elasticity of the flexible
partition. The outcome of this competition can be understood by
considering the linearized dynamics of the partition, characterized by
a Monge representation $h(x,t)$. Because the accumulation of
active particles is proportional to the curvature of a confining
interface~\cite{fily2014dynamicsSM,nikola2016active}, one expects
$\Delta P\propto \nabla^2 h$, leading to~\cite{nikola2016active}:
\begin{equation}\label{eq:hofx}
  \partial_t h(x,t)=(T-\kappa) \nabla^2 h(x,t)-\kappa_b \nabla^4 h(x,t)\;,
\end{equation}
where $T$ and $\kappa_b$ are the line tension and bending rigidity,
respectively, and $\kappa$ is a coefficient that measures the particle
activity. Equation~\eqref{eq:hofx} shows that for large activity, when
$\gamma>T$, a horizontal filament is unstable to fluctuations above a
characteristic length $\lambda\propto \sqrt{\kappa_b/(\kappa-T)}$. For
short filaments, one thus expects that the rigidity keeps the filament
straight, whereas long filaments are expected to be unstable. This has
indeed been observed both in simulations and
experiments~\cite{nikola2016active,junot2017active}, where the instability has been shown to coarsen and
lead to a deformation of the filament with a wavelength set by the
system size. For an unpinned passive flexible filament in an active medium, an even richer scenario has been
reported in both simulations and
experiments~\cite{nikola2016active,anderson2022polymer}. As the length
of the filament increases, the instability discussed above first leads to
a left-right asymmetry and the spontaneous formation of a
parachute-like structure~\cite{PolymerLooping,PolymerCollapse};
somewhat reminiscent of a sail blowing in the wind, except with the wind itself generated by the curving filament. 
The
pressure difference on the two sides of the filament then generates a
net propelling force and turns the filament into an emergent active
particle. Upon a further increase in length, a full period of the
unstable mode develops, leading to short-lived spontaneous
rotors. Finally, very long filaments are found to lead to folded
structures.

\section{Obstacles and localized inclusions}
\label{sec:obstacles}
\label{sec:inclusion}
\begin{figure}
\includegraphics{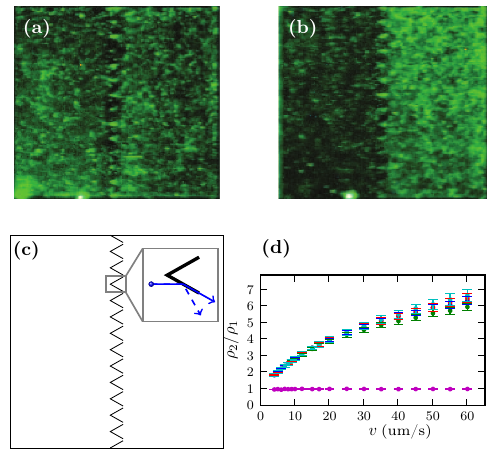}  
\if{\beginpgfgraphicnamed{Fig10}
        \begin{tikzpicture}[scale=.009]
      \begin{scope}[scale=2.5,xshift=200cm,yshift=180cm]
            \node (0,-0.2) {\includegraphics[totalheight=3.25cm]{GalajdaBacteria.png}};
              \draw[white] (-60,55) node {\bf (b)};
      \end{scope}

\begin{scope}[scale=2.5,xshift=0cm,yshift=180cm]
            \node (0,-0.2) {\includegraphics[totalheight=3.25cm]{GalajdaColloids.png}};
              \draw[white] (-60,55) node {\bf (a)};
      \end{scope}
      \begin{scope}[scale=2.5,xshift=180cm,yshift=-10cm]
            \node (0,-0.2) {\includegraphics[totalheight=2.75cm]{SimulationPaper4.pdf}};
              \draw (-60,80) node {\bf (d)};
              \draw (38,-58) node {\scriptsize (um/s)};

      \end{scope}
      
    \draw (-170,170) node {\bf (c)};
    \draw (-200,-200) -- (-200,200) -- (200,200) -- (200,-200) -- (-200,-200);
    \draw (-12,-200) -- (12,-200+13.5);
    \draw (12,-200+13.5+3.8) -- (-12,-200+27+3.8) -- (12,-200+27+13.5+3.8);
    \draw (12,-200+27+13.5+2*3.8) -- (-12,-200+2*27+2*3.8) -- (12,-200+2*27+13.5+2*3.8);
    \draw (12,-200+2*27+13.5+3*3.8) -- (-12,-200+3*27+3*3.8) -- (12,-200+3*27+13.5+3*3.8);
        \draw (12,-200+3*27+13.5+4*3.8) -- (-12,-200+4*27+4*3.8) -- (12,-200+4*27+13.5+4*3.8);
        \draw (12,-200+4*27+13.5+5*3.8) -- (-12,-200+5*27+5*3.8) -- (12,-200+5*27+13.5+5*3.8);
        \draw (12,-200+5*27+13.5+6*3.8) -- (-12,-200+6*27+6*3.8) -- (12,-200+6*27+13.5+6*3.8);
        \draw (12,-200+6*27+13.5+7*3.8) -- (-12,-200+7*27+7*3.8) -- (12,-200+7*27+13.5+7*3.8);
        \draw (12,-200+7*27+13.5+8*3.8) -- (-12,-200+8*27+8*3.8) -- (12,-200+8*27+13.5+8*3.8);
        \draw (12,-200+8*27+13.5+9*3.8) -- (-12,-200+9*27+9*3.8) -- (12,-200+9*27+13.5+9*3.8);
        \draw (12,-200+9*27+13.5+10*3.8) -- (-12,-200+10*27+10*3.8) -- (12,-200+10*27+13.5+10*3.8);
        \draw (12,-200+10*27+13.5+11*3.8) -- (-12,-200+11*27+11*3.8) -- (12,-200+11*27+13.5+11*3.8);
        \draw (12,-200+11*27+13.5+12*3.8) -- (-12,-200+12*27+12*3.8) -- (12,-200+12*27+13.5+12*3.8);
        \draw (-12,200) -- (12,200-13.5);
        \draw[thick,gray] (-20,93) rectangle +(37,30);
        \draw[gray,thick] (17,93)--(50cm-30,50cm-350);
        \draw[gray,thick] (17,123)--(50cm-30,50cm+125cm);
        \begin{scope}[scale=2.5,xshift=50cm,yshift=50cm]
        \draw[thick,gray] (-30,-35) rectangle +(56,55);
                  \draw[ultra thick] (12,-13.5) -- (-12,0) -- (12,13.5);
          \draw[blue,thick,->] (-24,-6.75) -- (0,-6.75) -- (24,-20.25);
          \shade[ball color=blue] (-24,-6.75) circle (2);
          \draw[blue,dashed,thick,->] (-24,-6.75) -- (0,-6.75) -- (13.24,-6.75-23.53);
          \shade[ball color=blue] (-24,-6.75) circle (2);
      \end{scope}

        \end{tikzpicture}
\endpgfgraphicnamed{Fig10}}\fi
\caption{\textbf{(a)}-\textbf{(c)} A uniform density of bacteria at the beginning of the experiment (a) becomes inhomogeneous at late times (b) in a 2d microfluidic chamber split into two compartments by an array of asymmetric obstacles, schematically depicted in (c). \textbf{(d)} Ratio between the densities of bacteria in the right ($\rho_2$) and left ($\rho_1$) sides of the cavity shown in (c), measured using numerical simulations of RTPs whose self-propulsion speed $v$ is varied. All dimensions match the experiment: the enclosures dimensions are $L\times H=400\mu\text{m}\times400\mu\text{m}$,
the arms of the funnels are 27$\mu$m long and their apex angle is
$\pi/3$, the funnels are separated by gaps 3.8$\mu$m wide. The
tumbling rate of the RTPs is set to $\alpha=1\text{s}^{-1}$. The
overlapping red, blue, green, and cyan symbols are obtained by varying
$H$ and the left and right chambers width, $L_L$ and $L_R$,
respectively, with $(L_R,L_L,H)\in
\{(200,200,400),(400,400,400),(200,200,800),(400,200,400)\}$. The vertical density of obstacles is kept constant. The rectification of bacterial density is thus independent of finite-size
effects. In all simulations, RTPs align with the walls upon
collision (solid line inside the inset of panel (c)). When the
RTPs instead experience specular reflection upon collision, as shown
by the dashed line inside the inset of panel (c), a uniform density
field is measured (magenta symbols). This is consistent with the fact
that collisions are the sole irreversible process in these
simulations. (a-b) were reproduced with permission from~\cite{galajda2007wall}. (c-d) were reproduced with permission from~\cite{tailleur2009sedimentation}.\label{fig:ratchets}\label{fig:funnel}}
\end{figure}

An even richer physics has been reported when considering the
mechanical interplay between active particles and obstacles
\textit{immersed} in active fluids. Arguably, the first observation
for this was the experimental study of \textit{asymmetric obstacles}
by Galajda and co-workers~\cite{galajda2007wall}, which demonstrated
that an array of V-shaped obstacles placed in a bacterial bath leads
to the accumulation of bacteria on one side of the array (see
Fig.~\ref{fig:ratchets}).  In this section, we discuss how a mechanical perspective
accounts for the induced organization of the bacterial fluid and how
to account for more general situations. In particular, we focus on the
universal aspects of the large-scale density modulation and current induced by
obstacles and not on the rich physics that can be observed in the near
field, at distances ${\cal O}( \ellp)$ from the obstacle, which has
attracted a lot of attention~\cite{Kaiser2012,Potiguar2014,Ni2015,ZaeifiYamchi2017,Yan2018,Wysocki2020,speck2021vorticity} and has already been reviewed~\cite{bechinger_active_2016}.

We start with the case of a single obstacle immersed in an infinite
active fluid and show how a multipole expansion allows one to predict the
far-field structure of density and current fields. A simple picture
emerges in which the local asymmetry of the obstacle induces a ratchet
current that mass-conservation turns into long-range density and
current modulations. Mathematically, the relation between the force
monopole exerted by the obstacle on the active fluid and the induced
current flow is identical to that between electrostatic dipoles and
fields. Next, we turn to consider the effect of boundary conditions
and show how an image theorem allows generalizing the results to
simple cases like periodic or closed boundary conditions. Then, we discuss how the above leads to long-range non-reciprocal interactions between inclusions in an active fluid. Finally, we
discuss the case of mobile inclusions and review recent works on
passive tracers and possible dynamics arising from interactions mediated by active baths.

\subsection{A single obstacle in an infinite system}
\label{sec:singleobj}

We consider $N$ non-interacting ABPs in the presence of an obstacle
modelled by an external potential $V(\br)$, localized on a compact
support near ${\bf r}=0$. In $d$ space dimensions, the average number
density of particles at position $\br$ with an orientation $\bfu$,
$\mathcal {P}(\br,\bfu)$, evolves according to:
\begin{align}\label{eq:FP}
	\partial_t \mathcal{P}({\bf r},\bfu)= &
        -\nabla\cdot\left[\mu f_{\rm p}\bfu \mathcal{P}-\mu\mathcal{P}\nabla
          V-D_{\rm t}\nabla
          \mathcal{P}\right]+D_{\rm r}\Delta_{\bfu}\mathcal{P}\;,
\end{align}
where $\Delta_{\bfu}$ is the spherical Laplacian. Integrating over the orientation $\bfu$ leads to Eqs.~\eqref{eq:dynamicsdensitylaw}
and~\eqref{eq:pres:current} with $V_{\rm w}$ replaced by $V$. Far away from
the obstacle, i.e. when $|\br|\gg \ellp$, the active dynamics is
diffusive at large scales so that $\bfJ \simeq \bfJ_D \equiv 
-D_{\rm eff} \nabla \rho$, where
$D_{\rm eff}=D_{\rm t}+(\mu f_{\rm p})^2/dD_{\rm r}$~\cite{cates_when_2013}. We then
introduce
\begin{equation}\label{eq:weirdJ}
  {\delta{\bfJ}}\equiv {\bf J}-\bfJ_D=\mu \bfF_a-\mu\rho \nabla V + (D_{\rm eff}-D_{\rm t})\nabla\rho\;,
\end{equation}
which measures the difference between $\bfJ$ and its diffusive
approximation. Intuitively, we expect $ {\delta{\bfJ}}$ to
vanish in the bulk and to be significant in the vicinity of the
obstacle, for $|\br|\lesssim \ellp$.  We can then rewrite the
conservation equation in the steady state, $\nabla \cdot {\bf J}=0$,
as
\begin{align}\label{eq:Poisson's eq}
	D_{\rm eff}\nabla^{2}\rho =\nabla \cdot {\delta{\bfJ}}({\bf r})\;.
\end{align}
Equation~\eqref{eq:Poisson's eq} is a Poisson equation for the density
field with a source term $\nabla\cdot  {\delta{\bfJ}}({\bf r})$ localized around
the obstacle. Its solution is
\begin{equation}\label{eq:rhoint}
	\rho({\bf r})=  \rho_b+\frac{1}{ D_{\rm eff}}\intop d^d \bfr'\,  G(\bfr,\bfr')\nabla \cdot   {\delta{\bfJ}}(\bfr') \;,
\end{equation}
where $G(\bfr,\bfr')$ is the Green's function of the Laplacian in $d$
space dimensions. To proceed, we employ a multipole expansion of
Eq.~\eqref{eq:rhoint} to determine the leading contribution in the
far-field limit, $|\bfr|\gg \ell_p$, using all the terms entering $\nabla
\cdot {\delta{\bfJ}}$. Using Eq.~\eqref{eq:pres:sigmaa}
and~\eqref{eq:weirdJ}, one sees that the leading order contribution is
given by $ {\delta{\bfJ}}(\br)\simeq -\mu\rho(\br)\nabla V
(\br)$. Introducing the force monopole
\begin{equation}\label{eq:forcemonopole}
  {\bf p} = - \int d^{d}\bfr'\, \rho(\bfr') \nabla V(\bfr')\;,
\end{equation}
and carrying out explicitly the multipole expansion
of Eq.~\eqref{eq:rhoint}, one then finds that the leading-order far-field
density and current are given by
\begin{align}\label{eq:rho d dimensions}
	\rho({\bf r})&= \rho_b + \frac{\beta_{\rm eff}}{S_d} \frac{{\bf r}\cdot{\bf p}}{r^d} +  \mathcal{O}\!\left(r^{-d}\right) \ , \\
	{\bf J}({\bf r})&= \frac{\mu}{S_d}\frac{d\left(\hat{\bf{r}}\cdot{\bf p}\right)\hat{\bf{r}}-{\bf p}}{r^{d}} + \mathcal{O}\!\left(r^{-(d+1)}\right) \ ,\label{eq:J d dimensions}
\end{align}
where $\beta_\mathrm{eff} \equiv 1/T_{\rm eff}=\mu / D_\mathrm{eff}$ and $S_d = (2\pi^{d/2})/\Gamma(d/2)$. As usual in multipole expansions, Eq.~\eqref{eq:rho d
  dimensions} is the solution of
\begin{align}\label{eq:Poisson's dipole}
  D_{\rm eff}\nabla^{2}\rho =\nabla \cdot [\mu\bfp \delta(\bfr)]\;.
\end{align}

We note that Eq.~\eqref{eq:rho d dimensions} predicts a universal
density modulation and current induced by an obstacle that exerts a
non-vanishing force monopole $\bfp$ on the active fluid. The
determination of $\bfp$, on the other hand, depends on the details of
the problem and requires an explicit derivation of the microscopic
structure of $\rho(\bfr)$ in the vicinity of the obstacle. This is, in
general, a very hard problem with very few exact
results~\cite{arnoulx2023run}. We also note that if $\bfp=0$, higher
orders in the multipole expansion have to be
considered~\cite{baek_generic_2018}. When the obstacle is spherical,
the density modulation and current vanish at all orders in the
multipole expansion.

The above derivation shows how forces exerted by obstacles lead to
large-scale ratchet currents. Since the pressure admits an equation of
state--- because the active force density can be written as the
divergence of a local stress tensor---there is an exact relation
between forces and current generated by obstacles. Integrating
Eq.~\eqref{eq:pres:current} over the full space indeed shows that
\begin{equation}
  \bfJ_{\rm tot}\equiv \int d^d \br\, \bfJ(\br)= \mu\bfp\;,\label{eq:currforce}
\end{equation}
where we have used that $\bfF_a(\br)=\nabla \cdot \sigma_a$ and the
divergence theorem to show that the contribution of the active force
vanishes when integrated over the full space.  This result is the
direct counterpart to Eq.~\eqref{eq:currentforcetot} for the case of an isolated obstacle.

While the above formulation pertained to noninteracting ABPs. The result
generalizes to homogeneous active fluids with  pairwise interparticle
forces~\cite{granek2020bodies}. 

\subsection{Finite systems and boundary conditions}\label{sec:finite}

So far, we have considered the case of an isolated object in an
infinite system and shown that asymmetric obstacles induce long-range
density modulations and currents. In turn, this implies that some care
has to be taken when finite systems of linear size $L$ are considered,
even when $L\gg \ellp$. We now discuss several such scenarios that
have been explored in the literature.

For periodic boundary conditions, when $L\gg \ellp$, one can show
that the system is equivalent to an infinite periodic lattice of
obstacles~\cite{granek2020bodies}. Interestingly, the density and
current fields, as well as the value of the force monopole ${\bf p}$,
differ between infinite and periodic systems by corrections of order
$\mathcal{O}(L^{-(d+2)})$~\cite{speck2021vorticity}.  For
$L\lesssim\ellp$, the far-field expansion is naturally
invalid. Numerical simulations and a scaling argument reveal that, for
$d=2$, the force monopole $\mathbf{p}$ grows as $\sim L^2$ until it
saturates at an asymptotic value for
$L\gg\ellp$~\cite{speck2021vorticity}.

Another scenario that was considered is the effect of confining flat
hard walls. The derivation of Sec.~\ref{sec:singleobj} can be extended
to this case---despite non-trivial boundary conditions---and, for a
single obstacle displaced by $\mathbf{X}=X\hat{\bf x}$ from a hard wall at
$x=0$, Eq.~\eqref{eq:rhoint} still holds in the far field of both the
obstacle and the wall~\cite{dor2022passive}. The Green's function $G$ is,
however, replaced by that of the Laplacian in a half-plane. To
leading order in the limit $|\mathbf{r}-\mathbf{X}|\gg\ellp$ and $X\gg\ellp$, the
solution is given by:
\begin{align}
\label{eq:rhowall}
    \rho(\mathbf{r})&=\rho_b+\frac{\beta_{\rm eff}}{S_d}\left[\frac{\left(\mathbf{r}-\mathbf{X}\right)\cdot\mathbf{p}}{\left|\mathbf{r}-\mathbf{X}\right|^d}+\frac{\left(\mathbf{r}-\mathbf{X}^*\right)\cdot\mathbf{p}^{*}}{\left|\mathbf{r}-\mathbf{X}^*\right|^d}\right]\nonumber\\
    &+\mathcal{O}\!\left(\left|\mathbf{r}-\mathbf{X}\right|^{-d},X^{-d}\right)\;.
\end{align}
where $\mathbf{p}^*$ and $\mathbf{X}^*$ are the images of $\mathbf{p}$
and $\mathbf{X}$ with respect to the wall, respectively. Note that Eq.~\eqref{eq:rhowall} holds with the force
monopole $\mathbf{p}$ given by its infinite-system value. Indeed, the corrections to $\mathbf{p}$ due to the image obstacle enter at order
$\mathcal{O}(X^{-(d-1)})$. Eq.~\eqref{eq:rhowall} was also
shown to hold when the obstacle is on the wall, so that
$\mathbf{X}=0$, with $\bfp$ parallel to the
wall~\cite{dor2022disordered}.  Finally, the derivation was also
carried out for a circular cavity, as will be discussed in
Sec.~\ref{sec:ptbreak}.

\begin{figure}
  \includegraphics[width=0.7\linewidth]{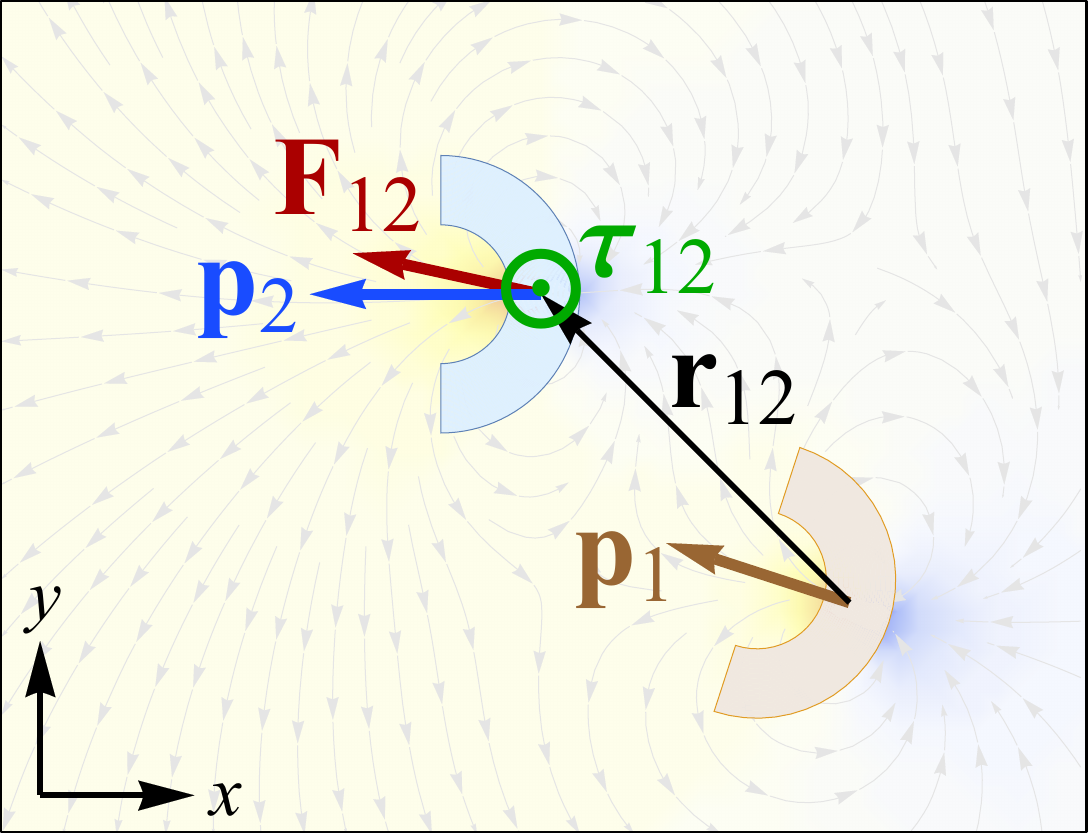} \caption{Schematic diagram of two interacting asymmetric
  passive obstacles.
  Obstacle $O^{(1)}$ (orange) is placed at $\mathbf{X}_1$ and obstacle $O^{(2)}$ (blue) is placed at $\mathbf{X}_2$ (black) and we denote by ${\bf r}_{12}$ their separation. Superimposed currents are shown in gray streamlines, and
  density modulations are shown with a blue-to-red
  colormap. Reproduced with permission from~\cite{granek2020bodies}
  }\label{fig:nonrec}
\end{figure}

\subsection{Non-reciprocal mediated interactions.}\label{sec:nonrec}
Since isolated obstacles create long-range density modulations, it is
natural to expect that several obstacles immersed in the same active
fluid will experience long-range mediated interactions. These turn out
to be \emph{non-reciprocal} and can be derived as follows. 
Consider two obstacles, $O^{(1)}$ and $O^{(2)}$, fixed at positions $\mathbf{X}^{(1)}$ and $\mathbf{X}^{(2)}$, respectively, and denote by $\bfr_{12}$ their separation 
(see
Fig.~\ref{fig:nonrec}). The effect of \OO{1} on \OO{2} can
be quantified by an emergent interaction force $\mathbf{F}_{12}$,
which can be identified as the net residual force exerted on \OO{2} due to the introduction of \OO{1} into the bath:
\begin{align}
\mathbf{F}_{12} & \equiv\mathbf{p}^0_{2}-\mathbf{p}_{2}\;,\label{eq:F12}
\end{align}
where $-\mathbf{p}_{k}\equiv\int d^{d}\mathbf{r}\, \rho(\mathbf{r})\nabla
V_{k}(\mathbf{r}-\mathbf{X}^{(k)})$ is the net force exerted by the
active bath on \OO{k}, and
$\mathbf{p}^0_{2}=\mathbf{p}_{2}|_{V_{1}=0}$ is the force on \OO{2} in
the absence of \OO{1}. Inspection of Eq.~\eqref{eq:forcemonopole}
shows that $\mathbf{F}_{12}$ is directly connected to the density
modulation induced by \OO{1} in the vicinity of \OO{2}. For large
separations $r_{12}$
, the leading
order contribution of this modulation is a local shift of the average
density:
\begin{equation}
      \delta \rho_1({\bf X}_2)=\frac{\beta_{\rm eff}}{S_d} \frac{{\bf r}_{12}\cdot{\bf p}_1}{r_{12}^d}+{\cal
        O}(r_{12}^{-d})\;.
\end{equation}
Due to the linearity of Eq.~\eqref{eq:rhoint}, the force monopole
exerted by \OO{2} is modified as
\begin{equation}\label{eq:shiftp2}
  {\bf p}_2 =  \frac{\rho_b+\delta \rho_1({\bf X}_2)}{\rho_b}{\bf p}^0_2\,.
\end{equation}
Inserting this into Eq.~\eqref{eq:F12} results in
\begin{align}
    \mathbf{F}_{12}&=-\frac{\beta_{\rm eff}}{ S_d} \frac{ {\bf r}_{12}\cdot{\bf p}_1^0} {\rho_b r_{12}^d}{\mathbf{p}_2^0}+\mathcal{O}(r_{12}^{-d})\,,\label{eq:drho1}
\end{align}
where we have used
$\mathbf{p}_k=\mathbf{p}_k^0+\mathcal{O}(r_{12}^{-(d-1)})$ and
$\mathbf{p}^0_{1}=\mathbf{p}_{1}|_{V_{2}=0}$. $\mathbf{F}_{21}$ can be
obtained by exchanging the indices $1 \leftrightarrow 2$ in
Eq. \eqref{eq:drho1}. Importantly, the equation implies that
$\mathbf{F}_{21}\neq-\mathbf{F}_{12}$, i.e. the bath-mediated
interactions are \textit{non-reciprocal}.

Note that the leading-order interactions require both obstacles to be
asymmetric so that ${\bf p}^0_k\neq 0$. Nonetheless, when \OO{2} is
symmetric, it still experiences a force from an asymmetric \OO{1}
because the latter induces a density gradient in the vicinity of
\OO{2}~\cite{baek_generic_2018}. Finally, we note that, while two isotropic obstacles
only experience short-range interactions, long-range interactions can
also emerge even if all ${\bf p}_{k}^0$ vanish. For example, rods
generate density modulations at order
$r_{12}^{-d}$~\cite{baek_generic_2018}.

So far, we have considered dilute active fluids, but these derivations
extend to the case of active bath particles subject to pairwise
interactions~\cite{granek2020bodies}. Furthermore, a similar expansion
has been derived for the interaction torque $\mathbf{n}_{12}$ between
the obstacles~\cite{baek_generic_2018}. The results also extend to
multiple obstacles, yielding additive interactions to leading order in
the far field.

\subsection{Mobile obstacles and dynamics}\label{sec:tracers}

\begin{figure}
\includegraphics{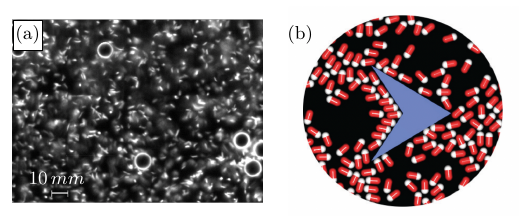}  
\if{\beginpgfgraphicnamed{Fig12ab}
 \begin{tikzpicture}[scale=.5]
  \draw (0,0) node {\includegraphics[width=0.49\linewidth]{WLexp.png}};
 \draw[color=black,fill=white] (-4.23,3.06) rectangle (-3.23,2.06);
 \draw (-3.73,2.56) node {(a)};
 \draw (-2.75,-2.3) node {\small \textcolor{white}{$10\,mm$}};
 \draw (9,0) node {\includegraphics[width=0.45\linewidth]{angelani.png}};
 \draw (5.5,2.56) node {(b)};
 \end{tikzpicture}
\endpgfgraphicnamed{Fig12ab}}\fi
\includegraphics{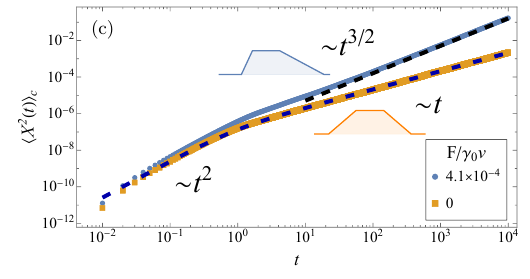}  
\if{\beginpgfgraphicnamed{Fig12c}
 \begin{tikzpicture}
 \draw (0,0) node {\includegraphics[width=\linewidth]{MSD.pdf} };
  \draw[color=white,fill=white] (-3,1.5) rectangle (-2.5,2.1);
 \draw (-2.8,1.85) node {(c)};
 \end{tikzpicture}
\endpgfgraphicnamed{Fig12c}}\fi
\caption{\textbf{(a)}
Passive polystyrene tracers immersed in a quasi-two-dimensional
  bacterial bath. Adapted with permission
  from~\cite{Wu2000}. \textbf{(b)} Simulation snapshot of a V-shaped passive
  tracer in a two-dimensional active bath. Holding the tracer
  orientation fixed leads to an average directed motion to the right
  with velocity $\propto-\mathbf{p}$. Reproduced with permission
  from~\cite{Angelani2010}. \textbf{(c)} Mean-square displacement for an adiabatic
  passive tracer in a one-dimensional active bath (microscopic
  simulations). Symmetric tracer (orange symbols): a transition from
  short-time ballistic motion to long-time diffusive is observed. Blue
  dashed line: the Wu-Libchaber model Eq.~\eqref{eq:WL}, with the
  exact calculation of the diffusivity $D_{\rm T}$ from the adiabatic
  theory (see text) and a fit leading to $\tau=0.41$. Asymmetric
  tracer (blue symbols) shows a long-time superdiffusion. Black dashed
  line: the prediction Eq.~\eqref{eq:sup} (no fitting parameters). All
  bath parameters are set to unity. Reproduced with permission
  from~\cite{Granek2022}.}\label{fig:tracer}
\end{figure}

So far, we have discussed the case in which obstacles are fixed in
space. However, there is also considerable interest in the dynamics
of mobile obstacles, referred to as passive tracers in active baths.
While a body of works focused on the effect of long-range hydrodynamic
interactions in wet active baths (see,
e.g. ~\cite{Chen2007,Leptos2009,Kurtuldu2011,Zaid2011,Thiffeault2015,Kurihara2017,Kanazawa2020}),
here, we focus on (dry) scalar active matter with, at most, short-range
interactions.

To write down the emerging equations of motion of a tracer, we
invoke an adiabatic limit where the motion of the tracer is so slow
that the statistics of the forces exerted by the bath are
indistinguishable from those exerted on a fixed tracer. In the
overdamped limit, the dynamics of the tracer can then be described using a
generalized Langevin equation
~\cite{Maes2020,Reichert2021,Reichert2021a,Granek2022,Feng2022,Shea2022}:
\begin{equation}
    \int_0^t dt'\,\gamma_{ij}(t-t')\dot{X}_j(t')=f_i (t)\;,\label{eq:GLE}
\end{equation}
where $\gamma_{ij}(t)$ is a memory kernel, $f_i (t)$ is a fluctuating
force, and summation over $j$ is implied. In Eq.~\eqref{eq:GLE}, the left-hand-side is the friction force exerted by the bath, i.e. the average contribution to the force due to the motion of the tracer, while the right hand side contains both the net force force exerted on a fixed tracer and the fluctuations of the force exerted by bath particles. Unlike in equilibrium, $\langle {\bf
f}(t) \rangle$ is generically non-zero for asymmetric tracers.

Within a
systematic adiabatic expansion~\cite{DAlessioJSTAT2016}, the memory
kernel $\gamma_{ij}(t)$ is given by an Agarwal-Kubo-type formula
\begin{align}
    \gamma_{ij}(t) &= \left\langle f_i (t)\partial_{X_j} \log P_{\rm s}(\{\mathbf{r}_k(0)-\mathbf{X},\mathbf{u}_k\})|_{\mathbf{X}=0}\right\rangle_c^{\rm s}\;,\label{eq:gamma}
\end{align}
where $\br_k$ and $\bfu_k$ are the positions and orientations of the
bath particles, $P_{\rm s}(\{\mathbf{r}_k-\mathbf{X},\mathbf{u}_k\})$
is their many-body steady-state distribution, and
$\langle\cdot\rangle^{\rm s}$ denotes an average with respect to an
ensemble in which the tracer is held fixed. The subscript $c$ indicates a connected correlation function, i.e., $\langle A B\rangle_c=\langle AB\rangle-\langle A\rangle\langle B\rangle$. 

Likewise, the statistics
of the fluctuating force $\mathbf{f}(t)$ are given by those of
$\sum_k\nabla V(\mathbf{r}_k-\mathbf{X})$ computed with the tracer
held fixed, and where $V$ the interaction potential between the tracer and the active particles.  For passive baths, the
Boltzmann distribution $P_{\rm
s}(\{\mathbf{r}_k-\mathbf{X},\mathbf{u}_k\})\propto e^{-\beta \sum_k
V(\mathbf{r}_k-\mathbf{X})}$ recovers from Eq.~\eqref{eq:gamma} the
fluctuation-dissipation relation $\gamma_{ij}(t)=\beta\langle f_i
(t)f_j (0)\rangle $. Out of equilibrium, the relation is generically
broken.

\subsubsection{Phenomenological description of symmetric tracers}
For isotropic tracers, it is common to use a heuristic suggestion by
Wu and Libchaber to describe their experiments on passive tracers
(see~\cite{Wu2000} and Fig.~\ref{fig:tracer}(a)). They postulated
$\gamma_{ij} = 2\Gamma_{\rm T} \delta(t-t') \delta_{ij}$ and $\langle
f_i(t)f_j(0)\rangle_c = D_{\rm T} \Gamma_{\rm T}^2
e^{-t/\tau_1}/\tau_1\delta_{ij}$, with $\tau_1$  a characteristic relaxation
time and $D_{\rm T}$ a diffusion
coefficient~\cite{Maggi2017a}. Namely, the motion of the tracer
behaves as an active particle with a mean-square displacement
$\langle \mathbf{X}^2(t)\rangle_c$ given by
\begin{align}
    \langle \mathbf{X}^2(t)\rangle_c&=2dD_{\rm T}[t-\tau_1(1-e^{-t/\tau_1})]\;.\label{eq:WL}
\end{align}
This describes a crossover between a short-time ballistic motion
($\sim t^2$) and a long-time diffusive motion ($\sim t$), that is
illustrated in Fig.~\ref{fig:tracer}(c). In the long-time limit, the
particle diffuses as $\langle \mathbf{X}^2(t)\rangle_c \sim 2dD_{\rm T}
t$.

\subsubsection{The Markovian approximation}

The long-time behavior can be obtained through more systematic
methods, starting from microscopic models and using Taylor's dispersion
theory~\cite{Burkholder2017,Burkholder2019a,Peng2022} or singular
perturbation methods~\cite{Solon2022,Jayaram2023}. Assuming that
$\gamma_{ij}$ and $\langle f_i(t)f_j(0)\rangle_c$ decay exponentially
in time, this leads to a Markovian approximation of Eq. \eqref{eq:GLE}
of the form
\begin{align}
     0&=-\Gamma_{ij}\dot{X}_j(t) + \xi_i(t)\;.\label{eq:Markovapp}
\end{align}
Here, ${\bm \xi} (t)$ is a Gaussian white noise that satisfies
    $\left\langle\xi_i (t) \xi_j(0)\right\rangle_c =2I_{ij}\delta(t)$
while the friction coefficient $\Gamma_{ij}$ and noise intensity $I_{ij}$ are given by the Green-Kubo relations
\begin{align}
    \Gamma_{ij} &= \int_0^\infty dt\, \gamma_{ij}(t)\;,\label{eq:GK1}\\
    I_{ij} &= \int_0^\infty dt\, \left\langle f_i (t)f_j (0) \right\rangle_c^{\rm s}\;.\label{eq:GK2}
\end{align}
This recovers the long-time diffusive behavior postulated by Wu and Libchaber.

Within the Markovian approximation and in the isotropic case where $I_{ij}=I\delta_{ij}$ and $\Gamma_{ij}=\Gamma_{\rm T}\delta_{ij}$, one can identify an effective tracer temperature by $T_{\rm T}=I/\Gamma_{\rm T}=D_{\rm T}\Gamma_{\rm T}$.
Interestingly, for hard-core tracers, the interpretation of
$T_{\rm T}$ extends to an effective fluctuation-dissipation relation
$\gamma_{ij}(t)\simeq T_{\rm T}^{-1}\langle f_i (t) f_j
(0)\rangle_c$ despite the system being out of equilibrium at the
microscopic scale~\cite{Solon2022}. We note, however, that in contrast
to equilibrium systems, the effective temperature depends on
the properties of the tracer. 

Note that, to model a tracer that is also in contact with a passive
(Markovian) bath, one adds a thermostat $-\Gamma_{\rm th}
\dot{X}_i(t)+\sqrt{2\Gamma_{\rm th} T_{\rm th}}\eta_i(t)$ to
Eqs.~\eqref{eq:GLE} and~\eqref{eq:Markovapp}. The effective
temperature then modifies straightforwardly
to~\cite{Granek2022,Solon2022}
\begin{align}
    T_{\rm T} &= \frac{I+\Gamma_{\rm th} T_{\rm th}}{\Gamma_{\rm T}+\Gamma_{\rm th}} = \frac{\Gamma_{\rm T}}{\Gamma_{\rm T}+\Gamma_{\rm th}} T_a + \frac{\Gamma_{\rm th}}{\Gamma_{\rm T}+\Gamma_{\rm th}} T_{\rm th}\;,\label{eq:Teff}
\end{align}
where $T_a \equiv I/\Gamma_{\rm T}$ is the active contribution.

The case of soft tracer brings in interesting differences, as was
recently shown in $d=1$~\cite{Granek2022}. First, the
fluctuation-dissipation relation only survives in the limit of a shallow wide
tracer, for which the system converges to an equilibrium distribution
$\rho(\mathbf{r})\propto e^{-V(\mathbf{r})/T_{\rm eff}}$ such that 
$T_a=T_{\rm eff}=D_{\rm eff}/\mu$. Then, for small tracers, the local
structure of the active bath around the tracer may enhance its motion
and lead to an effective \textit{negative friction coefficient}
$\Gamma_{\rm T}$~\cite{Granek2022}. Microscopic simulations and a
self-consistent calculation of the nonlinear tracer response reveal
that the negative friction induces a spontaneous symmetry breaking and a nonzero
average velocity in the limit
$\rho_b\rightarrow\infty$~\cite{Kim2023}. We note that, for hard
tracers in $d>1$, a related effect---termed ``swim
thinning''~\cite{Burkholder2019a}---was reported: the friction
experienced by the tracer is reduced by the active
bath~\cite{Burkholder2019a,Knezevic2021,Peng2022,Jayaram2023} but
negative frictions have not been reported so far in this case.

\subsubsection{Beyond the Markovian approximation}

It is well known that some care has to be taken when using a
Markovian description to describe the long-time behavior of passive
tracers. In particular, conservation laws lead to power-law decaying
time correlation functions, which might cause the integrals, Eqs. \eqref{eq:GK1}-\eqref{eq:GK2}, to diverge~\cite{VanBeijeren1982}. For dry scalar
active systems, the conserved density $\rho(\mathbf{r},t)$ is diffusive
and gives rise to slow hydrodynamic relaxation modes: for a system of
size $L \gg \ellp$, the relaxation time grows as $\sim L^2$. For an
infinite system and a fully symmetric tracer, the long-time tails of the force autocorrelation function and of $\gamma_{ij}$ have
been derived and shown to decay as $\sim
t^{-(d/2+1)}$~\cite{Granek2022,Solon2022,Feng2022}, identical to the
case of a tracer in passive diffusive
bath~\cite{Spohn1980,Hanna1981,VanBeijeren1982}. Then, the Green-Kubo
integrals in Eqs.~\eqref{eq:GK1}-\eqref{eq:GK2} converge in any
dimension, and the long-time diffusive behavior is unaffected.

Real tracers, however, always carry some degree of asymmetry, which
leads to very different behaviors. The underlying reason is that the
fluctuating force ${\bf f}(t)$ is non-zero on average, as discussed in
Sec.~\ref{sec:singleobj}. Indeed, in the adiabatic limit, $\langle {\bf
f}(t) \rangle = -{\bf p}$, which endows the tracer with an effective
self-propulsion.  Such dynamical rectification of random motion was
first observed numerically in~\cite{Angelani2010} (see
Fig.~\ref{fig:tracer}(b)) and later observed experimentally---in a wet
system---~\cite{Kaiser2014}. Recently, progress has been made in modeling the
motion of asymmetric tracers~\cite{Knezevic2020,Granek2022}.

The asymmetric tracer's dynamics has been derived from first
principles in $d=1$~\cite{Granek2022}, where Eq.~\eqref{eq:GLE}
continues to hold within an adiabatic expansion. Contrary to the case
of the symmetric tracer, where the long-time tails $\sim t^{-3/2}$
lead to finite Green-Kubo integrals and effective long-time diffusion,
the asymmetry shifts the decay to~\cite{Granek2022,Granek2023}
\begin{align}
    \left\langle f (t) f (0)\right\rangle_c = \frac{p^2}{\rho_b (4\pi D_{\rm eff} t)^{1/2}}+\mathcal{O}(t^{-3/2})\;,\label{eq:ff}\\
    \gamma(t) = \beta_{\rm eff} \left\langle f (t) f(0) \right\rangle_c + \mathcal{O}(t^{-3/2})\;,\label{eq:g}
\end{align}
rendering the Green-Kubo integrals infinite in $d=1$. As long as the
adiabatic approximation holds, one then predicts a superdiffusive behavior
with
\begin{align}
\langle X^2(t)\rangle_c &\sim \frac{4p^2}{3\rho_b\Gamma_{\rm th}^2\sqrt{\pi D_{\rm eff}}}t^{3/2}\;,\label{eq:sup}
\end{align}
which is illustrated in Fig.~\ref{fig:tracer}(c). Likewise, the
divergence in Eq.~\eqref{eq:GK1} is manifested in an apparent growth
of the friction coefficient, measured from the average force required
to tow the tracer at a constant velocity, as
\begin{align}
    \Gamma(t) &\sim \frac{\mu p^2}{\rho_b D_{\rm eff}}\left(\frac{t}{\pi D_{\rm eff}}\right)^{1/2}\;.
\end{align}
These anomalous properties demonstrate that ratchet effects may induce
a shift in the dynamical exponents caused by hydrodynamic modes.

Despite the theoretical advancement in the understanding of passive
tracer dynamics, various open questions remain unanswered. For
instance, the long-time tails in Eqs.~\eqref{eq:ff}-\eqref{eq:g} are
expected to yield logarithmic corrections to the diffusion of an
asymmetric tracer in $d=2$, a phenomenon yet to be confirmed. Most
importantly, exploring the dynamics in a controlled setting beyond the
adiabatic limit remains an outstanding technical challenge.

\subsection{Coupled dynamics of passive tracers: Non-reciprocal interactions and localization}\label{sec:ptbreak}

When several mobile tracers are embedded in an active bath, their
long-range interactions, described in Sec.~\ref{sec:nonrec}, lead to interesting
dynamical effects. Here we show how the non-reciprocal interactions
between the tracers lead to phase transitions and how the interaction
between a tracer and its boundary-induced image leads to a localization
transition.

\begin{figure}
\includegraphics{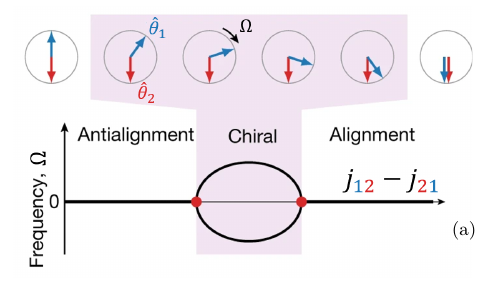}  
\if{\beginpgfgraphicnamed{Fig13a}
 \begin{tikzpicture}[scale=.5] \draw (0,0) node
  {\includegraphics[width=0.9\linewidth,trim=1 1 0 1,clip]{vitelli
  kuramoto.png}}; \draw (7.5,-3) node
  {(a)};
 \end{tikzpicture}
\endpgfgraphicnamed{Fig13a}}\fi
\includegraphics{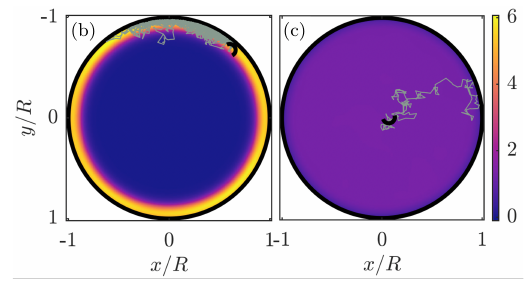}  
\if{\beginpgfgraphicnamed{Fig13b}
 \begin{tikzpicture}[scale=.5] \draw (0,0)
  node {\includegraphics[width=\linewidth,trim=1 1 1
  1,clip]{localization.png}}; \draw (-6.25,3.75) node {(b)}; \draw
  (0.9,3.75) node {(c)};
 \end{tikzpicture}
\endpgfgraphicnamed{Fig13b}}\fi
 \caption{\textbf{(a)}
  Non-reciprocal phase transitions. The non-reciprocal torque
  $\mathbf{n}_{12}=n_{12}\hat{\bf z}$ leads to an effective Kuramoto
  coupling $j_{12}\epsilon_{12}\propto n_{12}$, where $\epsilon_{ij}$ is the Levi-Civita symbol. The coupling asymmetry
  $j_{12}-j_{21}$ determines the steady-state frequency $\Omega$ and
  relative phase of $N\gg1$ identical copies of rotors $1$ and
  $2$. When $\Omega\neq0$, $PT$-symmetry is spontaneously
  broken. Reproduced with permission
  from~\cite{Fruchart2021}. \textbf{(b-c)} Localization transition of an asymmetric
  obstacle in a spherical cavity. Probability density and typical
  trajectories of a passive tracer in the active bath. Rotational
  friction is set to $\Gamma_r=10^{-2}$ (b) and $\Gamma_r=10^{-5}$
  (c). Particle speed: $v=10^{-2}$. All other parameters are set to
  unity. Reproduced with permission
  from~\protect\cite{dor2022passive}.  }\label{fig:vincenzo}
\end{figure}

Consider pinned asymmetric objects, each free to rotate along a fixed axis~(see
Fig.~\ref{fig:nonrec}). The density and current modulations induced by
the obstacles generate torques on each other, leading to a
Kuramoto-like dynamics of their
orientations~\cite{baek_generic_2018}. Analysis of this dynamics has
shown a transition between a phase where the obstacles' orientations
are locked and one in which they rotate in synchrony. While, for the
two-obstacle case studied in~\cite{baek_generic_2018}, one observes a
smooth crossover between these regimes, the many-body generalization
studied in~\cite{Fruchart2021} leads to \textit{bona fide} phase
transitions (see Fig.~\ref{fig:vincenzo}(a)). Unpinning the two rotors
was also shown to generate traveling bound
states~\cite{baek_generic_2018}. Similar bound states were
subsequently found for active particles in passive mass-conserving
baths~\cite{Dolai2022}. It remains an interesting open question
whether the many-body generalization of such states can lead to
flocking/anti-flocking, as was predicted by a non-reciprocal Vicsek
model~\cite{Fruchart2021} and a non-reciprocal active Ising model~\cite{Martin2023}.

Another interesting consequence of the non-reciprocal mediated
interaction can be observed for a \textit{single} asymmetric tracer in
a spherical cavity (see Fig.~\ref{fig:vincenzo}(b-c)). In this case,
the generalized method of images described in Section~\ref{sec:finite}
determines the density and current modulations as the sum of
contributions from the original tracer and its image upon a spherical
inversion~\cite{dor2022passive}. The non-reciprocal interactions
between the object and its image lead to a transition from a
steady-state distribution localized at the cavity wall to a
distribution localized at its center. We stress that such a phenomenon
is yet another signature of how different passive and active fluids
are. In a dilute passive diffusive fluid, an object cannot experience
any net force in a homogeneous system. To linear order, the sole force
${\bf f}$ it can experience is indeed proportional to $\nabla
\rho$. Consequently, in the steady state, $\nabla \cdot {\bf f}
\propto \nabla^2 \rho=0$ since the fluid is diffusive. In analogy to
Earnshaw’s theorem in electrostatics, this rules out the possibility
of a stable equilibrium for an object immersed in a passive diffusive
fluid~\cite{rohwer2020activated,dor2022passive}. Again, activity thus
leads to a completely different physics.

\begin{figure}[th!]
\includegraphics{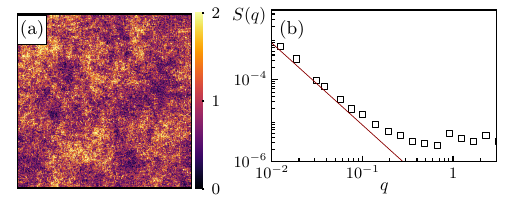}  
\if{\beginpgfgraphicnamed{Fig14}
 \begin{tikzpicture}
     \def \x{4.2}
     \draw (0,0) node {\includegraphics[totalheight=3.2cm]{Fig_disorder_config.pdf}};
     \draw[color=black, fill=white] (-1.7, 0.95) rectangle (-1.2, 1.45);
     \draw (-1.45, 0.95+0.25) node {(a)};
     \draw (\x,0) node {\includegraphics[totalheight=3.2cm]{Fig_disorder_Sq.pdf}};
     \draw (4.4-1.45, 0.95+0.25) node {(b)};
 \end{tikzpicture}
\endpgfgraphicnamed{Fig14}}\fi
\caption{Noninteracting
  active particles in the presence of a disordered
  potential. \textbf{(a)} Snapshot of the density field normalized by
  its average value $\rho({\bf r})/\rho_0$. \textbf{(b)} Structure
  factor (symbols) compared to $q^{-2}$ (red line), as predicted in
  Eq.~\eqref{eq:Sofqactivedisorder}. The results are obtained with RTPs on lattices with $v=4$, $\alpha = 1$, and the disorder potential is constructed by filling the space with square ramps with the size $\ell = 5$ and the height $\Delta V = 3.8$ in random locations and orientations. \label{fig:disorder}  }
\end{figure}

\section{Bulk disorder}
\label{sec:bulkdisorder}

As discussed in Section~\ref{sec:inclusion}, localized obstacles have
a far-reaching impact on the properties of active fluids. When
organized coherently, they can, for instance, act as pumps and
generate large-scale flows or density gradients (see
Fig.~\ref{fig:funnel}). Naturally, this raises the question as to
whether \textit{disordered} assemblies of obstacles can also impact
the properties of active fluids. In this colloquium, we discuss the
case of a quenched potential disorder, which corresponds to the
evolution of a system under a fixed realization of a random potential.

\begin{figure*}
\includegraphics{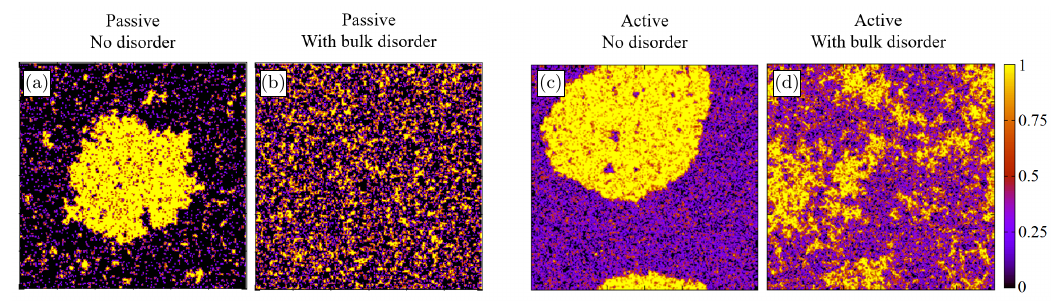}  
\if{\beginpgfgraphicnamed{Fig15}
  \begin{tikzpicture} \path (0,0) node
  {\includegraphics[width=500pt]{image001.png}}; \def\x{-8.6} \def\y{1} \def\w{8.7} \def\dy{.45} \def\dx{.45} \def\ddx{4} \def\xo{-6.75} \def\xgap{4} \def\yo{0.07} \def\wo{8.65} \path
  (\xo,\yo) node
  {\includegraphics[height=140pt]{img_passive_PS.png}}; \path
  (\xo+\xgap,\yo) node
  {\includegraphics[height=140pt]{img_passive_DIS.png}}; \path
  (\xo+\wo,\yo) node
  {\includegraphics[height=140pt]{img_active_PS.png}}; \path
  (\xo+\wo+\xgap+0.42,\yo) node
  {\includegraphics[height=140pt]{img_active_DIS.png}}; \draw[fill=white]
  (\x,\y) rectangle (\x+\dx,\y+\dy); \draw (\x+.5*\dx,\y+.5*\dy) node
  {(a)}; \draw[fill=white] (\x+\ddx,\y) rectangle
  (\x+\dx+\ddx,\y+\dy); \draw (\x+\ddx+.5*\dx,\y+.5*\dy) node
  {(b)}; \draw[fill=white] (\x+\w,\y) rectangle
  (\x+\w+\dx,\y+\dy); \draw (\x+\w+.5*\dx,\y+.5*\dy) node
  {(c)}; \draw[fill=white] (\x+\w+\ddx,\y) rectangle
  (\x+\w+\dx+\ddx,\y+\dy); \draw (\x+\w+\ddx+.5*\dx,\y+.5*\dy) node
  {(d)}; \end{tikzpicture}
\endpgfgraphicnamed{Fig15}}\fi

\caption{Comparison between equilibrium
    and active systems undergoing phase separation. A passive lattice
    gas with attractive interaction ({\bf a}), and an active lattice
    gas with repulsive interactions ({\bf c}), may both experience
    bulk phase separation. When a random potential is added to the
    passive system, the phase separation is suppressed, leading to a
    homogeneous phase with short-range correlations ({\bf b}). In the
    active case ({\bf d}), a scale-free distribution of particles
    replaces the bulk phase separation. Color encodes
    density. Reproduced with permission from~\cite{ro2021disorder}.
    }\label{fig:bulkdisorder}
\end{figure*}

For equilibrium systems, this question has attracted considerable
interest and many studies have been dedicated to the static and
dynamical properties of passive systems in the presence of random
potentials~\cite{imry_random-field_1975,aharony_lowering_1976,fisher_ising_1984,belanger_random-field_1983,imbrie_lower_1984,bricmont_lower_1987,glaus_correlations_1986}. Let
us first briefly review the results obtained in the passive case
before turning to the recent work on disordered scalar active matter. To
characterize the impact of disorder on a scalar system, it is natural
to consider density-density correlations, which are encoded in the
structure factor. In experiments, the structure factor can be measured
using scattering probes. For a system defined on a lattice of $N$
sites, it is computed as
\begin{equation}
    \overline{S({\bf q})}\equiv \frac{1}{N} \sum_{\br} \overline {\langle \phi(\br) \phi(0)\rangle} e^{i {\bf q}\cdot \br}\;,
\end{equation}
where $\phi(\br)\equiv \rho(\br)-\rho_0$ measures density
fluctuations, brackets stand for steady-state averages in a particular realization of disorder, and the
overline for an average over disorder realizations. In the
`high-temperature' homogeneous phase, the impact of disorder can be
computed using a Landau-Ginzburg field theory, which leads to the addition of a squared--Lorentzian form to the result in the absence of disorder~\cite{glaus_correlations_1986, kardar_2007}
\begin{equation}
\overline{S({\bf q})}\propto\frac{1}{({\bf q}^2+\ell_{\rm c}^{-2})^2}\;,
\end{equation}
where $\ell_{\rm c}$ is the correlation length of the density field. 
A natural question is then
whether impurities can impact the existence of the ordered state and the nature of the phase transition. The first question was addressed by
Imry and Ma some 50 years ago~\cite{imry_random-field_1975}, who showed that the
lower critical dimension below which the system does not admit an
ordered phase is $d_c=2$. Moreover, Aizenman and Wehr showed that the
system is also disordered at the marginal dimension of $d=2$~\cite{aizenman1989rounding}. This is illustrated in Fig.~\ref{fig:bulkdisorder}(a-b), where the ordered phase is destroyed by impurities and exhibits short-range correlations.

The active case is markedly distinct: In the homogeneous phase, the
disorder leads to a scale-free steady state with a structure factor
that diverges as~\cite{ro2021disorder}
\begin{equation}\label{eq:Sofqactivedisorder}
    \overline{S(\bq)}\underset{{q \to 0}}{\propto} \frac {1}{{\bf q}^2}\;.
\end{equation}
As shown in Fig.~\ref{fig:disorder}, fluctuations at small $q$ are
strongly enhanced compared to the passive case: The correlation length
is infinite and density-density correlations decay as
$\overline{\langle \phi (\br)\phi(0)\rangle} \propto r^{2-d}$ for
$d>2$, and logarithmically in $d=2$.

Beyond creating a long-range correlated fluid,  disorder also has a strong
impact on the existence of a long-range ordered phase. For a system undergoing
MIPS, disorder prevents phase separation in both 2 and 3
dimensions. Compared to the passive case, the lower critical
dimension is increased from $d_c=2$ to $d_c=4$. As shown in
  Fig.~\ref{fig:bulkdisorder}(c-d), the phase-separated state is
replaced by a frozen scale-free distribution with large-amplitude
density fluctuations.

\subsection{A simple physical picture}

\begin{figure}[b!]
 \includegraphics[width=.8\linewidth]{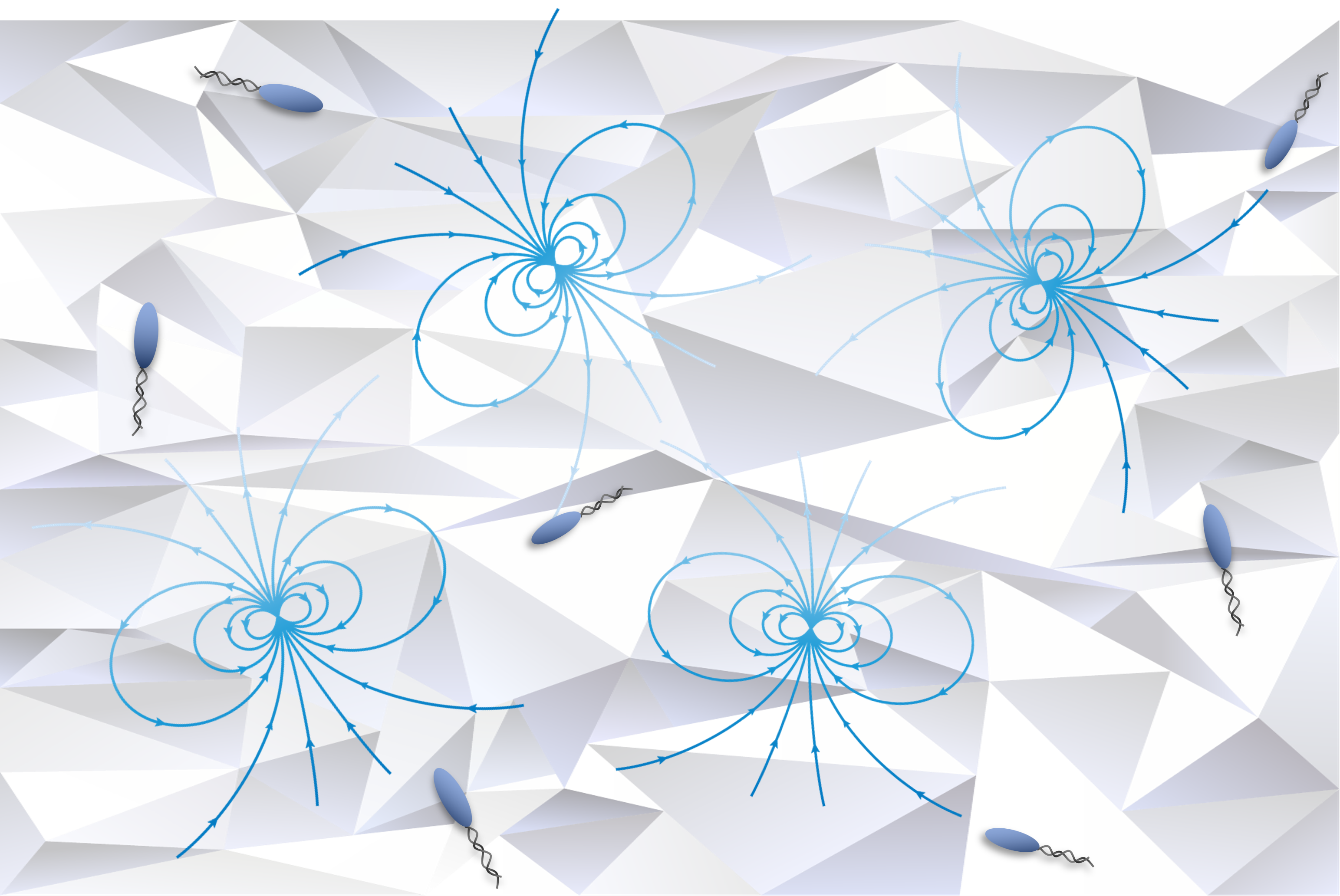} \caption{Schematic
  depiction of disorder acting as a random array of pumps for active
  particles, leading to a dipolar flow field.}\label{fig:dipole}
\end{figure}

To understand the difference between the active and passive cases, let
us first recall the results of Sec.~\ref{sec:inclusion}. At the
microscopic scale, a random potential is generically asymmetric and
thus, acting as a pump, it generates a ratchet current. As we now
argue, the net effect of a dilute collection of randomly placed pumps
is to generate density-density fluctuations consistent with
Eq.~\eqref{eq:Sofqactivedisorder} as depicted in
Fig.~\ref{fig:dipole}.

To see this explicitly, consider a dilute distribution of pumps whose
force density we denote by ${\bf f}({\bf r})$. Each pump acts as a current source that modifies the density according to Eq.~\eqref{eq:Poisson's dipole}. The contributions from the pumps add
up independently, resulting in the overall density modulation given by
\begin{equation} \label{Eq:phi}
  \langle \phi ({\bf r}) \rangle = \beta_\mathrm{eff} \int d^d{\bf r}'\, {\bf f}({\bf r}') \cdot \nabla_{\bf r} G({\bf r} - {\bf r}')~,
\end{equation}
where $G({\bf r} - {\bf r}')=G(\mathbf{r},\mathbf{r}')$ is the Green's function of the Laplacian. Note that for a single pump, ${\bf f}({\bf r})
= {\bf p} \delta^d({\bf r} - {\bf r}_0)$, Eq.~\eqref{Eq:phi}  gives back the solution Eq.~\eqref{eq:rho d dimensions} derived in Sec.~\ref{sec:singleobj} for an isolated asymmetric inclusion. In Fourier space, the
convolution in Eq.~\eqref{Eq:phi} takes the form\footnote{In
  what follows we use the Fourier convention: $f({\bf q}) =
  L^{-d/2} \int \mathrm{d}^d {\bf r}\, e^{-i {\bf q} \cdot
    {\bf r}} f({\bf r}) ~~~~ \mathrm{and} ~~~~ f({\bf r}) =
  L^{-d/2} \sum_{\bf q} e^{i {\bf q} \cdot {\bf r}} f({\bf
    q})$
    }
\begin{equation}
  \nonumber	\langle \phi ({\bf q}) \rangle =  \beta_{\rm eff}L^{d/2} i {\bf q} \cdot {\bf f}({\bf q}) G({\bf q} )~,
\end{equation}
where $G({\bf q}) = - L^{-d/2} q^{-2}$, $q = |{\bf q}|$ and $L$ is the system size. Assuming
Gaussian-distributed uncorrelated pumps with a typical strength
$\chi$, ${\bf f}({\bf q})$ is entirely characterized by
\begin{eqnarray}
  \nonumber	\overline{f_i ({\bf q})} &=& 0 ~,\\
  \nonumber	\overline{f_i ({\bf q}) f_j({\bf q}')} &=& \chi^2 \delta_{ij} \delta_{{\bf q}, -{\bf q}'}~\;.
\end{eqnarray}
For noninteracting particles, the structure factor satisfies
$\overline{ \langle \phi ({\bf q}) \phi (-{\bf q}) \rangle
}=\overline{ \langle \phi ({\bf q}) \rangle\langle \phi (-{\bf q})
  \rangle }$ so that
\begin{eqnarray}
  \nonumber	\overline{ \langle \phi ({\bf q})  \phi (-{\bf q}) \rangle } &=& \frac{\beta^2_\mathrm{eff}}{ q^4} \sum_{i,j=1}^d q_i q_j \overline{ f_i ({\bf q})  f_j (-{\bf q}) } \\
  \label{Eq:S}	&=& \frac{\beta^2_\mathrm{eff} \chi^2}{ q^2}~,  \label{Eq:chi}
\end{eqnarray}
which is indeed compatible with the 2d structure factor presented in
Fig.~\ref{fig:disorder}.

\subsection{Field-theoretical description}\label{sec:LFT1}
In order to account for the effects of interactions between the particles, it is useful to introduce a field-theoretical description by building up on the insights of the previous section. Since the quenched random potential translates, through the ratchet effect, into a quenched random forcing, one can start with a simple linear field theory of the form:
\begin{eqnarray}
\label{Eq:phi2} \frac{\partial}{\partial t} \phi({\bf r}, t) &=& - \nabla \cdot {\bf j}({\bf r}, t)~, \\
\label{Eq:j} {\bf j} ({\bf r}, t) &=& -  \nabla \mathfrak{u} [\phi] +  {\bf f}({\bf r}) + \sqrt{2D} \boldsymbol{\eta} ({\bf r}, t)~,
\end{eqnarray}
where $\phi({\bf r},t)$ describes the coarse-grained density
fluctuations, ${\bf j}({\bf r}, t)$ is the associated current, and
$\boldsymbol{\eta} ({\bf r}, t)$ is a centered Gaussian white noise
field such that $\langle \eta_i ({\bf r}, t) \eta_j ({\bf r'}, t') \rangle =\delta_{ij} \delta(t-t') \delta(\bfr-\bfr') $. The mobility is set to one and, similar to the previous
section, the force density satisfies $\overline{f_i ({\bf r})} = 0$
and $\overline{f_i ({\bf r}) f_j ({\bf r}')} = \sigma^2 \delta_{ij}
\delta^d({\bf r} - {\bf r}')$. At linear level in $\phi$ and to
leading order in a gradient expansion, the effective chemical
potential $\mathfrak{u}$ is of the form
\begin{equation} \label{Eq:g}
\mathfrak{u} [\phi({\bf r}, t)] =  u \phi({\bf r}, t) - K \nabla^2 \phi({\bf r}, t)~,
\end{equation} 
with $u, K >0$ to ensure stability.

\begin{figure}[b]
\includegraphics{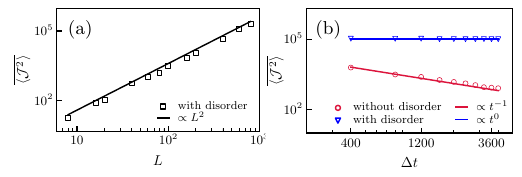}  
\if{\beginpgfgraphicnamed{Fig17}
  \begin{tikzpicture}
\def\x{-4.2}
    \draw (\x,0) node {\includegraphics[totalheight=2.78cm]{Fig_acc_current_scaling_J_L_v2.pdf}};
    \draw[color=white, fill=white] (\x-1.25,0.7) rectangle (\x-1.25+0.5,0.7+0.5);
    \draw(\x-1.25+0.2,0.7+0.3) node{\small (a)};
    \draw (0,0) node {\includegraphics[totalheight=2.78cm]{Fig_acc_current_scaling_J_t_v3.pdf}};
    \draw[color=white, fill=white] (-1.25,0.7) rectangle (-1.25+0.5,0.7+0.5);
    \draw(-1.25+0.2,0.7+0.3) node{\small (b)};
  \end{tikzpicture}
\endpgfgraphicnamed{Fig17}}\fi
    \caption{Scaling of the time-averaged total current with respect
      to, {\bf (a)} system size and, {\bf (b)} the measurement time
      interval. The second moment $\overline{ \langle {\cal J}^2
        \rangle }$ measured in simulation (symbols) is compared to the
      scaling trends predicted in Eq.~\eqref{eq:calJ_sq}
      (solid lines).  Parameters: $d = 2$, particle speed $v = 4$, the
      amplitude of the random potential $\Delta V = 3.8$, $\Delta t =
      4000$ in (a), $L = 800$ in (b), and averaged over 50
      realizations of disorder. }
    \label{fig:accJ}
\end{figure}

Before proceeding, let us highlight the difference between the current induced by the force density ${\bf f}({\bf r})$ and the Brownian noise $\boldsymbol{\eta} ({\bf r}, t)$. To do so, we consider the time-averaged total current, denoted as 
\begin{equation} \label{Eq:calJ}
    \bm{\mathcal{J}} = \frac{1}{\Delta t} \int_0^{\Delta t} dt \int d^d{\bf r} \, {\bf j}({\bf r}, t)\;,
\end{equation}
flowing through the whole system during the time interval $\Delta t$
in the presence of disorder and periodic boundary conditions. The
magnitude of $\bm{\mathcal{J}}$ can be characterized by averaging ${\cal J}^2$ over both noise and disorder. Direct algebra shows that this second moment is given by
\begin{equation} \label{eq:calJ_sq}
    \overline{ \langle {\cal J}^2 \rangle } = d \sigma^2 L^d + 2dD L^d \Delta t^{-1}\;.
\end{equation}
Equation~\eqref{eq:calJ_sq} demonstrates that although the scaling
with the system size is the same for both the random force and
noise-induced terms, the scaling with the time interval is drastically
different.  The quenched random force induces a net stationary
current, which remains finite as $\Delta t \to \infty$. By contrast,
the current fluctuations induced by the Brownian noise decay as
$\Delta t^{-1}$ once averaged over the time interval $\Delta t$.  To
verify this prediction, we measure $\bm{\mathcal{J}}$ by simulating
RTPs with and without disorder, subject to periodic boundary
conditions. The scaling of the second moment with the system size
shown in Fig.~\ref{fig:accJ}(a) indeed confirms
Eq.~\eqref{eq:calJ_sq}. In Fig.~\ref{fig:accJ}(b), we also show how
the second moment scales with the observation time interval, both with
and without disorder. In the absence of disorder, $\bm{\mathcal{J}}$
is dominated by the time-dependent noise, and its second moment indeed
shows a diffusive scaling $\overline{\langle {\cal J}^2 \rangle}
\propto \Delta t^{-1}$ (red symbols). In the presence of disorder, the
ballistic contribution of the random force dominates the long-time
scaling, as confirmed by our simulation (blue symbols). 

Let us now turn to the discussion of the density-density correlations.
A direct computation using Eqs.~\eqref{Eq:phi2}-\eqref{Eq:g} shows
that the structure factor of the (linearised) theory is given
by:
\begin{equation} \label{Eq:str_phi}
 S({\bf q}) = \frac{  \sigma^2}{  q^2 (u + Kq^2)^2} + \frac{D}{(u + K q^2)}~.
\end{equation}
We note that Eqs.~\eqref{Eq:str_phi} and~\eqref{Eq:chi} predict the
same small-$q$ behavior upon identifying $\sigma/u$ with $\chi
\beta_\mathrm{eff}$. To check the relevance of non-linearities, terms
of the form $s \phi^n({\bf r}, t)$ can be added to the effective
chemical potential $\mathfrak{u}$, and their relevance can be accessed
under diffusive scaling ${\bf r} \to b{\bf r}$ and $t \to b^2t$ in
Eqs.~\eqref{Eq:phi2}-\eqref{Eq:j}.  First, since the structure
factor satisfies $\overline{\phi ({\bf q}) \phi({\bf q}')} \propto
q^{-2} \delta^d({\bf q} + {\bf q}')$, the density modulations in real
space are rescaled as $\phi \to b^{1-d/2} \phi$. Accordingly, any
non-linear term transforms as $\phi^n \to b^{n(1-d/2)} \phi^n$, and
they are irrelevant for $d > 2$. The case $d=2$ is marginal, and it was
shown in~\cite{ro2021disorder} that the theory is self-consistent up
to length scales such that $\ell \ll \ell^\ast$ with $\ell^\ast \equiv
a \exp(\pi u^2 \rho_b^2 / \sigma^2 )$. Beyond such length scales, that
have never been explored numerically, a different behavior for $S({\bf
q})$ may emerge.

We now discuss how a generalization of the Imry-Ma argument predicts
the destruction of an ordered phase in dimensions below the lower
critical dimension of $d_c=4$. First, using a Helmholtz-Hodge
decomposition, we write the random force field as~\footnote{In the
  reported simulations, periodic boundary conditions were employed and
  the Helmholtz-Hodge decomposition also includes harmonic functions
  that are linear combinations of constant flows along the system
  axes. They do not impact the discussion of the generalized Imry Ma
  argument.}
\begin{equation}\label{eq:HH}
  {\bf f}({\bf r}) = -\nabla U({\bf r}) + \boldsymbol{\Xi} ({\bf
    r})\;,
\end{equation} where $U({\bf r})$
is an {\it effective potential} that differs from $V({\bf r})$ and the
vector field $\boldsymbol{\Xi} ({\bf r})$ satisfies $\nabla \cdot
\boldsymbol{\Xi}({\bf r}) = 0$. We note that $\boldsymbol{\Xi}$
impacts the current but not the dynamics of the density field or
its distribution. The statistics of ${\bf f}({\bf r})$ imply that
\begin{eqnarray}
  \overline{ U({\bf q}) U ({\bf q}') } &=& {\sigma^2}{q^{-2}}\label{eq:statU}
  \delta^d_{ {\bf q}, -{\bf q}' }\;,\\ \overline{ \Xi_i ({\bf q})
    \Xi_j ({\bf q}') } &=& \sigma^2 \left( \delta_{ij} - {q_i
    q_j'}/{q^{2}} \right) \delta^d_{ {\bf q}, -{\bf q}'
  }\;,\\ \overline{ U({\bf q}) \boldsymbol{\Xi} ({\bf q}') } &=& 0
  \;.
\end{eqnarray}
Because we are considering a linear theory and $\boldsymbol{\Xi}$ is
divergence-free, the dynamics of the density field is equivalent to an
equilibrium dynamics in a potential ${
  U(\bfr)}$~\cite{ao2004potential,kwon2005structure}. Inspection of
Eq.~\eqref{eq:statU} shows that $U({\bf r})$ behaves like a Gaussian
free field, equivalent to a random surface in $d=2$. Its deep and
scale-free minima account for the long-range correlations experienced
by the density field. Furthermore, we can now predict the lower
critical dimension by applying the standard Imry-Ma argument to an
equilibrium dynamics in the presence of the potential ${U}({\bf
  r})$~\cite{imry_random-field_1975,aharony_lowering_1976}. This
entails comparing the surface energy cost of overturning an ordered
domain of linear size $\ell$---given by $\gamma \ell^{d-1}$ with
$\gamma$ a surface-tension-like coefficient---to the bulk energy gain
of the domain due to a locally favorable disorder potential.  The
latter, given by $E(\ell) = \int_{\ell^d} \mathrm{d}^d {\bf r}'\,
\rho_0 U({\bf r}')$, has a typical value of $E(\ell)\propto
\sigma\rho_0\ell^{1+d/2}$, estimated by noting that the variance
satisfies $\overline{ E(\ell)^2 } = \sigma^2\rho_0^2 \ell^{d+2}$.  For
$d<4$, the surface energy cost is negligible on large enough length
scales, and the system cannot phase-separate into macroscopic domains.

We note that the field theory studied in
Eqs.~\eqref{Eq:phi}-\eqref{Eq:j} is equivalent to that describing
passive particles in a random force field. At the single-particle
level, this system has attracted some attention in the
past~\cite{sinai1983limiting,derrida1983velocity,derrida1982classical,fisher_random_1984,bouchaud1990classical}
and the analysis above extends these works to the many-body case. We
note that, for active particles in a random potential, the
one-dimensional dynamics can be shown to be equivalent to a Sinai
random walk~\cite{dor2019ramifications}. In higher dimensions, the
dynamics remains to be studied.

\section{Boundary disorder.}
\label{sec:boundarydisorder}

\begin{figure}
\includegraphics{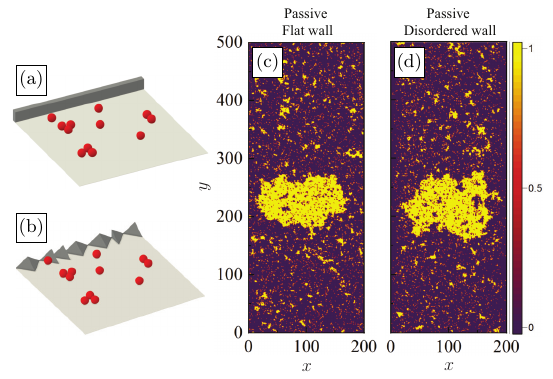}  
\if{\beginpgfgraphicnamed{Fig18}
\begin{tikzpicture}
    \draw (0,0) node {\includegraphics[totalheight=2.0cm]{Fig_schematic_passive_a.pdf}};
    \draw[color=black, fill=white] (-1.6,0.6) rectangle (-1.6+0.5,0.6+0.5);
    \draw(-1.6+0.25,0.6+0.25) node{(a)};
    \draw (0,-2.5) node {\includegraphics[totalheight=2.0cm]{Fig_schematic_passive_b.pdf}};
    \draw[color=black, fill=white] (-1.6,-2.5+0.6) rectangle (-1.6+0.5,-2.5+0.6+0.5);
    \draw(-1.6+0.25,-2.5+0.6+0.25) node{(b)};
    \draw (3,-1) node {\includegraphics[totalheight=6.2cm]{Fig_schematic_passive_c.pdf}};
    \draw[color=black, fill=white] (3.4-1,0.9) rectangle (3.4-1+0.5,0.9+0.5);
    \draw(3.4-1+0.25,0.9+0.25) node{(c)};
    \draw (6,-1) node {\includegraphics[totalheight=6.2cm]{Fig_schematic_passive_d.pdf}};
    \draw[color=black, fill=white] (5.8-1,0.9) rectangle (5.8-1+0.5,0.9+0.5);
    \draw(5.8-1+0.25,0.9+0.25) node{(d)};
\end{tikzpicture}
\endpgfgraphicnamed{Fig18}}\fi
\caption{Impact
  of flat {\bf (a)} and disordered {\bf (b)} walls on phase separation
  in passive systems.  In the presence of attractive interactions
  between the particles, simulations of a passive lattice gas at low
  temperature show phase separation in both settings (panels {\bf c}
  and {\bf d}). Color encodes density. Reproduced with permission
  from \cite{dor2022disordered}. } \label{fig:boundarydisorder}
\end{figure}

In equilibrium, boundaries and boundary conditions generically do not
alter bulk phase behaviors, due to the subextensive nature of their
contributions to the free energy.  This is illustrated in
Fig.~\ref{fig:boundarydisorder}, where replacing flat confining walls
with rough ones has no impact on bulk phase separation. An important
consequence is that, in simulations or theoretical analysis,
convenient boundary conditions (open or periodic) are often used to
study bulk properties in equilibrium statistical mechanics.

\begin{figure}
\includegraphics{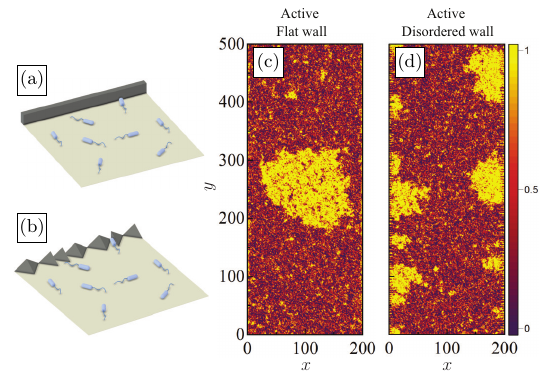}  
\if{\beginpgfgraphicnamed{Fig19}
\begin{tikzpicture}
    \draw (0,0) node {\includegraphics[totalheight=2.0cm]{Fig_schematic_active_a.pdf}};
    \draw[color=black, fill=white] (-1.6,0.6) rectangle (-1.6+0.5,0.6+0.5);
    \draw(-1.6+0.25,0.6+0.25) node{(a)};
    \draw (0,-2.5) node {\includegraphics[totalheight=2.0cm]{Fig_schematic_active_b.pdf}};
    \draw[color=black, fill=white] (-1.6,-2.5+0.6) rectangle (-1.6+0.5,-2.5+0.6+0.5);
    \draw(-1.6+0.25,-2.5+0.6+0.25) node{(b)};
    \draw (3,-1) node {\includegraphics[totalheight=6.2cm]{Fig_schematic_active_c.pdf}};
    \draw[color=black, fill=white] (3.4-1,0.9) rectangle (3.4-1+0.5,0.9+0.5);
    \draw(3.4-1+0.25,0.9+0.25) node{(c)};
    \draw (5.9,-1) node {\includegraphics[totalheight=6.2cm]{Fig_schematic_active_d.pdf}};
    \draw[color=black, fill=white] (5.8-1,0.9) rectangle (5.8-1+0.5,0.9+0.5);
    \draw(5.8-1+0.25,0.9+0.25) node{(d)};
\end{tikzpicture}
\endpgfgraphicnamed{Fig19}}\fi
\caption{Impact
  of flat {\bf (a)} and disordered {\bf (b)} walls on phase separation
  in active systems. In contrast to the passive case shown in
  Fig.~\ref{fig:boundarydisorder}, simulations of an active lattice
  gas {\bf (d)} show that the disordered boundary destroys the phase
  separation observed in the presence of a flat wall {\bf (c)},
  leading to a scale-free distribution of particles.  Color encodes
  density. Reproduced with permission
  from \cite{dor2022disordered}. } \label{fig:boundarydisorder_active}
\end{figure}

Rough boundary conditions can be implemented (as described earlier) through a wall potential 
$V(x,\bfr_\parallel)$, where $x$ is the coordinate normal to the wall
and $\bfr_\parallel$ is a ($d-1$)-dimensional vector parallel to the
wall. For example, $V(x,\bfr_\parallel)$ can be modeled by setting $V(x<0,\bfr_\parallel)=\infty$ and placing wedge-shaped asymmetric obstacles along the wall whose orientations are chosen randomly (see Figs.~\ref{fig:boundarydisorder}(b) and \ref{fig:boundarydisorder_active}(b) for a qualitative illustration). The obstacles then have a finite extent $x_{\rm w}$ in the $\hat{\bf x}$ direction.

Figure~\ref{fig:boundarydisorder_active} then shows a striking
difference with the passive case: the rough `disordered' walls destroy
bulk phase separation, leading to a nontrivial density
distribution. This was shown to hold even in the macroscopic limit
where the boundaries are sent to infinity. It shows that, even in the
thermodynamic limit, the bulk behaviors of active systems with closed
or periodic boundary conditions can be markedly
different~\cite{dor2022disordered}, in stark contrast with passive
systems (away from critical points). As we now discuss, the far-field
density modulations and currents induced by such a disordered wall and their effects on the bulk behavior can
be computed by modeling the disordered wall as a collection of force
monopoles randomly placed at $x=0$ and oriented along the
$\bfr_\parallel$ surface.

\subsection{A simple physical picture}
	
Again, we start by considering the dilute case and describe the force monopoles using a quenched Gaussian
random force density, ${\bf f}({\bf r}_\parallel, x)$, whose statistics
satisfy:
	\begin{align}\label{eq:stat}
		&\overline{f_i\left({\bf r}_\parallel,x\right)} = 0\ , \\
		&\overline{f_i\left({\bf r}_\parallel,x\right)f_j\left({\bf r}_\parallel\!',x'\right)} = 2 \chi^2\delta_{ij}^{\parallel}\delta(x)\delta(x') \delta^{(d-1)}\left({\bf r}_\parallel-{\bf r}_\parallel\!'\right)\,,\nonumber
	\end{align}
where $\chi$ sets the scale of the force density,
$\delta^\parallel_{ij}=1$ if $i=j \neq x$ and
$\delta^\parallel_{ij}=0$ otherwise.  Similar to the bulk disorder
case, the density modulations in the system satisfy
\begin{equation} \label{eq:density with dipole density}
  \langle \phi ({\bf r}) \rangle = \beta_\mathrm{eff} \int d^d {\bf r}'\, {\bf f}({\bf r}') \cdot \nabla_{\bf r} G_{\rm w}({\bf r} ,{\bf r}')~,
\end{equation}
but this time $G_{\rm w}({\bf r} , {\bf r}')$ is the Green's function
associated with a Poisson equation in a half-infinite system with
Neumann boundary conditions at $x=0$, see Sec. \ref{sec:finite}. Using this, one finds that the two-point correlation function
satisfies~\cite{dor2022disordered}:
	\begin{equation}\label{eq:rhorho}
		\overline{\langle \phi(x,{\bf r}_\parallel) \phi(x',{\bf r}_\parallel\!')}\rangle = \frac{2\beta_{\rm eff} \chi (x+x')}{S_d[(x+x')^2 + |\Delta{\bf r}_\parallel|^2]^\frac{d}{2}}\;,
	\end{equation}
with $S_d$ the $d$-dimensional solid angle and $\Delta{\bf
  r}_\parallel = {\bf r}_\parallel - {\bf r}_\parallel\!'$. This
equation shows that there are large-scale density modulations that
decay in amplitude but increase in range as one looks further from the
wall.

\begin{figure}
\includegraphics{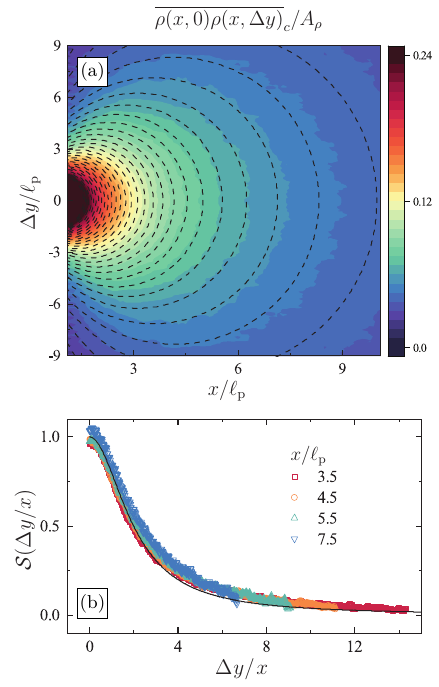}  
\if{\beginpgfgraphicnamed{Fig20}
    \begin{tikzpicture}
    \def\xl{-2.5}
        \draw (0,0) node {\includegraphics[width=7.0cm]{Fig_noint_cor_cm.pdf}};
        \draw[color=black, fill=white] (\xl,2) rectangle (\xl+0.5,2+0.5);
        \draw(\xl+0.25,2+0.25) node{(a)};
        \draw (-0.1,-5.8) node {\includegraphics[width=7.0cm]{Fig_noint_cor_collapse.pdf}};
        \draw[color=black, fill=white] (\xl,-5.5-1.5) rectangle (\xl+0.5,-5.5-1.5+0.5);
        \draw(\xl+0.25,-5.5-1.5+0.25) node{(b)};
    \end{tikzpicture}
\endpgfgraphicnamed{Fig20}}\fi
  \caption{Disorder-averaged two-point density correlation function
    of non-interacting RTPs in two dimensions in the presence of a
    disordered wall at $x=0$. {\bf (a)} The two-point correlation
    function as $x$ and $\Delta y$ are varied, calculated from
    simulations, is shown by the color map.  The value of $A_\rho
    \equiv \ellp S_d^{-1} (2 \beta_{
    \rm eff} \chi)^2$
    is obtained from a fit of the data to Eq.~\eqref{eq:rhorho}. 
    The theoretical prediction of Eq.~\eqref{eq:rhorho} is then used
    to produce dashed contour lines that match the levels of the color
    bar.  Both theory and simulations are normalized by $A_\rho$.
    {\bf (b)} A verification of the scaling form Eq.~\eqref{eq:C} for the
    density-density correlation function. The data shown in panel (a)
    for four different distances $x$ from the wall are collapsed onto
    a single curve, as predicted. Reproduced with permission from~\cite{dor2022disordered}.  }
  \label{fig:nonint_scaling}
\end{figure}

The heuristic prediction Eq.~\eqref{eq:rhorho} is verified numerically in
Fig.~\ref{fig:nonint_scaling} using microscopic simulations of
non-interacting particles in two space dimensions. To do so, the
prediction Eq.~\eqref{eq:rhorho} for $x=x'$ can be rewritten as:
\begin{align}
  \frac{\overline{\langle \phi(x, y) \phi(x,y+\Delta
      y)}\rangle}{\overline{\langle \phi(x, y)
      \phi(x,y)}\rangle} &= \frac{1}{1 + \left(
    \frac{\Delta y}{2x} \right)^2 }\equiv \mathcal{S}
  \left( \frac{\Delta y}{x} \right)\;, \label{eq:C}
\end{align}
which highlights that the correlations along $y$ increase
linearly with the distance $x$ to the wall.
	
\begin{figure}
  \centering
  \includegraphics[width=0.9\linewidth]{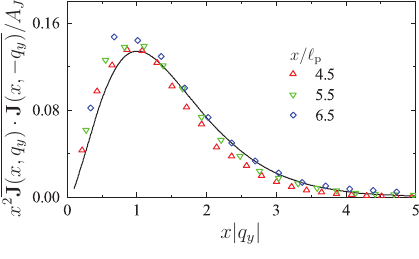}
  \caption{Fourier transform along the $\hat {\bf y}$ direction of the current-current correlation function measured at a distance $x$ from the wall and averaged over disorder.  The data are obtained for three values of $x$ and normalized by a factor $A_J \equiv 2 d (2 \pi)^{d-1} (\mu \chi)^2$. As predicted by the theory, the data can be collapsed onto a single curve, corresponding to Eq.~\eqref{eq:current correlations}, by properly scaling the abscissa and the ordinates. Reproduced with permission from~\cite{dor2022disordered}.
  }
  \label{fig:current_scaling}
\end{figure}

An illuminating way to understand these results is to consider the
currents generated in the system: the ratchet mechanism generates
local currents in the active fluid next to the wall. Since the number
of particles is conserved, this microscopic stirring develops into
large-scale eddies in the bulk. In fact, on large scales, the current
can be estimated as $ {\bf J}\left(x,{\bf r}_\parallel\right) \approx
-D_{\rm eff} \nabla \phi\left(x,{\bf r}_\parallel\right)$,
leading to:
\begin{align}\label{eq:current correlations}
  \overline{\langle {\bf J}\left(x,{\bf q_\parallel}\right) \cdot {\bf
      J}^*\left(x,{\bf k_\parallel}\right)\rangle} =& 
      \,2^d d \left(\mu
  \chi \right)^2 \left|{\bf q_\parallel}\right|^2 e^{-2\left|{\bf
      q_\parallel}\right| x}  \nonumber \\ & \times
  \pi^{d-1}\delta^{(d-1)}\!\left({\bf q_\parallel}+{\bf
    k_\parallel}\right)\;.
\end{align}
The current-current correlations first increase for small $|{\bf
  q_\parallel}|$ and are then exponentially suppressed for $|{\bf
  q_\parallel}|>x^{-1}$: like the density-density correlations, the
eddies have a transverse extent that grows linearly with the distance
$x$ to the wall, as verified in Fig.~\ref{fig:current_scaling} using a
scaling form similar to that of Eq.~\eqref{eq:C}.

\subsection{Linear field theory}\label{sec:LFT2}
To account for interactions, it is useful to follow
Sec.~\ref{sec:LFT1} and use the linear field
theory Eqs.~\eqref{Eq:phi2}-\eqref{Eq:j} to describe the evolution of the
density fluctuations $\phi(\bfr,t)$.  This time, however, the statistics
of the random force field ${\bf f}({\bf r})$ satisfy
\begin{align}
  {f}_x (x,{\bf r}_\parallel) =&\, 0\;, \label{eq:fstat} \\
  { \overline{{f}_i (x, {\bf r}_\parallel)} =}&{ \, 0\;,} \label{eq:fistat} \\
  \nonumber \overline{ {f}_i (x, {\bf r}_\parallel) {f}_j (x', {\bf r'}_\parallel ) } =&\, 2 \sigma^2 \delta_{ij} \delta(x) \delta(x') \delta^{(d-1)} ({\bf r}_\parallel - {\bf r'}_\parallel)\;,
\end{align}
where $i$ and $j$ label directions parallel to the wall. The amplitude
$\sigma$ of the random force density depends on the
average bulk density $\rho_b$ but is, to leading order, independent of
$\phi$. Direct algebra shows that this linear field theory predicts
density modulations and currents compatible with the
prediction Eqs.~\eqref{eq:C} and~\eqref{eq:current correlations} on large
length scales upon identifying $2 \chi \beta_{\rm eff}=\sigma/u$.

The self-consistency of the linear theory can be checked using a
scaling argument similar to what was done for the bulk disorder in
Sec.~\ref{sec:LFT1}. This time, non-linearities are found to be
irrelevant for $d>1$. We now discuss how the field theory allows
estimating the impact of boundary disorder on bulk phase separation in
active fluids.
	
\subsection{The effect of disordered boundaries on MIPS}\label{sec:Imry-Ma}
Again, one uses the Helmholtz-Hodge decomposition Eq.~\eqref{eq:HH} of the
random forcing to identify an effective potential $U(\bfr)$. Since,
the effective potential satisfies $\nabla^2 U({\bf r}) = - \nabla
\cdot {\bf f} ({\bf r})$, Eqs.~\eqref{eq:fstat}-\eqref{eq:fistat} imply
that the statistics of $U$ obey
\begin{align} 
  \overline{U({\bf r})} &= 0\;, \\
  \overline{U({\bf r}) U({\bf r}') } &= \frac{\sigma^2}{S_d} \frac{(x + x')}{\left[ (x + x')^2 + | \Delta {\bf r}_\parallel|^2 \right]^{d/2}}\;.\label{eq:U2}
\end{align}
This can then be used to compare the surface energy $\gamma
\ell^{d-1}$ of an ordered droplet of linear size $\ell$ to its bulk
energy $\int_{\ell^d} d^2 {\bfr}\, U(\bfr)$. Using Eq.~\eqref{eq:U2},
the latter is estimated to scale as $\ell^{(d+1)/2}$ so that the
surface energy is unable to stabilize a macroscopic droplet below a
lower critical dimension $d_c=3$. As shown in
Fig.~\ref{fig:boundarydisorder_active}, the phase-separated state is
indeed replaced by scale-free density modulations in $d=2$. This
result highlights how boundaries can play a much more important role
in active systems than in their passive counterparts.
	
\section{Conclusion, perspectives.}

In this colloquium, we have reviewed the anomalous mechanical
properties of dry scalar active systems that have attracted a lot of
attention recently. We have shown how the emergence of ratchet
currents and the lack of conservation of momentum lead to a wide
variety of phenomena, from the lack of an equation of state for pressure
to the destruction of bulk phase separation by boundary disorder, all of
which can be captured within a unifying perspective. These phenomena
endow active systems with properties that are strikingly different
from that of passive matter, whose consequences are only starting to
be explored.

Indeed, even basic concepts like surface tension are not
well understood and are intensely
debated~\cite{bialke2015negative,marconi2016pressure,paliwal2017non,hermann2019non,zakine2020surface,omar2020microscopic,lauersdorf2021phase,fausti2021capillary}. Furthermore,
much of the research so far has been based on theoretical studies of
minimal models. How the anomalous mechanical properties can be
measured or harnessed experimentally is an exciting research
direction~\cite{junot2017active}. Similarly, the far-field density and
current modulations induced by localized obstacles, as well as the
scale-free state induced by bulk and boundary disorder, are within
reach of modern experimental active-matter
systems~\cite{chardac2021emergence,bhattacharjee2019bacterial,takaha2023quasi}. Their
measurement is an ongoing challenge.

Finally, in this colloquium, we have focused on dry scalar active
matter to single out the interesting phenomena that are solely due to
the interplay between activity and mechanical forces. Other
situations, where more hydrodynamic fields are relevant, are bound to
lead to even richer physics. In particular, the role of obstacles,
boundaries and disorder in wet active matter, or in polar and nematic
active fluids, is a current frontier in the field.

\begin{acknowledgements}
  This colloquium is born out of many discussions and collaborations
  on the mechanical properties of scalar active systems that have
  involved many colleagues. In particular, we are grateful to Yongjoo
  Baek, Aparna Baskaran, Mike Cates, Adrian Daerr, Ydan Ben Dor,
  Yaouen Fily, Jordan Horowitz, Nikolai Nikola, Gianmarco Spera, Joakim Stenhammar, Ari
  Turner, Frédéric van Wijland, Raphaël Voituriez, Ruben Zakine, and
  Yongfeng Zhao. JT thanks ANR THEMA for financial support and MSC
  laboratory for hospitality. MK is supported by NSF grant DMR-2218849. YK, OG and SR acknowledge financial
  support from ISF (2038/21) and NSF/BSF (2022605).  OG also
  acknowledges support from the Adams Fellowship Program of the
  Israeli Academy of Science and Humanities.
\end{acknowledgements}

\bibliography{RMP_final}

\begin{thebibliography}{215}%
\makeatletter
\providecommand \@ifxundefined [1]{%
 \@ifx{#1\undefined}
}%
\providecommand \@ifnum [1]{%
 \ifnum #1\expandafter \@firstoftwo
 \else \expandafter \@secondoftwo
 \fi
}%
\providecommand \@ifx [1]{%
 \ifx #1\expandafter \@firstoftwo
 \else \expandafter \@secondoftwo
 \fi
}%
\providecommand \natexlab [1]{#1}%
\providecommand \enquote  [1]{``#1''}%
\providecommand \bibnamefont  [1]{#1}%
\providecommand \bibfnamefont [1]{#1}%
\providecommand \citenamefont [1]{#1}%
\providecommand \href@noop [0]{\@secondoftwo}%
\providecommand \href [0]{\begingroup \@sanitize@url \@href}%
\providecommand \@href[1]{\@@startlink{#1}\@@href}%
\providecommand \@@href[1]{\endgroup#1\@@endlink}%
\providecommand \@sanitize@url [0]{\catcode `\\12\catcode `\$12\catcode
  `\&12\catcode `\#12\catcode `\^12\catcode `\_12\catcode `\%12\relax}%
\providecommand \@@startlink[1]{}%
\providecommand \@@endlink[0]{}%
\providecommand \url  [0]{\begingroup\@sanitize@url \@url }%
\providecommand \@url [1]{\endgroup\@href {#1}{\urlprefix }}%
\providecommand \urlprefix  [0]{URL }%
\providecommand \Eprint [0]{\href }%
\providecommand \doibase [0]{https://doi.org/}%
\providecommand \selectlanguage [0]{\@gobble}%
\providecommand \bibinfo  [0]{\@secondoftwo}%
\providecommand \bibfield  [0]{\@secondoftwo}%
\providecommand \translation [1]{[#1]}%
\providecommand \BibitemOpen [0]{}%
\providecommand \bibitemStop [0]{}%
\providecommand \bibitemNoStop [0]{.\EOS\space}%
\providecommand \EOS [0]{\spacefactor3000\relax}%
\providecommand \BibitemShut  [1]{\csname bibitem#1\endcsname}%
\let\auto@bib@innerbib\@empty
\bibitem [{\citenamefont {Adachi}\ and\ \citenamefont
  {Kawaguchi}(2020)}]{adachi2020universality}%
  \BibitemOpen
  \bibfield  {author} {\bibinfo {author} {\bibnamefont {Adachi}, \bibfnamefont
  {Kyosuke}}, and\ \bibinfo {author} {\bibfnamefont {Kyogo}\ \bibnamefont
  {Kawaguchi}}} (\bibinfo {year} {2020}),\ \href@noop {} {\enquote {\bibinfo
  {title} {Universality of active and passive phase separation in a lattice
  model},}\ }\Eprint {https://arxiv.org/abs/2012.02517} {arXiv:2012.02517
  [cond-mat.stat-mech]} \BibitemShut {NoStop}%
\bibitem [{\citenamefont {Agranov}\ \emph {et~al.}(2021)\citenamefont
  {Agranov}, \citenamefont {Ro}, \citenamefont {Kafri},\ and\ \citenamefont
  {Lecomte}}]{agranov2021exact}%
  \BibitemOpen
  \bibfield  {author} {\bibinfo {author} {\bibnamefont {Agranov}, \bibfnamefont
  {Tal}}, \bibinfo {author} {\bibfnamefont {Sunghan}\ \bibnamefont {Ro}},
  \bibinfo {author} {\bibfnamefont {Yariv}\ \bibnamefont {Kafri}}, and\
  \bibinfo {author} {\bibfnamefont {Vivien}\ \bibnamefont {Lecomte}}} (\bibinfo
  {year} {2021}),\ \bibfield  {title} {\enquote {\bibinfo {title} {{Exact
  fluctuating hydrodynamics of active lattice gases—typical fluctuations}},}\
  }\href {https://doi.org/10.1088/1742-5468/ac1406} {\bibfield  {journal}
  {\bibinfo  {journal} {J. Stat. Mech. Theory Exp.}\ }\textbf {\bibinfo
  {volume} {2021}}~(\bibinfo {number} {8}),\ \bibinfo {pages}
  {083208}}\BibitemShut {NoStop}%
\bibitem [{\citenamefont {Aharony}\ \emph {et~al.}(1976)\citenamefont
  {Aharony}, \citenamefont {Imry},\ and\ \citenamefont
  {Ma}}]{aharony_lowering_1976}%
  \BibitemOpen
  \bibfield  {author} {\bibinfo {author} {\bibnamefont {Aharony}, \bibfnamefont
  {Amnon}}, \bibinfo {author} {\bibfnamefont {Yoseph}\ \bibnamefont {Imry}},
  and\ \bibinfo {author} {\bibfnamefont {Shang-keng}\ \bibnamefont {Ma}}}
  (\bibinfo {year} {1976}),\ \bibfield  {title} {\enquote {\bibinfo {title}
  {{Lowering of Dimensionality in Phase Transitions with Random Fields}},}\
  }\href {https://doi.org/10.1103/PhysRevLett.37.1364} {\bibfield  {journal}
  {\bibinfo  {journal} {Phys. Rev. Lett.}\ }\textbf {\bibinfo {volume}
  {37}}~(\bibinfo {number} {20}),\ \bibinfo {pages} {1364--1367}}\BibitemShut
  {NoStop}%
\bibitem [{\citenamefont {Ai}\ and\ \citenamefont
  {Wu}(2014)}]{ai2014transport}%
  \BibitemOpen
  \bibfield  {author} {\bibinfo {author} {\bibnamefont {Ai}, \bibfnamefont
  {Bao-Quan}}, and\ \bibinfo {author} {\bibfnamefont {Jian-Chun}\ \bibnamefont
  {Wu}}} (\bibinfo {year} {2014}),\ \bibfield  {title} {\enquote {\bibinfo
  {title} {{Transport of active ellipsoidal particles in ratchet
  potentials}},}\ }\href {https://doi.org/10.1063/1.4867283} {\bibfield
  {journal} {\bibinfo  {journal} {J. Chem. Phys.}\ }\textbf {\bibinfo {volume}
  {140}}~(\bibinfo {number} {9}),\ \bibinfo {pages} {094103}}\BibitemShut
  {NoStop}%
\bibitem [{\citenamefont {Aizenman}\ and\ \citenamefont
  {Wehr}(1989)}]{aizenman1989rounding}%
  \BibitemOpen
  \bibfield  {author} {\bibinfo {author} {\bibnamefont {Aizenman},
  \bibfnamefont {Michael}}, and\ \bibinfo {author} {\bibfnamefont {Jan}\
  \bibnamefont {Wehr}}} (\bibinfo {year} {1989}),\ \bibfield  {title} {\enquote
  {\bibinfo {title} {{Rounding of first-order phase transitions in systems with
  quenched disorder}},}\ }\href {https://doi.org/10.1103/PhysRevLett.62.2503}
  {\bibfield  {journal} {\bibinfo  {journal} {Phys. Rev. Lett.}\ }\textbf
  {\bibinfo {volume} {62}}~(\bibinfo {number} {21}),\ \bibinfo {pages}
  {2503--2506}}\BibitemShut {NoStop}%
\bibitem [{\citenamefont {Anderson}\ \emph {et~al.}(2022)\citenamefont
  {Anderson}, \citenamefont {Briand}, \citenamefont {Dauchot},\ and\
  \citenamefont {Fern{\'{a}}ndez-Nieves}}]{anderson2022polymer}%
  \BibitemOpen
  \bibfield  {author} {\bibinfo {author} {\bibnamefont {Anderson},
  \bibfnamefont {Caleb~J}}, \bibinfo {author} {\bibfnamefont {Guillaume}\
  \bibnamefont {Briand}}, \bibinfo {author} {\bibfnamefont {Olivier}\
  \bibnamefont {Dauchot}}, and\ \bibinfo {author} {\bibfnamefont {Alberto}\
  \bibnamefont {Fern{\'{a}}ndez-Nieves}}} (\bibinfo {year} {2022}),\ \bibfield
  {title} {\enquote {\bibinfo {title} {{Polymer-chain configurations in active
  and passive baths}},}\ }\href {https://doi.org/10.1103/PhysRevE.106.064606}
  {\bibfield  {journal} {\bibinfo  {journal} {Phys. Rev. E}\ }\textbf {\bibinfo
  {volume} {106}}~(\bibinfo {number} {6}),\ \bibinfo {pages}
  {064606}}\BibitemShut {NoStop}%
\bibitem [{\citenamefont {Angelani}\ \emph {et~al.}(2011)\citenamefont
  {Angelani}, \citenamefont {Costanzo},\ and\ \citenamefont {{Di
  Leonardo}}}]{angelani2011active}%
  \BibitemOpen
  \bibfield  {author} {\bibinfo {author} {\bibnamefont {Angelani},
  \bibfnamefont {L}}, \bibinfo {author} {\bibfnamefont {A.}~\bibnamefont
  {Costanzo}}, and\ \bibinfo {author} {\bibfnamefont {R.}~\bibnamefont {{Di
  Leonardo}}}} (\bibinfo {year} {2011}),\ \bibfield  {title} {\enquote
  {\bibinfo {title} {{Active ratchets}},}\ }\href
  {https://doi.org/10.1209/0295-5075/96/68002} {\bibfield  {journal} {\bibinfo
  {journal} {EPL}\ }\textbf {\bibinfo {volume} {96}}~(\bibinfo {number} {6}),\
  \bibinfo {pages} {68002}}\BibitemShut {NoStop}%
\bibitem [{\citenamefont {Angelani}\ and\ \citenamefont
  {Di~Leonardo}(2010)}]{Angelani2010}%
  \BibitemOpen
  \bibfield  {author} {\bibinfo {author} {\bibnamefont {Angelani},
  \bibfnamefont {Luca}}, and\ \bibinfo {author} {\bibfnamefont {Roberto}\
  \bibnamefont {Di~Leonardo}}} (\bibinfo {year} {2010}),\ \bibfield  {title}
  {\enquote {\bibinfo {title} {{Geometrically biased random walks in
  bacteria-driven micro-shuttles}},}\ }\href
  {https://doi.org/10.1088/1367-2630/12/11/113017} {\bibfield  {journal}
  {\bibinfo  {journal} {New J. Phys.}\ }\textbf {\bibinfo {volume}
  {12}}~(\bibinfo {number} {11}),\ \bibinfo {pages} {113017}}\BibitemShut
  {NoStop}%
\bibitem [{\citenamefont {Angelani}\ \emph {et~al.}(2009)\citenamefont
  {Angelani}, \citenamefont {Di~Leonardo},\ and\ \citenamefont
  {Ruocco}}]{angelani2009self}%
  \BibitemOpen
  \bibfield  {author} {\bibinfo {author} {\bibnamefont {Angelani},
  \bibfnamefont {Luca}}, \bibinfo {author} {\bibfnamefont {Roberto}\
  \bibnamefont {Di~Leonardo}}, and\ \bibinfo {author} {\bibfnamefont
  {Giancarlo}\ \bibnamefont {Ruocco}}} (\bibinfo {year} {2009}),\ \bibfield
  {title} {\enquote {\bibinfo {title} {Self-starting micromotors in a bacterial
  bath},}\ }\href@noop {} {\bibfield  {journal} {\bibinfo  {journal} {Physical
  review letters}\ }\textbf {\bibinfo {volume} {102}}~(\bibinfo {number} {4}),\
  \bibinfo {pages} {048104}}\BibitemShut {NoStop}%
\bibitem [{\citenamefont {Angelini}\ \emph {et~al.}(2011)\citenamefont
  {Angelini}, \citenamefont {Hannezo}, \citenamefont {Trepat}, \citenamefont
  {Marquez}, \citenamefont {Fredberg},\ and\ \citenamefont
  {Weitz}}]{angelini_glass-like_2011}%
  \BibitemOpen
  \bibfield  {author} {\bibinfo {author} {\bibnamefont {Angelini},
  \bibfnamefont {Thomas~E}}, \bibinfo {author} {\bibfnamefont {Edouard}\
  \bibnamefont {Hannezo}}, \bibinfo {author} {\bibfnamefont {Xavier}\
  \bibnamefont {Trepat}}, \bibinfo {author} {\bibfnamefont {Manuel}\
  \bibnamefont {Marquez}}, \bibinfo {author} {\bibfnamefont {Jeffrey~J.}\
  \bibnamefont {Fredberg}}, and\ \bibinfo {author} {\bibfnamefont {David~A.}\
  \bibnamefont {Weitz}}} (\bibinfo {year} {2011}),\ \bibfield  {title}
  {\enquote {\bibinfo {title} {{Glass-like dynamics of collective cell
  migration}},}\ }\href {https://doi.org/10.1073/pnas.1010059108} {\bibfield
  {journal} {\bibinfo  {journal} {Proc. Natl. Acad. Sci.}\ }\textbf {\bibinfo
  {volume} {108}}~(\bibinfo {number} {12}),\ \bibinfo {pages}
  {4714--4719}}\BibitemShut {NoStop}%
\bibitem [{\citenamefont {Ao}(2004)}]{ao2004potential}%
  \BibitemOpen
  \bibfield  {author} {\bibinfo {author} {\bibnamefont {Ao}, \bibfnamefont
  {Ping}}} (\bibinfo {year} {2004}),\ \bibfield  {title} {\enquote {\bibinfo
  {title} {{Potential in stochastic differential equations: novel
  construction}},}\ }\href {https://doi.org/10.1088/0305-4470/37/3/L01}
  {\bibfield  {journal} {\bibinfo  {journal} {J. Phys. A. Math. Gen.}\ }\textbf
  {\bibinfo {volume} {37}}~(\bibinfo {number} {3}),\ \bibinfo {pages}
  {L25--L30}}\BibitemShut {NoStop}%
\bibitem [{\citenamefont {Baek}\ \emph {et~al.}(2018)\citenamefont {Baek},
  \citenamefont {Solon}, \citenamefont {Xu}, \citenamefont {Nikola},\ and\
  \citenamefont {Kafri}}]{baek_generic_2018}%
  \BibitemOpen
  \bibfield  {author} {\bibinfo {author} {\bibnamefont {Baek}, \bibfnamefont
  {Yongjoo}}, \bibinfo {author} {\bibfnamefont {Alexandre~P.}\ \bibnamefont
  {Solon}}, \bibinfo {author} {\bibfnamefont {Xinpeng}\ \bibnamefont {Xu}},
  \bibinfo {author} {\bibfnamefont {Nikolai}\ \bibnamefont {Nikola}}, and\
  \bibinfo {author} {\bibfnamefont {Yariv}\ \bibnamefont {Kafri}}} (\bibinfo
  {year} {2018}),\ \bibfield  {title} {\enquote {\bibinfo {title} {{Generic
  Long-Range Interactions Between Passive Bodies in an Active Fluid}},}\ }\href
  {https://doi.org/10.1103/PhysRevLett.120.058002} {\bibfield  {journal}
  {\bibinfo  {journal} {Phys. Rev. Lett.}\ }\textbf {\bibinfo {volume}
  {120}}~(\bibinfo {number} {5}),\ \bibinfo {pages} {058002}}\BibitemShut
  {NoStop}%
\bibitem [{\citenamefont {Basu}\ \emph {et~al.}(2020)\citenamefont {Basu},
  \citenamefont {Majumdar}, \citenamefont {Rosso}, \citenamefont
  {Sabhapandit},\ and\ \citenamefont {Schehr}}]{basu2020exact}%
  \BibitemOpen
  \bibfield  {author} {\bibinfo {author} {\bibnamefont {Basu}, \bibfnamefont
  {Urna}}, \bibinfo {author} {\bibfnamefont {Satya~N.}\ \bibnamefont
  {Majumdar}}, \bibinfo {author} {\bibfnamefont {Alberto}\ \bibnamefont
  {Rosso}}, \bibinfo {author} {\bibfnamefont {Sanjib}\ \bibnamefont
  {Sabhapandit}}, and\ \bibinfo {author} {\bibfnamefont {Gr{\'{e}}gory}\
  \bibnamefont {Schehr}}} (\bibinfo {year} {2020}),\ \bibfield  {title}
  {\enquote {\bibinfo {title} {{Exact stationary state of a run-and-tumble
  particle with three internal states in a harmonic trap}},}\ }\href
  {https://doi.org/10.1088/1751-8121/ab6af0} {\bibfield  {journal} {\bibinfo
  {journal} {J. Phys. A Math. Theor.}\ }\textbf {\bibinfo {volume}
  {53}}~(\bibinfo {number} {9}),\ \bibinfo {pages} {09LT01}}\BibitemShut
  {NoStop}%
\bibitem [{\citenamefont {B{\"{a}}uerle}\ \emph {et~al.}(2018)\citenamefont
  {B{\"{a}}uerle}, \citenamefont {Fischer}, \citenamefont {Speck},\ and\
  \citenamefont {Bechinger}}]{bauerle_self-organization_2018}%
  \BibitemOpen
  \bibfield  {author} {\bibinfo {author} {\bibnamefont {B{\"{a}}uerle},
  \bibfnamefont {Tobias}}, \bibinfo {author} {\bibfnamefont {Andreas}\
  \bibnamefont {Fischer}}, \bibinfo {author} {\bibfnamefont {Thomas}\
  \bibnamefont {Speck}}, and\ \bibinfo {author} {\bibfnamefont {Clemens}\
  \bibnamefont {Bechinger}}} (\bibinfo {year} {2018}),\ \bibfield  {title}
  {\enquote {\bibinfo {title} {{Self-organization of active particles by quorum
  sensing rules}},}\ }\href {https://doi.org/10.1038/s41467-018-05675-7}
  {\bibfield  {journal} {\bibinfo  {journal} {Nat. Commun.}\ }\textbf {\bibinfo
  {volume} {9}}~(\bibinfo {number} {1}),\ \bibinfo {pages} {3232}}\BibitemShut
  {NoStop}%
\bibitem [{\citenamefont {Bechinger}\ \emph {et~al.}(2016)\citenamefont
  {Bechinger}, \citenamefont {{Di Leonardo}}, \citenamefont {L{\"{o}}wen},
  \citenamefont {Reichhardt}, \citenamefont {Volpe},\ and\ \citenamefont
  {Volpe}}]{bechinger_active_2016}%
  \BibitemOpen
  \bibfield  {author} {\bibinfo {author} {\bibnamefont {Bechinger},
  \bibfnamefont {Clemens}}, \bibinfo {author} {\bibfnamefont {Roberto}\
  \bibnamefont {{Di Leonardo}}}, \bibinfo {author} {\bibfnamefont {Hartmut}\
  \bibnamefont {L{\"{o}}wen}}, \bibinfo {author} {\bibfnamefont {Charles}\
  \bibnamefont {Reichhardt}}, \bibinfo {author} {\bibfnamefont {Giorgio}\
  \bibnamefont {Volpe}}, and\ \bibinfo {author} {\bibfnamefont {Giovanni}\
  \bibnamefont {Volpe}}} (\bibinfo {year} {2016}),\ \bibfield  {title}
  {\enquote {\bibinfo {title} {{Active Particles in Complex and Crowded
  Environments}},}\ }\href {https://doi.org/10.1103/RevModPhys.88.045006}
  {\bibfield  {journal} {\bibinfo  {journal} {Rev. Mod. Phys.}\ }\textbf
  {\bibinfo {volume} {88}}~(\bibinfo {number} {4}),\ \bibinfo {pages}
  {045006}}\BibitemShut {NoStop}%
\bibitem [{\citenamefont {van Beijeren}(1982)}]{VanBeijeren1982}%
  \BibitemOpen
  \bibfield  {author} {\bibinfo {author} {\bibnamefont {van Beijeren},
  \bibfnamefont {Henk}}} (\bibinfo {year} {1982}),\ \bibfield  {title}
  {\enquote {\bibinfo {title} {{Transport properties of stochastic Lorentz
  models}},}\ }\href {https://doi.org/10.1103/RevModPhys.54.195} {\bibfield
  {journal} {\bibinfo  {journal} {Rev. Mod. Phys.}\ }\textbf {\bibinfo {volume}
  {54}}~(\bibinfo {number} {1}),\ \bibinfo {pages} {195--234}}\BibitemShut
  {NoStop}%
\bibitem [{\citenamefont {Belanger}\ \emph {et~al.}(1983)\citenamefont
  {Belanger}, \citenamefont {King}, \citenamefont {Jaccarino},\ and\
  \citenamefont {Cardy}}]{belanger_random-field_1983}%
  \BibitemOpen
  \bibfield  {author} {\bibinfo {author} {\bibnamefont {Belanger},
  \bibfnamefont {D~P}}, \bibinfo {author} {\bibfnamefont {A.~R.}\ \bibnamefont
  {King}}, \bibinfo {author} {\bibfnamefont {V.}~\bibnamefont {Jaccarino}},
  and\ \bibinfo {author} {\bibfnamefont {J.~L.}\ \bibnamefont {Cardy}}}
  (\bibinfo {year} {1983}),\ \bibfield  {title} {\enquote {\bibinfo {title}
  {{Random-field critical behavior of a $d=3$ Ising system}},}\ }\href
  {https://doi.org/10.1103/PhysRevB.28.2522} {\bibfield  {journal} {\bibinfo
  {journal} {Phys. Rev. B}\ }\textbf {\bibinfo {volume} {28}}~(\bibinfo
  {number} {5}),\ \bibinfo {pages} {2522--2526}}\BibitemShut {NoStop}%
\bibitem [{\citenamefont {{Ben Dor}}\ \emph
  {et~al.}(2022{\natexlab{a}})\citenamefont {{Ben Dor}}, \citenamefont {Kafri},
  \citenamefont {Kardar},\ and\ \citenamefont {Tailleur}}]{dor2022passive}%
  \BibitemOpen
  \bibfield  {author} {\bibinfo {author} {\bibnamefont {{Ben Dor}},
  \bibfnamefont {Ydan}}, \bibinfo {author} {\bibfnamefont {Yariv}\ \bibnamefont
  {Kafri}}, \bibinfo {author} {\bibfnamefont {Mehran}\ \bibnamefont {Kardar}},
  and\ \bibinfo {author} {\bibfnamefont {Julien}\ \bibnamefont {Tailleur}}}
  (\bibinfo {year} {2022}{\natexlab{a}}),\ \bibfield  {title} {\enquote
  {\bibinfo {title} {{Passive objects in confined active fluids: A localization
  transition}},}\ }\href {https://doi.org/10.1103/PhysRevE.106.044604}
  {\bibfield  {journal} {\bibinfo  {journal} {Phys. Rev. E}\ }\textbf {\bibinfo
  {volume} {106}}~(\bibinfo {number} {4}),\ \bibinfo {pages}
  {044604}}\BibitemShut {NoStop}%
\bibitem [{\citenamefont {{Ben Dor}}\ \emph
  {et~al.}(2022{\natexlab{b}})\citenamefont {{Ben Dor}}, \citenamefont {Ro},
  \citenamefont {Kafri}, \citenamefont {Kardar},\ and\ \citenamefont
  {Tailleur}}]{dor2022disordered}%
  \BibitemOpen
  \bibfield  {author} {\bibinfo {author} {\bibnamefont {{Ben Dor}},
  \bibfnamefont {Ydan}}, \bibinfo {author} {\bibfnamefont {Sunghan}\
  \bibnamefont {Ro}}, \bibinfo {author} {\bibfnamefont {Yariv}\ \bibnamefont
  {Kafri}}, \bibinfo {author} {\bibfnamefont {Mehran}\ \bibnamefont {Kardar}},
  and\ \bibinfo {author} {\bibfnamefont {Julien}\ \bibnamefont {Tailleur}}}
  (\bibinfo {year} {2022}{\natexlab{b}}),\ \bibfield  {title} {\enquote
  {\bibinfo {title} {{Disordered boundaries destroy bulk phase separation in
  scalar active matter}},}\ }\href
  {https://doi.org/10.1103/PhysRevE.105.044603} {\bibfield  {journal} {\bibinfo
   {journal} {Phys. Rev. E}\ }\textbf {\bibinfo {volume} {105}}~(\bibinfo
  {number} {4}),\ \bibinfo {pages} {044603}}\BibitemShut {NoStop}%
\bibitem [{\citenamefont {{Ben Dor}}\ \emph {et~al.}(2019)\citenamefont {{Ben
  Dor}}, \citenamefont {Woillez}, \citenamefont {Kafri}, \citenamefont
  {Kardar},\ and\ \citenamefont {Solon}}]{dor2019ramifications}%
  \BibitemOpen
  \bibfield  {author} {\bibinfo {author} {\bibnamefont {{Ben Dor}},
  \bibfnamefont {Ydan}}, \bibinfo {author} {\bibfnamefont {Eric}\ \bibnamefont
  {Woillez}}, \bibinfo {author} {\bibfnamefont {Yariv}\ \bibnamefont {Kafri}},
  \bibinfo {author} {\bibfnamefont {Mehran}\ \bibnamefont {Kardar}}, and\
  \bibinfo {author} {\bibfnamefont {Alexandre~P.}\ \bibnamefont {Solon}}}
  (\bibinfo {year} {2019}),\ \bibfield  {title} {\enquote {\bibinfo {title}
  {{Ramifications of disorder on active particles in one dimension}},}\ }\href
  {https://doi.org/10.1103/PhysRevE.100.052610} {\bibfield  {journal} {\bibinfo
   {journal} {Phys. Rev. E}\ }\textbf {\bibinfo {volume} {100}}~(\bibinfo
  {number} {5}),\ \bibinfo {pages} {052610}}\BibitemShut {NoStop}%
\bibitem [{\citenamefont {Berg}(2004)}]{berg2004coli}%
  \BibitemOpen
  \bibfield  {author} {\bibinfo {author} {\bibnamefont {Berg}, \bibfnamefont
  {Howard~C}}} (\bibinfo {year} {2004}),\ \href@noop {} {\emph {\bibinfo
  {title} {{E. coli in Motion}}}}\ (\bibinfo  {publisher}
  {Springer})\BibitemShut {NoStop}%
\bibitem [{\citenamefont {Berthier}\ \emph {et~al.}(2017)\citenamefont
  {Berthier}, \citenamefont {Flenner},\ and\ \citenamefont
  {Szamel}}]{berthier_how_2017}%
  \BibitemOpen
  \bibfield  {author} {\bibinfo {author} {\bibnamefont {Berthier},
  \bibfnamefont {Ludovic}}, \bibinfo {author} {\bibfnamefont {Elijah}\
  \bibnamefont {Flenner}}, and\ \bibinfo {author} {\bibfnamefont {Grzegorz}\
  \bibnamefont {Szamel}}} (\bibinfo {year} {2017}),\ \bibfield  {title}
  {\enquote {\bibinfo {title} {{How active forces influence nonequilibrium
  glass transitions}},}\ }\href {https://doi.org/10.1088/1367-2630/aa914e}
  {\bibfield  {journal} {\bibinfo  {journal} {New J. Phys.}\ }\textbf {\bibinfo
  {volume} {19}}~(\bibinfo {number} {12}),\ \bibinfo {pages}
  {125006}}\BibitemShut {NoStop}%
\bibitem [{\citenamefont {Berthier}\ \emph {et~al.}(2019)\citenamefont
  {Berthier}, \citenamefont {Flenner},\ and\ \citenamefont
  {Szamel}}]{berthier_glassy_2019}%
  \BibitemOpen
  \bibfield  {author} {\bibinfo {author} {\bibnamefont {Berthier},
  \bibfnamefont {Ludovic}}, \bibinfo {author} {\bibfnamefont {Elijah}\
  \bibnamefont {Flenner}}, and\ \bibinfo {author} {\bibfnamefont {Grzegorz}\
  \bibnamefont {Szamel}}} (\bibinfo {year} {2019}),\ \bibfield  {title}
  {\enquote {\bibinfo {title} {{Glassy dynamics in dense systems of active
  particles}},}\ }\href {https://doi.org/10.1063/1.5093240} {\bibfield
  {journal} {\bibinfo  {journal} {J. Chem. Phys.}\ }\textbf {\bibinfo {volume}
  {150}}~(\bibinfo {number} {20}),\ \bibinfo {pages} {200901}}\BibitemShut
  {NoStop}%
\bibitem [{\citenamefont {Bhattacharjee}\ and\ \citenamefont
  {Datta}(2019)}]{bhattacharjee2019bacterial}%
  \BibitemOpen
  \bibfield  {author} {\bibinfo {author} {\bibnamefont {Bhattacharjee},
  \bibfnamefont {Tapomoy}}, and\ \bibinfo {author} {\bibfnamefont {Sujit~S.}\
  \bibnamefont {Datta}}} (\bibinfo {year} {2019}),\ \bibfield  {title}
  {\enquote {\bibinfo {title} {{Bacterial hopping and trapping in porous
  media}},}\ }\href {https://doi.org/10.1038/s41467-019-10115-1} {\bibfield
  {journal} {\bibinfo  {journal} {Nat. Commun.}\ }\textbf {\bibinfo {volume}
  {10}}~(\bibinfo {number} {1}),\ \bibinfo {pages} {2075}}\BibitemShut
  {NoStop}%
\bibitem [{\citenamefont {Bialk{\'{e}}}\ \emph {et~al.}(2015)\citenamefont
  {Bialk{\'{e}}}, \citenamefont {Siebert}, \citenamefont {L{\"{o}}wen},\ and\
  \citenamefont {Speck}}]{bialke2015negative}%
  \BibitemOpen
  \bibfield  {author} {\bibinfo {author} {\bibnamefont {Bialk{\'{e}}},
  \bibfnamefont {Julian}}, \bibinfo {author} {\bibfnamefont {Jonathan~T.}\
  \bibnamefont {Siebert}}, \bibinfo {author} {\bibfnamefont {Hartmut}\
  \bibnamefont {L{\"{o}}wen}}, and\ \bibinfo {author} {\bibfnamefont {Thomas}\
  \bibnamefont {Speck}}} (\bibinfo {year} {2015}),\ \bibfield  {title}
  {\enquote {\bibinfo {title} {{Negative Interfacial Tension in Phase-Separated
  Active Brownian Particles}},}\ }\href
  {https://doi.org/10.1103/PhysRevLett.115.098301} {\bibfield  {journal}
  {\bibinfo  {journal} {Phys. Rev. Lett.}\ }\textbf {\bibinfo {volume}
  {115}}~(\bibinfo {number} {9}),\ \bibinfo {pages} {098301}}\BibitemShut
  {NoStop}%
\bibitem [{\citenamefont {Bouchaud}\ \emph {et~al.}(1990)\citenamefont
  {Bouchaud}, \citenamefont {Comtet}, \citenamefont {Georges},\ and\
  \citenamefont {{Le Doussal}}}]{bouchaud1990classical}%
  \BibitemOpen
  \bibfield  {author} {\bibinfo {author} {\bibnamefont {Bouchaud},
  \bibfnamefont {J~P}}, \bibinfo {author} {\bibfnamefont {A.}~\bibnamefont
  {Comtet}}, \bibinfo {author} {\bibfnamefont {A.}~\bibnamefont {Georges}},
  and\ \bibinfo {author} {\bibfnamefont {P.}~\bibnamefont {{Le Doussal}}}}
  (\bibinfo {year} {1990}),\ \bibfield  {title} {\enquote {\bibinfo {title}
  {{Classical diffusion of a particle in a one-dimensional random force
  field}},}\ }\href {https://doi.org/10.1016/0003-4916(90)90043-N} {\bibfield
  {journal} {\bibinfo  {journal} {Ann. Phys. (N. Y).}\ }\textbf {\bibinfo
  {volume} {201}}~(\bibinfo {number} {2}),\ \bibinfo {pages}
  {285--341}}\BibitemShut {NoStop}%
\bibitem [{\citenamefont {Brenner}\ \emph {et~al.}(1998)\citenamefont
  {Brenner}, \citenamefont {Levitov},\ and\ \citenamefont
  {Budrene}}]{brenner1998physical}%
  \BibitemOpen
  \bibfield  {author} {\bibinfo {author} {\bibnamefont {Brenner}, \bibfnamefont
  {Michael~P}}, \bibinfo {author} {\bibfnamefont {Leonid~S.}\ \bibnamefont
  {Levitov}}, and\ \bibinfo {author} {\bibfnamefont {Elena~O.}\ \bibnamefont
  {Budrene}}} (\bibinfo {year} {1998}),\ \bibfield  {title} {\enquote {\bibinfo
  {title} {{Physical Mechanisms for Chemotactic Pattern Formation by
  Bacteria}},}\ }\href {https://doi.org/10.1016/S0006-3495(98)77880-4}
  {\bibfield  {journal} {\bibinfo  {journal} {Biophys. J.}\ }\textbf {\bibinfo
  {volume} {74}}~(\bibinfo {number} {4}),\ \bibinfo {pages}
  {1677--1693}}\BibitemShut {NoStop}%
\bibitem [{\citenamefont {Bricard}\ \emph {et~al.}(2013)\citenamefont
  {Bricard}, \citenamefont {Caussin}, \citenamefont {Desreumaux}, \citenamefont
  {Dauchot},\ and\ \citenamefont {Bartolo}}]{bricard2013emergence}%
  \BibitemOpen
  \bibfield  {author} {\bibinfo {author} {\bibnamefont {Bricard}, \bibfnamefont
  {Antoine}}, \bibinfo {author} {\bibfnamefont {Jean-Baptiste}\ \bibnamefont
  {Caussin}}, \bibinfo {author} {\bibfnamefont {Nicolas}\ \bibnamefont
  {Desreumaux}}, \bibinfo {author} {\bibfnamefont {Olivier}\ \bibnamefont
  {Dauchot}}, and\ \bibinfo {author} {\bibfnamefont {Denis}\ \bibnamefont
  {Bartolo}}} (\bibinfo {year} {2013}),\ \bibfield  {title} {\enquote {\bibinfo
  {title} {{Emergence of macroscopic directed motion in populations of motile
  colloids}},}\ }\href {https://doi.org/10.1038/nature12673} {\bibfield
  {journal} {\bibinfo  {journal} {Nature}\ }\textbf {\bibinfo {volume}
  {503}}~(\bibinfo {number} {7474}),\ \bibinfo {pages} {95--98}}\BibitemShut
  {NoStop}%
\bibitem [{\citenamefont {Bricmont}\ and\ \citenamefont
  {Kupiainen}(1987)}]{bricmont_lower_1987}%
  \BibitemOpen
  \bibfield  {author} {\bibinfo {author} {\bibnamefont {Bricmont},
  \bibfnamefont {J}}, and\ \bibinfo {author} {\bibfnamefont {A.}~\bibnamefont
  {Kupiainen}}} (\bibinfo {year} {1987}),\ \bibfield  {title} {\enquote
  {\bibinfo {title} {{Lower critical dimension for the random-field Ising
  model}},}\ }\href {https://doi.org/10.1103/PhysRevLett.59.1829} {\bibfield
  {journal} {\bibinfo  {journal} {Phys. Rev. Lett.}\ }\textbf {\bibinfo
  {volume} {59}}~(\bibinfo {number} {16}),\ \bibinfo {pages}
  {1829--1832}}\BibitemShut {NoStop}%
\bibitem [{\citenamefont {Burkholder}\ and\ \citenamefont
  {Brady}(2017)}]{Burkholder2017}%
  \BibitemOpen
  \bibfield  {author} {\bibinfo {author} {\bibnamefont {Burkholder},
  \bibfnamefont {Eric~W}}, and\ \bibinfo {author} {\bibfnamefont {John~F.}\
  \bibnamefont {Brady}}} (\bibinfo {year} {2017}),\ \bibfield  {title}
  {\enquote {\bibinfo {title} {{Tracer diffusion in active suspensions}},}\
  }\href {https://doi.org/10.1103/PhysRevE.95.052605} {\bibfield  {journal}
  {\bibinfo  {journal} {Phys. Rev. E}\ }\textbf {\bibinfo {volume}
  {95}}~(\bibinfo {number} {5}),\ \bibinfo {pages} {052605}}\BibitemShut
  {NoStop}%
\bibitem [{\citenamefont {Burkholder}\ and\ \citenamefont
  {Brady}(2019)}]{Burkholder2019a}%
  \BibitemOpen
  \bibfield  {author} {\bibinfo {author} {\bibnamefont {Burkholder},
  \bibfnamefont {Eric~W}}, and\ \bibinfo {author} {\bibfnamefont {John~F}\
  \bibnamefont {Brady}}} (\bibinfo {year} {2019}),\ \bibfield  {title}
  {\enquote {\bibinfo {title} {{Fluctuation-dissipation in active matter}},}\
  }\href {https://doi.org/10.1063/1.5081725} {\bibfield  {journal} {\bibinfo
  {journal} {J. Chem. Phys.}\ }\textbf {\bibinfo {volume} {150}}~(\bibinfo
  {number} {18}),\ \bibinfo {pages} {184901}}\BibitemShut {NoStop}%
\bibitem [{\citenamefont {Buttinoni}\ \emph {et~al.}(2013)\citenamefont
  {Buttinoni}, \citenamefont {Bialk{\'{e}}}, \citenamefont {K{\"{u}}mmel},
  \citenamefont {L{\"{o}}wen}, \citenamefont {Bechinger},\ and\ \citenamefont
  {Speck}}]{buttinoni_dynamical_2013}%
  \BibitemOpen
  \bibfield  {author} {\bibinfo {author} {\bibnamefont {Buttinoni},
  \bibfnamefont {Ivo}}, \bibinfo {author} {\bibfnamefont {Julian}\ \bibnamefont
  {Bialk{\'{e}}}}, \bibinfo {author} {\bibfnamefont {Felix}\ \bibnamefont
  {K{\"{u}}mmel}}, \bibinfo {author} {\bibfnamefont {Hartmut}\ \bibnamefont
  {L{\"{o}}wen}}, \bibinfo {author} {\bibfnamefont {Clemens}\ \bibnamefont
  {Bechinger}}, and\ \bibinfo {author} {\bibfnamefont {Thomas}\ \bibnamefont
  {Speck}}} (\bibinfo {year} {2013}),\ \bibfield  {title} {\enquote {\bibinfo
  {title} {{Dynamical Clustering and Phase Separation in Suspensions of
  Self-Propelled Colloidal Particles}},}\ }\href
  {https://doi.org/10.1103/PhysRevLett.110.238301} {\bibfield  {journal}
  {\bibinfo  {journal} {Phys. Rev. Lett.}\ }\textbf {\bibinfo {volume}
  {110}}~(\bibinfo {number} {23}),\ \bibinfo {pages} {238301}}\BibitemShut
  {NoStop}%
\bibitem [{\citenamefont {Caballero}\ \emph {et~al.}(2018)\citenamefont
  {Caballero}, \citenamefont {Nardini},\ and\ \citenamefont
  {Cates}}]{caballero2018bulk}%
  \BibitemOpen
  \bibfield  {author} {\bibinfo {author} {\bibnamefont {Caballero},
  \bibfnamefont {Fernando}}, \bibinfo {author} {\bibfnamefont {Cesare}\
  \bibnamefont {Nardini}}, and\ \bibinfo {author} {\bibfnamefont {Michael~E.}\
  \bibnamefont {Cates}}} (\bibinfo {year} {2018}),\ \bibfield  {title}
  {\enquote {\bibinfo {title} {{From bulk to microphase separation in scalar
  active matter: a perturbative renormalization group analysis}},}\ }\href
  {https://doi.org/10.1088/1742-5468/aaf321} {\bibfield  {journal} {\bibinfo
  {journal} {J. Stat. Mech. Theory Exp.}\ }\textbf {\bibinfo {volume}
  {2018}}~(\bibinfo {number} {12}),\ \bibinfo {pages} {123208}}\BibitemShut
  {NoStop}%
\bibitem [{\citenamefont {Caporusso}\ \emph {et~al.}(2020)\citenamefont
  {Caporusso}, \citenamefont {Digregorio}, \citenamefont {Levis}, \citenamefont
  {Cugliandolo},\ and\ \citenamefont
  {Gonnella}}]{caporusso_motility-induced_2020}%
  \BibitemOpen
  \bibfield  {author} {\bibinfo {author} {\bibnamefont {Caporusso},
  \bibfnamefont {Claudio~B}}, \bibinfo {author} {\bibfnamefont {Pasquale}\
  \bibnamefont {Digregorio}}, \bibinfo {author} {\bibfnamefont {Demian}\
  \bibnamefont {Levis}}, \bibinfo {author} {\bibfnamefont {Leticia~F.}\
  \bibnamefont {Cugliandolo}}, and\ \bibinfo {author} {\bibfnamefont
  {Giuseppe}\ \bibnamefont {Gonnella}}} (\bibinfo {year} {2020}),\ \bibfield
  {title} {\enquote {\bibinfo {title} {{Motility-Induced Microphase and
  Macrophase Separation in a Two-Dimensional Active Brownian Particle
  System}},}\ }\href {https://doi.org/10.1103/PhysRevLett.125.178004}
  {\bibfield  {journal} {\bibinfo  {journal} {Phys. Rev. Lett.}\ }\textbf
  {\bibinfo {volume} {125}}~(\bibinfo {number} {17}),\ \bibinfo {pages}
  {178004}}\BibitemShut {NoStop}%
\bibitem [{\citenamefont {Cates}\ and\ \citenamefont
  {Tailleur}(2013)}]{cates_when_2013}%
  \BibitemOpen
  \bibfield  {author} {\bibinfo {author} {\bibnamefont {Cates}, \bibfnamefont
  {M~E}}, and\ \bibinfo {author} {\bibfnamefont {J.}~\bibnamefont {Tailleur}}}
  (\bibinfo {year} {2013}),\ \bibfield  {title} {\enquote {\bibinfo {title}
  {{When are active Brownian particles and run-and-tumble particles equivalent?
  Consequences for motility-induced phase separation}},}\ }\href
  {https://doi.org/10.1209/0295-5075/101/20010} {\bibfield  {journal} {\bibinfo
   {journal} {EPL}\ }\textbf {\bibinfo {volume} {101}}~(\bibinfo {number}
  {2}),\ \bibinfo {pages} {20010}}\BibitemShut {NoStop}%
\bibitem [{\citenamefont {Cates}\ and\ \citenamefont
  {Tailleur}(2015)}]{cates2015motility}%
  \BibitemOpen
  \bibfield  {author} {\bibinfo {author} {\bibnamefont {Cates}, \bibfnamefont
  {Michael~E}}, and\ \bibinfo {author} {\bibfnamefont {Julien}\ \bibnamefont
  {Tailleur}}} (\bibinfo {year} {2015}),\ \bibfield  {title} {\enquote
  {\bibinfo {title} {{Motility-Induced Phase Separation}},}\ }\href
  {https://doi.org/10.1146/annurev-conmatphys-031214-014710} {\bibfield
  {journal} {\bibinfo  {journal} {Annu. Rev. Condens. Matter Phys.}\ }\textbf
  {\bibinfo {volume} {6}}~(\bibinfo {number} {1}),\ \bibinfo {pages}
  {219--244}}\BibitemShut {NoStop}%
\bibitem [{\citenamefont {Chardac}\ \emph {et~al.}(2021)\citenamefont
  {Chardac}, \citenamefont {Shankar}, \citenamefont {Marchetti},\ and\
  \citenamefont {Bartolo}}]{chardac2021emergence}%
  \BibitemOpen
  \bibfield  {author} {\bibinfo {author} {\bibnamefont {Chardac}, \bibfnamefont
  {Am{\'{e}}lie}}, \bibinfo {author} {\bibfnamefont {Suraj}\ \bibnamefont
  {Shankar}}, \bibinfo {author} {\bibfnamefont {M.~Cristina}\ \bibnamefont
  {Marchetti}}, and\ \bibinfo {author} {\bibfnamefont {Denis}\ \bibnamefont
  {Bartolo}}} (\bibinfo {year} {2021}),\ \bibfield  {title} {\enquote {\bibinfo
  {title} {{Emergence of dynamic vortex glasses in disordered polar active
  fluids}},}\ }\href {https://doi.org/10.1073/pnas.2018218118} {\bibfield
  {journal} {\bibinfo  {journal} {Proc. Natl. Acad. Sci.}\ }\textbf {\bibinfo
  {volume} {118}}~(\bibinfo {number} {10}),\ \bibinfo {pages}
  {e2018218118}}\BibitemShut {NoStop}%
\bibitem [{\citenamefont {Chat{\'{e}}}(2020)}]{chate_dry_2020}%
  \BibitemOpen
  \bibfield  {author} {\bibinfo {author} {\bibnamefont {Chat{\'{e}}},
  \bibfnamefont {Hugues}}} (\bibinfo {year} {2020}),\ \bibfield  {title}
  {\enquote {\bibinfo {title} {{Dry Aligning Dilute Active Matter}},}\ }\href
  {https://doi.org/10.1146/annurev-conmatphys-031119-050752} {\bibfield
  {journal} {\bibinfo  {journal} {Annu. Rev. Condens. Matter Phys.}\ }\textbf
  {\bibinfo {volume} {11}}~(\bibinfo {number} {1}),\ \bibinfo {pages}
  {189--212}}\BibitemShut {NoStop}%
\bibitem [{\citenamefont {Chen}\ \emph {et~al.}(2007)\citenamefont {Chen},
  \citenamefont {Lau}, \citenamefont {Hough}, \citenamefont {Islam},
  \citenamefont {Goulian}, \citenamefont {Lubensky},\ and\ \citenamefont
  {Yodh}}]{Chen2007}%
  \BibitemOpen
  \bibfield  {author} {\bibinfo {author} {\bibnamefont {Chen}, \bibfnamefont
  {D~T~N}}, \bibinfo {author} {\bibfnamefont {A.~W.~C.}\ \bibnamefont {Lau}},
  \bibinfo {author} {\bibfnamefont {L.~A.}\ \bibnamefont {Hough}}, \bibinfo
  {author} {\bibfnamefont {M.~F.}\ \bibnamefont {Islam}}, \bibinfo {author}
  {\bibfnamefont {M.}~\bibnamefont {Goulian}}, \bibinfo {author} {\bibfnamefont
  {T.~C.}\ \bibnamefont {Lubensky}}, and\ \bibinfo {author} {\bibfnamefont
  {A.~G.}\ \bibnamefont {Yodh}}} (\bibinfo {year} {2007}),\ \bibfield  {title}
  {\enquote {\bibinfo {title} {{Fluctuations and Rheology in Active Bacterial
  Suspensions}},}\ }\href {https://doi.org/10.1103/PhysRevLett.99.148302}
  {\bibfield  {journal} {\bibinfo  {journal} {Phys. Rev. Lett.}\ }\textbf
  {\bibinfo {volume} {99}}~(\bibinfo {number} {14}),\ \bibinfo {pages}
  {148302}}\BibitemShut {NoStop}%
\bibitem [{\citenamefont {Curatolo}\ \emph {et~al.}(2020)\citenamefont
  {Curatolo}, \citenamefont {Zhou}, \citenamefont {Zhao}, \citenamefont {Liu},
  \citenamefont {Daerr}, \citenamefont {Tailleur},\ and\ \citenamefont
  {Huang}}]{curatolo2020cooperative}%
  \BibitemOpen
  \bibfield  {author} {\bibinfo {author} {\bibnamefont {Curatolo},
  \bibfnamefont {A~I}}, \bibinfo {author} {\bibfnamefont {N.}~\bibnamefont
  {Zhou}}, \bibinfo {author} {\bibfnamefont {Y.}~\bibnamefont {Zhao}}, \bibinfo
  {author} {\bibfnamefont {C.}~\bibnamefont {Liu}}, \bibinfo {author}
  {\bibfnamefont {A.}~\bibnamefont {Daerr}}, \bibinfo {author} {\bibfnamefont
  {J.}~\bibnamefont {Tailleur}}, and\ \bibinfo {author} {\bibfnamefont
  {.J}~\bibnamefont {Huang}}} (\bibinfo {year} {2020}),\ \bibfield  {title}
  {\enquote {\bibinfo {title} {{Cooperative pattern formation in
  multi-component bacterial systems through reciprocal motility regulation}},}\
  }\href {https://doi.org/10.1038/s41567-020-0964-z} {\bibfield  {journal}
  {\bibinfo  {journal} {Nat. Phys.}\ }\textbf {\bibinfo {volume}
  {16}}~(\bibinfo {number} {11}),\ \bibinfo {pages} {1152--1157}}\BibitemShut
  {NoStop}%
\bibitem [{\citenamefont {D'Alessandro}\ \emph {et~al.}(2017)\citenamefont
  {D'Alessandro}, \citenamefont {Solon}, \citenamefont {Hayakawa},
  \citenamefont {Anjard}, \citenamefont {Detcheverry}, \citenamefont {Rieu},\
  and\ \citenamefont {Rivi{\`{e}}re}}]{dalessandro_contact_2017}%
  \BibitemOpen
  \bibfield  {author} {\bibinfo {author} {\bibnamefont {D'Alessandro},
  \bibfnamefont {Joseph}}, \bibinfo {author} {\bibfnamefont {Alexandre~P.}\
  \bibnamefont {Solon}}, \bibinfo {author} {\bibfnamefont {Yoshinori}\
  \bibnamefont {Hayakawa}}, \bibinfo {author} {\bibfnamefont {Christophe}\
  \bibnamefont {Anjard}}, \bibinfo {author} {\bibfnamefont {Fran{\c{c}}ois}\
  \bibnamefont {Detcheverry}}, \bibinfo {author} {\bibfnamefont {Jean-Paul}\
  \bibnamefont {Rieu}}, and\ \bibinfo {author} {\bibfnamefont {Charlotte}\
  \bibnamefont {Rivi{\`{e}}re}}} (\bibinfo {year} {2017}),\ \bibfield  {title}
  {\enquote {\bibinfo {title} {{Contact enhancement of locomotion in spreading
  cell colonies}},}\ }\href {https://doi.org/10.1038/nphys4180} {\bibfield
  {journal} {\bibinfo  {journal} {Nat. Phys.}\ }\textbf {\bibinfo {volume}
  {13}}~(\bibinfo {number} {10}),\ \bibinfo {pages} {999--1005}}\BibitemShut
  {NoStop}%
\bibitem [{\citenamefont {D'Alessio}\ \emph {et~al.}(2016)\citenamefont
  {D'Alessio}, \citenamefont {Kafri},\ and\ \citenamefont
  {Polkovnikov}}]{DAlessioJSTAT2016}%
  \BibitemOpen
  \bibfield  {author} {\bibinfo {author} {\bibnamefont {D'Alessio},
  \bibfnamefont {Luca}}, \bibinfo {author} {\bibfnamefont {Yariv}\ \bibnamefont
  {Kafri}}, and\ \bibinfo {author} {\bibfnamefont {Anatoli}\ \bibnamefont
  {Polkovnikov}}} (\bibinfo {year} {2016}),\ \bibfield  {title} {\enquote
  {\bibinfo {title} {{Negative mass corrections in a dissipative stochastic
  environment}},}\ }\href {https://doi.org/10.1088/1742-5468/2016/02/023105}
  {\bibfield  {journal} {\bibinfo  {journal} {J. Stat. Mech. Theory Exp.}\
  }\textbf {\bibinfo {volume} {2016}}~(\bibinfo {number} {2}),\ \bibinfo
  {pages} {023105}}\BibitemShut {NoStop}%
\bibitem [{\citenamefont {Daniels}\ \emph {et~al.}(2004)\citenamefont
  {Daniels}, \citenamefont {Vanderleyden},\ and\ \citenamefont
  {Michiels}}]{daniels_quorum_2004}%
  \BibitemOpen
  \bibfield  {author} {\bibinfo {author} {\bibnamefont {Daniels}, \bibfnamefont
  {Ruth}}, \bibinfo {author} {\bibfnamefont {Jos}\ \bibnamefont
  {Vanderleyden}}, and\ \bibinfo {author} {\bibfnamefont {Jan}\ \bibnamefont
  {Michiels}}} (\bibinfo {year} {2004}),\ \bibfield  {title} {\enquote
  {\bibinfo {title} {{Quorum sensing and swarming migration in bacteria}},}\
  }\href {https://doi.org/10.1016/j.femsre.2003.09.004} {\bibfield  {journal}
  {\bibinfo  {journal} {FEMS Microbiol. Rev.}\ }\textbf {\bibinfo {volume}
  {28}}~(\bibinfo {number} {3}),\ \bibinfo {pages} {261--289}}\BibitemShut
  {NoStop}%
\bibitem [{\citenamefont {Dean}(1996)}]{dean1996langevin}%
  \BibitemOpen
  \bibfield  {author} {\bibinfo {author} {\bibnamefont {Dean}, \bibfnamefont
  {David~S}}} (\bibinfo {year} {1996}),\ \bibfield  {title} {\enquote {\bibinfo
  {title} {{Langevin equation for the density of a system of interacting
  Langevin processes}},}\ }\href {https://doi.org/10.1088/0305-4470/29/24/001}
  {\bibfield  {journal} {\bibinfo  {journal} {J. Phys. A. Math. Gen.}\ }\textbf
  {\bibinfo {volume} {29}}~(\bibinfo {number} {24}),\ \bibinfo {pages}
  {L613--L617}}\BibitemShut {NoStop}%
\bibitem [{\citenamefont {Derrida}(1983)}]{derrida1983velocity}%
  \BibitemOpen
  \bibfield  {author} {\bibinfo {author} {\bibnamefont {Derrida}, \bibfnamefont
  {Bernard}}} (\bibinfo {year} {1983}),\ \bibfield  {title} {\enquote {\bibinfo
  {title} {{Velocity and diffusion constant of a periodic one-dimensional
  hopping model}},}\ }\href {https://doi.org/10.1007/BF01019492} {\bibfield
  {journal} {\bibinfo  {journal} {J. Stat. Phys.}\ }\textbf {\bibinfo {volume}
  {31}}~(\bibinfo {number} {3}),\ \bibinfo {pages} {433--450}}\BibitemShut
  {NoStop}%
\bibitem [{\citenamefont {Derrida}\ and\ \citenamefont
  {Pomeau}(1982)}]{derrida1982classical}%
  \BibitemOpen
  \bibfield  {author} {\bibinfo {author} {\bibnamefont {Derrida}, \bibfnamefont
  {Bernard}}, and\ \bibinfo {author} {\bibfnamefont {Y.}~\bibnamefont
  {Pomeau}}} (\bibinfo {year} {1982}),\ \bibfield  {title} {\enquote {\bibinfo
  {title} {{Classical Diffusion on a Random Chain}},}\ }\href
  {https://doi.org/10.1103/PhysRevLett.48.627} {\bibfield  {journal} {\bibinfo
  {journal} {Phys. Rev. Lett.}\ }\textbf {\bibinfo {volume} {48}}~(\bibinfo
  {number} {9}),\ \bibinfo {pages} {627--630}}\BibitemShut {NoStop}%
\bibitem [{\citenamefont {Deseigne}\ \emph {et~al.}(2010)\citenamefont
  {Deseigne}, \citenamefont {Dauchot},\ and\ \citenamefont
  {Chat{\'{e}}}}]{deseigne2010collective}%
  \BibitemOpen
  \bibfield  {author} {\bibinfo {author} {\bibnamefont {Deseigne},
  \bibfnamefont {Julien}}, \bibinfo {author} {\bibfnamefont {Olivier}\
  \bibnamefont {Dauchot}}, and\ \bibinfo {author} {\bibfnamefont {Hugues}\
  \bibnamefont {Chat{\'{e}}}}} (\bibinfo {year} {2010}),\ \bibfield  {title}
  {\enquote {\bibinfo {title} {{Collective Motion of Vibrated Polar Disks}},}\
  }\href {https://doi.org/10.1103/PhysRevLett.105.098001} {\bibfield  {journal}
  {\bibinfo  {journal} {Phys. Rev. Lett.}\ }\textbf {\bibinfo {volume}
  {105}}~(\bibinfo {number} {9}),\ \bibinfo {pages} {098001}}\BibitemShut
  {NoStop}%
\bibitem [{\citenamefont {{Di Leonardo}}\ \emph {et~al.}(2010)\citenamefont
  {{Di Leonardo}}, \citenamefont {Angelani}, \citenamefont {Dell'Arciprete},
  \citenamefont {Ruocco}, \citenamefont {Iebba}, \citenamefont {Schippa},
  \citenamefont {Conte}, \citenamefont {Mecarini}, \citenamefont {{De
  Angelis}},\ and\ \citenamefont {{Di Fabrizio}}}]{di2010bacterial}%
  \BibitemOpen
  \bibfield  {author} {\bibinfo {author} {\bibnamefont {{Di Leonardo}},
  \bibfnamefont {R}}, \bibinfo {author} {\bibfnamefont {L.}~\bibnamefont
  {Angelani}}, \bibinfo {author} {\bibfnamefont {D.}~\bibnamefont
  {Dell'Arciprete}}, \bibinfo {author} {\bibfnamefont {G.}~\bibnamefont
  {Ruocco}}, \bibinfo {author} {\bibfnamefont {V.}~\bibnamefont {Iebba}},
  \bibinfo {author} {\bibfnamefont {S.}~\bibnamefont {Schippa}}, \bibinfo
  {author} {\bibfnamefont {M.~P.}\ \bibnamefont {Conte}}, \bibinfo {author}
  {\bibfnamefont {F.}~\bibnamefont {Mecarini}}, \bibinfo {author}
  {\bibfnamefont {F.}~\bibnamefont {{De Angelis}}}, and\ \bibinfo {author}
  {\bibfnamefont {E.}~\bibnamefont {{Di Fabrizio}}}} (\bibinfo {year} {2010}),\
  \bibfield  {title} {\enquote {\bibinfo {title} {{Bacterial ratchet
  motors}},}\ }\href {https://doi.org/10.1073/pnas.0910426107} {\bibfield
  {journal} {\bibinfo  {journal} {Proc. Natl. Acad. Sci.}\ }\textbf {\bibinfo
  {volume} {107}}~(\bibinfo {number} {21}),\ \bibinfo {pages}
  {9541--9545}}\BibitemShut {NoStop}%
\bibitem [{\citenamefont {Digregorio}\ \emph {et~al.}(2018)\citenamefont
  {Digregorio}, \citenamefont {Levis}, \citenamefont {Suma}, \citenamefont
  {Cugliandolo}, \citenamefont {Gonnella},\ and\ \citenamefont
  {Pagonabarraga}}]{digregorio_full_2018}%
  \BibitemOpen
  \bibfield  {author} {\bibinfo {author} {\bibnamefont {Digregorio},
  \bibfnamefont {Pasquale}}, \bibinfo {author} {\bibfnamefont {Demian}\
  \bibnamefont {Levis}}, \bibinfo {author} {\bibfnamefont {Antonio}\
  \bibnamefont {Suma}}, \bibinfo {author} {\bibfnamefont {Leticia~F.}\
  \bibnamefont {Cugliandolo}}, \bibinfo {author} {\bibfnamefont {Giuseppe}\
  \bibnamefont {Gonnella}}, and\ \bibinfo {author} {\bibfnamefont {Ignacio}\
  \bibnamefont {Pagonabarraga}}} (\bibinfo {year} {2018}),\ \bibfield  {title}
  {\enquote {\bibinfo {title} {{Full Phase Diagram of Active Brownian Disks:
  From Melting to Motility-Induced Phase Separation}},}\ }\href
  {https://doi.org/10.1103/PhysRevLett.121.098003} {\bibfield  {journal}
  {\bibinfo  {journal} {Phys. Rev. Lett.}\ }\textbf {\bibinfo {volume}
  {121}}~(\bibinfo {number} {9}),\ \bibinfo {pages} {098003}}\BibitemShut
  {NoStop}%
\bibitem [{\citenamefont {Digregorio}\ \emph {et~al.}(2019)\citenamefont
  {Digregorio}, \citenamefont {Levis}, \citenamefont {Suma}, \citenamefont
  {Cugliandolo}, \citenamefont {Gonnella},\ and\ \citenamefont
  {Pagonabarraga}}]{digregorio20192d}%
  \BibitemOpen
  \bibfield  {author} {\bibinfo {author} {\bibnamefont {Digregorio},
  \bibfnamefont {Pasquale}}, \bibinfo {author} {\bibfnamefont {Demian}\
  \bibnamefont {Levis}}, \bibinfo {author} {\bibfnamefont {Antonio}\
  \bibnamefont {Suma}}, \bibinfo {author} {\bibfnamefont {Leticia~F.}\
  \bibnamefont {Cugliandolo}}, \bibinfo {author} {\bibfnamefont {Giuseppe}\
  \bibnamefont {Gonnella}}, and\ \bibinfo {author} {\bibfnamefont {Ignacio}\
  \bibnamefont {Pagonabarraga}}} (\bibinfo {year} {2019}),\ \bibfield  {title}
  {\enquote {\bibinfo {title} {{2D melting and motility induced phase
  separation in Active Brownian Hard Disks and Dumbbells}},}\ }in\ \href
  {https://doi.org/10.1088/1742-6596/1163/1/012073} {\emph {\bibinfo
  {booktitle} {Journal of Physics: Conference Series}}},\ Vol.\ \bibinfo
  {volume} {1163}\ (\bibinfo {organization} {IOP Publishing})\ p.\ \bibinfo
  {pages} {012073}\BibitemShut {NoStop}%
\bibitem [{\citenamefont {Dinelli}\ \emph {et~al.}(2022)\citenamefont
  {Dinelli}, \citenamefont {O'Byrne}, \citenamefont {Curatolo}, \citenamefont
  {Zhao}, \citenamefont {Sollich},\ and\ \citenamefont
  {Tailleur}}]{dinelli_non-reciprocity_2022}%
  \BibitemOpen
  \bibfield  {author} {\bibinfo {author} {\bibnamefont {Dinelli}, \bibfnamefont
  {Alberto}}, \bibinfo {author} {\bibfnamefont {Jérémy}\ \bibnamefont
  {O'Byrne}}, \bibinfo {author} {\bibfnamefont {Agnese}\ \bibnamefont
  {Curatolo}}, \bibinfo {author} {\bibfnamefont {Yongfeng}\ \bibnamefont
  {Zhao}}, \bibinfo {author} {\bibfnamefont {Peter}\ \bibnamefont {Sollich}},
  and\ \bibinfo {author} {\bibfnamefont {Julien}\ \bibnamefont {Tailleur}}}
  (\bibinfo {year} {2022}),\ \href@noop {} {\enquote {\bibinfo {title}
  {Non-reciprocity across scales in active mixtures},}\ }\Eprint
  {https://arxiv.org/abs/2203.07757} {arXiv:2203.07757 [cond-mat.stat-mech]}
  \BibitemShut {NoStop}%
\bibitem [{\citenamefont {Dittrich}\ \emph {et~al.}(2021)\citenamefont
  {Dittrich}, \citenamefont {Speck},\ and\ \citenamefont
  {Virnau}}]{dittrich2021critical}%
  \BibitemOpen
  \bibfield  {author} {\bibinfo {author} {\bibnamefont {Dittrich},
  \bibfnamefont {Florian}}, \bibinfo {author} {\bibfnamefont {Thomas}\
  \bibnamefont {Speck}}, and\ \bibinfo {author} {\bibfnamefont {Peter}\
  \bibnamefont {Virnau}}} (\bibinfo {year} {2021}),\ \bibfield  {title}
  {\enquote {\bibinfo {title} {{Critical behavior in active lattice models of
  motility-induced phase separation}},}\ }\href
  {https://doi.org/10.1140/epje/s10189-021-00058-1} {\bibfield  {journal}
  {\bibinfo  {journal} {Eur. Phys. J. E}\ }\textbf {\bibinfo {volume}
  {44}}~(\bibinfo {number} {4}),\ \bibinfo {pages} {53}}\BibitemShut {NoStop}%
\bibitem [{\citenamefont {Dolai}\ \emph {et~al.}(2022)\citenamefont {Dolai},
  \citenamefont {Krekels},\ and\ \citenamefont {Maes}}]{Dolai2022}%
  \BibitemOpen
  \bibfield  {author} {\bibinfo {author} {\bibnamefont {Dolai}, \bibfnamefont
  {Pritha}}, \bibinfo {author} {\bibfnamefont {Simon}\ \bibnamefont {Krekels}},
  and\ \bibinfo {author} {\bibfnamefont {Christian}\ \bibnamefont {Maes}}}
  (\bibinfo {year} {2022}),\ \bibfield  {title} {\enquote {\bibinfo {title}
  {{Inducing a bound state between active particles}},}\ }\href
  {https://doi.org/10.1103/PhysRevE.105.044605} {\bibfield  {journal} {\bibinfo
   {journal} {Phys. Rev. E}\ }\textbf {\bibinfo {volume} {105}}~(\bibinfo
  {number} {4}),\ \bibinfo {pages} {044605}}\BibitemShut {NoStop}%
\bibitem [{\citenamefont {Duan}\ \emph {et~al.}(2021)\citenamefont {Duan},
  \citenamefont {Mahault}, \citenamefont {Ma}, \citenamefont {Shi},\ and\
  \citenamefont {Chat{\'{e}}}}]{duan2021breakdown}%
  \BibitemOpen
  \bibfield  {author} {\bibinfo {author} {\bibnamefont {Duan}, \bibfnamefont
  {Yu}}, \bibinfo {author} {\bibfnamefont {Beno{\^{i}}t}\ \bibnamefont
  {Mahault}}, \bibinfo {author} {\bibfnamefont {Yu-qiang}\ \bibnamefont {Ma}},
  \bibinfo {author} {\bibfnamefont {Xia-qing}\ \bibnamefont {Shi}}, and\
  \bibinfo {author} {\bibfnamefont {Hugues}\ \bibnamefont {Chat{\'{e}}}}}
  (\bibinfo {year} {2021}),\ \bibfield  {title} {\enquote {\bibinfo {title}
  {{Breakdown of Ergodicity and Self-Averaging in Polar Flocks with Quenched
  Disorder}},}\ }\href {https://doi.org/10.1103/PhysRevLett.126.178001}
  {\bibfield  {journal} {\bibinfo  {journal} {Phys. Rev. Lett.}\ }\textbf
  {\bibinfo {volume} {126}}~(\bibinfo {number} {17}),\ \bibinfo {pages}
  {178001}}\BibitemShut {NoStop}%
\bibitem [{\citenamefont {Engebrecht}\ and\ \citenamefont
  {Silverman}(1984)}]{engebrecht1984lux}%
  \BibitemOpen
  \bibfield  {author} {\bibinfo {author} {\bibnamefont {Engebrecht},
  \bibfnamefont {Joanne}}, and\ \bibinfo {author} {\bibfnamefont {Michael}\
  \bibnamefont {Silverman}}} (\bibinfo {year} {1984}),\ \bibfield  {title}
  {\enquote {\bibinfo {title} {{Identification of genes and gene products
  necessary for bacterial bioluminescence.}}}\ }\href
  {https://doi.org/10.1073/pnas.81.13.4154} {\bibfield  {journal} {\bibinfo
  {journal} {Proc. Natl. Acad. Sci.}\ }\textbf {\bibinfo {volume}
  {81}}~(\bibinfo {number} {13}),\ \bibinfo {pages} {4154--4158}}\BibitemShut
  {NoStop}%
\bibitem [{\citenamefont {Falasco}\ \emph {et~al.}(2016)\citenamefont
  {Falasco}, \citenamefont {Baldovin}, \citenamefont {Kroy},\ and\
  \citenamefont {Baiesi}}]{Falasco:2016:NJP}%
  \BibitemOpen
  \bibfield  {author} {\bibinfo {author} {\bibnamefont {Falasco}, \bibfnamefont
  {G}}, \bibinfo {author} {\bibfnamefont {F.}~\bibnamefont {Baldovin}},
  \bibinfo {author} {\bibfnamefont {K.}~\bibnamefont {Kroy}}, and\ \bibinfo
  {author} {\bibfnamefont {M.}~\bibnamefont {Baiesi}}} (\bibinfo {year}
  {2016}),\ \bibfield  {title} {\enquote {\bibinfo {title} {{Mesoscopic virial
  equation for nonequilibrium statistical mechanics}},}\ }\href
  {https://doi.org/10.1088/1367-2630/18/9/093043} {\bibfield  {journal}
  {\bibinfo  {journal} {New J. Phys.}\ }\textbf {\bibinfo {volume}
  {18}}~(\bibinfo {number} {9}),\ \bibinfo {pages} {093043}}\BibitemShut
  {NoStop}%
\bibitem [{\citenamefont {Fausti}\ \emph {et~al.}(2021)\citenamefont {Fausti},
  \citenamefont {Tjhung}, \citenamefont {Cates},\ and\ \citenamefont
  {Nardini}}]{fausti2021capillary}%
  \BibitemOpen
  \bibfield  {author} {\bibinfo {author} {\bibnamefont {Fausti}, \bibfnamefont
  {G}}, \bibinfo {author} {\bibfnamefont {E.}~\bibnamefont {Tjhung}}, \bibinfo
  {author} {\bibfnamefont {M.~E.}\ \bibnamefont {Cates}}, and\ \bibinfo
  {author} {\bibfnamefont {C.}~\bibnamefont {Nardini}}} (\bibinfo {year}
  {2021}),\ \bibfield  {title} {\enquote {\bibinfo {title} {{Capillary
  Interfacial Tension in Active Phase Separation}},}\ }\href
  {https://doi.org/10.1103/PhysRevLett.127.068001} {\bibfield  {journal}
  {\bibinfo  {journal} {Phys. Rev. Lett.}\ }\textbf {\bibinfo {volume}
  {127}}~(\bibinfo {number} {6}),\ \bibinfo {pages} {068001}}\BibitemShut
  {NoStop}%
\bibitem [{\citenamefont {Feng}\ and\ \citenamefont {Hou}(2023)}]{Feng2022}%
  \BibitemOpen
  \bibfield  {author} {\bibinfo {author} {\bibnamefont {Feng}, \bibfnamefont
  {Mengkai}}, and\ \bibinfo {author} {\bibfnamefont {Zhonghuai}\ \bibnamefont
  {Hou}}} (\bibinfo {year} {2023}),\ \bibfield  {title} {\enquote {\bibinfo
  {title} {{Unraveling on kinesin acceleration in intracellular environments: A
  theory for active bath}},}\ }\href
  {https://doi.org/10.1103/PhysRevResearch.5.013206} {\bibfield  {journal}
  {\bibinfo  {journal} {Phys. Rev. Res.}\ }\textbf {\bibinfo {volume}
  {5}}~(\bibinfo {number} {1}),\ \bibinfo {pages} {013206}}\BibitemShut
  {NoStop}%
\bibitem [{\citenamefont {Fily}\ \emph {et~al.}(2014)\citenamefont {Fily},
  \citenamefont {Baskaran},\ and\ \citenamefont {Hagan}}]{fily2014dynamicsSM}%
  \BibitemOpen
  \bibfield  {author} {\bibinfo {author} {\bibnamefont {Fily}, \bibfnamefont
  {Yaouen}}, \bibinfo {author} {\bibfnamefont {Aparna}\ \bibnamefont
  {Baskaran}}, and\ \bibinfo {author} {\bibfnamefont {Michael~F.}\ \bibnamefont
  {Hagan}}} (\bibinfo {year} {2014}),\ \bibfield  {title} {\enquote {\bibinfo
  {title} {{Dynamics of self-propelled particles under strong confinement}},}\
  }\href {https://doi.org/10.1039/C4SM00975D} {\bibfield  {journal} {\bibinfo
  {journal} {Soft Matter}\ }\textbf {\bibinfo {volume} {10}}~(\bibinfo {number}
  {30}),\ \bibinfo {pages} {5609--5617}}\BibitemShut {NoStop}%
\bibitem [{\citenamefont {Fily}\ \emph {et~al.}(2015)\citenamefont {Fily},
  \citenamefont {Baskaran},\ and\ \citenamefont {Hagan}}]{fily2015dynamics}%
  \BibitemOpen
  \bibfield  {author} {\bibinfo {author} {\bibnamefont {Fily}, \bibfnamefont
  {Yaouen}}, \bibinfo {author} {\bibfnamefont {Aparna}\ \bibnamefont
  {Baskaran}}, and\ \bibinfo {author} {\bibfnamefont {Michael~F.}\ \bibnamefont
  {Hagan}}} (\bibinfo {year} {2015}),\ \bibfield  {title} {\enquote {\bibinfo
  {title} {{Dynamics and density distribution of strongly confined
  noninteracting nonaligning self-propelled particles in a nonconvex
  boundary}},}\ }\href {https://doi.org/10.1103/PhysRevE.91.012125} {\bibfield
  {journal} {\bibinfo  {journal} {Phys. Rev. E}\ }\textbf {\bibinfo {volume}
  {91}}~(\bibinfo {number} {1}),\ \bibinfo {pages} {012125}}\BibitemShut
  {NoStop}%
\bibitem [{\citenamefont {Fily}\ and\ \citenamefont
  {Marchetti}(2012)}]{fily_athermal_2012}%
  \BibitemOpen
  \bibfield  {author} {\bibinfo {author} {\bibnamefont {Fily}, \bibfnamefont
  {Yaouen}}, and\ \bibinfo {author} {\bibfnamefont {M.~Cristina}\ \bibnamefont
  {Marchetti}}} (\bibinfo {year} {2012}),\ \bibfield  {title} {\enquote
  {\bibinfo {title} {{Athermal Phase Separation of Self-Propelled Particles
  with No Alignment}},}\ }\href
  {https://doi.org/10.1103/PhysRevLett.108.235702} {\bibfield  {journal}
  {\bibinfo  {journal} {Phys. Rev. Lett.}\ }\textbf {\bibinfo {volume}
  {108}}~(\bibinfo {number} {23}),\ \bibinfo {pages} {235702}}\BibitemShut
  {NoStop}%
\bibitem [{\citenamefont {Fisher}(1984)}]{fisher_random_1984}%
  \BibitemOpen
  \bibfield  {author} {\bibinfo {author} {\bibnamefont {Fisher}, \bibfnamefont
  {Daniel~S}}} (\bibinfo {year} {1984}),\ \bibfield  {title} {\enquote
  {\bibinfo {title} {{Random walks in random environments}},}\ }\href
  {https://doi.org/10.1103/PhysRevA.30.960} {\bibfield  {journal} {\bibinfo
  {journal} {Phys. Rev. A}\ }\textbf {\bibinfo {volume} {30}}~(\bibinfo
  {number} {2}),\ \bibinfo {pages} {960--964}}\BibitemShut {NoStop}%
\bibitem [{\citenamefont {Fisher}\ \emph {et~al.}(1984)\citenamefont {Fisher},
  \citenamefont {Fr{\"{o}}hlich},\ and\ \citenamefont
  {Spencer}}]{fisher_ising_1984}%
  \BibitemOpen
  \bibfield  {author} {\bibinfo {author} {\bibnamefont {Fisher}, \bibfnamefont
  {Daniel~S}}, \bibinfo {author} {\bibfnamefont {J{\"{u}}rg}\ \bibnamefont
  {Fr{\"{o}}hlich}}, and\ \bibinfo {author} {\bibfnamefont {Thomas}\
  \bibnamefont {Spencer}}} (\bibinfo {year} {1984}),\ \bibfield  {title}
  {\enquote {\bibinfo {title} {{The Ising model in a random magnetic field}},}\
  }\href {https://doi.org/10.1007/BF01009445} {\bibfield  {journal} {\bibinfo
  {journal} {J. Stat. Phys.}\ }\textbf {\bibinfo {volume} {34}}~(\bibinfo
  {number} {5-6}),\ \bibinfo {pages} {863--870}}\BibitemShut {NoStop}%
\bibitem [{\citenamefont {Fodor}\ and\ \citenamefont
  {Marchetti}(2018)}]{fodor2018statistical}%
  \BibitemOpen
  \bibfield  {author} {\bibinfo {author} {\bibnamefont {Fodor}, \bibfnamefont
  {{\'{E}}tienne}}, and\ \bibinfo {author} {\bibfnamefont {Cristina}\
  \bibnamefont {Marchetti}}} (\bibinfo {year} {2018}),\ \bibfield  {title}
  {\enquote {\bibinfo {title} {{The statistical physics of active matter: From
  self-catalytic colloids to living cells}},}\ }\href
  {https://doi.org/10.1016/j.physa.2017.12.137} {\bibfield  {journal} {\bibinfo
   {journal} {Phys. A Stat. Mech. its Appl.}\ }\textbf {\bibinfo {volume}
  {504}},\ \bibinfo {pages} {106--120}}\BibitemShut {NoStop}%
\bibitem [{\citenamefont {Fodor}\ \emph {et~al.}(2016)\citenamefont {Fodor},
  \citenamefont {Nardini}, \citenamefont {Cates}, \citenamefont {Tailleur},
  \citenamefont {Visco},\ and\ \citenamefont {van Wijland}}]{fodor_how_2016}%
  \BibitemOpen
  \bibfield  {author} {\bibinfo {author} {\bibnamefont {Fodor}, \bibfnamefont
  {{\'{E}}tienne}}, \bibinfo {author} {\bibfnamefont {Cesare}\ \bibnamefont
  {Nardini}}, \bibinfo {author} {\bibfnamefont {Michael~E.}\ \bibnamefont
  {Cates}}, \bibinfo {author} {\bibfnamefont {Julien}\ \bibnamefont
  {Tailleur}}, \bibinfo {author} {\bibfnamefont {Paolo}\ \bibnamefont {Visco}},
  and\ \bibinfo {author} {\bibfnamefont {Fr{\'{e}}d{\'{e}}ric}\ \bibnamefont
  {van Wijland}}} (\bibinfo {year} {2016}),\ \bibfield  {title} {\enquote
  {\bibinfo {title} {{How Far from Equilibrium Is Active Matter?}}}\ }\href
  {https://doi.org/10.1103/PhysRevLett.117.038103} {\bibfield  {journal}
  {\bibinfo  {journal} {Phys. Rev. Lett.}\ }\textbf {\bibinfo {volume}
  {117}}~(\bibinfo {number} {3}),\ \bibinfo {pages} {038103}}\BibitemShut
  {NoStop}%
\bibitem [{\citenamefont {Fruchart}\ \emph {et~al.}(2021)\citenamefont
  {Fruchart}, \citenamefont {Hanai}, \citenamefont {Littlewood},\ and\
  \citenamefont {Vitelli}}]{Fruchart2021}%
  \BibitemOpen
  \bibfield  {author} {\bibinfo {author} {\bibnamefont {Fruchart},
  \bibfnamefont {Michel}}, \bibinfo {author} {\bibfnamefont {Ryo}\ \bibnamefont
  {Hanai}}, \bibinfo {author} {\bibfnamefont {Peter~B.}\ \bibnamefont
  {Littlewood}}, and\ \bibinfo {author} {\bibfnamefont {Vincenzo}\ \bibnamefont
  {Vitelli}}} (\bibinfo {year} {2021}),\ \bibfield  {title} {\enquote {\bibinfo
  {title} {{Non-reciprocal phase transitions}},}\ }\href
  {https://doi.org/10.1038/s41586-021-03375-9} {\bibfield  {journal} {\bibinfo
  {journal} {Nature}\ }\textbf {\bibinfo {volume} {592}}~(\bibinfo {number}
  {7854}),\ \bibinfo {pages} {363--369}}\BibitemShut {NoStop}%
\bibitem [{\citenamefont {Fuqua}\ \emph {et~al.}(1994)\citenamefont {Fuqua},
  \citenamefont {Winans},\ and\ \citenamefont {Greenberg}}]{fuqua1994quorum}%
  \BibitemOpen
  \bibfield  {author} {\bibinfo {author} {\bibnamefont {Fuqua}, \bibfnamefont
  {W~C}}, \bibinfo {author} {\bibfnamefont {S.~C.}\ \bibnamefont {Winans}},
  and\ \bibinfo {author} {\bibfnamefont {E.~P.}\ \bibnamefont {Greenberg}}}
  (\bibinfo {year} {1994}),\ \bibfield  {title} {\enquote {\bibinfo {title}
  {{Quorum sensing in bacteria: the LuxR-LuxI family of cell density-responsive
  transcriptional regulators}},}\ }\href
  {https://doi.org/10.1128/jb.176.2.269-275.1994} {\bibfield  {journal}
  {\bibinfo  {journal} {J. Bacteriol.}\ }\textbf {\bibinfo {volume}
  {176}}~(\bibinfo {number} {2}),\ \bibinfo {pages} {269--275}}\BibitemShut
  {NoStop}%
\bibitem [{\citenamefont {Galajda}\ \emph {et~al.}(2007)\citenamefont
  {Galajda}, \citenamefont {Keymer}, \citenamefont {Chaikin},\ and\
  \citenamefont {Austin}}]{galajda2007wall}%
  \BibitemOpen
  \bibfield  {author} {\bibinfo {author} {\bibnamefont {Galajda}, \bibfnamefont
  {Peter}}, \bibinfo {author} {\bibfnamefont {Juan}\ \bibnamefont {Keymer}},
  \bibinfo {author} {\bibfnamefont {Paul}\ \bibnamefont {Chaikin}}, and\
  \bibinfo {author} {\bibfnamefont {Robert}\ \bibnamefont {Austin}}} (\bibinfo
  {year} {2007}),\ \bibfield  {title} {\enquote {\bibinfo {title} {{A Wall of
  Funnels Concentrates Swimming Bacteria}},}\ }\href
  {https://doi.org/10.1128/JB.01033-07} {\bibfield  {journal} {\bibinfo
  {journal} {J. Bacteriol.}\ }\textbf {\bibinfo {volume} {189}}~(\bibinfo
  {number} {23}),\ \bibinfo {pages} {8704--8707}}\BibitemShut {NoStop}%
\bibitem [{\citenamefont {Garcia}\ \emph {et~al.}(2015)\citenamefont {Garcia},
  \citenamefont {Hannezo}, \citenamefont {Elgeti}, \citenamefont {Joanny},
  \citenamefont {Silberzan},\ and\ \citenamefont {Gov}}]{garcia_physics_2015}%
  \BibitemOpen
  \bibfield  {author} {\bibinfo {author} {\bibnamefont {Garcia}, \bibfnamefont
  {Simon}}, \bibinfo {author} {\bibfnamefont {Edouard}\ \bibnamefont
  {Hannezo}}, \bibinfo {author} {\bibfnamefont {Jens}\ \bibnamefont {Elgeti}},
  \bibinfo {author} {\bibfnamefont {Jean-Fran{\c{c}}ois}\ \bibnamefont
  {Joanny}}, \bibinfo {author} {\bibfnamefont {Pascal}\ \bibnamefont
  {Silberzan}}, and\ \bibinfo {author} {\bibfnamefont {Nir~S.}\ \bibnamefont
  {Gov}}} (\bibinfo {year} {2015}),\ \bibfield  {title} {\enquote {\bibinfo
  {title} {{Physics of active jamming during collective cellular motion in a
  monolayer}},}\ }\href {https://doi.org/10.1073/pnas.1510973112} {\bibfield
  {journal} {\bibinfo  {journal} {Proc. Natl. Acad. Sci.}\ }\textbf {\bibinfo
  {volume} {112}}~(\bibinfo {number} {50}),\ \bibinfo {pages}
  {15314--15319}}\BibitemShut {NoStop}%
\bibitem [{\citenamefont {Ginot}\ \emph {et~al.}(2018)\citenamefont {Ginot},
  \citenamefont {Solon}, \citenamefont {Kafri}, \citenamefont {Ybert},
  \citenamefont {Tailleur},\ and\ \citenamefont
  {Cottin-Bizonne}}]{ginot2018sedimentation}%
  \BibitemOpen
  \bibfield  {author} {\bibinfo {author} {\bibnamefont {Ginot}, \bibfnamefont
  {Felix}}, \bibinfo {author} {\bibfnamefont {Alexandre}\ \bibnamefont
  {Solon}}, \bibinfo {author} {\bibfnamefont {Yariv}\ \bibnamefont {Kafri}},
  \bibinfo {author} {\bibfnamefont {Christophe}\ \bibnamefont {Ybert}},
  \bibinfo {author} {\bibfnamefont {Julien}\ \bibnamefont {Tailleur}}, and\
  \bibinfo {author} {\bibfnamefont {Cecile}\ \bibnamefont {Cottin-Bizonne}}}
  (\bibinfo {year} {2018}),\ \bibfield  {title} {\enquote {\bibinfo {title}
  {{Sedimentation of self-propelled Janus colloids: polarization and
  pressure}},}\ }\href {https://doi.org/10.1088/1367-2630/aae732} {\bibfield
  {journal} {\bibinfo  {journal} {New J. Phys.}\ }\textbf {\bibinfo {volume}
  {20}}~(\bibinfo {number} {11}),\ \bibinfo {pages} {115001}}\BibitemShut
  {NoStop}%
\bibitem [{\citenamefont {Glaus}(1986)}]{glaus_correlations_1986}%
  \BibitemOpen
  \bibfield  {author} {\bibinfo {author} {\bibnamefont {Glaus}, \bibfnamefont
  {U}}} (\bibinfo {year} {1986}),\ \bibfield  {title} {\enquote {\bibinfo
  {title} {{Correlations in the two-dimensional random-field Ising model}},}\
  }\href {https://doi.org/10.1103/PhysRevB.34.3203} {\bibfield  {journal}
  {\bibinfo  {journal} {Phys. Rev. B}\ }\textbf {\bibinfo {volume}
  {34}}~(\bibinfo {number} {5}),\ \bibinfo {pages} {3203--3211}}\BibitemShut
  {NoStop}%
\bibitem [{\citenamefont {Granek}(2023)}]{Granek2023}%
  \BibitemOpen
  \bibfield  {author} {\bibinfo {author} {\bibnamefont {Granek}, \bibfnamefont
  {Omer}}} (\bibinfo {year} {2023}),\ \href@noop {} {\enquote {\bibinfo {title}
  {Universal fluctuations of local measurement in low-dimensional systems},}\
  }\Eprint {https://arxiv.org/abs/2308.08595} {arXiv:2308.08595
  [cond-mat.stat-mech]} \BibitemShut {NoStop}%
\bibitem [{\citenamefont {Granek}\ \emph {et~al.}(2020)\citenamefont {Granek},
  \citenamefont {Baek}, \citenamefont {Kafri},\ and\ \citenamefont
  {Solon}}]{granek2020bodies}%
  \BibitemOpen
  \bibfield  {author} {\bibinfo {author} {\bibnamefont {Granek}, \bibfnamefont
  {Omer}}, \bibinfo {author} {\bibfnamefont {Yongjoo}\ \bibnamefont {Baek}},
  \bibinfo {author} {\bibfnamefont {Yariv}\ \bibnamefont {Kafri}}, and\
  \bibinfo {author} {\bibfnamefont {Alexandre~P.}\ \bibnamefont {Solon}}}
  (\bibinfo {year} {2020}),\ \bibfield  {title} {\enquote {\bibinfo {title}
  {{Bodies in an interacting active fluid: far-field influence of a single body
  and interaction between two bodies}},}\ }\href
  {https://doi.org/10.1088/1742-5468/ab7f34} {\bibfield  {journal} {\bibinfo
  {journal} {J. Stat. Mech. Theory Exp.}\ }\textbf {\bibinfo {volume}
  {2020}}~(\bibinfo {number} {6}),\ \bibinfo {pages} {063211}}\BibitemShut
  {NoStop}%
\bibitem [{\citenamefont {Granek}\ \emph {et~al.}(2022)\citenamefont {Granek},
  \citenamefont {Kafri},\ and\ \citenamefont {Tailleur}}]{Granek2022}%
  \BibitemOpen
  \bibfield  {author} {\bibinfo {author} {\bibnamefont {Granek}, \bibfnamefont
  {Omer}}, \bibinfo {author} {\bibfnamefont {Yariv}\ \bibnamefont {Kafri}},
  and\ \bibinfo {author} {\bibfnamefont {Julien}\ \bibnamefont {Tailleur}}}
  (\bibinfo {year} {2022}),\ \bibfield  {title} {\enquote {\bibinfo {title}
  {{Anomalous Transport of Tracers in Active Baths}},}\ }\href
  {https://doi.org/10.1103/PhysRevLett.129.038001} {\bibfield  {journal}
  {\bibinfo  {journal} {Phys. Rev. Lett.}\ }\textbf {\bibinfo {volume}
  {129}}~(\bibinfo {number} {3}),\ \bibinfo {pages} {038001}}\BibitemShut
  {NoStop}%
\bibitem [{\citenamefont {Hammer}\ and\ \citenamefont
  {Bassler}(2003)}]{hammer_quorum_2003}%
  \BibitemOpen
  \bibfield  {author} {\bibinfo {author} {\bibnamefont {Hammer}, \bibfnamefont
  {Brian~K}}, and\ \bibinfo {author} {\bibfnamefont {Bonnie~L.}\ \bibnamefont
  {Bassler}}} (\bibinfo {year} {2003}),\ \bibfield  {title} {\enquote {\bibinfo
  {title} {{Quorum sensing controls biofilm formation in Vibrio cholerae}},}\
  }\href {https://doi.org/10.1046/j.1365-2958.2003.03688.x} {\bibfield
  {journal} {\bibinfo  {journal} {Mol. Microbiol.}\ }\textbf {\bibinfo {volume}
  {50}}~(\bibinfo {number} {1}),\ \bibinfo {pages} {101--104}}\BibitemShut
  {NoStop}%
\bibitem [{\citenamefont {H{\"{a}}nggi}\ and\ \citenamefont
  {Jung}(1994)}]{hanggi1994colored}%
  \BibitemOpen
  \bibfield  {author} {\bibinfo {author} {\bibnamefont {H{\"{a}}nggi},
  \bibfnamefont {Peter}}, and\ \bibinfo {author} {\bibfnamefont {Peter}\
  \bibnamefont {Jung}}} (\bibinfo {year} {1994}),\ \bibfield  {title} {\enquote
  {\bibinfo {title} {{Colored noise in dynamical systems}},}\ }\href@noop {}
  {\bibfield  {journal} {\bibinfo  {journal} {Adv. Chem. Phys.}\ }\textbf
  {\bibinfo {volume} {89}},\ \bibinfo {pages} {239--326}}\BibitemShut {NoStop}%
\bibitem [{\citenamefont {Hanna}\ \emph {et~al.}(1981)\citenamefont {Hanna},
  \citenamefont {Hess},\ and\ \citenamefont {Klein}}]{Hanna1981}%
  \BibitemOpen
  \bibfield  {author} {\bibinfo {author} {\bibnamefont {Hanna}, \bibfnamefont
  {S}}, \bibinfo {author} {\bibfnamefont {W.}~\bibnamefont {Hess}}, and\
  \bibinfo {author} {\bibfnamefont {R.}~\bibnamefont {Klein}}} (\bibinfo {year}
  {1981}),\ \bibfield  {title} {\enquote {\bibinfo {title} {{The velocity
  autocorrelation function of an overdamped Brownian system with hard-core
  intraction}},}\ }\href {https://doi.org/10.1088/0305-4470/14/12/004}
  {\bibfield  {journal} {\bibinfo  {journal} {J. Phys. A. Math. Gen.}\ }\textbf
  {\bibinfo {volume} {14}}~(\bibinfo {number} {12}),\ \bibinfo {pages}
  {L493--L498}}\BibitemShut {NoStop}%
\bibitem [{\citenamefont {Harder}\ \emph {et~al.}(2014)\citenamefont {Harder},
  \citenamefont {Valeriani},\ and\ \citenamefont {Cacciuto}}]{PolymerCollapse}%
  \BibitemOpen
  \bibfield  {author} {\bibinfo {author} {\bibnamefont {Harder}, \bibfnamefont
  {J}}, \bibinfo {author} {\bibfnamefont {C.}~\bibnamefont {Valeriani}}, and\
  \bibinfo {author} {\bibfnamefont {A.}~\bibnamefont {Cacciuto}}} (\bibinfo
  {year} {2014}),\ \bibfield  {title} {\enquote {\bibinfo {title}
  {{Activity-induced collapse and reexpansion of rigid polymers}},}\ }\href
  {https://doi.org/10.1103/PhysRevE.90.062312} {\bibfield  {journal} {\bibinfo
  {journal} {Phys. Rev. E}\ }\textbf {\bibinfo {volume} {90}}~(\bibinfo
  {number} {6}),\ \bibinfo {pages} {062312}}\BibitemShut {NoStop}%
\bibitem [{\citenamefont {Hennes}\ \emph {et~al.}(2014)\citenamefont {Hennes},
  \citenamefont {Wolff},\ and\ \citenamefont {Stark}}]{hennes2014self}%
  \BibitemOpen
  \bibfield  {author} {\bibinfo {author} {\bibnamefont {Hennes}, \bibfnamefont
  {Marc}}, \bibinfo {author} {\bibfnamefont {Katrin}\ \bibnamefont {Wolff}},
  and\ \bibinfo {author} {\bibfnamefont {Holger}\ \bibnamefont {Stark}}}
  (\bibinfo {year} {2014}),\ \bibfield  {title} {\enquote {\bibinfo {title}
  {{Self-Induced Polar Order of Active Brownian Particles in a Harmonic
  Trap}},}\ }\href {https://doi.org/10.1103/PhysRevLett.112.238104} {\bibfield
  {journal} {\bibinfo  {journal} {Phys. Rev. Lett.}\ }\textbf {\bibinfo
  {volume} {112}}~(\bibinfo {number} {23}),\ \bibinfo {pages}
  {238104}}\BibitemShut {NoStop}%
\bibitem [{\citenamefont {Hermann}\ \emph {et~al.}(2019)\citenamefont
  {Hermann}, \citenamefont {de~las Heras},\ and\ \citenamefont
  {Schmidt}}]{hermann2019non}%
  \BibitemOpen
  \bibfield  {author} {\bibinfo {author} {\bibnamefont {Hermann}, \bibfnamefont
  {Sophie}}, \bibinfo {author} {\bibfnamefont {Daniel}\ \bibnamefont {de~las
  Heras}}, and\ \bibinfo {author} {\bibfnamefont {Matthias}\ \bibnamefont
  {Schmidt}}} (\bibinfo {year} {2019}),\ \bibfield  {title} {\enquote {\bibinfo
  {title} {{Non-negative Interfacial Tension in Phase-Separated Active Brownian
  Particles}},}\ }\href {https://doi.org/10.1103/PhysRevLett.123.268002}
  {\bibfield  {journal} {\bibinfo  {journal} {Phys. Rev. Lett.}\ }\textbf
  {\bibinfo {volume} {123}}~(\bibinfo {number} {26}),\ \bibinfo {pages}
  {268002}}\BibitemShut {NoStop}%
\bibitem [{\citenamefont {Howse}\ \emph {et~al.}(2007)\citenamefont {Howse},
  \citenamefont {Jones}, \citenamefont {Ryan}, \citenamefont {Gough},
  \citenamefont {Vafabakhsh},\ and\ \citenamefont
  {Golestanian}}]{howse2007self}%
  \BibitemOpen
  \bibfield  {author} {\bibinfo {author} {\bibnamefont {Howse}, \bibfnamefont
  {Jonathan~R}}, \bibinfo {author} {\bibfnamefont {Richard A.~L.}\ \bibnamefont
  {Jones}}, \bibinfo {author} {\bibfnamefont {Anthony~J.}\ \bibnamefont
  {Ryan}}, \bibinfo {author} {\bibfnamefont {Tim}\ \bibnamefont {Gough}},
  \bibinfo {author} {\bibfnamefont {Reza}\ \bibnamefont {Vafabakhsh}}, and\
  \bibinfo {author} {\bibfnamefont {Ramin}\ \bibnamefont {Golestanian}}}
  (\bibinfo {year} {2007}),\ \bibfield  {title} {\enquote {\bibinfo {title}
  {{Self-Motile Colloidal Particles: From Directed Propulsion to Random
  Walk}},}\ }\href {https://doi.org/10.1103/PhysRevLett.99.048102} {\bibfield
  {journal} {\bibinfo  {journal} {Phys. Rev. Lett.}\ }\textbf {\bibinfo
  {volume} {99}}~(\bibinfo {number} {4}),\ \bibinfo {pages}
  {048102}}\BibitemShut {NoStop}%
\bibitem [{\citenamefont {Ilker}\ and\ \citenamefont
  {Joanny}(2020)}]{ilker_phase_2020}%
  \BibitemOpen
  \bibfield  {author} {\bibinfo {author} {\bibnamefont {Ilker}, \bibfnamefont
  {Efe}}, and\ \bibinfo {author} {\bibfnamefont {Jean-Fran{\c{c}}ois}\
  \bibnamefont {Joanny}}} (\bibinfo {year} {2020}),\ \bibfield  {title}
  {\enquote {\bibinfo {title} {{Phase separation and nucleation in mixtures of
  particles with different temperatures}},}\ }\href
  {https://doi.org/10.1103/PhysRevResearch.2.023200} {\bibfield  {journal}
  {\bibinfo  {journal} {Phys. Rev. Res.}\ }\textbf {\bibinfo {volume}
  {2}}~(\bibinfo {number} {2}),\ \bibinfo {pages} {023200}}\BibitemShut
  {NoStop}%
\bibitem [{\citenamefont {Imbrie}(1984)}]{imbrie_lower_1984}%
  \BibitemOpen
  \bibfield  {author} {\bibinfo {author} {\bibnamefont {Imbrie}, \bibfnamefont
  {John~Z}}} (\bibinfo {year} {1984}),\ \bibfield  {title} {\enquote {\bibinfo
  {title} {{Lower Critical Dimension of the Random-Field Ising Model}},}\
  }\href {https://doi.org/10.1103/PhysRevLett.53.1747} {\bibfield  {journal}
  {\bibinfo  {journal} {Phys. Rev. Lett.}\ }\textbf {\bibinfo {volume}
  {53}}~(\bibinfo {number} {18}),\ \bibinfo {pages} {1747--1750}}\BibitemShut
  {NoStop}%
\bibitem [{\citenamefont {Imry}\ and\ \citenamefont
  {Ma}(1975)}]{imry_random-field_1975}%
  \BibitemOpen
  \bibfield  {author} {\bibinfo {author} {\bibnamefont {Imry}, \bibfnamefont
  {Yoseph}}, and\ \bibinfo {author} {\bibfnamefont {Shang-keng}\ \bibnamefont
  {Ma}}} (\bibinfo {year} {1975}),\ \bibfield  {title} {\enquote {\bibinfo
  {title} {{Random-Field Instability of the Ordered State of Continuous
  Symmetry}},}\ }\href {https://doi.org/10.1103/PhysRevLett.35.1399} {\bibfield
   {journal} {\bibinfo  {journal} {Phys. Rev. Lett.}\ }\textbf {\bibinfo
  {volume} {35}}~(\bibinfo {number} {21}),\ \bibinfo {pages}
  {1399--1401}}\BibitemShut {NoStop}%
\bibitem [{\citenamefont {Jayaram}\ and\ \citenamefont
  {Speck}(2023)}]{Jayaram2023}%
  \BibitemOpen
  \bibfield  {author} {\bibinfo {author} {\bibnamefont {Jayaram}, \bibfnamefont
  {Ashreya}}, and\ \bibinfo {author} {\bibfnamefont {Thomas}\ \bibnamefont
  {Speck}}} (\bibinfo {year} {2023}),\ \bibfield  {title} {\enquote {\bibinfo
  {title} {{Effective dynamics and fluctuations of a trapped probe moving in a
  fluid of active hard discs}},}\ }\href
  {https://doi.org/10.1209/0295-5075/acdf1a} {\bibfield  {journal} {\bibinfo
  {journal} {EPL}\ }\textbf {\bibinfo {volume} {143}}~(\bibinfo {number} {1}),\
  \bibinfo {pages} {17005}}\BibitemShut {NoStop}%
\bibitem [{\citenamefont {Junot}\ \emph {et~al.}(2017)\citenamefont {Junot},
  \citenamefont {Briand}, \citenamefont {Ledesma-Alonso},\ and\ \citenamefont
  {Dauchot}}]{junot2017active}%
  \BibitemOpen
  \bibfield  {author} {\bibinfo {author} {\bibnamefont {Junot}, \bibfnamefont
  {G}}, \bibinfo {author} {\bibfnamefont {G.}~\bibnamefont {Briand}}, \bibinfo
  {author} {\bibfnamefont {R.}~\bibnamefont {Ledesma-Alonso}}, and\ \bibinfo
  {author} {\bibfnamefont {O.}~\bibnamefont {Dauchot}}} (\bibinfo {year}
  {2017}),\ \bibfield  {title} {\enquote {\bibinfo {title} {{Active versus
  Passive Hard Disks against a Membrane: Mechanical Pressure and
  Instability}},}\ }\href {https://doi.org/10.1103/PhysRevLett.119.028002}
  {\bibfield  {journal} {\bibinfo  {journal} {Phys. Rev. Lett.}\ }\textbf
  {\bibinfo {volume} {119}}~(\bibinfo {number} {2}),\ \bibinfo {pages}
  {028002}}\BibitemShut {NoStop}%
\bibitem [{\citenamefont {Kaiser}\ \emph {et~al.}(2012)\citenamefont {Kaiser},
  \citenamefont {Wensink},\ and\ \citenamefont {L{\"{o}}wen}}]{Kaiser2012}%
  \BibitemOpen
  \bibfield  {author} {\bibinfo {author} {\bibnamefont {Kaiser}, \bibfnamefont
  {A}}, \bibinfo {author} {\bibfnamefont {H.~H.}\ \bibnamefont {Wensink}}, and\
  \bibinfo {author} {\bibfnamefont {H}~\bibnamefont {L{\"{o}}wen}}} (\bibinfo
  {year} {2012}),\ \bibfield  {title} {\enquote {\bibinfo {title} {{How to
  Capture Active Particles}},}\ }\href
  {https://doi.org/10.1103/PhysRevLett.108.268307} {\bibfield  {journal}
  {\bibinfo  {journal} {Phys. Rev. Lett.}\ }\textbf {\bibinfo {volume}
  {108}}~(\bibinfo {number} {26}),\ \bibinfo {pages} {268307}}\BibitemShut
  {NoStop}%
\bibitem [{\citenamefont {Kaiser}\ \emph {et~al.}(2014)\citenamefont {Kaiser},
  \citenamefont {Peshkov}, \citenamefont {Sokolov}, \citenamefont {ten Hagen},
  \citenamefont {L{\"{o}}wen},\ and\ \citenamefont {Aranson}}]{Kaiser2014}%
  \BibitemOpen
  \bibfield  {author} {\bibinfo {author} {\bibnamefont {Kaiser}, \bibfnamefont
  {Andreas}}, \bibinfo {author} {\bibfnamefont {Anton}\ \bibnamefont
  {Peshkov}}, \bibinfo {author} {\bibfnamefont {Andrey}\ \bibnamefont
  {Sokolov}}, \bibinfo {author} {\bibfnamefont {Borge}\ \bibnamefont {ten
  Hagen}}, \bibinfo {author} {\bibfnamefont {Hartmut}\ \bibnamefont
  {L{\"{o}}wen}}, and\ \bibinfo {author} {\bibfnamefont {Igor~S.}\ \bibnamefont
  {Aranson}}} (\bibinfo {year} {2014}),\ \bibfield  {title} {\enquote {\bibinfo
  {title} {{Transport Powered by Bacterial Turbulence}},}\ }\href
  {https://doi.org/10.1103/PhysRevLett.112.158101} {\bibfield  {journal}
  {\bibinfo  {journal} {Phys. Rev. Lett.}\ }\textbf {\bibinfo {volume}
  {112}}~(\bibinfo {number} {15}),\ \bibinfo {pages} {158101}}\BibitemShut
  {NoStop}%
\bibitem [{\citenamefont {Kanazawa}\ \emph {et~al.}(2020)\citenamefont
  {Kanazawa}, \citenamefont {Sano}, \citenamefont {Cairoli},\ and\
  \citenamefont {Baule}}]{Kanazawa2020}%
  \BibitemOpen
  \bibfield  {author} {\bibinfo {author} {\bibnamefont {Kanazawa},
  \bibfnamefont {Kiyoshi}}, \bibinfo {author} {\bibfnamefont {Tomohiko~G.}\
  \bibnamefont {Sano}}, \bibinfo {author} {\bibfnamefont {Andrea}\ \bibnamefont
  {Cairoli}}, and\ \bibinfo {author} {\bibfnamefont {Adrian}\ \bibnamefont
  {Baule}}} (\bibinfo {year} {2020}),\ \bibfield  {title} {\enquote {\bibinfo
  {title} {{Loopy L{\'{e}}vy flights enhance tracer diffusion in active
  suspensions}},}\ }\href {https://doi.org/10.1038/s41586-020-2086-2}
  {\bibfield  {journal} {\bibinfo  {journal} {Nature}\ }\textbf {\bibinfo
  {volume} {579}}~(\bibinfo {number} {7799}),\ \bibinfo {pages}
  {364--367}}\BibitemShut {NoStop}%
\bibitem [{\citenamefont {Kardar}(2007)}]{kardar_2007}%
  \BibitemOpen
  \bibfield  {author} {\bibinfo {author} {\bibnamefont {Kardar}, \bibfnamefont
  {Mehran}}} (\bibinfo {year} {2007}),\ \href
  {https://doi.org/10.1017/CBO9780511815881} {\emph {\bibinfo {title}
  {{Statistical Physics of Fields}}}}\ (\bibinfo  {publisher} {Cambridge
  University Press})\BibitemShut {NoStop}%
\bibitem [{\citenamefont {Kim}\ \emph {et~al.}(2023)\citenamefont {Kim},
  \citenamefont {Choe},\ and\ \citenamefont {Baek}}]{Kim2023}%
  \BibitemOpen
  \bibfield  {author} {\bibinfo {author} {\bibnamefont {Kim}, \bibfnamefont
  {Ki-Won}}, \bibinfo {author} {\bibfnamefont {Yunsik}\ \bibnamefont {Choe}},
  and\ \bibinfo {author} {\bibfnamefont {Yongjoo}\ \bibnamefont {Baek}}}
  (\bibinfo {year} {2023}),\ \href@noop {} {\enquote {\bibinfo {title} {Generic
  symmetry-breaking motility in active fluids},}\ }\Eprint
  {https://arxiv.org/abs/2304.01645} {arXiv:2304.01645 [cond-mat.soft]}
  \BibitemShut {NoStop}%
\bibitem [{\citenamefont {Klongvessa}\ \emph {et~al.}(2019)\citenamefont
  {Klongvessa}, \citenamefont {Ginot}, \citenamefont {Ybert}, \citenamefont
  {Cottin-Bizonne},\ and\ \citenamefont {Leocmach}}]{klongvessa_active_2019}%
  \BibitemOpen
  \bibfield  {author} {\bibinfo {author} {\bibnamefont {Klongvessa},
  \bibfnamefont {Natsuda}}, \bibinfo {author} {\bibfnamefont {F{\'{e}}lix}\
  \bibnamefont {Ginot}}, \bibinfo {author} {\bibfnamefont {Christophe}\
  \bibnamefont {Ybert}}, \bibinfo {author} {\bibfnamefont {C{\'{e}}cile}\
  \bibnamefont {Cottin-Bizonne}}, and\ \bibinfo {author} {\bibfnamefont
  {Mathieu}\ \bibnamefont {Leocmach}}} (\bibinfo {year} {2019}),\ \bibfield
  {title} {\enquote {\bibinfo {title} {{Active glass: {Ergodicity} breaking
  dramatically affects response to self-propulsion}},}\ }\href
  {https://doi.org/10.1103/PhysRevLett.123.248004} {\bibfield  {journal}
  {\bibinfo  {journal} {Phys. Rev. Lett.}\ }\textbf {\bibinfo {volume}
  {123}}~(\bibinfo {number} {24}),\ \bibinfo {pages} {248004}}\BibitemShut
  {NoStop}%
\bibitem [{\citenamefont {Kne{\v{z}}evi{\'{c}}}\ \emph
  {et~al.}(2021)\citenamefont {Kne{\v{z}}evi{\'{c}}}, \citenamefont
  {{Avil{\'{e}}s Podgurski}},\ and\ \citenamefont {Stark}}]{Knezevic2021}%
  \BibitemOpen
  \bibfield  {author} {\bibinfo {author} {\bibnamefont {Kne{\v{z}}evi{\'{c}}},
  \bibfnamefont {Milo{\v{s}}}}, \bibinfo {author} {\bibfnamefont {Luisa~E}\
  \bibnamefont {{Avil{\'{e}}s Podgurski}}}, and\ \bibinfo {author}
  {\bibfnamefont {Holger}\ \bibnamefont {Stark}}} (\bibinfo {year} {2021}),\
  \bibfield  {title} {\enquote {\bibinfo {title} {{Oscillatory active
  microrheology of active suspensions}},}\ }\href
  {https://doi.org/10.1038/s41598-021-02103-7} {\bibfield  {journal} {\bibinfo
  {journal} {Sci. Rep.}\ }\textbf {\bibinfo {volume} {11}}~(\bibinfo {number}
  {1}),\ \bibinfo {pages} {22706}}\BibitemShut {NoStop}%
\bibitem [{\citenamefont {Kne{\v{z}}evi{\'{c}}}\ and\ \citenamefont
  {Stark}(2020)}]{Knezevic2020}%
  \BibitemOpen
  \bibfield  {author} {\bibinfo {author} {\bibnamefont {Kne{\v{z}}evi{\'{c}}},
  \bibfnamefont {Milo{\v{s}}}}, and\ \bibinfo {author} {\bibfnamefont {Holger}\
  \bibnamefont {Stark}}} (\bibinfo {year} {2020}),\ \bibfield  {title}
  {\enquote {\bibinfo {title} {{Effective Langevin equations for a polar tracer
  in an active bath}},}\ }\href {https://doi.org/10.1088/1367-2630/abc91e}
  {\bibfield  {journal} {\bibinfo  {journal} {New J. Phys.}\ }\textbf {\bibinfo
  {volume} {22}}~(\bibinfo {number} {11}),\ \bibinfo {pages}
  {113025}}\BibitemShut {NoStop}%
\bibitem [{\citenamefont {Kourbane-Houssene}\ \emph {et~al.}(2018)\citenamefont
  {Kourbane-Houssene}, \citenamefont {Erignoux}, \citenamefont {Bodineau},\
  and\ \citenamefont {Tailleur}}]{kourbane2018exact}%
  \BibitemOpen
  \bibfield  {author} {\bibinfo {author} {\bibnamefont {Kourbane-Houssene},
  \bibfnamefont {Mourtaza}}, \bibinfo {author} {\bibfnamefont {Cl{\'{e}}ment}\
  \bibnamefont {Erignoux}}, \bibinfo {author} {\bibfnamefont {Thierry}\
  \bibnamefont {Bodineau}}, and\ \bibinfo {author} {\bibfnamefont {Julien}\
  \bibnamefont {Tailleur}}} (\bibinfo {year} {2018}),\ \bibfield  {title}
  {\enquote {\bibinfo {title} {{Exact Hydrodynamic Description of Active
  Lattice Gases}},}\ }\href {https://doi.org/10.1103/PhysRevLett.120.268003}
  {\bibfield  {journal} {\bibinfo  {journal} {Phys. Rev. Lett.}\ }\textbf
  {\bibinfo {volume} {120}}~(\bibinfo {number} {26}),\ \bibinfo {pages}
  {268003}}\BibitemShut {NoStop}%
\bibitem [{\citenamefont {Kurihara}\ \emph {et~al.}(2017)\citenamefont
  {Kurihara}, \citenamefont {Aridome}, \citenamefont {Ayade}, \citenamefont
  {Zaid},\ and\ \citenamefont {Mizuno}}]{Kurihara2017}%
  \BibitemOpen
  \bibfield  {author} {\bibinfo {author} {\bibnamefont {Kurihara},
  \bibfnamefont {Takashi}}, \bibinfo {author} {\bibfnamefont {Msato}\
  \bibnamefont {Aridome}}, \bibinfo {author} {\bibfnamefont {Heev}\
  \bibnamefont {Ayade}}, \bibinfo {author} {\bibfnamefont {Irwin}\ \bibnamefont
  {Zaid}}, and\ \bibinfo {author} {\bibfnamefont {Daisuke}\ \bibnamefont
  {Mizuno}}} (\bibinfo {year} {2017}),\ \bibfield  {title} {\enquote {\bibinfo
  {title} {{Non-Gaussian limit fluctuations in active swimmer suspensions}},}\
  }\href {https://doi.org/10.1103/PhysRevE.95.030601} {\bibfield  {journal}
  {\bibinfo  {journal} {Phys. Rev. E}\ }\textbf {\bibinfo {volume}
  {95}}~(\bibinfo {number} {3}),\ \bibinfo {pages} {030601}}\BibitemShut
  {NoStop}%
\bibitem [{\citenamefont {Kurtuldu}\ \emph {et~al.}(2011)\citenamefont
  {Kurtuldu}, \citenamefont {Guasto}, \citenamefont {Johnson},\ and\
  \citenamefont {Gollub}}]{Kurtuldu2011}%
  \BibitemOpen
  \bibfield  {author} {\bibinfo {author} {\bibnamefont {Kurtuldu},
  \bibfnamefont {H{\"{u}}seyin}}, \bibinfo {author} {\bibfnamefont
  {Jeffrey~S.}\ \bibnamefont {Guasto}}, \bibinfo {author} {\bibfnamefont
  {Karl~A.}\ \bibnamefont {Johnson}}, and\ \bibinfo {author} {\bibfnamefont
  {J.~P.}\ \bibnamefont {Gollub}}} (\bibinfo {year} {2011}),\ \bibfield
  {title} {\enquote {\bibinfo {title} {{Enhancement of biomixing by swimming
  algal cells in two-dimensional films}},}\ }\href
  {https://doi.org/10.1073/pnas.1107046108} {\bibfield  {journal} {\bibinfo
  {journal} {Proc. Natl. Acad. Sci.}\ }\textbf {\bibinfo {volume}
  {108}}~(\bibinfo {number} {26}),\ \bibinfo {pages}
  {10391--10395}}\BibitemShut {NoStop}%
\bibitem [{\citenamefont {Kwon}\ \emph {et~al.}(2005)\citenamefont {Kwon},
  \citenamefont {Ao},\ and\ \citenamefont {Thouless}}]{kwon2005structure}%
  \BibitemOpen
  \bibfield  {author} {\bibinfo {author} {\bibnamefont {Kwon}, \bibfnamefont
  {Chulan}}, \bibinfo {author} {\bibfnamefont {Ping}\ \bibnamefont {Ao}}, and\
  \bibinfo {author} {\bibfnamefont {David~J.}\ \bibnamefont {Thouless}}}
  (\bibinfo {year} {2005}),\ \bibfield  {title} {\enquote {\bibinfo {title}
  {{Structure of stochastic dynamics near fixed points}},}\ }\href
  {https://doi.org/10.1073/pnas.0506347102} {\bibfield  {journal} {\bibinfo
  {journal} {Proc. Natl. Acad. Sci.}\ }\textbf {\bibinfo {volume}
  {102}}~(\bibinfo {number} {37}),\ \bibinfo {pages}
  {13029--13033}}\BibitemShut {NoStop}%
\bibitem [{\citenamefont {Lauersdorf}\ \emph {et~al.}(2021)\citenamefont
  {Lauersdorf}, \citenamefont {Kolb}, \citenamefont {Moradi}, \citenamefont
  {Nazockdast},\ and\ \citenamefont {Klotsa}}]{lauersdorf2021phase}%
  \BibitemOpen
  \bibfield  {author} {\bibinfo {author} {\bibnamefont {Lauersdorf},
  \bibfnamefont {Nicholas}}, \bibinfo {author} {\bibfnamefont {Thomas}\
  \bibnamefont {Kolb}}, \bibinfo {author} {\bibfnamefont {Moslem}\ \bibnamefont
  {Moradi}}, \bibinfo {author} {\bibfnamefont {Ehssan}\ \bibnamefont
  {Nazockdast}}, and\ \bibinfo {author} {\bibfnamefont {Daphne}\ \bibnamefont
  {Klotsa}}} (\bibinfo {year} {2021}),\ \bibfield  {title} {\enquote {\bibinfo
  {title} {{Phase behavior and surface tension of soft active Brownian
  particles}},}\ }\href {https://doi.org/10.1039/D1SM00350J} {\bibfield
  {journal} {\bibinfo  {journal} {Soft Matter}\ }\textbf {\bibinfo {volume}
  {17}}~(\bibinfo {number} {26}),\ \bibinfo {pages} {6337--6351}}\BibitemShut
  {NoStop}%
\bibitem [{\citenamefont {Leptos}\ \emph {et~al.}(2009)\citenamefont {Leptos},
  \citenamefont {Guasto}, \citenamefont {Gollub}, \citenamefont {Pesci},\ and\
  \citenamefont {Goldstein}}]{Leptos2009}%
  \BibitemOpen
  \bibfield  {author} {\bibinfo {author} {\bibnamefont {Leptos}, \bibfnamefont
  {Kyriacos~C}}, \bibinfo {author} {\bibfnamefont {Jeffrey~S.}\ \bibnamefont
  {Guasto}}, \bibinfo {author} {\bibfnamefont {J.~P.}\ \bibnamefont {Gollub}},
  \bibinfo {author} {\bibfnamefont {Adriana~I.}\ \bibnamefont {Pesci}}, and\
  \bibinfo {author} {\bibfnamefont {Raymond~E.}\ \bibnamefont {Goldstein}}}
  (\bibinfo {year} {2009}),\ \bibfield  {title} {\enquote {\bibinfo {title}
  {{Dynamics of Enhanced Tracer Diffusion in Suspensions of Swimming Eukaryotic
  Microorganisms}},}\ }\href {https://doi.org/10.1103/PhysRevLett.103.198103}
  {\bibfield  {journal} {\bibinfo  {journal} {Phys. Rev. Lett.}\ }\textbf
  {\bibinfo {volume} {103}}~(\bibinfo {number} {19}),\ \bibinfo {pages}
  {198103}}\BibitemShut {NoStop}%
\bibitem [{\citenamefont {Levis}\ and\ \citenamefont
  {Berthier}(2015)}]{levis_single-particle_2015}%
  \BibitemOpen
  \bibfield  {author} {\bibinfo {author} {\bibnamefont {Levis}, \bibfnamefont
  {Demian}}, and\ \bibinfo {author} {\bibfnamefont {Ludovic}\ \bibnamefont
  {Berthier}}} (\bibinfo {year} {2015}),\ \bibfield  {title} {\enquote
  {\bibinfo {title} {{From single-particle to collective effective temperatures
  in an active fluid of self-propelled particles}},}\ }\href
  {https://doi.org/10.1209/0295-5075/111/60006} {\bibfield  {journal} {\bibinfo
   {journal} {EPL}\ }\textbf {\bibinfo {volume} {111}}~(\bibinfo {number}
  {6}),\ \bibinfo {pages} {60006}}\BibitemShut {NoStop}%
\bibitem [{\citenamefont {Li}\ and\ \citenamefont
  {Zhang}(2013)}]{li_asymmetric_2013}%
  \BibitemOpen
  \bibfield  {author} {\bibinfo {author} {\bibnamefont {Li}, \bibfnamefont
  {He}}, and\ \bibinfo {author} {\bibfnamefont {H.~P.}\ \bibnamefont {Zhang}}}
  (\bibinfo {year} {2013}),\ \bibfield  {title} {\enquote {\bibinfo {title}
  {{Asymmetric gear rectifies random robot motion}},}\ }\href
  {https://doi.org/10.1209/0295-5075/102/50007} {\bibfield  {journal} {\bibinfo
   {journal} {EPL}\ }\textbf {\bibinfo {volume} {102}}~(\bibinfo {number}
  {5}),\ \bibinfo {pages} {50007}}\BibitemShut {NoStop}%
\bibitem [{\citenamefont {van~der Linden}\ \emph {et~al.}(2019)\citenamefont
  {van~der Linden}, \citenamefont {Alexander}, \citenamefont {Aarts},\ and\
  \citenamefont {Dauchot}}]{van2019interrupted}%
  \BibitemOpen
  \bibfield  {author} {\bibinfo {author} {\bibnamefont {van~der Linden},
  \bibfnamefont {Marjolein~N}}, \bibinfo {author} {\bibfnamefont {Lachlan~C.}\
  \bibnamefont {Alexander}}, \bibinfo {author} {\bibfnamefont {Dirk G. A.~L.}\
  \bibnamefont {Aarts}}, and\ \bibinfo {author} {\bibfnamefont {Olivier}\
  \bibnamefont {Dauchot}}} (\bibinfo {year} {2019}),\ \bibfield  {title}
  {\enquote {\bibinfo {title} {{Interrupted Motility Induced Phase Separation
  in Aligning Active Colloids}},}\ }\href
  {https://doi.org/10.1103/PhysRevLett.123.098001} {\bibfield  {journal}
  {\bibinfo  {journal} {Phys. Rev. Lett.}\ }\textbf {\bibinfo {volume}
  {123}}~(\bibinfo {number} {9}),\ \bibinfo {pages} {098001}}\BibitemShut
  {NoStop}%
\bibitem [{\citenamefont {Liu}\ \emph {et~al.}(2011)\citenamefont {Liu},
  \citenamefont {Fu}, \citenamefont {Liu}, \citenamefont {Ren}, \citenamefont
  {Chau}, \citenamefont {Li}, \citenamefont {Xiang}, \citenamefont {Zeng},
  \citenamefont {Chen}, \citenamefont {Tang}, \citenamefont {Lenz},
  \citenamefont {Cui}, \citenamefont {Huang}, \citenamefont {Hwa},\ and\
  \citenamefont {Huang}}]{liu2011sequential}%
  \BibitemOpen
  \bibfield  {author} {\bibinfo {author} {\bibnamefont {Liu}, \bibfnamefont
  {Chenli}}, \bibinfo {author} {\bibfnamefont {Xiongfei}\ \bibnamefont {Fu}},
  \bibinfo {author} {\bibfnamefont {Lizhong}\ \bibnamefont {Liu}}, \bibinfo
  {author} {\bibfnamefont {Xiaojing}\ \bibnamefont {Ren}}, \bibinfo {author}
  {\bibfnamefont {Carlos~K.L.}\ \bibnamefont {Chau}}, \bibinfo {author}
  {\bibfnamefont {Sihong}\ \bibnamefont {Li}}, \bibinfo {author} {\bibfnamefont
  {Lu}~\bibnamefont {Xiang}}, \bibinfo {author} {\bibfnamefont {Hualing}\
  \bibnamefont {Zeng}}, \bibinfo {author} {\bibfnamefont {Guanhua}\
  \bibnamefont {Chen}}, \bibinfo {author} {\bibfnamefont {Lei-Han}\
  \bibnamefont {Tang}}, \bibinfo {author} {\bibfnamefont {Peter}\ \bibnamefont
  {Lenz}}, \bibinfo {author} {\bibfnamefont {Xiaodong}\ \bibnamefont {Cui}},
  \bibinfo {author} {\bibfnamefont {Wei}\ \bibnamefont {Huang}}, \bibinfo
  {author} {\bibfnamefont {Terence}\ \bibnamefont {Hwa}}, and\ \bibinfo
  {author} {\bibfnamefont {Jian-Dong}\ \bibnamefont {Huang}}} (\bibinfo {year}
  {2011}),\ \bibfield  {title} {\enquote {\bibinfo {title} {{Sequential
  Establishment of Stripe Patterns in an Expanding Cell Population}},}\ }\href
  {https://doi.org/10.1126/science.1209042} {\bibfield  {journal} {\bibinfo
  {journal} {Science}\ }\textbf {\bibinfo {volume} {334}}~(\bibinfo {number}
  {6053}),\ \bibinfo {pages} {238--241}}\BibitemShut {NoStop}%
\bibitem [{\citenamefont {Liu}\ \emph {et~al.}(2019)\citenamefont {Liu},
  \citenamefont {Patch}, \citenamefont {Bahar}, \citenamefont {Yllanes},
  \citenamefont {Welch}, \citenamefont {Marchetti}, \citenamefont
  {Thutupalli},\ and\ \citenamefont {Shaevitz}}]{liu2019self}%
  \BibitemOpen
  \bibfield  {author} {\bibinfo {author} {\bibnamefont {Liu}, \bibfnamefont
  {Guannan}}, \bibinfo {author} {\bibfnamefont {Adam}\ \bibnamefont {Patch}},
  \bibinfo {author} {\bibfnamefont {Fatmag{\"{u}}l}\ \bibnamefont {Bahar}},
  \bibinfo {author} {\bibfnamefont {David}\ \bibnamefont {Yllanes}}, \bibinfo
  {author} {\bibfnamefont {Roy~D.}\ \bibnamefont {Welch}}, \bibinfo {author}
  {\bibfnamefont {M.~Cristina}\ \bibnamefont {Marchetti}}, \bibinfo {author}
  {\bibfnamefont {Shashi}\ \bibnamefont {Thutupalli}}, and\ \bibinfo {author}
  {\bibfnamefont {Joshua~W.}\ \bibnamefont {Shaevitz}}} (\bibinfo {year}
  {2019}),\ \bibfield  {title} {\enquote {\bibinfo {title} {{Self-driven phase
  transitions drive Myxococcus xanthus fruiting body formation}},}\ }\href
  {https://doi.org/10.1103/PhysRevLett.122.248102} {\bibfield  {journal}
  {\bibinfo  {journal} {Phys. Rev. Lett.}\ }\textbf {\bibinfo {volume}
  {122}}~(\bibinfo {number} {24}),\ \bibinfo {pages} {248102}}\BibitemShut
  {NoStop}%
\bibitem [{\citenamefont {Maes}(2020)}]{Maes2020}%
  \BibitemOpen
  \bibfield  {author} {\bibinfo {author} {\bibnamefont {Maes}, \bibfnamefont
  {Christian}}} (\bibinfo {year} {2020}),\ \bibfield  {title} {\enquote
  {\bibinfo {title} {{Fluctuating Motion in an Active Environment}},}\ }\href
  {https://doi.org/10.1103/PhysRevLett.125.208001} {\bibfield  {journal}
  {\bibinfo  {journal} {Phys. Rev. Lett.}\ }\textbf {\bibinfo {volume}
  {125}}~(\bibinfo {number} {20}),\ \bibinfo {pages} {208001}}\BibitemShut
  {NoStop}%
\bibitem [{\citenamefont {Maggi}\ \emph {et~al.}(2022)\citenamefont {Maggi},
  \citenamefont {Gnan}, \citenamefont {Paoluzzi}, \citenamefont {Zaccarelli},\
  and\ \citenamefont {Crisanti}}]{maggi_critical_2022}%
  \BibitemOpen
  \bibfield  {author} {\bibinfo {author} {\bibnamefont {Maggi}, \bibfnamefont
  {Claudio}}, \bibinfo {author} {\bibfnamefont {Nicoletta}\ \bibnamefont
  {Gnan}}, \bibinfo {author} {\bibfnamefont {Matteo}\ \bibnamefont {Paoluzzi}},
  \bibinfo {author} {\bibfnamefont {Emanuela}\ \bibnamefont {Zaccarelli}}, and\
  \bibinfo {author} {\bibfnamefont {Andrea}\ \bibnamefont {Crisanti}}}
  (\bibinfo {year} {2022}),\ \bibfield  {title} {\enquote {\bibinfo {title}
  {{Critical active dynamics is captured by a colored-noise driven field
  theory}},}\ }\href {https://doi.org/10.1038/s42005-022-00830-5} {\bibfield
  {journal} {\bibinfo  {journal} {Commun. Phys.}\ }\textbf {\bibinfo {volume}
  {5}}~(\bibinfo {number} {1}),\ \bibinfo {pages} {55}}\BibitemShut {NoStop}%
\bibitem [{\citenamefont {Maggi}\ \emph {et~al.}(2015)\citenamefont {Maggi},
  \citenamefont {Marconi}, \citenamefont {Gnan},\ and\ \citenamefont {{Di
  Leonardo}}}]{maggi2015multidimensional}%
  \BibitemOpen
  \bibfield  {author} {\bibinfo {author} {\bibnamefont {Maggi}, \bibfnamefont
  {Claudio}}, \bibinfo {author} {\bibfnamefont {Umberto Marini~Bettolo}\
  \bibnamefont {Marconi}}, \bibinfo {author} {\bibfnamefont {Nicoletta}\
  \bibnamefont {Gnan}}, and\ \bibinfo {author} {\bibfnamefont {Roberto}\
  \bibnamefont {{Di Leonardo}}}} (\bibinfo {year} {2015}),\ \bibfield  {title}
  {\enquote {\bibinfo {title} {{Multidimensional stationary probability
  distribution for interacting active particles}},}\ }\href
  {https://doi.org/10.1038/srep10742} {\bibfield  {journal} {\bibinfo
  {journal} {Sci. Rep.}\ }\textbf {\bibinfo {volume} {5}}~(\bibinfo {number}
  {1}),\ \bibinfo {pages} {10742}}\BibitemShut {NoStop}%
\bibitem [{\citenamefont {Maggi}\ \emph {et~al.}(2017)\citenamefont {Maggi},
  \citenamefont {Paoluzzi}, \citenamefont {Angelani},\ and\ \citenamefont {{Di
  Leonardo}}}]{Maggi2017a}%
  \BibitemOpen
  \bibfield  {author} {\bibinfo {author} {\bibnamefont {Maggi}, \bibfnamefont
  {Claudio}}, \bibinfo {author} {\bibfnamefont {Matteo}\ \bibnamefont
  {Paoluzzi}}, \bibinfo {author} {\bibfnamefont {Luca}\ \bibnamefont
  {Angelani}}, and\ \bibinfo {author} {\bibfnamefont {Roberto}\ \bibnamefont
  {{Di Leonardo}}}} (\bibinfo {year} {2017}),\ \bibfield  {title} {\enquote
  {\bibinfo {title} {{Memory-less response and violation of the
  fluctuation-dissipation theorem in colloids suspended in an active bath}},}\
  }\href {https://doi.org/10.1038/s41598-017-17900-2} {\bibfield  {journal}
  {\bibinfo  {journal} {Sci. Rep.}\ }\textbf {\bibinfo {volume} {7}}~(\bibinfo
  {number} {1}),\ \bibinfo {pages} {17588}}\BibitemShut {NoStop}%
\bibitem [{\citenamefont {Maggi}\ \emph {et~al.}(2021)\citenamefont {Maggi},
  \citenamefont {Paoluzzi}, \citenamefont {Crisanti}, \citenamefont
  {Zaccarelli},\ and\ \citenamefont {Gnan}}]{maggi_universality_2021}%
  \BibitemOpen
  \bibfield  {author} {\bibinfo {author} {\bibnamefont {Maggi}, \bibfnamefont
  {Claudio}}, \bibinfo {author} {\bibfnamefont {Matteo}\ \bibnamefont
  {Paoluzzi}}, \bibinfo {author} {\bibfnamefont {Andrea}\ \bibnamefont
  {Crisanti}}, \bibinfo {author} {\bibfnamefont {Emanuela}\ \bibnamefont
  {Zaccarelli}}, and\ \bibinfo {author} {\bibfnamefont {Nicoletta}\
  \bibnamefont {Gnan}}} (\bibinfo {year} {2021}),\ \bibfield  {title} {\enquote
  {\bibinfo {title} {{Universality class of the motility-induced critical point
  in large scale off-lattice simulations of active particles}},}\ }\href
  {https://doi.org/10.1039/D0SM02162H} {\bibfield  {journal} {\bibinfo
  {journal} {Soft Matter}\ }\textbf {\bibinfo {volume} {17}}~(\bibinfo {number}
  {14}),\ \bibinfo {pages} {3807--3812}}\BibitemShut {NoStop}%
\bibitem [{\citenamefont {Malakar}\ \emph {et~al.}(2020)\citenamefont
  {Malakar}, \citenamefont {Das}, \citenamefont {Kundu}, \citenamefont
  {Kumar},\ and\ \citenamefont {Dhar}}]{malakar2020steady}%
  \BibitemOpen
  \bibfield  {author} {\bibinfo {author} {\bibnamefont {Malakar}, \bibfnamefont
  {Kanaya}}, \bibinfo {author} {\bibfnamefont {Arghya}\ \bibnamefont {Das}},
  \bibinfo {author} {\bibfnamefont {Anupam}\ \bibnamefont {Kundu}}, \bibinfo
  {author} {\bibfnamefont {K.~Vijay}\ \bibnamefont {Kumar}}, and\ \bibinfo
  {author} {\bibfnamefont {Abhishek}\ \bibnamefont {Dhar}}} (\bibinfo {year}
  {2020}),\ \bibfield  {title} {\enquote {\bibinfo {title} {{Steady state of an
  active Brownian particle in a two-dimensional harmonic trap}},}\ }\href
  {https://doi.org/10.1103/PhysRevE.101.022610} {\bibfield  {journal} {\bibinfo
   {journal} {Phys. Rev. E}\ }\textbf {\bibinfo {volume} {101}}~(\bibinfo
  {number} {2}),\ \bibinfo {pages} {022610}}\BibitemShut {NoStop}%
\bibitem [{\citenamefont {Mallory}\ \emph
  {et~al.}(2014{\natexlab{a}})\citenamefont {Mallory}, \citenamefont
  {{\v{S}}ari{\'{c}}}, \citenamefont {Valeriani},\ and\ \citenamefont
  {Cacciuto}}]{mallory2014anomalous}%
  \BibitemOpen
  \bibfield  {author} {\bibinfo {author} {\bibnamefont {Mallory}, \bibfnamefont
  {S~A}}, \bibinfo {author} {\bibfnamefont {A.}~\bibnamefont
  {{\v{S}}ari{\'{c}}}}, \bibinfo {author} {\bibfnamefont {C.}~\bibnamefont
  {Valeriani}}, and\ \bibinfo {author} {\bibfnamefont {A.}~\bibnamefont
  {Cacciuto}}} (\bibinfo {year} {2014}{\natexlab{a}}),\ \bibfield  {title}
  {\enquote {\bibinfo {title} {{Anomalous thermomechanical properties of a
  self-propelled colloidal fluid}},}\ }\href
  {https://doi.org/10.1103/PhysRevE.89.052303} {\bibfield  {journal} {\bibinfo
  {journal} {Phys. Rev. E}\ }\textbf {\bibinfo {volume} {89}}~(\bibinfo
  {number} {5}),\ \bibinfo {pages} {052303}}\BibitemShut {NoStop}%
\bibitem [{\citenamefont {Mallory}\ \emph
  {et~al.}(2014{\natexlab{b}})\citenamefont {Mallory}, \citenamefont
  {Valeriani},\ and\ \citenamefont {Cacciuto}}]{mallory2014curvature}%
  \BibitemOpen
  \bibfield  {author} {\bibinfo {author} {\bibnamefont {Mallory}, \bibfnamefont
  {S~A}}, \bibinfo {author} {\bibfnamefont {C.}~\bibnamefont {Valeriani}}, and\
  \bibinfo {author} {\bibfnamefont {A.}~\bibnamefont {Cacciuto}}} (\bibinfo
  {year} {2014}{\natexlab{b}}),\ \bibfield  {title} {\enquote {\bibinfo {title}
  {{Curvature-induced activation of a passive tracer in an active bath}},}\
  }\href {https://doi.org/10.1103/PhysRevE.90.032309} {\bibfield  {journal}
  {\bibinfo  {journal} {Phys. Rev. E}\ }\textbf {\bibinfo {volume}
  {90}}~(\bibinfo {number} {3}),\ \bibinfo {pages} {032309}}\BibitemShut
  {NoStop}%
\bibitem [{\citenamefont {Marchetti}\ \emph {et~al.}(2013)\citenamefont
  {Marchetti}, \citenamefont {Joanny}, \citenamefont {Ramaswamy}, \citenamefont
  {Liverpool}, \citenamefont {Prost}, \citenamefont {Rao},\ and\ \citenamefont
  {Simha}}]{marchetti2013hydrodynamics}%
  \BibitemOpen
  \bibfield  {author} {\bibinfo {author} {\bibnamefont {Marchetti},
  \bibfnamefont {M~C}}, \bibinfo {author} {\bibfnamefont {J.~F.}\ \bibnamefont
  {Joanny}}, \bibinfo {author} {\bibfnamefont {S.}~\bibnamefont {Ramaswamy}},
  \bibinfo {author} {\bibfnamefont {T.~B.}\ \bibnamefont {Liverpool}}, \bibinfo
  {author} {\bibfnamefont {J.}~\bibnamefont {Prost}}, \bibinfo {author}
  {\bibfnamefont {Madan}\ \bibnamefont {Rao}}, and\ \bibinfo {author}
  {\bibfnamefont {R.~Aditi}\ \bibnamefont {Simha}}} (\bibinfo {year} {2013}),\
  \bibfield  {title} {\enquote {\bibinfo {title} {{Hydrodynamics of soft active
  matter}},}\ }\href {https://doi.org/10.1103/RevModPhys.85.1143} {\bibfield
  {journal} {\bibinfo  {journal} {Rev. Mod. Phys.}\ }\textbf {\bibinfo {volume}
  {85}}~(\bibinfo {number} {3}),\ \bibinfo {pages} {1143--1189}}\BibitemShut
  {NoStop}%
\bibitem [{\citenamefont {{Marini Bettolo Marconi}}\ \emph
  {et~al.}(2016)\citenamefont {{Marini Bettolo Marconi}}, \citenamefont
  {Maggi},\ and\ \citenamefont {Melchionna}}]{marconi2016pressure}%
  \BibitemOpen
  \bibfield  {author} {\bibinfo {author} {\bibnamefont {{Marini Bettolo
  Marconi}}, \bibfnamefont {Umberto}}, \bibinfo {author} {\bibfnamefont
  {Claudio}\ \bibnamefont {Maggi}}, and\ \bibinfo {author} {\bibfnamefont
  {Simone}\ \bibnamefont {Melchionna}}} (\bibinfo {year} {2016}),\ \bibfield
  {title} {\enquote {\bibinfo {title} {{Pressure and surface tension of an
  active simple liquid: a comparison between kinetic, mechanical and
  free-energy based approaches}},}\ }\href {https://doi.org/10.1039/C6SM00667A}
  {\bibfield  {journal} {\bibinfo  {journal} {Soft Matter}\ }\textbf {\bibinfo
  {volume} {12}}~(\bibinfo {number} {26}),\ \bibinfo {pages}
  {5727--5738}}\BibitemShut {NoStop}%
\bibitem [{\citenamefont {Martin}\ \emph {et~al.}(2021)\citenamefont {Martin},
  \citenamefont {O'Byrne}, \citenamefont {Cates}, \citenamefont {Fodor},
  \citenamefont {Nardini}, \citenamefont {Tailleur},\ and\ \citenamefont {van
  Wijland}}]{martin_statistical_2021}%
  \BibitemOpen
  \bibfield  {author} {\bibinfo {author} {\bibnamefont {Martin}, \bibfnamefont
  {David}}, \bibinfo {author} {\bibfnamefont {J{\'{e}}r{\'{e}}my}\ \bibnamefont
  {O'Byrne}}, \bibinfo {author} {\bibfnamefont {Michael~E.}\ \bibnamefont
  {Cates}}, \bibinfo {author} {\bibfnamefont {{\'{E}}tienne}\ \bibnamefont
  {Fodor}}, \bibinfo {author} {\bibfnamefont {Cesare}\ \bibnamefont {Nardini}},
  \bibinfo {author} {\bibfnamefont {Julien}\ \bibnamefont {Tailleur}}, and\
  \bibinfo {author} {\bibfnamefont {Fr{\'{e}}d{\'{e}}ric}\ \bibnamefont {van
  Wijland}}} (\bibinfo {year} {2021}),\ \bibfield  {title} {\enquote {\bibinfo
  {title} {{Statistical mechanics of active Ornstein-Uhlenbeck particles}},}\
  }\href {https://doi.org/10.1103/PhysRevE.103.032607} {\bibfield  {journal}
  {\bibinfo  {journal} {Phys. Rev. E}\ }\textbf {\bibinfo {volume}
  {103}}~(\bibinfo {number} {3}),\ \bibinfo {pages} {032607}}\BibitemShut
  {NoStop}%
\bibitem [{\citenamefont {Martin}\ \emph {et~al.}(2023)\citenamefont {Martin},
  \citenamefont {Seara}, \citenamefont {Avni}, \citenamefont {Fruchart},\ and\
  \citenamefont {Vitelli}}]{Martin2023}%
  \BibitemOpen
  \bibfield  {author} {\bibinfo {author} {\bibnamefont {Martin}, \bibfnamefont
  {David}}, \bibinfo {author} {\bibfnamefont {Daniel}\ \bibnamefont {Seara}},
  \bibinfo {author} {\bibfnamefont {Yael}\ \bibnamefont {Avni}}, \bibinfo
  {author} {\bibfnamefont {Michel}\ \bibnamefont {Fruchart}}, and\ \bibinfo
  {author} {\bibfnamefont {Vincenzo}\ \bibnamefont {Vitelli}}} (\bibinfo {year}
  {2023}),\ \href@noop {} {\enquote {\bibinfo {title} {An exact model for the
  transition to collective motion in nonreciprocal active matter},}\ }\Eprint
  {https://arxiv.org/abs/2307.08251} {arXiv:2307.08251 [cond-mat.stat-mech]}
  \BibitemShut {NoStop}%
\bibitem [{\citenamefont {Massana-Cid}\ \emph {et~al.}(2022)\citenamefont
  {Massana-Cid}, \citenamefont {Maggi}, \citenamefont {Frangipane},\ and\
  \citenamefont {{Di Leonardo}}}]{massana-cid_rectification_2022}%
  \BibitemOpen
  \bibfield  {author} {\bibinfo {author} {\bibnamefont {Massana-Cid},
  \bibfnamefont {Helena}}, \bibinfo {author} {\bibfnamefont {Claudio}\
  \bibnamefont {Maggi}}, \bibinfo {author} {\bibfnamefont {Giacomo}\
  \bibnamefont {Frangipane}}, and\ \bibinfo {author} {\bibfnamefont {Roberto}\
  \bibnamefont {{Di Leonardo}}}} (\bibinfo {year} {2022}),\ \bibfield  {title}
  {\enquote {\bibinfo {title} {{Rectification and confinement of photokinetic
  bacteria in an optical feedback loop}},}\ }\href
  {https://doi.org/10.1038/s41467-022-30201-1} {\bibfield  {journal} {\bibinfo
  {journal} {Nat. Commun.}\ }\textbf {\bibinfo {volume} {13}}~(\bibinfo
  {number} {1}),\ \bibinfo {pages} {2740}}\BibitemShut {NoStop}%
\bibitem [{\citenamefont {Miller}\ and\ \citenamefont
  {Bassler}(2001)}]{miller_quorum_2001}%
  \BibitemOpen
  \bibfield  {author} {\bibinfo {author} {\bibnamefont {Miller}, \bibfnamefont
  {Melissa~B}}, and\ \bibinfo {author} {\bibfnamefont {Bonnie~L.}\ \bibnamefont
  {Bassler}}} (\bibinfo {year} {2001}),\ \bibfield  {title} {\enquote {\bibinfo
  {title} {{Quorum Sensing in Bacteria}},}\ }\href
  {https://doi.org/10.1146/annurev.micro.55.1.165} {\bibfield  {journal}
  {\bibinfo  {journal} {Annu. Rev. Microbiol.}\ }\textbf {\bibinfo {volume}
  {55}}~(\bibinfo {number} {1}),\ \bibinfo {pages} {165--199}}\BibitemShut
  {NoStop}%
\bibitem [{\citenamefont {Mognetti}\ \emph {et~al.}(2013)\citenamefont
  {Mognetti}, \citenamefont {{\v{S}}ari{\'{c}}}, \citenamefont
  {Angioletti-Uberti}, \citenamefont {Cacciuto}, \citenamefont {Valeriani},\
  and\ \citenamefont {Frenkel}}]{mognetti2013living}%
  \BibitemOpen
  \bibfield  {author} {\bibinfo {author} {\bibnamefont {Mognetti},
  \bibfnamefont {B~M}}, \bibinfo {author} {\bibfnamefont {A.}~\bibnamefont
  {{\v{S}}ari{\'{c}}}}, \bibinfo {author} {\bibfnamefont {S.}~\bibnamefont
  {Angioletti-Uberti}}, \bibinfo {author} {\bibfnamefont {A.}~\bibnamefont
  {Cacciuto}}, \bibinfo {author} {\bibfnamefont {C.}~\bibnamefont {Valeriani}},
  and\ \bibinfo {author} {\bibfnamefont {D.}~\bibnamefont {Frenkel}}} (\bibinfo
  {year} {2013}),\ \bibfield  {title} {\enquote {\bibinfo {title} {{Living
  Clusters and Crystals from Low-Density Suspensions of Active Colloids}},}\
  }\href {https://doi.org/10.1103/PhysRevLett.111.245702} {\bibfield  {journal}
  {\bibinfo  {journal} {Phys. Rev. Lett.}\ }\textbf {\bibinfo {volume}
  {111}}~(\bibinfo {number} {24}),\ \bibinfo {pages} {245702}}\BibitemShut
  {NoStop}%
\bibitem [{\citenamefont {Morin}\ \emph {et~al.}(2017)\citenamefont {Morin},
  \citenamefont {Desreumaux}, \citenamefont {Caussin},\ and\ \citenamefont
  {Bartolo}}]{morin2017distortion}%
  \BibitemOpen
  \bibfield  {author} {\bibinfo {author} {\bibnamefont {Morin}, \bibfnamefont
  {Alexandre}}, \bibinfo {author} {\bibfnamefont {Nicolas}\ \bibnamefont
  {Desreumaux}}, \bibinfo {author} {\bibfnamefont {Jean-Baptiste}\ \bibnamefont
  {Caussin}}, and\ \bibinfo {author} {\bibfnamefont {Denis}\ \bibnamefont
  {Bartolo}}} (\bibinfo {year} {2017}),\ \bibfield  {title} {\enquote {\bibinfo
  {title} {{Distortion and destruction of colloidal flocks in disordered
  environments}},}\ }\href {https://doi.org/10.1038/nphys3903} {\bibfield
  {journal} {\bibinfo  {journal} {Nat. Phys.}\ }\textbf {\bibinfo {volume}
  {13}}~(\bibinfo {number} {1}),\ \bibinfo {pages} {63--67}}\BibitemShut
  {NoStop}%
\bibitem [{\citenamefont {Nardini}\ \emph {et~al.}(2017)\citenamefont
  {Nardini}, \citenamefont {Fodor}, \citenamefont {Tjhung}, \citenamefont {van
  Wijland}, \citenamefont {Tailleur},\ and\ \citenamefont
  {Cates}}]{nardini_entropy_2017}%
  \BibitemOpen
  \bibfield  {author} {\bibinfo {author} {\bibnamefont {Nardini}, \bibfnamefont
  {Cesare}}, \bibinfo {author} {\bibfnamefont {{\'{E}}tienne}\ \bibnamefont
  {Fodor}}, \bibinfo {author} {\bibfnamefont {Elsen}\ \bibnamefont {Tjhung}},
  \bibinfo {author} {\bibfnamefont {Fr{\'{e}}d{\'{e}}ric}\ \bibnamefont {van
  Wijland}}, \bibinfo {author} {\bibfnamefont {Julien}\ \bibnamefont
  {Tailleur}}, and\ \bibinfo {author} {\bibfnamefont {Michael~E.}\ \bibnamefont
  {Cates}}} (\bibinfo {year} {2017}),\ \bibfield  {title} {\enquote {\bibinfo
  {title} {{Entropy Production in Field Theories without Time-Reversal
  Symmetry: Quantifying the Non-Equilibrium Character of Active Matter}},}\
  }\href {https://doi.org/10.1103/PhysRevX.7.021007} {\bibfield  {journal}
  {\bibinfo  {journal} {Phys. Rev. X}\ }\textbf {\bibinfo {volume}
  {7}}~(\bibinfo {number} {2}),\ \bibinfo {pages} {021007}}\BibitemShut
  {NoStop}%
\bibitem [{\citenamefont {Nealson}\ \emph {et~al.}(1970)\citenamefont
  {Nealson}, \citenamefont {Platt},\ and\ \citenamefont
  {Hastings}}]{nealson1970luminescence}%
  \BibitemOpen
  \bibfield  {author} {\bibinfo {author} {\bibnamefont {Nealson}, \bibfnamefont
  {Kenneth~H}}, \bibinfo {author} {\bibfnamefont {Terry}\ \bibnamefont
  {Platt}}, and\ \bibinfo {author} {\bibfnamefont {J.~Woodland}\ \bibnamefont
  {Hastings}}} (\bibinfo {year} {1970}),\ \bibfield  {title} {\enquote
  {\bibinfo {title} {{Cellular Control of the Synthesis and Activity of the
  Bacterial Luminescent System}},}\ }\href
  {https://doi.org/10.1128/jb.104.1.313-322.1970} {\bibfield  {journal}
  {\bibinfo  {journal} {J. Bacteriol.}\ }\textbf {\bibinfo {volume}
  {104}}~(\bibinfo {number} {1}),\ \bibinfo {pages} {313--322}}\BibitemShut
  {NoStop}%
\bibitem [{\citenamefont {Ni}\ \emph {et~al.}(2015)\citenamefont {Ni},
  \citenamefont {{Cohen Stuart}},\ and\ \citenamefont {Bolhuis}}]{Ni2015}%
  \BibitemOpen
  \bibfield  {author} {\bibinfo {author} {\bibnamefont {Ni}, \bibfnamefont
  {Ran}}, \bibinfo {author} {\bibfnamefont {Martien~A.}\ \bibnamefont {{Cohen
  Stuart}}}, and\ \bibinfo {author} {\bibfnamefont {Peter~G.}\ \bibnamefont
  {Bolhuis}}} (\bibinfo {year} {2015}),\ \bibfield  {title} {\enquote {\bibinfo
  {title} {{Tunable Long Range Forces Mediated by Self-Propelled Colloidal Hard
  Spheres}},}\ }\href {https://doi.org/10.1103/PhysRevLett.114.018302}
  {\bibfield  {journal} {\bibinfo  {journal} {Phys. Rev. Lett.}\ }\textbf
  {\bibinfo {volume} {114}}~(\bibinfo {number} {1}),\ \bibinfo {pages}
  {018302}}\BibitemShut {NoStop}%
\bibitem [{\citenamefont {Nikola}\ \emph {et~al.}(2016)\citenamefont {Nikola},
  \citenamefont {Solon}, \citenamefont {Kafri}, \citenamefont {Kardar},
  \citenamefont {Tailleur},\ and\ \citenamefont
  {Voituriez}}]{nikola2016active}%
  \BibitemOpen
  \bibfield  {author} {\bibinfo {author} {\bibnamefont {Nikola}, \bibfnamefont
  {Nikolai}}, \bibinfo {author} {\bibfnamefont {Alexandre~P.}\ \bibnamefont
  {Solon}}, \bibinfo {author} {\bibfnamefont {Yariv}\ \bibnamefont {Kafri}},
  \bibinfo {author} {\bibfnamefont {Mehran}\ \bibnamefont {Kardar}}, \bibinfo
  {author} {\bibfnamefont {Julien}\ \bibnamefont {Tailleur}}, and\ \bibinfo
  {author} {\bibfnamefont {Rapha{\"{e}}l}\ \bibnamefont {Voituriez}}} (\bibinfo
  {year} {2016}),\ \bibfield  {title} {\enquote {\bibinfo {title} {{Active
  Particles with Soft and Curved Walls: Equation of State, Ratchets, and
  Instabilities}},}\ }\href {https://doi.org/10.1103/PhysRevLett.117.098001}
  {\bibfield  {journal} {\bibinfo  {journal} {Phys. Rev. Lett.}\ }\textbf
  {\bibinfo {volume} {117}}~(\bibinfo {number} {9}),\ \bibinfo {pages}
  {098001}}\BibitemShut {NoStop}%
\bibitem [{\citenamefont {O'Byrne}\ and\ \citenamefont
  {Tailleur}(2020)}]{o2020lamellar}%
  \BibitemOpen
  \bibfield  {author} {\bibinfo {author} {\bibnamefont {O'Byrne}, \bibfnamefont
  {J}}, and\ \bibinfo {author} {\bibfnamefont {J.}~\bibnamefont {Tailleur}}}
  (\bibinfo {year} {2020}),\ \bibfield  {title} {\enquote {\bibinfo {title}
  {{Lamellar to Micellar Phases and Beyond: When Tactic Active Systems Admit
  Free Energy Functionals}},}\ }\href
  {https://doi.org/10.1103/PhysRevLett.125.208003} {\bibfield  {journal}
  {\bibinfo  {journal} {Phys. Rev. Lett.}\ }\textbf {\bibinfo {volume}
  {125}}~(\bibinfo {number} {20}),\ \bibinfo {pages} {208003}}\BibitemShut
  {NoStop}%
\bibitem [{\citenamefont {O'Byrne}\ \emph {et~al.}(2023)\citenamefont
  {O'Byrne}, \citenamefont {Solon}, \citenamefont {Tailleur},\ and\
  \citenamefont {Zhao}}]{kurzthaler2022out}%
  \BibitemOpen
  \bibfield  {author} {\bibinfo {author} {\bibnamefont {O'Byrne}, \bibfnamefont
  {Jérémy}}, \bibinfo {author} {\bibfnamefont {Alexandre}\ \bibnamefont
  {Solon}}, \bibinfo {author} {\bibfnamefont {Julien}\ \bibnamefont
  {Tailleur}}, and\ \bibinfo {author} {\bibfnamefont {Yongfeng}\ \bibnamefont
  {Zhao}}} (\bibinfo {year} {2023}),\ \bibfield  {title} {\enquote {\bibinfo
  {title} {An introduction to motility-induced phase separation},}\ }in\
  \href@noop {} {\emph {\bibinfo {booktitle} {Out-of-equilibrium Soft Matter:
  Active Fluids}}},\ \bibinfo {series and number} {Soft Matter Series},\
  \bibinfo {editor} {edited by\ \bibinfo {editor} {\bibfnamefont {Christina}\
  \bibnamefont {Kurzthaler}}, \bibinfo {editor} {\bibfnamefont {Luigi}\
  \bibnamefont {Gentile}}, \ and\ \bibinfo {editor} {\bibfnamefont {Howard~A.}\
  \bibnamefont {Stone}}},\ Chap.~\bibinfo {chapter} {4}\ (\bibinfo  {publisher}
  {Royal Society of Chemistry})\BibitemShut {NoStop}%
\bibitem [{\citenamefont {Omar}\ \emph {et~al.}(2023)\citenamefont {Omar},
  \citenamefont {Row}, \citenamefont {Mallory},\ and\ \citenamefont
  {Brady}}]{omar2023mechanical}%
  \BibitemOpen
  \bibfield  {author} {\bibinfo {author} {\bibnamefont {Omar}, \bibfnamefont
  {Ahmad~K}}, \bibinfo {author} {\bibfnamefont {Hyeongjoo}\ \bibnamefont
  {Row}}, \bibinfo {author} {\bibfnamefont {Stewart~A.}\ \bibnamefont
  {Mallory}}, and\ \bibinfo {author} {\bibfnamefont {John~F.}\ \bibnamefont
  {Brady}}} (\bibinfo {year} {2023}),\ \bibfield  {title} {\enquote {\bibinfo
  {title} {{Mechanical theory of nonequilibrium coexistence and
  motility-induced phase separation}},}\ }\href
  {https://doi.org/10.1073/pnas.2219900120} {\bibfield  {journal} {\bibinfo
  {journal} {Proc. Natl. Acad. Sci.}\ }\textbf {\bibinfo {volume}
  {120}}~(\bibinfo {number} {18}),\ \bibinfo {pages} {e2219900120}}\BibitemShut
  {NoStop}%
\bibitem [{\citenamefont {Omar}\ \emph {et~al.}(2020)\citenamefont {Omar},
  \citenamefont {Wang},\ and\ \citenamefont {Brady}}]{omar2020microscopic}%
  \BibitemOpen
  \bibfield  {author} {\bibinfo {author} {\bibnamefont {Omar}, \bibfnamefont
  {Ahmad~K}}, \bibinfo {author} {\bibfnamefont {Zhen-Gang}\ \bibnamefont
  {Wang}}, and\ \bibinfo {author} {\bibfnamefont {John~F.}\ \bibnamefont
  {Brady}}} (\bibinfo {year} {2020}),\ \bibfield  {title} {\enquote {\bibinfo
  {title} {{Microscopic origins of the swim pressure and the anomalous surface
  tension of active matter}},}\ }\href
  {https://doi.org/10.1103/PhysRevE.101.012604} {\bibfield  {journal} {\bibinfo
   {journal} {Phys. Rev. E}\ }\textbf {\bibinfo {volume} {101}}~(\bibinfo
  {number} {1}),\ \bibinfo {pages} {012604}}\BibitemShut {NoStop}%
\bibitem [{\citenamefont {Palacci}\ \emph {et~al.}(2013)\citenamefont
  {Palacci}, \citenamefont {Sacanna}, \citenamefont {Steinberg}, \citenamefont
  {Pine},\ and\ \citenamefont {Chaikin}}]{palacci2013living}%
  \BibitemOpen
  \bibfield  {author} {\bibinfo {author} {\bibnamefont {Palacci}, \bibfnamefont
  {Jeremie}}, \bibinfo {author} {\bibfnamefont {Stefano}\ \bibnamefont
  {Sacanna}}, \bibinfo {author} {\bibfnamefont {Asher~Preska}\ \bibnamefont
  {Steinberg}}, \bibinfo {author} {\bibfnamefont {David~J.}\ \bibnamefont
  {Pine}}, and\ \bibinfo {author} {\bibfnamefont {Paul~M.}\ \bibnamefont
  {Chaikin}}} (\bibinfo {year} {2013}),\ \bibfield  {title} {\enquote {\bibinfo
  {title} {{Living Crystals of Light-Activated Colloidal Surfers}},}\ }\href
  {https://doi.org/10.1126/science.1230020} {\bibfield  {journal} {\bibinfo
  {journal} {Science}\ }\textbf {\bibinfo {volume} {339}}~(\bibinfo {number}
  {6122}),\ \bibinfo {pages} {936--940}}\BibitemShut {NoStop}%
\bibitem [{\citenamefont {Paliwal}\ \emph {et~al.}(2017)\citenamefont
  {Paliwal}, \citenamefont {Prymidis}, \citenamefont {Filion},\ and\
  \citenamefont {Dijkstra}}]{paliwal2017non}%
  \BibitemOpen
  \bibfield  {author} {\bibinfo {author} {\bibnamefont {Paliwal}, \bibfnamefont
  {Siddharth}}, \bibinfo {author} {\bibfnamefont {Vasileios}\ \bibnamefont
  {Prymidis}}, \bibinfo {author} {\bibfnamefont {Laura}\ \bibnamefont
  {Filion}}, and\ \bibinfo {author} {\bibfnamefont {Marjolein}\ \bibnamefont
  {Dijkstra}}} (\bibinfo {year} {2017}),\ \bibfield  {title} {\enquote
  {\bibinfo {title} {{Non-equilibrium surface tension of the vapour-liquid
  interface of active Lennard-Jones particles}},}\ }\href
  {https://doi.org/10.1063/1.4989764} {\bibfield  {journal} {\bibinfo
  {journal} {J. Chem. Phys.}\ }\textbf {\bibinfo {volume} {147}}~(\bibinfo
  {number} {8}),\ \bibinfo {pages} {84902}}\BibitemShut {NoStop}%
\bibitem [{\citenamefont {Paliwal}\ \emph {et~al.}(2018)\citenamefont
  {Paliwal}, \citenamefont {Rodenburg}, \citenamefont {van Roij},\ and\
  \citenamefont {Dijkstra}}]{paliwal2018chemical}%
  \BibitemOpen
  \bibfield  {author} {\bibinfo {author} {\bibnamefont {Paliwal}, \bibfnamefont
  {Siddharth}}, \bibinfo {author} {\bibfnamefont {Jeroen}\ \bibnamefont
  {Rodenburg}}, \bibinfo {author} {\bibfnamefont {Ren{\'{e}}}\ \bibnamefont
  {van Roij}}, and\ \bibinfo {author} {\bibfnamefont {Marjolein}\ \bibnamefont
  {Dijkstra}}} (\bibinfo {year} {2018}),\ \bibfield  {title} {\enquote
  {\bibinfo {title} {{Chemical potential in active systems: predicting phase
  equilibrium from bulk equations of state?}}}\ }\href
  {https://doi.org/10.1088/1367-2630/aa9b4d} {\bibfield  {journal} {\bibinfo
  {journal} {New J. Phys.}\ }\textbf {\bibinfo {volume} {20}}~(\bibinfo
  {number} {1}),\ \bibinfo {pages} {015003}}\BibitemShut {NoStop}%
\bibitem [{\citenamefont {Partridge}\ and\ \citenamefont
  {Lee}(2019)}]{partridge_critical_2019}%
  \BibitemOpen
  \bibfield  {author} {\bibinfo {author} {\bibnamefont {Partridge},
  \bibfnamefont {Benjamin}}, and\ \bibinfo {author} {\bibfnamefont {Chiu~Fan}\
  \bibnamefont {Lee}}} (\bibinfo {year} {2019}),\ \bibfield  {title} {\enquote
  {\bibinfo {title} {{Critical Motility-Induced Phase Separation Belongs to the
  Ising Universality Class}},}\ }\href
  {https://doi.org/10.1103/PhysRevLett.123.068002} {\bibfield  {journal}
  {\bibinfo  {journal} {Phys. Rev. Lett.}\ }\textbf {\bibinfo {volume}
  {123}}~(\bibinfo {number} {6}),\ \bibinfo {pages} {068002}}\BibitemShut
  {NoStop}%
\bibitem [{\citenamefont {Peng}\ and\ \citenamefont {Brady}(2022)}]{Peng2022}%
  \BibitemOpen
  \bibfield  {author} {\bibinfo {author} {\bibnamefont {Peng}, \bibfnamefont
  {Zhiwei}}, and\ \bibinfo {author} {\bibfnamefont {John~F.}\ \bibnamefont
  {Brady}}} (\bibinfo {year} {2022}),\ \bibfield  {title} {\enquote {\bibinfo
  {title} {{Forced microrheology of active colloids}},}\ }\href
  {https://doi.org/10.1122/8.0000504} {\bibfield  {journal} {\bibinfo
  {journal} {Journal of Rheology}\ }\textbf {\bibinfo {volume} {66}}~(\bibinfo
  {number} {5}),\ \bibinfo {pages} {955--972}}\BibitemShut {NoStop}%
\bibitem [{\citenamefont {de~Pirey}\ and\ \citenamefont {van
  Wijland}(2023)}]{arnoulx2023run}%
  \BibitemOpen
  \bibfield  {author} {\bibinfo {author} {\bibnamefont {de~Pirey},
  \bibfnamefont {Thibaut~Arnoulx}}, and\ \bibinfo {author} {\bibfnamefont
  {Frédéric}\ \bibnamefont {van Wijland}}} (\bibinfo {year} {2023}),\
  \href@noop {} {\enquote {\bibinfo {title} {A run-and-tumble particle around a
  spherical obstacle: steady-state distribution far-from-equilibrium},}\
  }\Eprint {https://arxiv.org/abs/2303.00331} {arXiv:2303.00331
  [cond-mat.stat-mech]} \BibitemShut {NoStop}%
\bibitem [{\citenamefont {Potiguar}\ \emph {et~al.}(2014)\citenamefont
  {Potiguar}, \citenamefont {Farias},\ and\ \citenamefont
  {Ferreira}}]{Potiguar2014}%
  \BibitemOpen
  \bibfield  {author} {\bibinfo {author} {\bibnamefont {Potiguar},
  \bibfnamefont {Fabricio~Q}}, \bibinfo {author} {\bibfnamefont {G.~A.}\
  \bibnamefont {Farias}}, and\ \bibinfo {author} {\bibfnamefont {W.~P.}\
  \bibnamefont {Ferreira}}} (\bibinfo {year} {2014}),\ \bibfield  {title}
  {\enquote {\bibinfo {title} {{Self-propelled particle transport in regular
  arrays of rigid asymmetric obstacles}},}\ }\href
  {https://doi.org/10.1103/PhysRevE.90.012307} {\bibfield  {journal} {\bibinfo
  {journal} {Phys. Rev. E}\ }\textbf {\bibinfo {volume} {90}}~(\bibinfo
  {number} {1}),\ \bibinfo {pages} {012307}}\BibitemShut {NoStop}%
\bibitem [{\citenamefont {Ramaswamy}(2010)}]{ramaswamy2010mechanics}%
  \BibitemOpen
  \bibfield  {author} {\bibinfo {author} {\bibnamefont {Ramaswamy},
  \bibfnamefont {Sriram}}} (\bibinfo {year} {2010}),\ \bibfield  {title}
  {\enquote {\bibinfo {title} {{The Mechanics and Statistics of Active
  Matter}},}\ }\href {https://doi.org/10.1146/annurev-conmatphys-070909-104101}
  {\bibfield  {journal} {\bibinfo  {journal} {Annu. Rev. Condens. Matter
  Phys.}\ }\textbf {\bibinfo {volume} {1}}~(\bibinfo {number} {1}),\ \bibinfo
  {pages} {323--345}}\BibitemShut {NoStop}%
\bibitem [{\citenamefont {Redner}\ \emph
  {et~al.}(2013{\natexlab{a}})\citenamefont {Redner}, \citenamefont
  {Baskaran},\ and\ \citenamefont {Hagan}}]{redner2013reentrant}%
  \BibitemOpen
  \bibfield  {author} {\bibinfo {author} {\bibnamefont {Redner}, \bibfnamefont
  {Gabriel~S}}, \bibinfo {author} {\bibfnamefont {Aparna}\ \bibnamefont
  {Baskaran}}, and\ \bibinfo {author} {\bibfnamefont {Michael~F.}\ \bibnamefont
  {Hagan}}} (\bibinfo {year} {2013}{\natexlab{a}}),\ \bibfield  {title}
  {\enquote {\bibinfo {title} {{Reentrant phase behavior in active colloids
  with attraction}},}\ }\href {https://doi.org/10.1103/PhysRevE.88.012305}
  {\bibfield  {journal} {\bibinfo  {journal} {Phys. Rev. E}\ }\textbf {\bibinfo
  {volume} {88}}~(\bibinfo {number} {1}),\ \bibinfo {pages}
  {12305}}\BibitemShut {NoStop}%
\bibitem [{\citenamefont {Redner}\ \emph
  {et~al.}(2013{\natexlab{b}})\citenamefont {Redner}, \citenamefont {Hagan},\
  and\ \citenamefont {Baskaran}}]{redner_structure_2013}%
  \BibitemOpen
  \bibfield  {author} {\bibinfo {author} {\bibnamefont {Redner}, \bibfnamefont
  {Gabriel~S}}, \bibinfo {author} {\bibfnamefont {Michael~F.}\ \bibnamefont
  {Hagan}}, and\ \bibinfo {author} {\bibfnamefont {Aparna}\ \bibnamefont
  {Baskaran}}} (\bibinfo {year} {2013}{\natexlab{b}}),\ \bibfield  {title}
  {\enquote {\bibinfo {title} {{Structure and Dynamics of a Phase-Separating
  Active Colloidal Fluid}},}\ }\href
  {https://doi.org/10.1103/PhysRevLett.110.055701} {\bibfield  {journal}
  {\bibinfo  {journal} {Phys. Rev. Lett.}\ }\textbf {\bibinfo {volume}
  {110}}~(\bibinfo {number} {5}),\ \bibinfo {pages} {055701}}\BibitemShut
  {NoStop}%
\bibitem [{\citenamefont {Reichert}\ \emph {et~al.}(2021)\citenamefont
  {Reichert}, \citenamefont {Granz},\ and\ \citenamefont
  {Voigtmann}}]{Reichert2021}%
  \BibitemOpen
  \bibfield  {author} {\bibinfo {author} {\bibnamefont {Reichert},
  \bibfnamefont {Julian}}, \bibinfo {author} {\bibfnamefont {Leon~F}\
  \bibnamefont {Granz}}, and\ \bibinfo {author} {\bibfnamefont {Thomas}\
  \bibnamefont {Voigtmann}}} (\bibinfo {year} {2021}),\ \bibfield  {title}
  {\enquote {\bibinfo {title} {{Transport coefficients in dense active Brownian
  particle systems: mode-coupling theory and simulation results}},}\ }\href
  {https://doi.org/10.1140/epje/s10189-021-00039-4} {\bibfield  {journal}
  {\bibinfo  {journal} {Eur. Phys. J. E}\ }\textbf {\bibinfo {volume}
  {44}}~(\bibinfo {number} {3}),\ \bibinfo {pages} {27}}\BibitemShut {NoStop}%
\bibitem [{\citenamefont {Reichert}\ and\ \citenamefont
  {Voigtmann}(2021)}]{Reichert2021a}%
  \BibitemOpen
  \bibfield  {author} {\bibinfo {author} {\bibnamefont {Reichert},
  \bibfnamefont {Julian}}, and\ \bibinfo {author} {\bibfnamefont {Thomas}\
  \bibnamefont {Voigtmann}}} (\bibinfo {year} {2021}),\ \bibfield  {title}
  {\enquote {\bibinfo {title} {{Tracer dynamics in crowded active-particle
  suspensions}},}\ }\href {https://doi.org/10.1039/D1SM01092A} {\bibfield
  {journal} {\bibinfo  {journal} {Soft Matter}\ }\textbf {\bibinfo {volume}
  {17}}~(\bibinfo {number} {46}),\ \bibinfo {pages} {10492--10504}}\BibitemShut
  {NoStop}%
\bibitem [{\citenamefont {Reichhardt}\ and\ \citenamefont
  {Reichhardt}(2013)}]{reichhardt2013active}%
  \BibitemOpen
  \bibfield  {author} {\bibinfo {author} {\bibnamefont {Reichhardt},
  \bibfnamefont {C}}, and\ \bibinfo {author} {\bibfnamefont {C.~J.~Olson}\
  \bibnamefont {Reichhardt}}} (\bibinfo {year} {2013}),\ \bibfield  {title}
  {\enquote {\bibinfo {title} {{Active matter ratchets with an external
  drift}},}\ }\href {https://doi.org/10.1103/PhysRevE.88.062310} {\bibfield
  {journal} {\bibinfo  {journal} {Phys. Rev. E}\ }\textbf {\bibinfo {volume}
  {88}}~(\bibinfo {number} {6}),\ \bibinfo {pages} {062310}}\BibitemShut
  {NoStop}%
\bibitem [{\citenamefont {Ro}\ \emph {et~al.}(2021)\citenamefont {Ro},
  \citenamefont {Kafri}, \citenamefont {Kardar},\ and\ \citenamefont
  {Tailleur}}]{ro2021disorder}%
  \BibitemOpen
  \bibfield  {author} {\bibinfo {author} {\bibnamefont {Ro}, \bibfnamefont
  {Sunghan}}, \bibinfo {author} {\bibfnamefont {Yariv}\ \bibnamefont {Kafri}},
  \bibinfo {author} {\bibfnamefont {Mehran}\ \bibnamefont {Kardar}}, and\
  \bibinfo {author} {\bibfnamefont {Julien}\ \bibnamefont {Tailleur}}}
  (\bibinfo {year} {2021}),\ \bibfield  {title} {\enquote {\bibinfo {title}
  {{Disorder-Induced Long-Ranged Correlations in Scalar Active Matter}},}\
  }\href {https://doi.org/10.1103/PhysRevLett.126.048003} {\bibfield  {journal}
  {\bibinfo  {journal} {Phys. Rev. Lett.}\ }\textbf {\bibinfo {volume}
  {126}}~(\bibinfo {number} {4}),\ \bibinfo {pages} {048003}}\BibitemShut
  {NoStop}%
\bibitem [{\citenamefont {Rodenburg}\ \emph {et~al.}(2018)\citenamefont
  {Rodenburg}, \citenamefont {Paliwal}, \citenamefont {de~Jager}, \citenamefont
  {Bolhuis}, \citenamefont {Dijkstra},\ and\ \citenamefont {van
  Roij}}]{rodenburg2018ratchet}%
  \BibitemOpen
  \bibfield  {author} {\bibinfo {author} {\bibnamefont {Rodenburg},
  \bibfnamefont {Jeroen}}, \bibinfo {author} {\bibfnamefont {Siddharth}\
  \bibnamefont {Paliwal}}, \bibinfo {author} {\bibfnamefont {Marjolein}\
  \bibnamefont {de~Jager}}, \bibinfo {author} {\bibfnamefont {Peter~G.}\
  \bibnamefont {Bolhuis}}, \bibinfo {author} {\bibfnamefont {Marjolein}\
  \bibnamefont {Dijkstra}}, and\ \bibinfo {author} {\bibfnamefont {Ren{\'{e}}}\
  \bibnamefont {van Roij}}} (\bibinfo {year} {2018}),\ \bibfield  {title}
  {\enquote {\bibinfo {title} {{Ratchet-induced variations in bulk states of an
  active ideal gas}},}\ }\href {https://doi.org/10.1063/1.5048698} {\bibfield
  {journal} {\bibinfo  {journal} {J. Chem. Phys.}\ }\textbf {\bibinfo {volume}
  {149}}~(\bibinfo {number} {17}),\ \bibinfo {pages} {174910}}\BibitemShut
  {NoStop}%
\bibitem [{\citenamefont {Rohwer}\ \emph {et~al.}(2020)\citenamefont {Rohwer},
  \citenamefont {Kardar},\ and\ \citenamefont
  {Kr{\"{u}}ger}}]{rohwer2020activated}%
  \BibitemOpen
  \bibfield  {author} {\bibinfo {author} {\bibnamefont {Rohwer}, \bibfnamefont
  {Christian~M}}, \bibinfo {author} {\bibfnamefont {Mehran}\ \bibnamefont
  {Kardar}}, and\ \bibinfo {author} {\bibfnamefont {Matthias}\ \bibnamefont
  {Kr{\"{u}}ger}}} (\bibinfo {year} {2020}),\ \bibfield  {title} {\enquote
  {\bibinfo {title} {{Activated diffusiophoresis}},}\ }\href
  {https://doi.org/10.1063/1.5139017} {\bibfield  {journal} {\bibinfo
  {journal} {J. Chem. Phys.}\ }\textbf {\bibinfo {volume} {152}}~(\bibinfo
  {number} {8}),\ \bibinfo {pages} {084109}}\BibitemShut {NoStop}%
\bibitem [{\citenamefont {Romanczuk}\ \emph {et~al.}(2012)\citenamefont
  {Romanczuk}, \citenamefont {B{\"{a}}r}, \citenamefont {Ebeling},
  \citenamefont {Lindner},\ and\ \citenamefont
  {Schimansky-Geier}}]{Romanczuk2012b}%
  \BibitemOpen
  \bibfield  {author} {\bibinfo {author} {\bibnamefont {Romanczuk},
  \bibfnamefont {P}}, \bibinfo {author} {\bibfnamefont {M.}~\bibnamefont
  {B{\"{a}}r}}, \bibinfo {author} {\bibfnamefont {W.}~\bibnamefont {Ebeling}},
  \bibinfo {author} {\bibfnamefont {B.}~\bibnamefont {Lindner}}, and\ \bibinfo
  {author} {\bibfnamefont {L.}~\bibnamefont {Schimansky-Geier}}} (\bibinfo
  {year} {2012}),\ \bibfield  {title} {\enquote {\bibinfo {title} {{Active
  Brownian particles}},}\ }\href {https://doi.org/10.1140/epjst/e2012-01529-y}
  {\bibfield  {journal} {\bibinfo  {journal} {Eur. Phys. J. Spec. Top.}\
  }\textbf {\bibinfo {volume} {202}}~(\bibinfo {number} {1}),\ \bibinfo {pages}
  {1--162}}\BibitemShut {NoStop}%
\bibitem [{\citenamefont {Saha}\ \emph {et~al.}(2020)\citenamefont {Saha},
  \citenamefont {Agudo-Canalejo},\ and\ \citenamefont
  {Golestanian}}]{saha_scalar_2020}%
  \BibitemOpen
  \bibfield  {author} {\bibinfo {author} {\bibnamefont {Saha}, \bibfnamefont
  {Suropriya}}, \bibinfo {author} {\bibfnamefont {Jaime}\ \bibnamefont
  {Agudo-Canalejo}}, and\ \bibinfo {author} {\bibfnamefont {Ramin}\
  \bibnamefont {Golestanian}}} (\bibinfo {year} {2020}),\ \bibfield  {title}
  {\enquote {\bibinfo {title} {{Scalar Active Mixtures: The Nonreciprocal
  Cahn-Hilliard Model}},}\ }\href {https://doi.org/10.1103/PhysRevX.10.041009}
  {\bibfield  {journal} {\bibinfo  {journal} {Phys. Rev. X}\ }\textbf {\bibinfo
  {volume} {10}}~(\bibinfo {number} {4}),\ \bibinfo {pages}
  {041009}}\BibitemShut {NoStop}%
\bibitem [{\citenamefont {Saha}\ \emph {et~al.}(2014)\citenamefont {Saha},
  \citenamefont {Golestanian},\ and\ \citenamefont
  {Ramaswamy}}]{saha2014clusters}%
  \BibitemOpen
  \bibfield  {author} {\bibinfo {author} {\bibnamefont {Saha}, \bibfnamefont
  {Suropriya}}, \bibinfo {author} {\bibfnamefont {Ramin}\ \bibnamefont
  {Golestanian}}, and\ \bibinfo {author} {\bibfnamefont {Sriram}\ \bibnamefont
  {Ramaswamy}}} (\bibinfo {year} {2014}),\ \bibfield  {title} {\enquote
  {\bibinfo {title} {{Clusters, asters, and collective oscillations in
  chemotactic colloids}},}\ }\href {https://doi.org/10.1103/PhysRevE.89.062316}
  {\bibfield  {journal} {\bibinfo  {journal} {Phys. Rev. E}\ }\textbf {\bibinfo
  {volume} {89}}~(\bibinfo {number} {6}),\ \bibinfo {pages}
  {062316}}\BibitemShut {NoStop}%
\bibitem [{\citenamefont {Sandford}\ \emph {et~al.}(2017)\citenamefont
  {Sandford}, \citenamefont {Grosberg},\ and\ \citenamefont
  {Joanny}}]{sandford2017pressure}%
  \BibitemOpen
  \bibfield  {author} {\bibinfo {author} {\bibnamefont {Sandford},
  \bibfnamefont {Cato}}, \bibinfo {author} {\bibfnamefont {Alexander~Y.}\
  \bibnamefont {Grosberg}}, and\ \bibinfo {author} {\bibfnamefont
  {Jean-Fran{\c{c}}ois}\ \bibnamefont {Joanny}}} (\bibinfo {year} {2017}),\
  \bibfield  {title} {\enquote {\bibinfo {title} {{Pressure and flow of
  exponentially self-correlated active particles}},}\ }\href
  {https://doi.org/10.1103/PhysRevE.96.052605} {\bibfield  {journal} {\bibinfo
  {journal} {Phys. Rev. E}\ }\textbf {\bibinfo {volume} {96}}~(\bibinfo
  {number} {5}),\ \bibinfo {pages} {052605}}\BibitemShut {NoStop}%
\bibitem [{\citenamefont {Schmidt}\ \emph {et~al.}(2021)\citenamefont
  {Schmidt}, \citenamefont {{\v{S}}{\'{i}}pov{\'{a}}-Jungov{\'{a}}},
  \citenamefont {K{\"{a}}ll}, \citenamefont {W{\"{u}}rger},\ and\ \citenamefont
  {Volpe}}]{schmidt2021non}%
  \BibitemOpen
  \bibfield  {author} {\bibinfo {author} {\bibnamefont {Schmidt}, \bibfnamefont
  {Falko}}, \bibinfo {author} {\bibfnamefont {Hana}\ \bibnamefont
  {{\v{S}}{\'{i}}pov{\'{a}}-Jungov{\'{a}}}}, \bibinfo {author} {\bibfnamefont
  {Mikael}\ \bibnamefont {K{\"{a}}ll}}, \bibinfo {author} {\bibfnamefont
  {Alois}\ \bibnamefont {W{\"{u}}rger}}, and\ \bibinfo {author} {\bibfnamefont
  {Giovanni}\ \bibnamefont {Volpe}}} (\bibinfo {year} {2021}),\ \bibfield
  {title} {\enquote {\bibinfo {title} {{Non-equilibrium properties of an active
  nanoparticle in a harmonic potential}},}\ }\href
  {https://doi.org/10.1038/s41467-021-22187-z} {\bibfield  {journal} {\bibinfo
  {journal} {Nat. Commun.}\ }\textbf {\bibinfo {volume} {12}}~(\bibinfo
  {number} {1}),\ \bibinfo {pages} {1902}}\BibitemShut {NoStop}%
\bibitem [{\citenamefont {Schnitzer}(1993)}]{schnitzer1993theory}%
  \BibitemOpen
  \bibfield  {author} {\bibinfo {author} {\bibnamefont {Schnitzer},
  \bibfnamefont {Mark~J}}} (\bibinfo {year} {1993}),\ \bibfield  {title}
  {\enquote {\bibinfo {title} {{Theory of continuum random walks and
  application to chemotaxis}},}\ }\href
  {https://doi.org/10.1103/PhysRevE.48.2553} {\bibfield  {journal} {\bibinfo
  {journal} {Phys. Rev. E}\ }\textbf {\bibinfo {volume} {48}}~(\bibinfo
  {number} {4}),\ \bibinfo {pages} {2553--2568}}\BibitemShut {NoStop}%
\bibitem [{\citenamefont {Sep{\'{u}}lveda}\ \emph {et~al.}(2013)\citenamefont
  {Sep{\'{u}}lveda}, \citenamefont {Petitjean}, \citenamefont {Cochet},
  \citenamefont {Grasland-Mongrain}, \citenamefont {Silberzan},\ and\
  \citenamefont {Hakim}}]{sepulveda2013collective}%
  \BibitemOpen
  \bibfield  {author} {\bibinfo {author} {\bibnamefont {Sep{\'{u}}lveda},
  \bibfnamefont {N{\'{e}}stor}}, \bibinfo {author} {\bibfnamefont {Laurence}\
  \bibnamefont {Petitjean}}, \bibinfo {author} {\bibfnamefont {Olivier}\
  \bibnamefont {Cochet}}, \bibinfo {author} {\bibfnamefont {Erwan}\
  \bibnamefont {Grasland-Mongrain}}, \bibinfo {author} {\bibfnamefont {Pascal}\
  \bibnamefont {Silberzan}}, and\ \bibinfo {author} {\bibfnamefont {Vincent}\
  \bibnamefont {Hakim}}} (\bibinfo {year} {2013}),\ \bibfield  {title}
  {\enquote {\bibinfo {title} {{Collective Cell Motion in an Epithelial Sheet
  Can Be Quantitatively Described by a Stochastic Interacting Particle
  Model}},}\ }\href {https://doi.org/10.1371/journal.pcbi.1002944} {\bibfield
  {journal} {\bibinfo  {journal} {PLoS Comput. Biol.}\ }\textbf {\bibinfo
  {volume} {9}}~(\bibinfo {number} {3}),\ \bibinfo {pages}
  {e1002944}}\BibitemShut {NoStop}%
\bibitem [{\citenamefont {Ses{\'{e}}-Sansa}\ \emph {et~al.}(2018)\citenamefont
  {Ses{\'{e}}-Sansa}, \citenamefont {Pagonabarraga},\ and\ \citenamefont
  {Levis}}]{sese2018velocity}%
  \BibitemOpen
  \bibfield  {author} {\bibinfo {author} {\bibnamefont {Ses{\'{e}}-Sansa},
  \bibfnamefont {E}}, \bibinfo {author} {\bibfnamefont {I.}~\bibnamefont
  {Pagonabarraga}}, and\ \bibinfo {author} {\bibfnamefont {D.}~\bibnamefont
  {Levis}}} (\bibinfo {year} {2018}),\ \bibfield  {title} {\enquote {\bibinfo
  {title} {{Velocity alignment promotes motility-induced phase separation}},}\
  }\href {https://doi.org/10.1209/0295-5075/124/30004} {\bibfield  {journal}
  {\bibinfo  {journal} {EPL}\ }\textbf {\bibinfo {volume} {124}}~(\bibinfo
  {number} {3}),\ \bibinfo {pages} {30004}}\BibitemShut {NoStop}%
\bibitem [{\citenamefont {Shea}\ \emph {et~al.}(2022)\citenamefont {Shea},
  \citenamefont {Jung},\ and\ \citenamefont {Schmid}}]{Shea2022}%
  \BibitemOpen
  \bibfield  {author} {\bibinfo {author} {\bibnamefont {Shea}, \bibfnamefont
  {Jeanine}}, \bibinfo {author} {\bibfnamefont {Gerhard}\ \bibnamefont {Jung}},
  and\ \bibinfo {author} {\bibfnamefont {Friederike}\ \bibnamefont {Schmid}}}
  (\bibinfo {year} {2022}),\ \bibfield  {title} {\enquote {\bibinfo {title}
  {{Passive probe particle in an active bath: can we tell it is out of
  equilibrium?}}}\ }\href {https://doi.org/10.1039/D2SM00905F} {\bibfield
  {journal} {\bibinfo  {journal} {Soft Matter}\ }\textbf {\bibinfo {volume}
  {18}}~(\bibinfo {number} {36}),\ \bibinfo {pages} {6965--6973}}\BibitemShut
  {NoStop}%
\bibitem [{\citenamefont {Shi}\ \emph {et~al.}(2020)\citenamefont {Shi},
  \citenamefont {Fausti}, \citenamefont {Chat{\'{e}}}, \citenamefont
  {Nardini},\ and\ \citenamefont {Solon}}]{shi_self-organized_2020}%
  \BibitemOpen
  \bibfield  {author} {\bibinfo {author} {\bibnamefont {Shi}, \bibfnamefont
  {Xia-qing}}, \bibinfo {author} {\bibfnamefont {Giordano}\ \bibnamefont
  {Fausti}}, \bibinfo {author} {\bibfnamefont {Hugues}\ \bibnamefont
  {Chat{\'{e}}}}, \bibinfo {author} {\bibfnamefont {Cesare}\ \bibnamefont
  {Nardini}}, and\ \bibinfo {author} {\bibfnamefont {Alexandre}\ \bibnamefont
  {Solon}}} (\bibinfo {year} {2020}),\ \bibfield  {title} {\enquote {\bibinfo
  {title} {{Self-Organized Critical Coexistence Phase in Repulsive Active
  Particles}},}\ }\href {https://doi.org/10.1103/PhysRevLett.125.168001}
  {\bibfield  {journal} {\bibinfo  {journal} {Phys. Rev. Lett.}\ }\textbf
  {\bibinfo {volume} {125}}~(\bibinfo {number} {16}),\ \bibinfo {pages}
  {168001}}\BibitemShut {NoStop}%
\bibitem [{\citenamefont {Shin}\ \emph {et~al.}(2015)\citenamefont {Shin},
  \citenamefont {Cherstvy}, \citenamefont {Kim},\ and\ \citenamefont
  {Metzler}}]{PolymerLooping}%
  \BibitemOpen
  \bibfield  {author} {\bibinfo {author} {\bibnamefont {Shin}, \bibfnamefont
  {Jaeoh}}, \bibinfo {author} {\bibfnamefont {Andrey~G.}\ \bibnamefont
  {Cherstvy}}, \bibinfo {author} {\bibfnamefont {Won~Kyu}\ \bibnamefont {Kim}},
  and\ \bibinfo {author} {\bibfnamefont {Ralf}\ \bibnamefont {Metzler}}}
  (\bibinfo {year} {2015}),\ \bibfield  {title} {\enquote {\bibinfo {title}
  {{Facilitation of polymer looping and giant polymer diffusivity in crowded
  solutions of active particles}},}\ }\href
  {https://doi.org/10.1088/1367-2630/17/11/113008} {\bibfield  {journal}
  {\bibinfo  {journal} {New J. Phys.}\ }\textbf {\bibinfo {volume}
  {17}}~(\bibinfo {number} {11}),\ \bibinfo {pages} {113008}}\BibitemShut
  {NoStop}%
\bibitem [{\citenamefont {Siebert}\ \emph {et~al.}(2018)\citenamefont
  {Siebert}, \citenamefont {Dittrich}, \citenamefont {Schmid}, \citenamefont
  {Binder}, \citenamefont {Speck},\ and\ \citenamefont
  {Virnau}}]{siebert_critical_2018}%
  \BibitemOpen
  \bibfield  {author} {\bibinfo {author} {\bibnamefont {Siebert}, \bibfnamefont
  {Jonathan~Tammo}}, \bibinfo {author} {\bibfnamefont {Florian}\ \bibnamefont
  {Dittrich}}, \bibinfo {author} {\bibfnamefont {Friederike}\ \bibnamefont
  {Schmid}}, \bibinfo {author} {\bibfnamefont {Kurt}\ \bibnamefont {Binder}},
  \bibinfo {author} {\bibfnamefont {Thomas}\ \bibnamefont {Speck}}, and\
  \bibinfo {author} {\bibfnamefont {Peter}\ \bibnamefont {Virnau}}} (\bibinfo
  {year} {2018}),\ \bibfield  {title} {\enquote {\bibinfo {title} {{Critical
  behavior of active Brownian particles}},}\ }\href
  {https://doi.org/10.1103/PhysRevE.98.030601} {\bibfield  {journal} {\bibinfo
  {journal} {Phys. Rev. E}\ }\textbf {\bibinfo {volume} {98}}~(\bibinfo
  {number} {3}),\ \bibinfo {pages} {030601}}\BibitemShut {NoStop}%
\bibitem [{\citenamefont {Sinai}(1983)}]{sinai1983limiting}%
  \BibitemOpen
  \bibfield  {author} {\bibinfo {author} {\bibnamefont {Sinai}, \bibfnamefont
  {Ya~G}}} (\bibinfo {year} {1983}),\ \bibfield  {title} {\enquote {\bibinfo
  {title} {{The Limiting Behavior of a One-Dimensional Random Walk in a Random
  Medium}},}\ }\href {https://doi.org/10.1137/1127028} {\bibfield  {journal}
  {\bibinfo  {journal} {Theory Probab. Its Appl.}\ }\textbf {\bibinfo {volume}
  {27}}~(\bibinfo {number} {2}),\ \bibinfo {pages} {256--268}}\BibitemShut
  {NoStop}%
\bibitem [{\citenamefont {Smith}\ \emph {et~al.}(2022)\citenamefont {Smith},
  \citenamefont {{Le Doussal}}, \citenamefont {Majumdar},\ and\ \citenamefont
  {Schehr}}]{smith2022exact}%
  \BibitemOpen
  \bibfield  {author} {\bibinfo {author} {\bibnamefont {Smith}, \bibfnamefont
  {Naftali~R}}, \bibinfo {author} {\bibfnamefont {Pierre}\ \bibnamefont {{Le
  Doussal}}}, \bibinfo {author} {\bibfnamefont {Satya~N.}\ \bibnamefont
  {Majumdar}}, and\ \bibinfo {author} {\bibfnamefont {Gr{\'{e}}gory}\
  \bibnamefont {Schehr}}} (\bibinfo {year} {2022}),\ \bibfield  {title}
  {\enquote {\bibinfo {title} {{Exact position distribution of a harmonically
  confined run-and-tumble particle in two dimensions}},}\ }\href
  {https://doi.org/10.1103/PhysRevE.106.054133} {\bibfield  {journal} {\bibinfo
   {journal} {Phys. Rev. E}\ }\textbf {\bibinfo {volume} {106}}~(\bibinfo
  {number} {5}),\ \bibinfo {pages} {054133}}\BibitemShut {NoStop}%
\bibitem [{\citenamefont {Sokolov}\ \emph {et~al.}(2010)\citenamefont
  {Sokolov}, \citenamefont {Apodaca}, \citenamefont {Grzybowski},\ and\
  \citenamefont {Aranson}}]{sokolov2010swimming}%
  \BibitemOpen
  \bibfield  {author} {\bibinfo {author} {\bibnamefont {Sokolov}, \bibfnamefont
  {Andrey}}, \bibinfo {author} {\bibfnamefont {Mario~M.}\ \bibnamefont
  {Apodaca}}, \bibinfo {author} {\bibfnamefont {Bartosz~A.}\ \bibnamefont
  {Grzybowski}}, and\ \bibinfo {author} {\bibfnamefont {Igor~S.}\ \bibnamefont
  {Aranson}}} (\bibinfo {year} {2010}),\ \bibfield  {title} {\enquote {\bibinfo
  {title} {{Swimming bacteria power microscopic gears}},}\ }\href
  {https://doi.org/10.1073/pnas.0913015107} {\bibfield  {journal} {\bibinfo
  {journal} {Proc. Natl. Acad. Sci.}\ }\textbf {\bibinfo {volume}
  {107}}~(\bibinfo {number} {3}),\ \bibinfo {pages} {969--974}}\BibitemShut
  {NoStop}%
\bibitem [{\citenamefont {Solon}\ \emph
  {et~al.}(2015{\natexlab{a}})\citenamefont {Solon}, \citenamefont {Cates},\
  and\ \citenamefont {Tailleur}}]{solon2015active}%
  \BibitemOpen
  \bibfield  {author} {\bibinfo {author} {\bibnamefont {Solon}, \bibfnamefont
  {A~P}}, \bibinfo {author} {\bibfnamefont {M.~E.}\ \bibnamefont {Cates}}, and\
  \bibinfo {author} {\bibfnamefont {J.}~\bibnamefont {Tailleur}}} (\bibinfo
  {year} {2015}{\natexlab{a}}),\ \bibfield  {title} {\enquote {\bibinfo {title}
  {{Active brownian particles and run-and-tumble particles: A comparative
  study}},}\ }\href {https://doi.org/10.1140/epjst/e2015-02457-0} {\bibfield
  {journal} {\bibinfo  {journal} {Eur. Phys. J. Spec. Top.}\ }\textbf {\bibinfo
  {volume} {224}}~(\bibinfo {number} {7}),\ \bibinfo {pages}
  {1231--1262}}\BibitemShut {NoStop}%
\bibitem [{\citenamefont {Solon}\ \emph
  {et~al.}(2015{\natexlab{b}})\citenamefont {Solon}, \citenamefont {Fily},
  \citenamefont {Baskaran}, \citenamefont {Cates}, \citenamefont {Kafri},
  \citenamefont {Kardar},\ and\ \citenamefont {Tailleur}}]{solon2015pressure}%
  \BibitemOpen
  \bibfield  {author} {\bibinfo {author} {\bibnamefont {Solon}, \bibfnamefont
  {A~P}}, \bibinfo {author} {\bibfnamefont {Y.}~\bibnamefont {Fily}}, \bibinfo
  {author} {\bibfnamefont {A.}~\bibnamefont {Baskaran}}, \bibinfo {author}
  {\bibfnamefont {M.~E.}\ \bibnamefont {Cates}}, \bibinfo {author}
  {\bibfnamefont {Y.}~\bibnamefont {Kafri}}, \bibinfo {author} {\bibfnamefont
  {M.}~\bibnamefont {Kardar}}, and\ \bibinfo {author} {\bibfnamefont
  {J.}~\bibnamefont {Tailleur}}} (\bibinfo {year} {2015}{\natexlab{b}}),\
  \bibfield  {title} {\enquote {\bibinfo {title} {{Pressure is not a state
  function for generic active fluids}},}\ }\href
  {https://doi.org/10.1038/nphys3377} {\bibfield  {journal} {\bibinfo
  {journal} {Nat. Phys.}\ }\textbf {\bibinfo {volume} {11}}~(\bibinfo {number}
  {8}),\ \bibinfo {pages} {673--678}}\BibitemShut {NoStop}%
\bibitem [{\citenamefont {Solon}\ and\ \citenamefont
  {Horowitz}(2022)}]{Solon2022}%
  \BibitemOpen
  \bibfield  {author} {\bibinfo {author} {\bibnamefont {Solon}, \bibfnamefont
  {Alexandre}}, and\ \bibinfo {author} {\bibfnamefont {Jordan~M.}\ \bibnamefont
  {Horowitz}}} (\bibinfo {year} {2022}),\ \bibfield  {title} {\enquote
  {\bibinfo {title} {{On the Einstein relation between mobility and diffusion
  coefficient in an active bath}},}\ }\href
  {https://doi.org/10.1088/1751-8121/ac5d82} {\bibfield  {journal} {\bibinfo
  {journal} {J. Phys. A Math. Theor.}\ }\textbf {\bibinfo {volume}
  {55}}~(\bibinfo {number} {18}),\ \bibinfo {pages} {184002}}\BibitemShut
  {NoStop}%
\bibitem [{\citenamefont {Solon}\ \emph
  {et~al.}(2018{\natexlab{a}})\citenamefont {Solon}, \citenamefont
  {Stenhammar}, \citenamefont {Cates}, \citenamefont {Kafri},\ and\
  \citenamefont {Tailleur}}]{solon_generalized_2018-1}%
  \BibitemOpen
  \bibfield  {author} {\bibinfo {author} {\bibnamefont {Solon}, \bibfnamefont
  {Alexandre~P}}, \bibinfo {author} {\bibfnamefont {Joakim}\ \bibnamefont
  {Stenhammar}}, \bibinfo {author} {\bibfnamefont {Michael~E.}\ \bibnamefont
  {Cates}}, \bibinfo {author} {\bibfnamefont {Yariv}\ \bibnamefont {Kafri}},
  and\ \bibinfo {author} {\bibfnamefont {Julien}\ \bibnamefont {Tailleur}}}
  (\bibinfo {year} {2018}{\natexlab{a}}),\ \bibfield  {title} {\enquote
  {\bibinfo {title} {{Generalized thermodynamics of motility-induced phase
  separation: phase equilibria, Laplace pressure, and change of ensembles}},}\
  }\href {https://doi.org/10.1088/1367-2630/aaccdd} {\bibfield  {journal}
  {\bibinfo  {journal} {New J. Phys.}\ }\textbf {\bibinfo {volume}
  {20}}~(\bibinfo {number} {7}),\ \bibinfo {pages} {075001}}\BibitemShut
  {NoStop}%
\bibitem [{\citenamefont {Solon}\ \emph
  {et~al.}(2018{\natexlab{b}})\citenamefont {Solon}, \citenamefont
  {Stenhammar}, \citenamefont {Cates}, \citenamefont {Kafri},\ and\
  \citenamefont {Tailleur}}]{solon_generalized_2018}%
  \BibitemOpen
  \bibfield  {author} {\bibinfo {author} {\bibnamefont {Solon}, \bibfnamefont
  {Alexandre~P}}, \bibinfo {author} {\bibfnamefont {Joakim}\ \bibnamefont
  {Stenhammar}}, \bibinfo {author} {\bibfnamefont {Michael~E.}\ \bibnamefont
  {Cates}}, \bibinfo {author} {\bibfnamefont {Yariv}\ \bibnamefont {Kafri}},
  and\ \bibinfo {author} {\bibfnamefont {Julien}\ \bibnamefont {Tailleur}}}
  (\bibinfo {year} {2018}{\natexlab{b}}),\ \bibfield  {title} {\enquote
  {\bibinfo {title} {{Generalized thermodynamics of phase equilibria in scalar
  active matter}},}\ }\href {https://doi.org/10.1103/PhysRevE.97.020602}
  {\bibfield  {journal} {\bibinfo  {journal} {Phys. Rev. E}\ }\textbf {\bibinfo
  {volume} {97}}~(\bibinfo {number} {2}),\ \bibinfo {pages}
  {020602}}\BibitemShut {NoStop}%
\bibitem [{\citenamefont {Solon}\ \emph
  {et~al.}(2015{\natexlab{c}})\citenamefont {Solon}, \citenamefont
  {Stenhammar}, \citenamefont {Wittkowski}, \citenamefont {Kardar},
  \citenamefont {Kafri}, \citenamefont {Cates},\ and\ \citenamefont
  {Tailleur}}]{Solon_interactions}%
  \BibitemOpen
  \bibfield  {author} {\bibinfo {author} {\bibnamefont {Solon}, \bibfnamefont
  {Alexandre~P}}, \bibinfo {author} {\bibfnamefont {Joakim}\ \bibnamefont
  {Stenhammar}}, \bibinfo {author} {\bibfnamefont {Raphael}\ \bibnamefont
  {Wittkowski}}, \bibinfo {author} {\bibfnamefont {Mehran}\ \bibnamefont
  {Kardar}}, \bibinfo {author} {\bibfnamefont {Yariv}\ \bibnamefont {Kafri}},
  \bibinfo {author} {\bibfnamefont {Michael~E.}\ \bibnamefont {Cates}}, and\
  \bibinfo {author} {\bibfnamefont {Julien}\ \bibnamefont {Tailleur}}}
  (\bibinfo {year} {2015}{\natexlab{c}}),\ \bibfield  {title} {\enquote
  {\bibinfo {title} {{Pressure and Phase Equilibria in Interacting Active
  Brownian Spheres}},}\ }\href {https://doi.org/10.1103/PhysRevLett.114.198301}
  {\bibfield  {journal} {\bibinfo  {journal} {Phys. Rev. Lett.}\ }\textbf
  {\bibinfo {volume} {114}}~(\bibinfo {number} {19}),\ \bibinfo {pages}
  {198301}}\BibitemShut {NoStop}%
\bibitem [{\citenamefont {Speck}(2021)}]{speck2021coexistence}%
  \BibitemOpen
  \bibfield  {author} {\bibinfo {author} {\bibnamefont {Speck}, \bibfnamefont
  {Thomas}}} (\bibinfo {year} {2021}),\ \bibfield  {title} {\enquote {\bibinfo
  {title} {{Coexistence of active Brownian disks: van der Waals theory and
  analytical results}},}\ }\href {https://doi.org/10.1103/PhysRevE.103.012607}
  {\bibfield  {journal} {\bibinfo  {journal} {Phys. Rev. E}\ }\textbf {\bibinfo
  {volume} {103}}~(\bibinfo {number} {1}),\ \bibinfo {pages}
  {012607}}\BibitemShut {NoStop}%
\bibitem [{\citenamefont {Speck}(2022)}]{Speck2022rg}%
  \BibitemOpen
  \bibfield  {author} {\bibinfo {author} {\bibnamefont {Speck}, \bibfnamefont
  {Thomas}}} (\bibinfo {year} {2022}),\ \bibfield  {title} {\enquote {\bibinfo
  {title} {{Critical behavior of active Brownian particles: Connection to field
  theories}},}\ }\href {https://doi.org/10.1103/PhysRevE.105.064601} {\bibfield
   {journal} {\bibinfo  {journal} {Phys. Rev. E}\ }\textbf {\bibinfo {volume}
  {105}}~(\bibinfo {number} {6}),\ \bibinfo {pages} {064601}}\BibitemShut
  {NoStop}%
\bibitem [{\citenamefont {Speck}\ and\ \citenamefont
  {Jayaram}(2021)}]{speck2021vorticity}%
  \BibitemOpen
  \bibfield  {author} {\bibinfo {author} {\bibnamefont {Speck}, \bibfnamefont
  {Thomas}}, and\ \bibinfo {author} {\bibfnamefont {Ashreya}\ \bibnamefont
  {Jayaram}}} (\bibinfo {year} {2021}),\ \bibfield  {title} {\enquote {\bibinfo
  {title} {{Vorticity Determines the Force on Bodies Immersed in Active
  Fluids}},}\ }\href {https://doi.org/10.1103/PhysRevLett.126.138002}
  {\bibfield  {journal} {\bibinfo  {journal} {Phys. Rev. Lett.}\ }\textbf
  {\bibinfo {volume} {126}}~(\bibinfo {number} {13}),\ \bibinfo {pages}
  {138002}}\BibitemShut {NoStop}%
\bibitem [{\citenamefont {Spera}\ \emph {et~al.}(2023)\citenamefont {Spera},
  \citenamefont {Duclut}, \citenamefont {Durand},\ and\ \citenamefont
  {Tailleur}}]{spera2023nematic}%
  \BibitemOpen
  \bibfield  {author} {\bibinfo {author} {\bibnamefont {Spera}, \bibfnamefont
  {Gianmarco}}, \bibinfo {author} {\bibfnamefont {Charlie}\ \bibnamefont
  {Duclut}}, \bibinfo {author} {\bibfnamefont {Marc}\ \bibnamefont {Durand}},
  and\ \bibinfo {author} {\bibfnamefont {Julien}\ \bibnamefont {Tailleur}}}
  (\bibinfo {year} {2023}),\ \href@noop {} {\enquote {\bibinfo {title} {Nematic
  torques in scalar active matter: when fluctuations favor polar order and
  persistence},}\ }\Eprint {https://arxiv.org/abs/2301.02568} {arXiv:2301.02568
  [cond-mat.stat-mech]} \BibitemShut {NoStop}%
\bibitem [{\citenamefont {Spohn}(1980)}]{Spohn1980}%
  \BibitemOpen
  \bibfield  {author} {\bibinfo {author} {\bibnamefont {Spohn}, \bibfnamefont
  {Herbert}}} (\bibinfo {year} {1980}),\ \bibfield  {title} {\enquote {\bibinfo
  {title} {{Kinetic equations from Hamiltonian dynamics: Markovian limits}},}\
  }\href {https://doi.org/10.1103/RevModPhys.52.569} {\bibfield  {journal}
  {\bibinfo  {journal} {Rev. Mod. Phys.}\ }\textbf {\bibinfo {volume}
  {52}}~(\bibinfo {number} {3}),\ \bibinfo {pages} {569--615}}\BibitemShut
  {NoStop}%
\bibitem [{\citenamefont {Stenhammar}(2021)}]{stenhammar2021introduction}%
  \BibitemOpen
  \bibfield  {author} {\bibinfo {author} {\bibnamefont {Stenhammar},
  \bibfnamefont {Joakim}}} (\bibinfo {year} {2021}),\ \href@noop {} {\enquote
  {\bibinfo {title} {An introduction to motility-induced phase separation},}\
  }\Eprint {https://arxiv.org/abs/2112.05024} {arXiv:2112.05024
  [cond-mat.soft]} \BibitemShut {NoStop}%
\bibitem [{\citenamefont {Stenhammar}\ \emph {et~al.}(2014)\citenamefont
  {Stenhammar}, \citenamefont {Marenduzzo}, \citenamefont {Allen},\ and\
  \citenamefont {Cates}}]{stenhammar2014phase}%
  \BibitemOpen
  \bibfield  {author} {\bibinfo {author} {\bibnamefont {Stenhammar},
  \bibfnamefont {Joakim}}, \bibinfo {author} {\bibfnamefont {Davide}\
  \bibnamefont {Marenduzzo}}, \bibinfo {author} {\bibfnamefont {Rosalind~J.}\
  \bibnamefont {Allen}}, and\ \bibinfo {author} {\bibfnamefont {Michael~E.}\
  \bibnamefont {Cates}}} (\bibinfo {year} {2014}),\ \bibfield  {title}
  {\enquote {\bibinfo {title} {{Phase behaviour of active Brownian particles:
  the role of dimensionality}},}\ }\href {https://doi.org/10.1039/C3SM52813H}
  {\bibfield  {journal} {\bibinfo  {journal} {Soft Matter}\ }\textbf {\bibinfo
  {volume} {10}}~(\bibinfo {number} {10}),\ \bibinfo {pages}
  {1489--1499}}\BibitemShut {NoStop}%
\bibitem [{\citenamefont {Stenhammar}\ \emph {et~al.}(2015)\citenamefont
  {Stenhammar}, \citenamefont {Wittkowski}, \citenamefont {Marenduzzo},\ and\
  \citenamefont {Cates}}]{stenhammar_activity-induced_2015}%
  \BibitemOpen
  \bibfield  {author} {\bibinfo {author} {\bibnamefont {Stenhammar},
  \bibfnamefont {Joakim}}, \bibinfo {author} {\bibfnamefont {Raphael}\
  \bibnamefont {Wittkowski}}, \bibinfo {author} {\bibfnamefont {Davide}\
  \bibnamefont {Marenduzzo}}, and\ \bibinfo {author} {\bibfnamefont
  {Michael~E.}\ \bibnamefont {Cates}}} (\bibinfo {year} {2015}),\ \bibfield
  {title} {\enquote {\bibinfo {title} {{Activity-Induced Phase Separation and
  Self-Assembly in Mixtures of Active and Passive Particles}},}\ }\href
  {https://doi.org/10.1103/PhysRevLett.114.018301} {\bibfield  {journal}
  {\bibinfo  {journal} {Phys. Rev. Lett.}\ }\textbf {\bibinfo {volume}
  {114}}~(\bibinfo {number} {1}),\ \bibinfo {pages} {018301}}\BibitemShut
  {NoStop}%
\bibitem [{\citenamefont {Stramer}\ and\ \citenamefont
  {Mayor}(2017)}]{stramer2017mechanisms}%
  \BibitemOpen
  \bibfield  {author} {\bibinfo {author} {\bibnamefont {Stramer}, \bibfnamefont
  {Brian}}, and\ \bibinfo {author} {\bibfnamefont {Roberto}\ \bibnamefont
  {Mayor}}} (\bibinfo {year} {2017}),\ \bibfield  {title} {\enquote {\bibinfo
  {title} {{Mechanisms and in vivo functions of contact inhibition of
  locomotion}},}\ }\href {https://doi.org/10.1038/nrm.2016.118} {\bibfield
  {journal} {\bibinfo  {journal} {Nat. Rev. Mol. Cell Biol.}\ }\textbf
  {\bibinfo {volume} {18}}~(\bibinfo {number} {1}),\ \bibinfo {pages}
  {43--55}}\BibitemShut {NoStop}%
\bibitem [{\citenamefont {Szamel}(2014)}]{szamel2014self}%
  \BibitemOpen
  \bibfield  {author} {\bibinfo {author} {\bibnamefont {Szamel}, \bibfnamefont
  {Grzegorz}}} (\bibinfo {year} {2014}),\ \bibfield  {title} {\enquote
  {\bibinfo {title} {{Self-propelled particle in an external potential:
  Existence of an effective temperature}},}\ }\href
  {https://doi.org/10.1103/PhysRevE.90.012111} {\bibfield  {journal} {\bibinfo
  {journal} {Phys. Rev. E}\ }\textbf {\bibinfo {volume} {90}}~(\bibinfo
  {number} {1}),\ \bibinfo {pages} {012111}}\BibitemShut {NoStop}%
\bibitem [{\citenamefont {Tailleur}\ and\ \citenamefont
  {Cates}(2008)}]{tailleur2008statistical}%
  \BibitemOpen
  \bibfield  {author} {\bibinfo {author} {\bibnamefont {Tailleur},
  \bibfnamefont {J}}, and\ \bibinfo {author} {\bibfnamefont {M.~E.}\
  \bibnamefont {Cates}}} (\bibinfo {year} {2008}),\ \bibfield  {title}
  {\enquote {\bibinfo {title} {{Statistical Mechanics of Interacting
  Run-and-Tumble Bacteria}},}\ }\href
  {https://doi.org/10.1103/PhysRevLett.100.218103} {\bibfield  {journal}
  {\bibinfo  {journal} {Phys. Rev. Lett.}\ }\textbf {\bibinfo {volume}
  {100}}~(\bibinfo {number} {21}),\ \bibinfo {pages} {218103}}\BibitemShut
  {NoStop}%
\bibitem [{\citenamefont {Tailleur}\ and\ \citenamefont
  {Cates}(2009)}]{tailleur2009sedimentation}%
  \BibitemOpen
  \bibfield  {author} {\bibinfo {author} {\bibnamefont {Tailleur},
  \bibfnamefont {J}}, and\ \bibinfo {author} {\bibfnamefont {M.~E.}\
  \bibnamefont {Cates}}} (\bibinfo {year} {2009}),\ \bibfield  {title}
  {\enquote {\bibinfo {title} {{Sedimentation, trapping, and rectification of
  dilute bacteria}},}\ }\href {https://doi.org/10.1209/0295-5075/86/60002}
  {\bibfield  {journal} {\bibinfo  {journal} {EPL}\ }\textbf {\bibinfo {volume}
  {86}}~(\bibinfo {number} {6}),\ \bibinfo {pages} {60002}}\BibitemShut
  {NoStop}%
\bibitem [{\citenamefont {Tailleur}\ \emph {et~al.}(2022)\citenamefont
  {Tailleur}, \citenamefont {Gompper}, \citenamefont {Marchetti}, \citenamefont
  {Yeomans},\ and\ \citenamefont {Salomon}}]{tailleur2022active}%
  \BibitemOpen
  \bibinfo {editor} {\bibnamefont {Tailleur}, \bibfnamefont {Julien}}, \bibinfo
  {editor} {\bibfnamefont {Gerhard}\ \bibnamefont {Gompper}}, \bibinfo {editor}
  {\bibfnamefont {M.~Cristina}\ \bibnamefont {Marchetti}}, \bibinfo {editor}
  {\bibfnamefont {Julia~M.}\ \bibnamefont {Yeomans}}, and\ \bibinfo {editor}
  {\bibfnamefont {Christophe}\ \bibnamefont {Salomon}},\ Eds. (\bibinfo {year}
  {2022}),\ \href@noop {} {\emph {\bibinfo {title} {{Active Matter and
  Nonequilibrium Statistical Physics: Lecture Notes of the Les Houches Summer
  School: Volume 112, September 2018}}}},\ Vol.\ \bibinfo {volume} {112}\
  (\bibinfo  {publisher} {Oxford University Press})\BibitemShut {NoStop}%
\bibitem [{\citenamefont {Takaha}\ and\ \citenamefont
  {Nishiguchi}(2023)}]{takaha2023quasi}%
  \BibitemOpen
  \bibfield  {author} {\bibinfo {author} {\bibnamefont {Takaha}, \bibfnamefont
  {Yuki}}, and\ \bibinfo {author} {\bibfnamefont {Daiki}\ \bibnamefont
  {Nishiguchi}}} (\bibinfo {year} {2023}),\ \bibfield  {title} {\enquote
  {\bibinfo {title} {{Quasi-two-dimensional bacterial swimming around pillars:
  Enhanced trapping efficiency and curvature dependence}},}\ }\href
  {https://doi.org/10.1103/PhysRevE.107.014602} {\bibfield  {journal} {\bibinfo
   {journal} {Phys. Rev. E}\ }\textbf {\bibinfo {volume} {107}}~(\bibinfo
  {number} {1}),\ \bibinfo {pages} {014602}}\BibitemShut {NoStop}%
\bibitem [{\citenamefont {Takatori}\ and\ \citenamefont
  {Brady}(2015)}]{takatori2015towards}%
  \BibitemOpen
  \bibfield  {author} {\bibinfo {author} {\bibnamefont {Takatori},
  \bibfnamefont {S~C}}, and\ \bibinfo {author} {\bibfnamefont {J.~F.}\
  \bibnamefont {Brady}}} (\bibinfo {year} {2015}),\ \bibfield  {title}
  {\enquote {\bibinfo {title} {{Towards a thermodynamics of active matter}},}\
  }\href {https://doi.org/10.1103/PhysRevE.91.032117} {\bibfield  {journal}
  {\bibinfo  {journal} {Phys. Rev. E}\ }\textbf {\bibinfo {volume}
  {91}}~(\bibinfo {number} {3}),\ \bibinfo {pages} {032117}}\BibitemShut
  {NoStop}%
\bibitem [{\citenamefont {Takatori}\ \emph {et~al.}(2014)\citenamefont
  {Takatori}, \citenamefont {Yan},\ and\ \citenamefont
  {Brady}}]{takatori2014swim}%
  \BibitemOpen
  \bibfield  {author} {\bibinfo {author} {\bibnamefont {Takatori},
  \bibfnamefont {S~C}}, \bibinfo {author} {\bibfnamefont {W.}~\bibnamefont
  {Yan}}, and\ \bibinfo {author} {\bibfnamefont {J.~F.}\ \bibnamefont {Brady}}}
  (\bibinfo {year} {2014}),\ \bibfield  {title} {\enquote {\bibinfo {title}
  {{Swim Pressure: Stress Generation in Active Matter}},}\ }\href
  {https://doi.org/10.1103/PhysRevLett.113.028103} {\bibfield  {journal}
  {\bibinfo  {journal} {Phys. Rev. Lett.}\ }\textbf {\bibinfo {volume}
  {113}}~(\bibinfo {number} {2}),\ \bibinfo {pages} {028103}}\BibitemShut
  {NoStop}%
\bibitem [{\citenamefont {Takatori}\ \emph {et~al.}(2016)\citenamefont
  {Takatori}, \citenamefont {{De Dier}}, \citenamefont {Vermant},\ and\
  \citenamefont {Brady}}]{takatori2016acoustic}%
  \BibitemOpen
  \bibfield  {author} {\bibinfo {author} {\bibnamefont {Takatori},
  \bibfnamefont {Sho~C}}, \bibinfo {author} {\bibfnamefont {Raf}\ \bibnamefont
  {{De Dier}}}, \bibinfo {author} {\bibfnamefont {Jan}\ \bibnamefont
  {Vermant}}, and\ \bibinfo {author} {\bibfnamefont {John~F.}\ \bibnamefont
  {Brady}}} (\bibinfo {year} {2016}),\ \bibfield  {title} {\enquote {\bibinfo
  {title} {{Acoustic trapping of active matter}},}\ }\href
  {https://doi.org/10.1038/ncomms10694} {\bibfield  {journal} {\bibinfo
  {journal} {Nat. Commun.}\ }\textbf {\bibinfo {volume} {7}}~(\bibinfo {number}
  {1}),\ \bibinfo {pages} {10694}}\BibitemShut {NoStop}%
\bibitem [{\citenamefont {Theurkauff}\ \emph {et~al.}(2012)\citenamefont
  {Theurkauff}, \citenamefont {Cottin-Bizonne}, \citenamefont {Palacci},
  \citenamefont {Ybert},\ and\ \citenamefont
  {Bocquet}}]{theurkauff2012dynamic}%
  \BibitemOpen
  \bibfield  {author} {\bibinfo {author} {\bibnamefont {Theurkauff},
  \bibfnamefont {I}}, \bibinfo {author} {\bibfnamefont {C.}~\bibnamefont
  {Cottin-Bizonne}}, \bibinfo {author} {\bibfnamefont {J.}~\bibnamefont
  {Palacci}}, \bibinfo {author} {\bibfnamefont {C.}~\bibnamefont {Ybert}}, and\
  \bibinfo {author} {\bibfnamefont {L.}~\bibnamefont {Bocquet}}} (\bibinfo
  {year} {2012}),\ \bibfield  {title} {\enquote {\bibinfo {title} {{Dynamic
  Clustering in Active Colloidal Suspensions with Chemical Signaling}},}\
  }\href {https://doi.org/10.1103/PhysRevLett.108.268303} {\bibfield  {journal}
  {\bibinfo  {journal} {Phys. Rev. Lett.}\ }\textbf {\bibinfo {volume}
  {108}}~(\bibinfo {number} {26}),\ \bibinfo {pages} {268303}}\BibitemShut
  {NoStop}%
\bibitem [{\citenamefont {Thiffeault}(2015)}]{Thiffeault2015}%
  \BibitemOpen
  \bibfield  {author} {\bibinfo {author} {\bibnamefont {Thiffeault},
  \bibfnamefont {Jean-Luc}}} (\bibinfo {year} {2015}),\ \bibfield  {title}
  {\enquote {\bibinfo {title} {{Distribution of particle displacements due to
  swimming microorganisms}},}\ }\href
  {https://doi.org/10.1103/PhysRevE.92.023023} {\bibfield  {journal} {\bibinfo
  {journal} {Phys. Rev. E}\ }\textbf {\bibinfo {volume} {92}}~(\bibinfo
  {number} {2}),\ \bibinfo {pages} {023023}}\BibitemShut {NoStop}%
\bibitem [{\citenamefont {Thompson}\ \emph {et~al.}(2011)\citenamefont
  {Thompson}, \citenamefont {Tailleur}, \citenamefont {Cates},\ and\
  \citenamefont {Blythe}}]{thompson_lattice_2011}%
  \BibitemOpen
  \bibfield  {author} {\bibinfo {author} {\bibnamefont {Thompson},
  \bibfnamefont {A~G}}, \bibinfo {author} {\bibfnamefont {J.}~\bibnamefont
  {Tailleur}}, \bibinfo {author} {\bibfnamefont {M.~E.}\ \bibnamefont {Cates}},
  and\ \bibinfo {author} {\bibfnamefont {R.~A.}\ \bibnamefont {Blythe}}}
  (\bibinfo {year} {2011}),\ \bibfield  {title} {\enquote {\bibinfo {title}
  {{Lattice models of nonequilibrium bacterial dynamics}},}\ }\href
  {https://doi.org/10.1088/1742-5468/2011/02/P02029} {\bibfield  {journal}
  {\bibinfo  {journal} {J. Stat. Mech. Theory Exp.}\ }\textbf {\bibinfo
  {volume} {2011}}~(\bibinfo {number} {02}),\ \bibinfo {pages}
  {P02029}}\BibitemShut {NoStop}%
\bibitem [{\citenamefont {Thutupalli}\ \emph {et~al.}(2011)\citenamefont
  {Thutupalli}, \citenamefont {Seemann},\ and\ \citenamefont
  {Herminghaus}}]{thutupalli_swarming_2011}%
  \BibitemOpen
  \bibfield  {author} {\bibinfo {author} {\bibnamefont {Thutupalli},
  \bibfnamefont {Shashi}}, \bibinfo {author} {\bibfnamefont {Ralf}\
  \bibnamefont {Seemann}}, and\ \bibinfo {author} {\bibfnamefont {Stephan}\
  \bibnamefont {Herminghaus}}} (\bibinfo {year} {2011}),\ \bibfield  {title}
  {\enquote {\bibinfo {title} {{Swarming behavior of simple model
  squirmers}},}\ }\href {https://doi.org/10.1088/1367-2630/13/7/073021}
  {\bibfield  {journal} {\bibinfo  {journal} {New J. Phys.}\ }\textbf {\bibinfo
  {volume} {13}}~(\bibinfo {number} {7}),\ \bibinfo {pages}
  {073021}}\BibitemShut {NoStop}%
\bibitem [{\citenamefont {Tjhung}\ \emph {et~al.}(2018)\citenamefont {Tjhung},
  \citenamefont {Nardini},\ and\ \citenamefont {Cates}}]{tjhung_cluster_2018}%
  \BibitemOpen
  \bibfield  {author} {\bibinfo {author} {\bibnamefont {Tjhung}, \bibfnamefont
  {Elsen}}, \bibinfo {author} {\bibfnamefont {Cesare}\ \bibnamefont {Nardini}},
  and\ \bibinfo {author} {\bibfnamefont {Michael~E.}\ \bibnamefont {Cates}}}
  (\bibinfo {year} {2018}),\ \bibfield  {title} {\enquote {\bibinfo {title}
  {{Cluster Phases and Bubbly Phase Separation in Active Fluids: Reversal of
  the Ostwald Process}},}\ }\href {https://doi.org/10.1103/PhysRevX.8.031080}
  {\bibfield  {journal} {\bibinfo  {journal} {Phys. Rev. X}\ }\textbf {\bibinfo
  {volume} {8}}~(\bibinfo {number} {3}),\ \bibinfo {pages}
  {031080}}\BibitemShut {NoStop}%
\bibitem [{\citenamefont {Toner}\ \emph {et~al.}(2018)\citenamefont {Toner},
  \citenamefont {Guttenberg},\ and\ \citenamefont
  {Tu}}]{toner2018hydrodynamic}%
  \BibitemOpen
  \bibfield  {author} {\bibinfo {author} {\bibnamefont {Toner}, \bibfnamefont
  {John}}, \bibinfo {author} {\bibfnamefont {Nicholas}\ \bibnamefont
  {Guttenberg}}, and\ \bibinfo {author} {\bibfnamefont {Yuhai}\ \bibnamefont
  {Tu}}} (\bibinfo {year} {2018}),\ \bibfield  {title} {\enquote {\bibinfo
  {title} {{Hydrodynamic theory of flocking in the presence of quenched
  disorder}},}\ }\href {https://doi.org/10.1103/PhysRevE.98.062604} {\bibfield
  {journal} {\bibinfo  {journal} {Phys. Rev. E}\ }\textbf {\bibinfo {volume}
  {98}}~(\bibinfo {number} {6}),\ \bibinfo {pages} {062604}}\BibitemShut
  {NoStop}%
\bibitem [{\citenamefont {Toner}\ \emph {et~al.}(2005)\citenamefont {Toner},
  \citenamefont {Tu},\ and\ \citenamefont
  {Ramaswamy}}]{toner_hydrodynamics_2005}%
  \BibitemOpen
  \bibfield  {author} {\bibinfo {author} {\bibnamefont {Toner}, \bibfnamefont
  {John}}, \bibinfo {author} {\bibfnamefont {Yuhai}\ \bibnamefont {Tu}}, and\
  \bibinfo {author} {\bibfnamefont {Sriram}\ \bibnamefont {Ramaswamy}}}
  (\bibinfo {year} {2005}),\ \bibfield  {title} {\enquote {\bibinfo {title}
  {{Hydrodynamics and phases of flocks}},}\ }\href
  {https://doi.org/10.1016/j.aop.2005.04.011} {\bibfield  {journal} {\bibinfo
  {journal} {Annals of Physics}\ }\textbf {\bibinfo {volume} {318}}~(\bibinfo
  {number} {1}),\ \bibinfo {pages} {170--244}}\BibitemShut {NoStop}%
\bibitem [{\citenamefont {Tsou}\ and\ \citenamefont
  {Zhu}(2010)}]{tsou2010virulence}%
  \BibitemOpen
  \bibfield  {author} {\bibinfo {author} {\bibnamefont {Tsou}, \bibfnamefont
  {Amy~M}}, and\ \bibinfo {author} {\bibfnamefont {Jun}\ \bibnamefont {Zhu}}}
  (\bibinfo {year} {2010}),\ \bibfield  {title} {\enquote {\bibinfo {title}
  {{Quorum Sensing Negatively Regulates Hemolysin Transcriptionally and
  Posttranslationally in Vibrio cholerae}},}\ }\href
  {https://doi.org/10.1128/IAI.00590-09} {\bibfield  {journal} {\bibinfo
  {journal} {Infect. Immun.}\ }\textbf {\bibinfo {volume} {78}}~(\bibinfo
  {number} {1}),\ \bibinfo {pages} {461--467}}\BibitemShut {NoStop}%
\bibitem [{\citenamefont {Verma}\ and\ \citenamefont
  {Miyashiro}(2013)}]{verma2013quorum}%
  \BibitemOpen
  \bibfield  {author} {\bibinfo {author} {\bibnamefont {Verma}, \bibfnamefont
  {Subhash}}, and\ \bibinfo {author} {\bibfnamefont {Tim}\ \bibnamefont
  {Miyashiro}}} (\bibinfo {year} {2013}),\ \bibfield  {title} {\enquote
  {\bibinfo {title} {{Quorum Sensing in the Squid-Vibrio Symbiosis}},}\ }\href
  {https://doi.org/10.3390/ijms140816386} {\bibfield  {journal} {\bibinfo
  {journal} {Int. J. Mol. Sci.}\ }\textbf {\bibinfo {volume} {14}}~(\bibinfo
  {number} {8}),\ \bibinfo {pages} {16386--16401}}\BibitemShut {NoStop}%
\bibitem [{\citenamefont {Vicsek}\ \emph {et~al.}(1995)\citenamefont {Vicsek},
  \citenamefont {Czir{\'{o}}k}, \citenamefont {Ben-Jacob}, \citenamefont
  {Cohen},\ and\ \citenamefont {Shochet}}]{vicsek1995novel}%
  \BibitemOpen
  \bibfield  {author} {\bibinfo {author} {\bibnamefont {Vicsek}, \bibfnamefont
  {Tam{\'{a}}s}}, \bibinfo {author} {\bibfnamefont {Andr{\'{a}}s}\ \bibnamefont
  {Czir{\'{o}}k}}, \bibinfo {author} {\bibfnamefont {Eshel}\ \bibnamefont
  {Ben-Jacob}}, \bibinfo {author} {\bibfnamefont {Inon}\ \bibnamefont {Cohen}},
  and\ \bibinfo {author} {\bibfnamefont {Ofer}\ \bibnamefont {Shochet}}}
  (\bibinfo {year} {1995}),\ \bibfield  {title} {\enquote {\bibinfo {title}
  {{Novel Type of Phase Transition in a System of Self-Driven Particles}},}\
  }\href {https://doi.org/10.1103/PhysRevLett.75.1226} {\bibfield  {journal}
  {\bibinfo  {journal} {Phys. Rev. Lett.}\ }\textbf {\bibinfo {volume}
  {75}}~(\bibinfo {number} {6}),\ \bibinfo {pages} {1226--1229}}\BibitemShut
  {NoStop}%
\bibitem [{\citenamefont {Weber}\ \emph {et~al.}(2016)\citenamefont {Weber},
  \citenamefont {Weber},\ and\ \citenamefont {Frey}}]{weber_binary_2016}%
  \BibitemOpen
  \bibfield  {author} {\bibinfo {author} {\bibnamefont {Weber}, \bibfnamefont
  {Simon~N}}, \bibinfo {author} {\bibfnamefont {Christoph~A.}\ \bibnamefont
  {Weber}}, and\ \bibinfo {author} {\bibfnamefont {Erwin}\ \bibnamefont
  {Frey}}} (\bibinfo {year} {2016}),\ \bibfield  {title} {\enquote {\bibinfo
  {title} {{Binary Mixtures of Particles with Different Diffusivities
  Demix}},}\ }\href {https://doi.org/10.1103/PhysRevLett.116.058301} {\bibfield
   {journal} {\bibinfo  {journal} {Phys. Rev. Lett.}\ }\textbf {\bibinfo
  {volume} {116}}~(\bibinfo {number} {5}),\ \bibinfo {pages}
  {058301}}\BibitemShut {NoStop}%
\bibitem [{\citenamefont {Whitelam}\ \emph {et~al.}(2018)\citenamefont
  {Whitelam}, \citenamefont {Klymko},\ and\ \citenamefont
  {Mandal}}]{whitelam2018phase}%
  \BibitemOpen
  \bibfield  {author} {\bibinfo {author} {\bibnamefont {Whitelam},
  \bibfnamefont {Stephen}}, \bibinfo {author} {\bibfnamefont {Katherine}\
  \bibnamefont {Klymko}}, and\ \bibinfo {author} {\bibfnamefont {Dibyendu}\
  \bibnamefont {Mandal}}} (\bibinfo {year} {2018}),\ \bibfield  {title}
  {\enquote {\bibinfo {title} {{Phase separation and large deviations of
  lattice active matter}},}\ }\href
  {https://pubs.aip.org/jcp/article/148/15/154902/195348/Phase-separation-and-large-deviations-of-lattice}
  {\bibfield  {journal} {\bibinfo  {journal} {J. Chem. Phys.}\ }\textbf
  {\bibinfo {volume} {148}}~(\bibinfo {number} {15})}\BibitemShut {NoStop}%
\bibitem [{\citenamefont {Winkler}\ \emph {et~al.}(2015)\citenamefont
  {Winkler}, \citenamefont {Wysocki},\ and\ \citenamefont
  {Gompper}}]{Winkler2015SoftMatter}%
  \BibitemOpen
  \bibfield  {author} {\bibinfo {author} {\bibnamefont {Winkler}, \bibfnamefont
  {Roland~G}}, \bibinfo {author} {\bibfnamefont {Adam}\ \bibnamefont
  {Wysocki}}, and\ \bibinfo {author} {\bibfnamefont {Gerhard}\ \bibnamefont
  {Gompper}}} (\bibinfo {year} {2015}),\ \bibfield  {title} {\enquote {\bibinfo
  {title} {{Virial pressure in systems of spherical active Brownian
  particles}},}\ }\href {https://doi.org/10.1039/C5SM01412C} {\bibfield
  {journal} {\bibinfo  {journal} {Soft Matter}\ }\textbf {\bibinfo {volume}
  {11}}~(\bibinfo {number} {33}),\ \bibinfo {pages} {6680--6691}}\BibitemShut
  {NoStop}%
\bibitem [{\citenamefont {Wittkowski}\ \emph {et~al.}(2017)\citenamefont
  {Wittkowski}, \citenamefont {Stenhammar},\ and\ \citenamefont
  {Cates}}]{wittkowski_nonequilibrium_2017}%
  \BibitemOpen
  \bibfield  {author} {\bibinfo {author} {\bibnamefont {Wittkowski},
  \bibfnamefont {Raphael}}, \bibinfo {author} {\bibfnamefont {Joakim}\
  \bibnamefont {Stenhammar}}, and\ \bibinfo {author} {\bibfnamefont
  {Michael~E.}\ \bibnamefont {Cates}}} (\bibinfo {year} {2017}),\ \bibfield
  {title} {\enquote {\bibinfo {title} {{Nonequilibrium dynamics of mixtures of
  active and passive colloidal particles}},}\ }\href
  {https://doi.org/10.1088/1367-2630/aa8195} {\bibfield  {journal} {\bibinfo
  {journal} {New J. Phys.}\ }\textbf {\bibinfo {volume} {19}}~(\bibinfo
  {number} {10}),\ \bibinfo {pages} {105003}}\BibitemShut {NoStop}%
\bibitem [{\citenamefont {Wittkowski}\ \emph {et~al.}(2014)\citenamefont
  {Wittkowski}, \citenamefont {Tiribocchi}, \citenamefont {Stenhammar},
  \citenamefont {Allen}, \citenamefont {Marenduzzo},\ and\ \citenamefont
  {Cates}}]{wittkowski_scalar_2014}%
  \BibitemOpen
  \bibfield  {author} {\bibinfo {author} {\bibnamefont {Wittkowski},
  \bibfnamefont {Raphael}}, \bibinfo {author} {\bibfnamefont {Adriano}\
  \bibnamefont {Tiribocchi}}, \bibinfo {author} {\bibfnamefont {Joakim}\
  \bibnamefont {Stenhammar}}, \bibinfo {author} {\bibfnamefont {Rosalind~J.}\
  \bibnamefont {Allen}}, \bibinfo {author} {\bibfnamefont {Davide}\
  \bibnamefont {Marenduzzo}}, and\ \bibinfo {author} {\bibfnamefont
  {Michael~E.}\ \bibnamefont {Cates}}} (\bibinfo {year} {2014}),\ \bibfield
  {title} {\enquote {\bibinfo {title} {{Scalar $\phi^4$ field theory for
  active-particle phase separation}},}\ }\href
  {https://doi.org/10.1038/ncomms5351} {\bibfield  {journal} {\bibinfo
  {journal} {Nat. Commun.}\ }\textbf {\bibinfo {volume} {5}}~(\bibinfo {number}
  {1}),\ \bibinfo {pages} {4351}}\BibitemShut {NoStop}%
\bibitem [{\citenamefont {Wittmann}\ \emph
  {et~al.}(2017{\natexlab{a}})\citenamefont {Wittmann}, \citenamefont {Maggi},
  \citenamefont {Sharma}, \citenamefont {Scacchi}, \citenamefont {Brader},\
  and\ \citenamefont {{Marini Bettolo Marconi}}}]{wittmann2017effective1}%
  \BibitemOpen
  \bibfield  {author} {\bibinfo {author} {\bibnamefont {Wittmann},
  \bibfnamefont {Ren{\'{e}}}}, \bibinfo {author} {\bibfnamefont {Claudio}\
  \bibnamefont {Maggi}}, \bibinfo {author} {\bibfnamefont {A.}~\bibnamefont
  {Sharma}}, \bibinfo {author} {\bibfnamefont {A.}~\bibnamefont {Scacchi}},
  \bibinfo {author} {\bibfnamefont {Joseph~M.}\ \bibnamefont {Brader}}, and\
  \bibinfo {author} {\bibfnamefont {U.}~\bibnamefont {{Marini Bettolo
  Marconi}}}} (\bibinfo {year} {2017}{\natexlab{a}}),\ \bibfield  {title}
  {\enquote {\bibinfo {title} {{Effective equilibrium states in the
  colored-noise model for active matter I. Pairwise forces in the Fox and
  unified colored noise approximations}},}\ }\href
  {https://doi.org/10.1088/1742-5468/aa8c1f} {\bibfield  {journal} {\bibinfo
  {journal} {J. Stat. Mech. Theory Exp.}\ }\textbf {\bibinfo {volume}
  {2017}}~(\bibinfo {number} {11}),\ \bibinfo {pages} {113207}}\BibitemShut
  {NoStop}%
\bibitem [{\citenamefont {Wittmann}\ \emph
  {et~al.}(2017{\natexlab{b}})\citenamefont {Wittmann}, \citenamefont
  {Marconi}, \citenamefont {Maggi},\ and\ \citenamefont
  {Brader}}]{wittmann2017effective2}%
  \BibitemOpen
  \bibfield  {author} {\bibinfo {author} {\bibnamefont {Wittmann},
  \bibfnamefont {Ren{\'{e}}}}, \bibinfo {author} {\bibfnamefont
  {U.~Marini~Bettolo}\ \bibnamefont {Marconi}}, \bibinfo {author}
  {\bibfnamefont {C.}~\bibnamefont {Maggi}}, and\ \bibinfo {author}
  {\bibfnamefont {J.~M.}\ \bibnamefont {Brader}}} (\bibinfo {year}
  {2017}{\natexlab{b}}),\ \bibfield  {title} {\enquote {\bibinfo {title}
  {{Effective equilibrium states in the colored-noise model for active matter
  II. A unified framework for phase equilibria, structure and mechanical
  properties}},}\ }\href {https://doi.org/10.1088/1742-5468/aa8c37} {\bibfield
  {journal} {\bibinfo  {journal} {J. Stat. Mech. Theory Exp.}\ }\textbf
  {\bibinfo {volume} {2017}}~(\bibinfo {number} {11}),\ \bibinfo {pages}
  {113208}}\BibitemShut {NoStop}%
\bibitem [{\citenamefont {Woillez}\ \emph
  {et~al.}(2020{\natexlab{a}})\citenamefont {Woillez}, \citenamefont {Kafri},\
  and\ \citenamefont {Gov}}]{woillez2020active}%
  \BibitemOpen
  \bibfield  {author} {\bibinfo {author} {\bibnamefont {Woillez}, \bibfnamefont
  {Eric}}, \bibinfo {author} {\bibfnamefont {Yariv}\ \bibnamefont {Kafri}},
  and\ \bibinfo {author} {\bibfnamefont {Nir~S.}\ \bibnamefont {Gov}}}
  (\bibinfo {year} {2020}{\natexlab{a}}),\ \bibfield  {title} {\enquote
  {\bibinfo {title} {{Active Trap Model}},}\ }\href
  {https://doi.org/10.1103/PhysRevLett.124.118002} {\bibfield  {journal}
  {\bibinfo  {journal} {Phys. Rev. Lett.}\ }\textbf {\bibinfo {volume}
  {124}}~(\bibinfo {number} {11}),\ \bibinfo {pages} {118002}}\BibitemShut
  {NoStop}%
\bibitem [{\citenamefont {Woillez}\ \emph
  {et~al.}(2020{\natexlab{b}})\citenamefont {Woillez}, \citenamefont {Kafri},\
  and\ \citenamefont {Lecomte}}]{woillez2020nonlocal}%
  \BibitemOpen
  \bibfield  {author} {\bibinfo {author} {\bibnamefont {Woillez}, \bibfnamefont
  {Eric}}, \bibinfo {author} {\bibfnamefont {Yariv}\ \bibnamefont {Kafri}},
  and\ \bibinfo {author} {\bibfnamefont {Vivien}\ \bibnamefont {Lecomte}}}
  (\bibinfo {year} {2020}{\natexlab{b}}),\ \bibfield  {title} {\enquote
  {\bibinfo {title} {{Nonlocal stationary probability distributions and escape
  rates for an active Ornstein–Uhlenbeck particle}},}\ }\href
  {https://doi.org/10.1088/1742-5468/ab7e2e} {\bibfield  {journal} {\bibinfo
  {journal} {J. Stat. Mech. Theory Exp.}\ }\textbf {\bibinfo {volume}
  {2020}}~(\bibinfo {number} {6}),\ \bibinfo {pages} {063204}}\BibitemShut
  {NoStop}%
\bibitem [{\citenamefont {Wu}\ and\ \citenamefont {Libchaber}(2000)}]{Wu2000}%
  \BibitemOpen
  \bibfield  {author} {\bibinfo {author} {\bibnamefont {Wu}, \bibfnamefont
  {Xiao-Lun}}, and\ \bibinfo {author} {\bibfnamefont {Albert}\ \bibnamefont
  {Libchaber}}} (\bibinfo {year} {2000}),\ \bibfield  {title} {\enquote
  {\bibinfo {title} {{Particle Diffusion in a Quasi-Two-Dimensional Bacterial
  Bath}},}\ }\href {https://doi.org/10.1103/PhysRevLett.84.3017} {\bibfield
  {journal} {\bibinfo  {journal} {Phys. Rev. Lett.}\ }\textbf {\bibinfo
  {volume} {84}}~(\bibinfo {number} {13}),\ \bibinfo {pages}
  {3017--3020}}\BibitemShut {NoStop}%
\bibitem [{\citenamefont {Wysocki}\ and\ \citenamefont
  {Rieger}(2020)}]{Wysocki2020}%
  \BibitemOpen
  \bibfield  {author} {\bibinfo {author} {\bibnamefont {Wysocki}, \bibfnamefont
  {Adam}}, and\ \bibinfo {author} {\bibfnamefont {Heiko}\ \bibnamefont
  {Rieger}}} (\bibinfo {year} {2020}),\ \bibfield  {title} {\enquote {\bibinfo
  {title} {{Capillary Action in Scalar Active Matter}},}\ }\href
  {https://doi.org/10.1103/PhysRevLett.124.048001} {\bibfield  {journal}
  {\bibinfo  {journal} {Phys. Rev. Lett.}\ }\textbf {\bibinfo {volume}
  {124}}~(\bibinfo {number} {4}),\ \bibinfo {pages} {048001}}\BibitemShut
  {NoStop}%
\bibitem [{\citenamefont {Wysocki}\ \emph {et~al.}(2014)\citenamefont
  {Wysocki}, \citenamefont {Winkler},\ and\ \citenamefont
  {Gompper}}]{wysocki2014cooperative}%
  \BibitemOpen
  \bibfield  {author} {\bibinfo {author} {\bibnamefont {Wysocki}, \bibfnamefont
  {Adam}}, \bibinfo {author} {\bibfnamefont {Roland~G.}\ \bibnamefont
  {Winkler}}, and\ \bibinfo {author} {\bibfnamefont {Gerhard}\ \bibnamefont
  {Gompper}}} (\bibinfo {year} {2014}),\ \bibfield  {title} {\enquote {\bibinfo
  {title} {{Cooperative motion of active Brownian spheres in three-dimensional
  dense suspensions}},}\ }\href {https://doi.org/10.1209/0295-5075/105/48004}
  {\bibfield  {journal} {\bibinfo  {journal} {EPL}\ }\textbf {\bibinfo {volume}
  {105}}~(\bibinfo {number} {4}),\ \bibinfo {pages} {48004}}\BibitemShut
  {NoStop}%
\bibitem [{\citenamefont {Wysocki}\ \emph {et~al.}(2016)\citenamefont
  {Wysocki}, \citenamefont {Winkler},\ and\ \citenamefont
  {Gompper}}]{wysocki_propagating_2016}%
  \BibitemOpen
  \bibfield  {author} {\bibinfo {author} {\bibnamefont {Wysocki}, \bibfnamefont
  {Adam}}, \bibinfo {author} {\bibfnamefont {Roland~G.}\ \bibnamefont
  {Winkler}}, and\ \bibinfo {author} {\bibfnamefont {Gerhard}\ \bibnamefont
  {Gompper}}} (\bibinfo {year} {2016}),\ \bibfield  {title} {\enquote {\bibinfo
  {title} {{Propagating interfaces in mixtures of active and passive Brownian
  particles}},}\ }\href {https://doi.org/10.1088/1367-2630/aa529d} {\bibfield
  {journal} {\bibinfo  {journal} {New J. Phys.}\ }\textbf {\bibinfo {volume}
  {18}}~(\bibinfo {number} {12}),\ \bibinfo {pages} {123030}}\BibitemShut
  {NoStop}%
\bibitem [{\citenamefont {Yan}\ and\ \citenamefont
  {Brady}(2015)}]{yan2015force}%
  \BibitemOpen
  \bibfield  {author} {\bibinfo {author} {\bibnamefont {Yan}, \bibfnamefont
  {Wen}}, and\ \bibinfo {author} {\bibfnamefont {John~F.}\ \bibnamefont
  {Brady}}} (\bibinfo {year} {2015}),\ \bibfield  {title} {\enquote {\bibinfo
  {title} {{The force on a boundary in active matter}},}\ }\href
  {https://doi.org/10.1017/jfm.2015.621} {\bibfield  {journal} {\bibinfo
  {journal} {J. Fluid Mech.}\ }\textbf {\bibinfo {volume} {785}},\ \bibinfo
  {pages} {R1}}\BibitemShut {NoStop}%
\bibitem [{\citenamefont {Yan}\ and\ \citenamefont {Brady}(2018)}]{Yan2018}%
  \BibitemOpen
  \bibfield  {author} {\bibinfo {author} {\bibnamefont {Yan}, \bibfnamefont
  {Wen}}, and\ \bibinfo {author} {\bibfnamefont {John~F.}\ \bibnamefont
  {Brady}}} (\bibinfo {year} {2018}),\ \bibfield  {title} {\enquote {\bibinfo
  {title} {{The curved kinetic boundary layer of active matter}},}\ }\href
  {https://doi.org/10.1039/C7SM01643C} {\bibfield  {journal} {\bibinfo
  {journal} {Soft Matter}\ }\textbf {\bibinfo {volume} {14}}~(\bibinfo {number}
  {2}),\ \bibinfo {pages} {279--290}}\BibitemShut {NoStop}%
\bibitem [{\citenamefont {Yang}\ \emph {et~al.}(2014)\citenamefont {Yang},
  \citenamefont {Manning},\ and\ \citenamefont
  {Marchetti}}]{yang2014aggregation}%
  \BibitemOpen
  \bibfield  {author} {\bibinfo {author} {\bibnamefont {Yang}, \bibfnamefont
  {Xingbo}}, \bibinfo {author} {\bibfnamefont {M.~Lisa}\ \bibnamefont
  {Manning}}, and\ \bibinfo {author} {\bibfnamefont {M.~Cristina}\ \bibnamefont
  {Marchetti}}} (\bibinfo {year} {2014}),\ \bibfield  {title} {\enquote
  {\bibinfo {title} {{Aggregation and segregation of confined active
  particles}},}\ }\href {https://doi.org/10.1039/C4SM00927D} {\bibfield
  {journal} {\bibinfo  {journal} {Soft Matter}\ }\textbf {\bibinfo {volume}
  {10}}~(\bibinfo {number} {34}),\ \bibinfo {pages} {6477--6484}}\BibitemShut
  {NoStop}%
\bibitem [{\citenamefont {You}\ \emph {et~al.}(2020)\citenamefont {You},
  \citenamefont {Baskaran},\ and\ \citenamefont
  {Marchetti}}]{you2020nonreciprocity}%
  \BibitemOpen
  \bibfield  {author} {\bibinfo {author} {\bibnamefont {You}, \bibfnamefont
  {Zhihong}}, \bibinfo {author} {\bibfnamefont {Aparna}\ \bibnamefont
  {Baskaran}}, and\ \bibinfo {author} {\bibfnamefont {M.~Cristina}\
  \bibnamefont {Marchetti}}} (\bibinfo {year} {2020}),\ \bibfield  {title}
  {\enquote {\bibinfo {title} {{Nonreciprocity as a generic route to traveling
  states}},}\ }\href {https://doi.org/10.1073/pnas.2010318117} {\bibfield
  {journal} {\bibinfo  {journal} {Proc. Natl. Acad. Sci.}\ }\textbf {\bibinfo
  {volume} {117}}~(\bibinfo {number} {33}),\ \bibinfo {pages}
  {19767--19772}}\BibitemShut {NoStop}%
\bibitem [{\citenamefont {{Zaeifi Yamchi}}\ and\ \citenamefont
  {Naji}(2017)}]{ZaeifiYamchi2017}%
  \BibitemOpen
  \bibfield  {author} {\bibinfo {author} {\bibnamefont {{Zaeifi Yamchi}},
  \bibfnamefont {Mahdi}}, and\ \bibinfo {author} {\bibfnamefont {Ali}\
  \bibnamefont {Naji}}} (\bibinfo {year} {2017}),\ \bibfield  {title} {\enquote
  {\bibinfo {title} {{Effective interactions between inclusions in an active
  bath}},}\ }\href {https://doi.org/10.1063/1.5001505} {\bibfield  {journal}
  {\bibinfo  {journal} {J. Chem. Phys.}\ }\textbf {\bibinfo {volume}
  {147}}~(\bibinfo {number} {19}),\ \bibinfo {pages} {194901}}\BibitemShut
  {NoStop}%
\bibitem [{\citenamefont {Zaid}\ \emph {et~al.}(2011)\citenamefont {Zaid},
  \citenamefont {Dunkel},\ and\ \citenamefont {Yeomans}}]{Zaid2011}%
  \BibitemOpen
  \bibfield  {author} {\bibinfo {author} {\bibnamefont {Zaid}, \bibfnamefont
  {Irwin~M}}, \bibinfo {author} {\bibfnamefont {J{\"{o}}rn}\ \bibnamefont
  {Dunkel}}, and\ \bibinfo {author} {\bibfnamefont {Julia~M.}\ \bibnamefont
  {Yeomans}}} (\bibinfo {year} {2011}),\ \bibfield  {title} {\enquote {\bibinfo
  {title} {{L{\'{e}}vy fluctuations and mixing in dilute suspensions of algae
  and bacteria}},}\ }\href {https://doi.org/10.1098/rsif.2010.0545} {\bibfield
  {journal} {\bibinfo  {journal} {J. R. Soc. Interface}\ }\textbf {\bibinfo
  {volume} {8}}~(\bibinfo {number} {62}),\ \bibinfo {pages}
  {1314--1331}}\BibitemShut {NoStop}%
\bibitem [{\citenamefont {Zakine}\ \emph {et~al.}(2020)\citenamefont {Zakine},
  \citenamefont {Zhao}, \citenamefont {Kne{\v{z}}evi{\'{c}}}, \citenamefont
  {Daerr}, \citenamefont {Kafri}, \citenamefont {Tailleur},\ and\ \citenamefont
  {van Wijland}}]{zakine2020surface}%
  \BibitemOpen
  \bibfield  {author} {\bibinfo {author} {\bibnamefont {Zakine}, \bibfnamefont
  {R}}, \bibinfo {author} {\bibfnamefont {Y.}~\bibnamefont {Zhao}}, \bibinfo
  {author} {\bibfnamefont {M.}~\bibnamefont {Kne{\v{z}}evi{\'{c}}}}, \bibinfo
  {author} {\bibfnamefont {A.}~\bibnamefont {Daerr}}, \bibinfo {author}
  {\bibfnamefont {Y.}~\bibnamefont {Kafri}}, \bibinfo {author} {\bibfnamefont
  {J.}~\bibnamefont {Tailleur}}, and\ \bibinfo {author} {\bibfnamefont
  {F.}~\bibnamefont {van Wijland}}} (\bibinfo {year} {2020}),\ \bibfield
  {title} {\enquote {\bibinfo {title} {{Surface Tensions between Active Fluids
  and Solid Interfaces: Bare vs Dressed}},}\ }\href
  {https://doi.org/10.1103/PhysRevLett.124.248003} {\bibfield  {journal}
  {\bibinfo  {journal} {Phys. Rev. Lett.}\ }\textbf {\bibinfo {volume}
  {124}}~(\bibinfo {number} {24}),\ \bibinfo {pages} {248003}}\BibitemShut
  {NoStop}%
\bibitem [{\citenamefont {Zhang}\ \emph {et~al.}(2021)\citenamefont {Zhang},
  \citenamefont {Alert}, \citenamefont {Yan}, \citenamefont {Wingreen},\ and\
  \citenamefont {Granick}}]{zhang2021active}%
  \BibitemOpen
  \bibfield  {author} {\bibinfo {author} {\bibnamefont {Zhang}, \bibfnamefont
  {Jie}}, \bibinfo {author} {\bibfnamefont {Ricard}\ \bibnamefont {Alert}},
  \bibinfo {author} {\bibfnamefont {Jing}\ \bibnamefont {Yan}}, \bibinfo
  {author} {\bibfnamefont {Ned~S.}\ \bibnamefont {Wingreen}}, and\ \bibinfo
  {author} {\bibfnamefont {Steve}\ \bibnamefont {Granick}}} (\bibinfo {year}
  {2021}),\ \bibfield  {title} {\enquote {\bibinfo {title} {{Active phase
  separation by turning towards regions of higher density}},}\ }\href
  {https://doi.org/10.1038/s41567-021-01238-8} {\bibfield  {journal} {\bibinfo
  {journal} {Nat. Phys.}\ }\textbf {\bibinfo {volume} {17}}~(\bibinfo {number}
  {8}),\ \bibinfo {pages} {961--967}}\BibitemShut {NoStop}%
\bibitem [{\citenamefont {Zhao}\ \emph {et~al.}(2023)\citenamefont {Zhao},
  \citenamefont {Ko\ifmmode~\check{s}\else \v{s}\fi{}mrlj},\ and\ \citenamefont
  {Datta}}]{Hongbo2023}%
  \BibitemOpen
  \bibfield  {author} {\bibinfo {author} {\bibnamefont {Zhao}, \bibfnamefont
  {Hongbo}}, \bibinfo {author} {\bibfnamefont {Andrej}\ \bibnamefont
  {Ko\ifmmode~\check{s}\else \v{s}\fi{}mrlj}}, and\ \bibinfo {author}
  {\bibfnamefont {Sujit~S.}\ \bibnamefont {Datta}}} (\bibinfo {year} {2023}),\
  \bibfield  {title} {\enquote {\bibinfo {title} {Chemotactic motility-induced
  phase separation},}\ }\href {https://doi.org/10.1103/PhysRevLett.131.118301}
  {\bibfield  {journal} {\bibinfo  {journal} {Phys. Rev. Lett.}\ }\textbf
  {\bibinfo {volume} {131}},\ \bibinfo {pages} {118301}}\BibitemShut {NoStop}%
\end{thebibliography}%

\end{document}